%% file: main.tex
\documentclass[twoside,openright,frontopenright]{ip3thesis}
\pdfoutput=1
\input{text/preamble}
\usepackage{subfiles}

\begin{document}
\title{Automated inclusion of QED corrections in Monte Carlo event generators}
% \subtitle{An optional subtitle}
\author{Lois Flower}
\researchgroup{Institute for Particle Physics Phenomenology}
\maketitlepage*

\begin{abstract}
In this thesis, we present automated, process-independent methods 
for the calculation of QED real radiative corrections. We review 
the construction of a parton shower based on Catani-Seymour 
dipole subtraction, and thus detail the implementation of a QED 
parton shower. We validate the predictions made by the shower 
against the YFS soft-photon resummation, and discuss the 
algorithmic choices made. We then present results for the 
production of a Higgs boson at the LHC and its decay 
to leptons, showing that the interleaved QCD+QED parton 
shower predicts distributions in excellent agreement with 
the YFS approach.
We then study the \MCatNLO method for matching a next-to-leading 
order calculation with a parton shower. Showing that the method 
preserves its accuracy for the case of QED corrections and of 
mixed QCD and QED corrections, we present the QCD+QED \MCatNLO
method. Validating the method against both the YFS resummation 
and the QED parton shower, we find very good agreement. 
Finally, we present an extension to the YFS soft-photon 
resummation, in which we use a one-step parton shower to 
resum the logarithms associated with charged particle 
pair production. Throughout this thesis we also discuss the 
impact of dressed lepton definitions on observables.
The methods presented in this thesis are made available in a 
public Monte Carlo event generator and analysis framework.
\end{abstract}

\disableprotrusion
\tableofcontents*
\listoffigures
% \listoftables
\enableprotrusion

\chapter*{List of Abbreviations}
\addcontentsline{toc}{chapter}{List of Abbreviations}

\begin{acronym}
    % \acro{ATLAS}{\textit{A Toroidal \acs*{LHC} Apparatus}}
    % \acro{CMS}{\textit{Compact Muon Solenoid}}
    \acro{CS}{Catani-Seymour}
    \acro{DGLAP}{Dokshitser-Gribov-Lipatov-Altarelli-Parisi}
    % \acro{FC}{full-colour}
    \acro{FKS}{Frixione-Kunszt-Signer}
    \acro{KLN}{Kinoshita-Lee-Nauenberg}
    \acro{BLM}{Brodsky-Lepage-Mackenzie}
    % \acro{LC}{leading-colour}
    \acro{LHC}{Large Hadron Collider}
    \acro{LO}{leading order}
    \acro{LL}{leading-logarithmic}
    \acro{MC}{Monte Carlo}
    \acro{ME}{matrix element}
	\acroplural{ME}[MEs]{matrix elements}
    % \acro{N3LO}[N\textsuperscript{3}\acs*{LO}]{(next-to-)\textsuperscript{3}leading order}
    % \acro{N4LO}[N\textsuperscript{4}\acs*{LO}]{(next-to-)\textsuperscript{4}leading order}
    % \acro{NkLL}[N\textsuperscript{k}LL]{(next-to-)\textsuperscript{k}leading log}
    % \acro{NkLO}[N\textsuperscript{k}\acs*{LO}]{(next-to-)\textsuperscript{k}leading order}
    % \acro{NkLP}[N\textsuperscript{k}LP]{(next-to-)\textsuperscript{k}leading power}
    \acro{NLO}{next-to-leading order}
    \acro{NNLO}{next-to-next-to-leading order}
    \acro{PDF}{parton distribution function}
	\acroplural{PDF}[PDFs]{parton distribution functions}
    \acro{QCD}{quantum chromodynamics}
	\acro{EW}{electroweak}
    \acro{QED}{quantum electrodynamics}
	\acro{EM}{electromagnetic}
	\acro{YFS}{Yennie-Frautschi-Suura}
    % \acro{SHERPA}[\code{Sherpa}]{\textit{Simulation of High Energy Reactions of Particles}}
    \acro{SM}{Standard Model}
    \acro{HEFT}{Higgs effective field theory}
    \acro{IR}{infrared}
    \acrodef{IR-safe}{infrared-safe}
    \acro{UV}{ultraviolet}
	\acro{PS}{parton shower}
    \acro{OSSF}{opposite-sign charge same-flavour}
	% \acro{FF}{final-final}
	% \acro{FI}{final-initial}
	% \acro{IF}{initial-final}
	% \acro{II}{initial-initial}

    \acroindefinite{IR}{an}{an}
    \acroindefinite{IR-safe}{an}{an}
    \acroindefinite{UV}{a}{an}
    \acroindefinite{MC}{an}{a}
	\acroindefinite{FKS}{an}{a}
    \acroindefinite{FF}{an}{a}
	\acroindefinite{FI}{an}{a}
	\acroindefinite{IF}{an}{an}
	\acroindefinite{II}{an}{an}
	\acroindefinite{EW}{an}{an}
	\acroindefinite{LC}{an}{a}
    \acroindefinite{NLO}{an}{a}
    \acroindefinite{LO}{an}{a}
\end{acronym}
\pagebreak

\begin{declaration*}
	The work in this thesis is based on research carried out in the Department of
	Physics at Durham University. No part of this thesis has been
	submitted elsewhere for any degree or qualification.
    All original work is my own, carried out in collaboration with 
    my supervisor, Dr Marek Sch\"{o}nherr. Chapters \ref{chapter:qedps}
    and \ref{chapter:mcatnlo} of this thesis are based on ongoing 
    joint research to be published. Chapter \ref{chapter:yfs} is 
    based on ref. \cite{Flower:2022iew}.
\end{declaration*}

\begin{acknowledgements*}
I would like to begin by expressing my deepest gratitude 
to my supervisor, Marek Sch\"{o}nherr, for all his support 
throughout my PhD, for 
endless help with \Sherpa, and for putting his trust in me.
I am also grateful to everyone in the \Sherpa 
collaboration for being such lovely people to work with
and for valuable physics discussions.

A very special thanks go to Elliot Fox, Hitham Hassan, 
Ery McPartland and Malina Rosca, for proofreading.
They, and the whole community at IPPP, were invaluable 
sources of companionship and support, especially coming 
out of the Covid-19 pandemic. Of course, the IPPP would 
not function without Trudy Forster and Joanne Bentham, 
who helped me with countless admin tasks and made 
everything run smoothly.

I would like to thank all my housemates over the years I 
spent in Durham for providing good company and many hours 
of board games. Thank you to my family for checking in 
with me often, and for all the practical help and advice.
Finally, I would like to thank Joe Callow, for always 
being there for me.
\end{acknowledgements*}

\begin{epigraph*}
    But if you must and you can, then there's no excuse.
	\source{The Amber Spyglass}{Philip Pullman}
\end{epigraph*}

%\dedicationtext{This thesis is dedicated to}
% \begin{dedication*}
% %
% 	Someone special
% 	\also[and]{that other guy}
% %
% \end{dedication*}

\cleardoublepage

\subfile{text/intro}

\subfile{text/QEDps}

\subfile{text/mcatnlo}

\subfile{text/yfs}

\subfile{text/conclusions}

%\subfile{background}
%\subfile{paper1}
%\subfile{paper2}
%\subfile{paper3}

\appendix

\subfile{text/appendix1}

%\subfile{appendix2}

\bibliographystyle{JHEP}
\bibliography{journal}

\end{document}

%% file: text/preamble.tex
\usepackage[utf8]{inputenc} % utf8 input good for non-ascii characters
\usepackage[T1]{fontenc} % Good for hyphenating non-ascii characters
\usepackage{lmodern}
\usepackage[british]{babel}
\usepackage[nottoc]{tocbibind} % Includes bibliography in table of contents, but not the table of contents itself
\usepackage[figuresright]{rotating} % Allows for sideways figures and tables
\usepackage[hidelinks,pdfusetitle]{hyperref} % hidelinks removes the boxes around links when viewed on screen, pdfusetitle option includes the thesis title and author in the PDF metadata
\usepackage{cite} % Combine multiple citations in range

\usepackage[shrink=10,stretch=10]{microtype}
\makeatletter
\@ifpackageloaded{microtype}{%
	\providecommand{\disableprotrusion}{\microtypesetup{protrusion=false}}
	\providecommand{\enableprotrusion}{\microtypesetup{protrusion=true}}
}{%
	\providecommand{\disableprotrusion}{}
	\providecommand{\enableprotrusion}{}
}
\makeatother

\usepackage{booktabs} % Better tables

\usepackage[margin=15mm,hang]{caption}
\usepackage{subcaption}
\usepackage{perpage} % Allows to reset counters on each page
\MakePerPage{footnote} % Resets footnote counters on each page

\usepackage[printonlyused,withpage]{acronym}

\graphicspath{{img/}}

\newcommand\numberthis{\addtocounter{equation}{1}\tag{\theequation}}

\newenvironment{itemize*}%
{\begin{itemize}%
	\setlength{\itemsep}{0pt}%
	\setlength{\parskip}{0pt}}%
{\end{itemize}}
\newenvironment{enumerate*}%
{\begin{enumerate}%
	\setlength{\itemsep}{0pt}%
	\setlength{\parskip}{0pt}}%
{\end{enumerate}}

\usepackage{amsmath,amssymb}
\numberwithin{equation}{section}
\allowdisplaybreaks % Allow display equations to break over different pages
\usepackage{amssymb}
\usepackage{bm} % More bold maths symbols
\usepackage{bbm} % More blackboard bold characters, via \mathbbm. Mostly for blackboard bold 1
\usepackage{xfrac} % Nice small fractions
\usepackage{array} % Allows us to define custom column types for tables
\newcolumntype{L}{>{$}l<{$}} % a left aligned maths column type

\usepackage{cleveref} % Add automatic reference type text via \cref command
\usepackage{xcolor}

\usepackage[italic]{hepnames} % Add particle name macros (e.g. \PBs)
\usepackage{braket} % Adds Dirac bra-ket notation
\usepackage{slashed} % Adds Dirac slash notation
\usepackage{siunitx} % Add support for units
\DeclareSIUnit\fb{\femto\barn}

\makeatletter
\providecommand*{\diff}%
{\@ifnextchar^{\DIfF}{\DIfF^{}}}
\def\DIfF^#1{%
\mathop{\mathrm{\mathstrut d}}%
\nolimits^{#1}\gobblespace}
\def\gobblespace{%
\futurelet\diffarg\opspace}
\def\opspace{%
\let\DiffSpace\!%
\ifx\diffarg(%
\let\DiffSpace\relax
\else
\ifx\diffarg[%
\let\DiffSpace\relax
\else
\ifx\diffarg\{%
\let\DiffSpace\relax
\fi\fi\fi\DiffSpace}

\providecommand{\intdphi}[1]{\ensuremath{\int\diff\Phi_{#1}\,}}
\providecommand{\Oangle}[1]{\ensuremath{\langle O\rangle^\mathrm{#1}}}

\newcommand{\ie}{\textit{i.e.\@}\xspace}

\newcommand{\see}{\textit{see\@}\xspace}

\newcommand{\SMCatNLO}{S--M\protect\scalebox{0.8}{C}@N\protect\scalebox{0.8}{LO}\xspace}
\newcommand{\MCatNLO}{M\protect\scalebox{0.8}{C}@N\protect\scalebox{0.8}{LO}\xspace}

\newcommand{\Powheg}{P\protect\scalebox{0.8}{OWHEG}\xspace}

\newcommand{\Herwig}{H\protect\scalebox{0.8}{ERWIG}\xspace}

\newcommand{\Madgraph}{M\protect\scalebox{0.8}{AD}G\protect\scalebox{0.8}{RAPH}\xspace}

\newcommand{\Pythia}{P\protect\scalebox{0.8}{YTHIA}\xspace}

\newcommand{\Photos}{P\protect\scalebox{0.8}{HOTOS}\xspace}

\newcommand{\Horace}{H\protect\scalebox{0.8}{ORACE}\xspace}
\newcommand{\Winhac}{W\protect\scalebox{0.8}{INHAC}\xspace}
\newcommand{\Zinhac}{Z\protect\scalebox{0.8}{INHAC}\xspace}
\newcommand{\KoralW}{K\protect\scalebox{0.8}{ORAL}W\xspace}
\newcommand{\YFSWW}{Y\protect\scalebox{0.8}{FS}WW\xspace}
\newcommand{\KKMC}{K\protect\scalebox{0.8}{KMC}\xspace}
\newcommand{\BHLumi}{BHL\protect\scalebox{0.8}{UMI}\xspace}
\newcommand{\BHWide}{BHW\protect\scalebox{0.8}{IDE}\xspace}
\newcommand{\Rady}{R\protect\scalebox{0.8}{ADY}\xspace}
\newcommand{\BabaYaga}{B\protect\scalebox{0.8}{ABA}Y\protect\scalebox{0.8}{AGA}\xspace}
\newcommand{\LHAPDF}{L\protect\scalebox{0.8}{HAPDF}\xspace}

\newcommand{\Rivet}{R\protect\scalebox{0.8}{IVET}\xspace}

\newcommand{\OpenLoops}{O\protect\scalebox{0.8}{PEN}L\protect\scalebox{0.8}{OOPS}\xspace}

\newcommand{\Sherpa}{S\protect\scalebox{0.8}{HERPA}\xspace}
\newcommand{\Comix}{C\protect\scalebox{0.8}{OMIX}\xspace}

\newcommand{\Amegic}{A\protect\scalebox{0.8}{MEGIC}\xspace}

\newcommand{\Ahadic}{A\protect\scalebox{0.8}{HADIC}\xspace}

\newcommand{\Photons}{P\protect\scalebox{0.8}{HOTONS}\xspace}
\newcommand{\Tevatron}{Tevatron\xspace}

\long\def\symbolfootnote[#1]#2{\begingroup%
\def\thefootnote{\fnsymbol{footnote}}\footnote[#1]{#2}\endgroup}

\newcommand{\abs}[1]{\left| #1\right|}

\newcommand{\im}{\imath}
\newcommand{\jm}{\jmath}
\newcommand{\ijt}{{\widetilde{\im\hspace*{-1pt}\jm}}}

\newcommand{\kt}{{\tilde{k}}}
\newcommand{\ajt}{{\widetilde{a\;\!\!\jm}}}
\newcommand{\at}{{\tilde{a}}}
\newcommand{\bt}{{\tilde{b}}}

\newcommand{\ffbar}{\mathrm{f}\bar{\mathrm{f}}}

\newcommand{\done}{{\rm d}}
\newcommand{\order}{\mathcal{O}}

\newcommand{\bea}{\begin{eqnarray}}
\newcommand{\eea}{\end{eqnarray}}
\newcommand{\bi}{\begin{itemize}}
\newcommand{\ei}{\end{itemize}}

\newcommand*{\TeV}{\ensuremath{\text{Te\kern -0.1em V}}}
\newcommand*{\GeV}{\ensuremath{\text{Ge\kern -0.1em V}}}
\newcommand*{\MeV}{\ensuremath{\text{Me\kern -0.1em V}}}
\newcommand*{\keV}{\ensuremath{\text{ke\kern -0.1em V}}}
\newcommand*{\eV}{\ensuremath{\text{e\kern -0.1em V}}}

\newcommand{\tstart}{\ensuremath{t_\text{start}}\xspace}
\newcommand{\qbar}{\ensuremath{\bar{q}}\xspace}
\newcommand{\kT}{\ensuremath{{\mathrm{k}_\mathrm{T}}}\xspace}
\newcommand{\kTsq}{\ensuremath{{\mathrm{k}_\mathrm{T}^2}}\xspace}

\newcommand{\dRdress}{\ensuremath{\Delta \Theta_\text{dress}}}
\newcommand{\fdress}{\ensuremath{\mathrm{f}_\text{dress}}}
\newcommand{\ellp}{\ensuremath{\ell_\mathrm{p}}}
\newcommand{\ells}{\ensuremath{\mathrm{f}_\mathrm{s}}}

\newcolumntype{C}{>{\centering\arraybackslash}p{0.14\textwidth}}

\newlength{\unitcharwidth}
\newcommand{\changed}[1]{{\color{blue}#1}}
\renewcommand{\changed}[1]{{\color{black}#1}}

%% file: text/intro.tex
\chapter{Introduction}
\label{chapter:intro}

The \ac{SM} of particle physics is a triumph of modern physics, 
being our most precisely tested theory of nature. Over a huge 
range of energy and distance scales, we are able to describe 
the interactions of particles at collider experiments 
with incredible precision. By 
utilising a number of mathematical and computational methods, 
ever more precise predictions can be made. At the time of writing, 
no discovery of physics beyond the \ac{SM} has definitively been 
made, although it is known that it must exist (there is evidence 
for the existence of dark matter, dark energy and 
massive neutrinos). More concretely, the recent measurements of 
the muon anomalous magnetic moment $(g-2)_\mu$ 
\cite{Muong-2:2006rrc,Muong-2:2023cdq} 
are in conflict with the \ac{SM} prediction \cite{Aoyama:2020ynm}.
The uncertainty on the prediction is dominated by the hadronic 
vacuum polarisation, which can be determined using various 
experiments, such as low-energy $e^+ e^-$ collisions or 
muon-electron scattering.
To confirm the disagreement, and an unquestionable signal 
of new physics coupling to the \ac{SM}, or to resolve 
the tension, we require higher precision predictions of 
these experiments.
For this reason, it is imperative that we 
continue to improve our understanding of the \ac{SM} itself.

Modern particle colliders such as the \ac{LHC} and electron-positron
`$b$ factories' are entering a precision era where the 
experimental error on many measurements is smaller than the 
theoretical error on the \ac{SM} prediction. In addition 
to this success of current experiments, there are multiple 
proposals for a future lepton-lepton collider, which would 
provide the most precise measurements yet of the \ac{SM}, and 
any physics beyond the \ac{SM}. While the 
success of the engineering involved is to be celebrated, this 
fact presents a challenge to the theory and phenomenology community. 
The last 20 years have seen a plethora of new methods for 
increasing theoretical precision, in many cases in a largely 
automated and process-independent way. 

Many of these automated methods for producing precise theoretical 
predictions fall under the category of \ac{MC} statistical
methods. They are implemented in \ac{MC} event generators,
codes which simulate the physics involved in 
particle collisions and produce descriptions of the final-state 
particles, called events. These events can be treated as 
similar enough to real collider final states and can 
be used to model the detector response to certain physics 
scenarios in experiments. Events can also be analysed to 
produce cross sections and kinematical distributions, to 
be compared to data. Due to the public nature of these codes 
and their vast applicability, \ac{MC} event generators are 
a cornerstone of modern particle physics. As a result, their 
physics models are constantly under development to address 
the aforementioned challenge to theoretical precision. 

In this thesis we concern ourselves with improving the 
precision of \ac{MC} event generator predictions by 
implementing new methods to describe electromagnetic 
radiative corrections in an automated, process-independent 
manner. We describe each method and present results to 
validate the description, before presenting new 
phenomenological results. In chapter \ref{chapter:qedps}
we introduce the \ac{QED} dipole parton shower, in chapter 
\ref{chapter:mcatnlo} we describe the process of 
matching the shower to \iac{NLO}
calculation, and in chapter \ref{chapter:yfs} we introduce 
the \ac{YFS} soft-photon resummation and describe its extension to 
charged particle pair production. Finally, in 
chapter \ref{chapter:conclusions}, we conclude and reflect on the 
impact of all three methods, and give an outlook on 
future developments in this area.

\section{The Standard Model and perturbation theory} \label{sec:intro:SM}

The \ac{SM} of particle physics is an SU(3)$_C \times$ SU(2)$_L 
\times$ U(1)$_Y$ gauge theory. \Ac{QCD} is a Yang-Mills 
theory which describes quarks, gluons (the gauge bosons of the strong force) 
and the hadrons they make up. It exhibits confinement \cite{Wilson:1974sk} 
and asymptotic freedom \cite{Gross:1973id,Politzer:1973fx}.
The former means that objects with a colour charge, notably quarks and gluons, 
cannot be observed in isolation because the energy required to separate them 
exceeds the energy required for pair production and the formation of colourless 
hadrons. The latter allows strong interactions to be calculated using perturbation 
theory when the energy scale of the interaction is sufficiently high. The strong 
coupling constant $\alpha_s = g_s^2/4\pi$ has a pole at the confinement scale,
$\Lambda_{\mathrm{QCD}} = 210 \pm 14$ \MeV\, in the modified minimal-subtraction
($\overline{\mathrm{MS}}$)
scheme with 5 light quark flavours \cite{ParticleDataGroup:2018ovx}.
At energies higher than a few GeV, 
$\alpha_s < 1$ and therefore perturbation theory can be employed. Below this scale 
non-perturbative methods, such as lattice \ac{QCD}, must be used. 
In practice, however, since 
lattice \ac{QCD} is still a developing technology,
the solution is usually to use other non-perturbative 
models with free parameters extracted from data. 
For initial-state hadrons we make phenomenologically-motivated 
Ans\"{a}tze for \acp{PDF} and fit these to data. Similarly, 
to understand the production of final-state hadrons we must 
model hadronisation. There are various 
physically-motivated methods to model hadronisation, 
including the Lund string model \cite{Andersson:1983ia} and the 
cluster model \cite{Winter:2003tt}.
Of the whole \ac{SM}, only quarks, antiquarks 
and gluons carry colour charge and hence feel the strong force. 
However, as the strongest interaction in nature, a huge amount of 
effort has gone into developing methods to allow us to model 
\ac{QCD} ever more precisely.
% Another 
% feature of \ac{QCD} is that gluons are massless and the up, down and strange quark 
% masses are very small compared to even hadronic scales.
% As a result the massless approximation is often taken for the three lightest 
% quarks, and when the relevant energy scale is far above the mass of the charm or 
% bottom quark these can also be taken to be massless. The massless approximation 
% cannot be taken for the top quark as its mass $m_t =$ 175 GeV is higher than the 
% \ac{EW} scale $Q_{\mathrm{EW}} \approx m_Z  \approx m_W \approx$ 100 GeV. 

At high energies, the remaining SU(2)$_L \times$ U(1)$_Y$ group 
describes the \ac{EW} theory. This symmetry 
is spontaneously broken in nature, leaving an unbroken 
U(1)$_\text{QED}$ \ac{EM} symmetry and massive weak gauge 
bosons, as well as a massive Higgs boson. The low-energy 
\ac{EM} force is described by \ac{QED}.
The gauge boson of \ac{QED} is the photon, 
which has no electromagnetic charge 
(contrary to gluons which carry colour charge). All charged particles feel the 
\ac{EM} force: quarks, the charged leptons $e,
\mu$ and $\tau$, and the charged gauge bosons $W^\pm$. 
The \ac{EM} force decreases with distance and hence increases with the 
energy of the interaction, unlike the strong force. The 
electroweak coupling constant $\alpha = e^2/4\pi$, also known as 
the fine structure constant, is of order $10^{-2}$ and has a 
slow running, so the theory remains perturbative over the whole 
range of energies available in experiments.
The weak sector 
contains the massive $Z$ and $W$ gauge bosons and the Higgs boson. 
Due to the high energies required to produce these bosons on-shell, 
when discussing their production and decay we will usually refer 
to the unified electroweak interaction instead of the low-energy 
weak and \ac{EM} interactions.

In this thesis we will be primarily concerned with the \ac{QED} sector of the \ac{SM},
as well as its interplay with \ac{QCD}. We will propose, motivate and present results 
for methods to bridge the gap between perturbative calculations and 
experimental observations. As far as possible, these methods will be 
process-independent, automated and publicly available.

\subsection{Perturbation theory, renormalisation and cancellation 
of infrared singularities}

The primary technique of quantum field theory calculations is perturbation theory, 
in which any interactions are assumed to be a small correction to the free theory 
and so a series expansion in the coupling constant can be performed. The first 
interacting term of this series is referred to as \ac{LO}.
We define a Feynman diagram without loops as a tree-level
diagram. While for many quantities the the \ac{LO} contribution 
is tree-level, the two terms are not synonymous.
The corresponding \ac{ME} is often referred to as the Born 
\ac{ME} to distinguish it from higher-order corrections.
\Ac{NLO} calculations generally include diagrams which involve 
loops. The 
momentum flowing through the loop particles is not determined by momentum 
conservation so must be integrated over all possible values. Some of these loop 
integrals are divergent due to the requirement to integrate the magnitude of the 
loop momentum up to infinity. 
\changed{However, by regularising the integrals, it can be 
seen that the divergences can be absorbed into redefinitions of 
Lagrangian parameters and all integrals become \ac{UV} finite.}
% These ultraviolet divergences can be removed using 
% renormalisation since they correspond to parameters in the Lagrangian which are 
% unobservable and hence unphysical. Loop integrals can then be calculated using 
% dimensional regularisation. 

In general, calculations of terms in the perturbative 
series up to a given order in powers of the relevant coupling ($\alpha$ or 
$\alpha_s$ or a mix of the two) are called `fixed-order'. When ordering 
a perturbative expansion in a mixture of $\alpha$ and $\alpha_s$, a rule of 
thumb for the relative magnitudes of the terms is $\alpha \approx \alpha_s^2$.
Care must be taken not to double-count contributions, since 
\iac{EW} correction to \iac{QCD}-induced process 
and \iac{QCD} correction to \iac{EW}-induced process
may be identical. 

After renormalisation, higher-order calculations still lead to 
divergences. The emission of a soft or collinear photon or gluon,
or a collinear massless fermion, is associated with a so-called \ac{IR} 
singularity in the cross section due to the phase space integration. 
Separately, the \ac{NLO} correction involving exchange of a virtual 
(internal)
photon or gluon to a tree-level process leads to \iac{IR} 
singularity in the integration over the loop momentum. The solution 
to both these problems lies in abandoning the attempt to calculate 
them separately. At \ac{NLO} in the cross section, we must consider 
both the interference of the Born \ac{ME} with the one-loop \ac{ME}
and the real-emission \ac{ME} squared. With the use of regularisation, 
it can be seen that the two singularities cancel exactly. This is 
the \ac{KLN} theorem \cite{Kinoshita:1962ur,Lee:1964is}.

This cancellation of \ac{IR} singularities occurs not only for 
the total cross section, but for a range of observables. We 
will define such observables, where \ac{IR} divergences exactly 
cancel, as \ac{IR-safe}. More precisely, \iac{IR-safe} 
observable $O$ satisfies the two conditions 
\begin{equation}
  \begin{aligned}
    &O_{n+1}(\dots,p_i,\dots) \xrightarrow{p_i \to 0} 
    O_n(\dots) \qquad\qquad\qquad\qquad\quad\, \text{soft-safe,}\\
    &O_{n+1}(\dots,p_i,p_j,\dots) \xrightarrow{p_i \to Cp_j} 
    O_n(\dots,(p_i+p_j),\dots) \quad \text{collinear-safe,}\\
  \end{aligned}
\end{equation}
where `soft' means low-energy, such that the soft limit is the limit of zero 
energy; if two particles are exactly collinear they have parallel momenta 
$p_i = C p_j$
and the collinear limit is that of zero separation angle.
Examples of \ac{IR-safe} observables 
include transverse momenta and energies of jets or suitably 
dressed leptons, and geometrical observables such as thrust, but do not 
include observables such as the total number of photons in the 
final state.

In this section we have discussed perturbation theory as a tool 
to obtain fixed-order predictions for \ac{IR-safe} observables.
However, fixed-order calculations are not always sufficient to get a good description 
of the physical process. There can exist very significant contributions even 
from large powers of the coupling when these are multiplied by logarithms of 
very large or very small quantities, such as a ratio of energies or masses. Calculations 
which are not fixed-order use resummation of leading logarithms to obtain a better 
approximation of the quantity under consideration. This can be extended to next-to-leading 
logarithms, and so on in the same way as fixed-order calculations. Where the size of 
logarithms in the perturbative series is controlled and small, a resummed calculation 
will not achieve significantly higher precision than a fixed-order one.

\subsection{The YFS soft-photon resummation} \label{sec:intro:yfs}

The work of Yennie, Frautschi and Suura \cite{Yennie:1961ad} 
describes \changed{the removal of} the \ac{IR} singularities of \ac{QED} to all orders
by reordering the perturbative
expansion of a scattering or decay \ac{ME}. 
This is achieved by separating the \ac{IR} divergences from the finite 
remainders, to all orders. The IR divergent terms form a series which can 
be exponentiated, amounting to a resummation of soft-photon logarithms
in the enhanced real and virtual regions. This leaves a perturbative expansion
in \ac{IR}-finite, hard photons (both real emissions and virtual exchanges).
The \ac{YFS} approach considers all charged 
particles of the theory 
to be massive, and as a consequence only singularities associated with 
soft-photon emission are present.

In the \ac{YFS} description, photons are emitted coherently from 
the charged multipole through the eikonal radiator function. This 
is in contrast to parton shower descriptions of photon radiation, 
which usually consider emission from one specific emitter particle 
or from a charge dipole. The photons produced by \iac{YFS} 
implementation are also not ordered in any specific kinematical 
variable, but do have \iac{IR} cutoff on their energy - all 
radiation below this energy is resummed with the virtual corrections.
The \ac{YFS} resummation has the advantage that, when the 
perturbative expansion is reordered, what remains after all 
soft-photon singularities are resummed is \changed{the complete} perturbative expansion.
This means that \changed{\ac{IR}}-subtracted real-correct \acp{ME} 
can be calculated and combined with the resummation to yield 
fixed-order-matched results.

The \ac{YFS} resummation is widely used in \ac{MC}
event generators to describe photon radiation. An implementation 
\changed{of the algorithm was first included in the event generator 
\KKMC \cite{Jadach:1988gb,Jadach:1999vf}, 
and later in \Herwig \cite{Hamilton:2006xz},
\Sherpa \cite{Schonherr:2008av,Krauss:2018djz}, 
\BHLumi \cite{Jadach:1991by}, \BHWide \cite{Jadach:1995nk},
\Winhac/\Zinhac \cite{Placzek:2003zg}, and \KoralW/\YFSWW \cite{Jadach:2001mp}.}
The inclusion of the algorithm in general-purpose event 
generators allows it to be used in conjunction with \ac{QCD} 
parton showers, hadronisation and other non-perturbative effects.
However, the \ac{YFS} resummation is not the only approach to 
modelling \ac{QED} radiative corrections. One can also use 
a collinear resummation approach and supplement this with 
soft effects. A prominent example of a successful tool for 
\ac{QED} final-state radiation which \changed{uses a bespoke 
algorithm for exponentiation of multiple soft-photon emissions}
is \Photos \cite{Barberio:1990ms, Barberio:1993qi,Golonka:2005pn,Davidson:2010ew}.
\changed{The \Photos algorithm is similar to that of \ac{YFS}, but was 
developed independently.}

\section{Monte Carlo event generators} \label{sec:intro:mceg}

The modern workhorses of particle physics are event generators. These 
programs translate \ac{ME} calculations into `events', collections 
of particles with defined momenta, which can be treated almost
identically to 
those seen in experimental detectors. An event generator usually consists 
of multiple modules which simulate different effects: parton showers, 
decays, hadronisation, \acp{PDF}, beam structure, and sometimes multiple 
interactions and the underlying event. Event generators carry out cross section 
calculations using \ac{MC} integration techniques. There are many 
excellent reviews of MC event generators in the literature, for example 
refs. \cite{Hoche:2014rga,Sjostrand:2009ad,Campbell:2022qmc}.

\subsection{Factorisation of the cross section} \label{sec:intro:factorisation}

Many of the techniques carried out by event generators rely on the 
factorisation of the cross section into parts characterised by 
different energy scales. At high scales, we have the partonic 
cross section, which can be 
calculated using perturbation theory. But the long-distance objects 
which we measure are dependent on low-energy physics.
Thus, we must describe how the initial state that 
enters the hard scattering is produced and how the observed final state 
emerges from the hard scattering. We assume that the relevant processes 
involved at these two energy scales factorise.
The factorisation Ansatz is 
\cite{Bodwin:1984hc,Collins:1985ue}

\begin{equation} \label{eq:intro:factorisation}
  \sigma_{A B \to X} = \sum_{a \in A} \sum_{b \in B}
  \int \diff x_a \int \diff x_b \, f_a^{A}(x_a,\mu_F^2)
  f_b^{B}(x_b,\mu_F^2) \int \diff \Phi_{ab \to X} 
  \frac{\diff \hat{\sigma}_{ab}(\Phi_{ab \to X},\mu_F^2)}
  {\diff \Phi_{ab \to X}}
\end{equation}
where $A,B$ are the incoming beam particles (hadrons or 
leptons), $a,b$ are their constituent flavours (quarks, 
gluons, leptons and/or photons), $x_a$ is the momentum fraction 
carried by constituent $a$, and $f_a^A$ is the Parton Distribution Function
of constituent $a$ in particle $A$. \changed{Additionally,} 
$\mu_F^2$ is the factorisation 
scale, with units of energy squared.
$\hat{\sigma}_{ab \to X}$ is 
the partonic cross section for the production of a final state 
$X$, calculated using perturbation theory.

\begin{figure}
  \centering
  \includegraphics[width=0.7\textwidth]{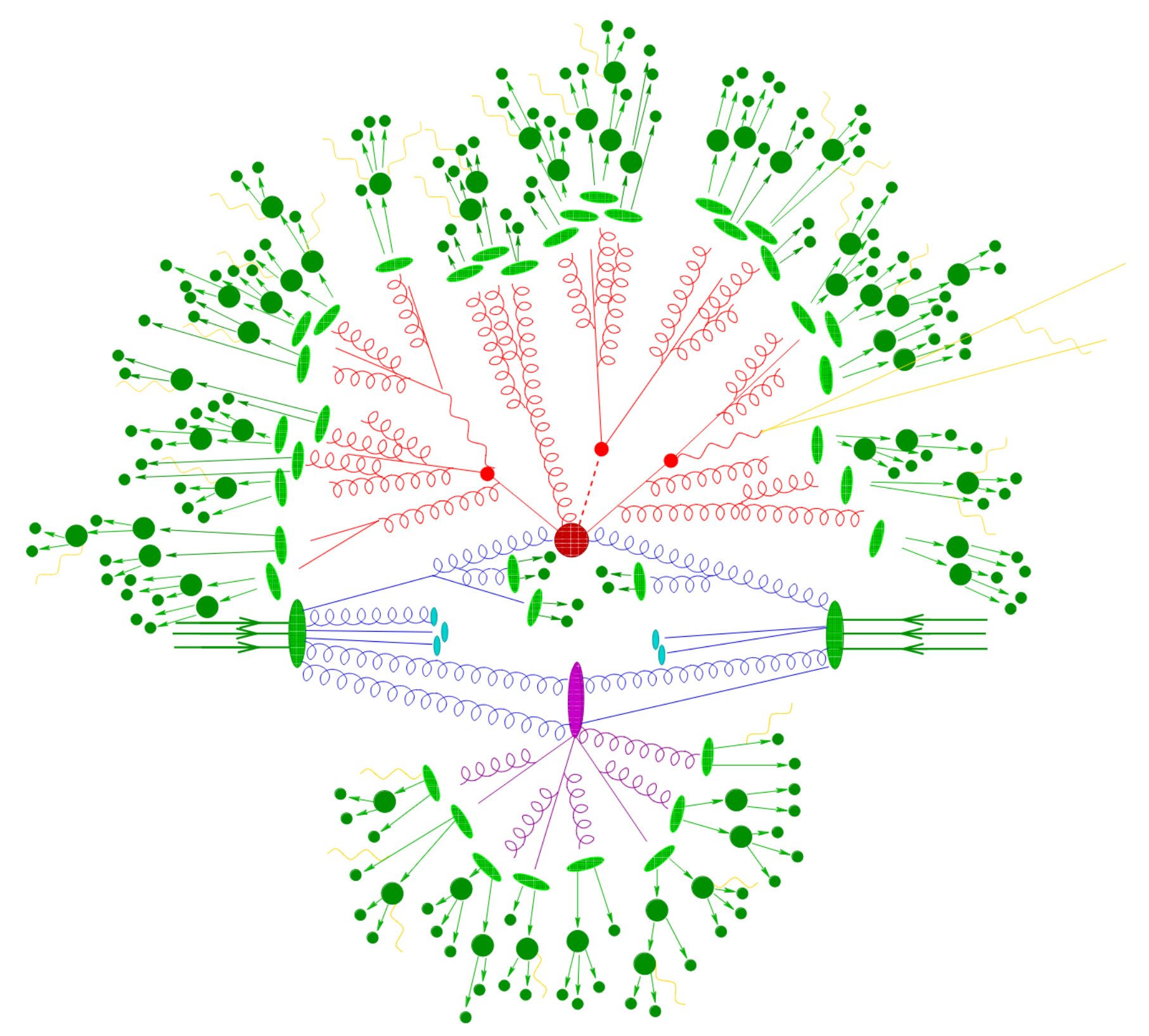}
  \caption[A pictorial representation of a $t\bar{t}H$ event at a 
  hadron collider, as represented in the event generator 
  \Sherpa.]{A pictorial representation of a $t\bar{t}H$ event at a 
  hadron collider, as represented in the event generator 
  \Sherpa \cite{Gleisberg:2008ta}. The beams are resolved into partons (blue) 
  which radiate before entering either the hard interaction 
  (red blob) or secondary softer interactions (purple).
  The heavy final state, the Higgs and top quarks, decay 
  (smaller red blobs) and further QCD radiation is produced 
  by the parton shower (red). Hadronisation is shown in 
  light green and the produced hadrons then decay (dark green).
  QED radiation is shown in yellow and can occur at all 
  scales.
  \label{fig:intro:SherpaEventPic}
  }
\end{figure}

Note that the cross section in eq.\ \eqref{eq:intro:factorisation}
is the inclusive cross section: it specifies the cross section 
for the production of $X$ in any kinematical configuration 
and accompanied by any number of additional particles. However, 
the way that \ac{MC} event generators work is to generate 
an event according to eq.\ \eqref{eq:intro:factorisation} 
containing exactly $X$ as the final state, and then augment 
this event with additional particles. By using Markov 
processes to produce these additional particles, probability is 
conserved, as well as four-momentum, respecting the inclusivity
requirement. This is shown schematically in fig.\ 
\ref{fig:intro:SherpaEventPic}.
For this thesis the relevant Markov process, 
applicable at hadron and lepton colliders, is the parton shower,
which we introduce in section \ref{sec:intro:ps}.

\subsection{The parton shower} \label{sec:intro:ps}

The property that a real-emission \ac{ME} factorises 
into the Born \ac{ME} multiplied by a universal 
splitting function in the collinear limit leads to 
the derivation of evolution equations for the \acp{PDF},
since these must obey the strict collinear limit.
The dependence of the \acp{PDF} on the factorisation 
scale is given by the \ac{DGLAP} equation
\cite{Gribov:1972ri,Lipatov:1974qm,Dokshitzer:1977sg,Altarelli:1977zs},
\begin{equation} \label{eq:intro:dglap}
  \frac{\diff f_a^A(x,\mu_F^2)}{\diff \log{(\mu_F^2)}}
   = \sum_{b \in A} \int_x^1 \frac{\diff z}{z} 
   \frac{\alpha}{2\pi} \hat{P}_{b\to a}(z) f_b^A(x/z,\mu_F^2),
\end{equation}
where the $\hat{P}_{b\to a}(z)$ are the regularised 
Altarelli-Parisi splitting functions. \changed{These describe
the collinear splitting of particle $b$ into particle $a$ in 
\ac{QCD} or \ac{QED} \cite{Altarelli:1977zs}}.
This leads to an 
interpretation of the factorisation Ansatz as separating 
unresolved parton branchings (absorbed into the definition 
of the \acp{PDF}) and resolved parton branchings, which 
are generated by a parton shower using variants of the 
Altarelli-Parisi splitting functions. The crossover point 
between the two is related to the factorisation scale.

Repeatedly generating parton branchings according to 
eq.\ \eqref{eq:intro:dglap} produces new final-state particles 
in a Markov process which we call initial-state radiation.
The final-state radiation will be generated in an analogous 
manner, without considering the \acp{PDF} and instead 
considering physical cutoff scales for the evolution, such 
as $\Lambda_{\text{QCD}}$ for \ac{QCD} radiation or 
photon resolution scales for \ac{QED} radiation.

\subsection{Next-to-leading order calculations and matching}
\label{sec:intro:mceg:NLO}

\ac{NLO} calculations of cross sections generally involve interference terms between 
loop diagrams (virtual corrections) and the Born \ac{ME}, and emissions of 
undetectable particles off external legs (real corrections). \Ac{IR} divergences 
appear in both cases, when the propagator of a virtual particle goes on-shell or 
when a real emission becomes infinitely soft or collinear to another particle. 
In the soft or collinear limit, the 
particle becomes undetectable. There is no divergence associated with the soft 
limit of a fermion, but photons and gluons have this divergence and the collinear 
limit of any massless emission is associated with a divergence. 
As described above, to ensure the cancellation of these \ac{IR} singularities 
by the \ac{KLN} theorem, we must consider only \ac{IR-safe} observables.
\changed{When the integrals are regularised, for example by moving to 
$d=4-2\epsilon$ dimensions, the cancellation of singularities is 
apparent, and the $\epsilon \to 0$ limit can be safely taken.
Nevertheless, the need to carry out integrals in $d$ 
dimensions to facilitate the cancellation prevents numerical 
integration of \ac{NLO} quantities.}

In order to carry out \ac{NLO} calculations as described in sec. 
\ref{sec:intro:SM} in an event generator, we must ensure that the 
real and virtual contributions are separately finite. 
There are two main families of methods to achieve this: phase-space 
slicing, and subtraction. In this thesis we will exclusively 
consider subtraction methods. \changed{To define \iac{NLO} 
subtraction, we add or remove particles from the Born 
configuration using a dipole splitting picture. The notation is detailed
in sec. \ref{sec:intro:notation} below; briefly, a tilde is used 
to denote quantities pre-splitting (in this case, relating to 
the Born configuration) while unmodified quantities relate to 
the real-emission configuration. We introduce a notation for mapping 
from the real-emission phase space $\Phi_{n+1}$ to the Born phase space $\Phi_n$ \cite{Hoche:2010pf},}
\begin{equation} \label{eq:intro:PSmapping}
  b_{ij,k}(\{a\}) = \begin{cases}
    \{f\} \cup \{f_{\ijt}\} \setminus \{f_i,f_j\} \\
    \{\vec{p}\} \to \{\vec{\tilde{p}}\},
  \end{cases}
\end{equation}
where $\{a\} = \{a_1,a_2,\dots a_n\}$ denotes the particle configuration, 
$\{f\}$ is the set of flavours and $\{\vec{p}\}$ is the set of 
four-momenta. 
\changed{This mapping removes the emitter particle $\ijt$ and adds 
its splitting products $i$ and $j$. The inverse is a 
mapping from the Born phase space 
to the real-emission phase space,}
\begin{equation} \label{eq:intro:PSmappingInverse}
  r_{\ijt,\kt}(\changed{f_j},\Phi_{R|B}^{ij,k},\changed{\{\tilde{a}\}}) = 
  \begin{cases}
    \changed{\{\tilde{f}\}} \cup \{f_i,f_j\} \setminus \{f_\ijt\} \\
    \{\vec{\tilde{p}}\} \to \{\vec{p}\},
  \end{cases}
\end{equation}
\changed{where $\Phi_{R|B}^{ij,k}$ is the factorised 
one-particle phase space for an emission $\ijt (\kt) \to i j (k)$,
and $f_j$ is the emitted particle flavour. Of course, only 
those splittings which are allowed by the theory are considered.
In the soft limit, $p_j \to 0$, both these mappings must 
leave all non-soft particles unchanged. Similarly, in 
the collinear limit, only the emitter particle may be changed.}

\changed{These mappings allow us to write down the factorisation 
of the squared matrix element in the infrared limits. 
In the collinear limit,}
\begin{equation}
  |\mathcal{M}_{n+1}|^2(\{a\}) \longrightarrow \frac{8\pi 
\alpha}{2 p_i p_j} \mathcal{M}_n(b_{ij,k}(\{a\}))
\otimes \hat{P}_{\ijt \to i}(z) \otimes \mathcal{M}^*_n(b_{ij,k}(\{a\}))
\end{equation}
where the tensor products indicate spin correlations. In \ac{QED},
these spin correlations only exist for photon splittings 
into fermions, which do not have a true collinear singularity in 
any case. In this thesis 
we will use spin-averaged splitting functions and reduce this to 
an ordinary product. In the soft limit, a similar factorisation 
holds where the Altarelli-Parisi splitting functions divided by 
the virtuality $2p_i p_j$ 
are replaced by an eikonal 
\begin{equation}
  \mathcal{S}_{ij,k} = \frac{p_i p_k}{(p_i p_j)(p_j p_k)}
\end{equation}
accompanied by appropriate colour or charge correlators for \ac{QCD} or 
\ac{QED} evolution respectively. These are the ingredients needed to 
define a subtraction scheme, to allow us to calculate the parts of 
\iac{NLO} calculation individually without encountering divergences.

Here we will use the \ac{CS} subtraction method, 
which is built \changed{on dipole factorisation}. The subtraction
terms include both soft and collinear singularities, obtained by partial 
fractioning the eikonals \cite{Catani:1996vz,Catani:2002hc}. 
The dipole subtraction terms are denoted by $D^S_{ij,k}$ and their 
integrals by $I^S_{\ijt,\kt}$. Then an observable $O$ has 
expectation value 
\begin{align*} \label{eq:intro:nlo}
  \langle O \rangle^{\text{NLO}} =& \sum_{\{\tilde{f}\}} \int \diff \Phi_n\left( \{\tilde{f}\} \right)
  \left( B\left( \changed{\{\tilde{a}\}} \right) + \tilde{V}\left( \changed{\{\tilde{a}\}} \right) + 
  \sum_{\ijt,\kt} I^S_{\ijt,\kt}\left( \changed{\{\tilde{a}\}} \right) \right)
  O\left( \{\tilde{p}\} \right) \\
  &+ \sum_{\{f\}} \int \diff \Phi_{n+1}\left( \{f\} \right) 
  \left( R_{n+1}\left( \{a\} \right) O\left( \{p\} \right) - \sum_{ij,k} D^S_{ij,k}\left( \{a\} \right) 
  O(b_{ij,k}\left( \{p\} \right)) \right) \numberthis
\end{align*}
where $\tilde{V}$, in addition to the one-loop \acp{ME}
interfered with the Born contribution,
contains the collinear counterterm. It is common to redefine 
$I^S$ to only cancel the divergences from 1-loop diagrams, and 
absorb the collinear counterterm and its cancellation into 
separate $K$ and $P$ terms. This simplifies the evaluation in 4 dimensions 
and separates the phase space dependence of the terms. $R_{n+1}$ is 
the real \ac{ME} squared.

When combining \iac{NLO} calculation with a particle-generating Markov
process such as a parton shower, care must be taken to avoid double-counting 
the first emission. Further details will be discussed in chapter 
\ref{chapter:mcatnlo}, but the main difficulty lies in exponentiating only 
the logarithmically enhanced soft and collinear parts of the subtraction 
terms. This requires defining suitable starting scale for the parton shower, while 
treating hard emissions as \ac{IR}-regular. Many matching methods have 
been proposed and implemented, including \Powheg \cite{Nason:2004rx,Frixione:2007vw},
\MCatNLO \cite{Frixione:2002ik} and {\sc KrkNlo} \cite{Jadach:2015mza}.

\subsection{Summary}

The Standard Model of particle physics has been hugely successful in 
describing a wide range of physics at collider experiments. Modern 
efforts focus on using perturbation theory and resummation 
techniques to increase the precision of \ac{SM} predictions. Monte 
Carlo event generators are one aspect of this effort, implementing
automated methods to make precise predictions for many
processes and observables. In this thesis, we will detail the 
implementation of new methods to describe \ac{QED} radiative 
corrections in the event generator \Sherpa \cite{Gleisberg:2008ta}.

% The \Sherpa modules which are relevant here 
% are two automated \ac{ME} generators, \Amegic and \Comix, 
% a phase space generator and integrator \textsc{Phasic}, a dipole parton shower 
% \CSS based on the \ac{CS} subtraction, an \MCatNLO implementation, 
% and a soft-photon resummation generator \Photons.

\section{Notation} \label{sec:intro:notation}

Throughout this thesis we will use notation consistent with the \Sherpa 
parton shower and \ac{NLO} matching publications, introduced in refs.
\cite{Schumann:2007mg} and \cite{Hoeche:2011fd}.

A dipole parton splitting process is denoted $\ijt \, (\kt) \to i j \, (k)$,
where $\ijt$ is the emitter parton and $\kt$ is the spectator parton.
A tilde denotes quantities before the splitting, as ordered 
in a parton shower ordering variable, and $i$ and $j$ are the 
splitting products.
The corresponding splitting function is denoted 
$S_{\mathrm{f}_\ijt (\kt)\to \mathrm{f}_i \mathrm{f}_j (k)}$,
where $\mathrm{f}_i$ is the flavour of parton $i$. Fermions are denoted 
$f$ in general, or $q = \{u,d,c,s,t,b\}$ for the quarks, $\ell = \{e,\mu,\tau\}$
for the charged leptons and $\nu = \{\nu_e,\nu_\mu,\nu_\tau\}$ for the 
neutrinos. Neutrinos are considered massless throughout this thesis. 
Scalars are denoted $s$ and photons are denoted $\gamma$. 
Momenta, masses and other quantities in 
splitting processes are labelled with the splitting index.

For \ac{NLO} calculations, the \ac{LO} contribution to the squared \ac{ME}
is denoted $B$, such that the total cross section $\sigma = \int 
\diff \Phi \, B$, where $\Phi$ is the 
Lorentz-invariant phase space. Similarly, the real correction is represented 
by $R$ and the virtual by $V$.

The strong coupling constant $\alpha_s$ is always considered to run with 
the scale, usually denoted by its argument. The \ac{EW} coupling constant 
$\alpha$ depends on the input scheme used in the calculation. Unless 
otherwise specified, for \ac{ME} calculations we use a fixed $\alpha$ 
in the $G_\mu$ scheme.

%% file: text/QEDps.tex
\chapter{QED parton showers}
\label{chapter:qedps}

\section{Introduction to parton showers} \label{sec:qedps:intro}

One of the standard tools, almost the defining feature, of any 
\ac{MC} event generator is \iac{QCD} parton shower. 
Early parton showers were developed in the 
1980s due to the need to describe the large number of final-state 
particles produced in the newer, higher-energy particle colliders.
The process of hadronisation depends on the quantum properties 
of the hadronising parton ensemble, necessitating a modelling of 
the whole partonic final state up to timescales comparable to 
hadronisation. A parton shower provides this by modelling the 
perturbative splitting of quarks and gluons over a wide range 
of energy scales using a Markov process. By solving 
variants of the \ac{DGLAP} equations for parton splittings of the 
initial and final states of a hard process, the evolution between 
the hard scattering partons and the long-distance states is bridged 
probabilistically. The first parton showers were built 
on the \ac{LL} collinear approximation of \ac{QCD} 
\acp{ME} for the decay of off-shell partons, ordered in 
the virtuality of the partons \cite{Fox:1979ag,Odorico:1980gg,
Kajantie:1980we,Gottschalk:1982yt}. Initially, there was no 
description of momentum conservation in the shower since
the process of hadronisation involved a smooth momentum 
convolution into the momenta of hadrons. 
In addition to collinear logarithms, soft logarithms were 
first resummed in a parton shower in the work of Marchesini and 
Webber \cite{Marchesini:1983bm,Webber:1983if}.
This work also introduced angular ordering, ensuring the 
inclusion of colour coherence effects. This was developed into 
a full-fledged parton shower in the event generator \Herwig
\cite{Marchesini:1991ch}. For the case of the initial state, an 
important development in \ac{QCD} parton showers was the advent 
of backward evolution \cite{Sjostrand:1985xi}. This led to the 
interleaved initial- and final-state \ac{QCD} shower 
in many modern showers, pioneered by \Pythia \cite{Sjostrand:2006za}.
While early work on parton showers was intended to replace the 
use of higher-order \acp{ME}, it became clear 
that even at high-energy 
colliders, the soft- and collinearly-approximated parton shower 
was not sufficient to describe the hard wide-angle radiation. Thus 
a method to merge \ac{LO} multileg \acp{ME} with 
the parton shower was developed \cite{Catani:2001cc,Lonnblad:2001iq},
which became the cornerstone of the event generator \Sherpa 
\cite{Gleisberg:2003xi}.

Despite the flourishing development of \ac{QCD} parton showers 
over the last 40 years, the need for precise description of 
\ac{QED} radiation has been much less pronounced and so \ac{QED} showers 
are not ubiquitous in event generators. 
\Photos \cite{Barberio:1990ms, Barberio:1993qi,Golonka:2005pn,Davidson:2010ew}
was the first process-independent tool developed for modelling 
\ac{QED} final-state radiation. \Herwig has \iac{QED}
parton shower which includes all spin correlation effects
\cite{Masouminia:2021kne}. \Pythia also includes 
a \ac{QED} shower which treats each $f\bar{f}$ pair as an 
independent dipole, thus approximating the correct 
multipole soft limit for photon emissions.
The parton shower in \Sherpa has also been previously updated to 
include \ac{QED} splittings in order to model hard photon production
\cite{Hoeche:2009xc}. However, until this work, only positive 
dipoles have been included, leading to an overestimate of the photon 
radiation produced in the soft limit. 
In addition, photon splittings into fermions were previously not 
modelled correctly in the dipole picture. This has little 
effect on observables where a hard photon is selected, 
but can have non-negligible effects on reconstructed 
resonances, especially when dressed leptons are used 
to reconstruct them. 
In this chapter, we will detail the modifications
made to the \ac{QED} parton shower in \Sherpa.

We begin by outlining the ingredients behind a parton shower 
algorithm. We follow the description given in ref. 
\cite{Hoche:2014rga}, which, like the \ac{QED} parton shower 
we will outline here, is built on the \ac{CS} dipole 
subtraction.

First, we motivate an initial-state parton shower as the 
ability to model initial-state radiation, not inclusively 
via \acp{PDF}, but exclusively such that actual particles 
are produced. 
Since any parton shower must have \iac{IR} cutoff, 
however, the \ac{DGLAP}
equations must be modified for initial-state radiation.
% We require four-momentum conservation 
% after each splitting, so we use the Sudakov decomposition 
% \cite{Sudakov:1954sw}
% \begin{equation} \label{eq:qedps:SudakovDecomposition}
%   p_{i}^\mu = z p^\mu_\ijt + \frac{-k_\perp^2}{z} \frac{n_\mu}{2np_\ijt}
%   + k_\perp^\mu, \quad p_j^\mu = (1-z)p^\mu_\ijt + \frac{-k_\perp^2}{1-z}
%   \frac{n^\mu}{2np_\ijt} - k_\perp^\mu,
% \end{equation}
% where $n^\mu$ is a lightlike reference vector, usually identified with 
% the spectator momentum $p_\kt$. $k_\perp$ is the spacelike transverse 
% momentum, $p_\ijt k_\perp = n k_\perp = 0$.
% $z$ is the light-cone momentum fraction
% which, if identified with the energy fraction in the \ac{DGLAP} equations 
% (eq. \eqref{eq:intro:dglap}), suggests that we can model initial-state 
% radiation using the \ac{DGLAP} equations with the unregularised 
% Altarelli-Parisi functions.
The parton shower \ac{IR} cutoff means that 
the plus-prescription and $\delta$-function part are no longer needed. 
This would amount to including real-emission corrections to the 
cross section in the \ac{LL} approximation. 
This is the approach taken in 
many \ac{QED} parton showers for $e^+ e^-$ colliders, for example in the 
event generator \BabaYaga \cite{Balossini:2006wc,Balossini:2008xr,
CarloniCalame:2000pz,CarloniCalame:2001ny,CarloniCalame:2003yt}.
For the case of parton showers at hadron colliders, however, the 
approach taken is to respect unitarity of the \ac{LO} cross section.
This results in a probabilistic interpretation of the splitting 
functions, at least in the leading-colour limit of \iac{QCD} parton shower.
To achieve unitarity, an extra term is added to the \ac{DGLAP} equations 
\changed{(eq.\ \eqref{eq:intro:dglap})} to 
represent the virtual corrections in the restricted phase space, 
\begin{align*} \label{eq:qedps:modiDGLAP}
  \frac{\diff f_a^A(x,t)}{\diff \log{t}}
   = &\sum_{b \in A} \int_x^{z_\text{max}} \frac{\diff z}{z} 
   \frac{\alpha}{2\pi} P_{b\to a}(z) f_b^A(x/z,t) \\
   &- f_a^A(x,t) \sum_{b \in A} \int_{z_\text{min}}^{z_\text{max}}
   \diff z \frac{\alpha}{2\pi} \frac{1}{2} P_{a\to b}(z), \numberthis
\end{align*}
\changed{where the second term represents the probability for removing 
a particle of flavour $a$ through its splitting to other flavours.
The factor $1/2$ is due to the symmetry in the splitting variable $z$ 
and its complement $1-z$.
In this equation,} we have also written $t$ instead of the factorisation scale 
$\mu_F^2$. This is the parton shower evolution variable, 
which will be defined later.

A final-state parton shower, or more generally one which 
can model initial-final interference as well as radiation 
from the initial and final states, does not need to replicate 
\ac{PDF} evolution, but is built from the same collinear 
factorisation picture which will be detailed in the next section. 

In chapter \ref{chapter:intro}, we introduced a mapping from a Born 
particle configuration to that of a real-emission event, 
eq.\ \eqref{eq:intro:PSmappingInverse}. This mapping is not uniquely 
defined \changed{for each splitter-spectator pair},
unlike its inverse. It depends not only on the three 
kinematical variables needed to describe the momentum of the newly 
created particle, but also on its flavour (although the latter is often 
fixed by the Born particle configuration). The parton shower uses 
the \ac{MC} methods described in this section to select these variables
while respecting the underlying evolution equations.

Key to the parton shower algorithm is the `hit and miss' provided by the 
veto algorithm described below. The process begins at some scale $t'$ where
all the possible splitter partons are iterated over, then all the possible 
spectators. An overestimate for the splitting is calculated and a new scale 
$t$ is computed. The new scale is then accepted with a probability which 
corrects for the overestimate. If the splitting is accepted, the new 
particle is created and flavours, colours and kinematics are updated. 
The whole process is repeated at the new lower scale until some \ac{IR} 
cutoff $t_c$, needed to regulate the divergence of the splitting functions.
This cutoff is usually taken around the hadronisation scale for \ac{QCD} 
showers, since quarks and gluons would no longer be meaningful 
degrees of freedom below this scale. For \ac{QED} evolution, a natural 
cutoff for photon splittings is provided by the mass of the produced 
fermions, and for photon emissions the cutoff should be chosen 
low enough to not affect the observables of interest.

In this chapter, we describe the construction of \iac{QED} parton 
shower, starting from the \ac{CS} dipole subtraction scheme 
for \ac{NLO} electroweak corrections. In the construction we take 
inspiration from \ac{QCD} parton showers based on \ac{CS} subtraction.
In section \ref{sec:qedps:veto} we show that the veto algorithm 
reproduces the desired integral of the probability distribution.
We present results from the updated 
\ac{QED} shower in \Sherpa in section \ref{sec:qedps:FSresults}.
Before concluding and providing an outlook in section 
\ref{sec:qedps:conclusion}, we detail the additional considerations 
required in the case of initial-state radiation at an $e^+ e^-$ 
collider in section \ref{sec:qedps:ISmethods}.

%%%%%%%%%%%%%%%%%%%%%%%%%%%%%%%%%%%%%%%%%%%%%%%%%%%%%%%%%%%%%%%%%%

\section{Construction of a QED parton shower} \label{sec:qedps:construction}

As motivated in section \ref{sec:intro:ps}, parton showers based 
on \ac{CS} dipole subtraction have been highly successful 
phenomenological tools for \ac{QCD}. In this section we 
describe the construction of a \ac{QED} parton shower from the 
\ac{CS} dipole subtraction terms.

Eq.\ \eqref{eq:intro:nlo} for the expectation value of 
\iac{IR-safe} observable at \ac{NLO} in the \ac{CS} dipole 
subtraction formalism is repeated here for convenience,
\begin{align*} \label{eq:qedps:nlo}
  \langle O \rangle^{\text{NLO}} =& \sum_{\{\tilde{f}\}} \int \diff \Phi_n\left( \{\tilde{f}\} \right)
  \left( B\left( \changed{\{\tilde{a}\}} \right) + \tilde{V}\left( \changed{\{\tilde{a}\}} \right) + 
  \sum_{\ijt,\kt} I^S_{\ijt,\kt}\left( \changed{\{\tilde{a}\}} \right) \right)
  O\left( \{\tilde{p}\} \right) \\
  &+ \sum_{\{f\}} \int \diff \Phi_{n+1}\left( \{p\} \right) 
  \left( R_{n+1}\left( \{a\} \right) O\left( \{p\} \right) - \sum_{ij,k} D^S_{ij,k}\left( \{a\} \right) 
  O(b_{ij,k}\left( \{p\} \right)) \right).
\end{align*}

To construct a parton shower from the subtraction terms \changed{$D^S_{ij,k}$}, 
we start from the factorisation of the $(n+1)$-particle phase space
\changed{in the \ac{IR} limit},
\begin{equation}
  \diff \Phi_{n+1}\, D^S_{ij,k}(\{a\})
  \changed{\xrightarrow{\text{soft or collinear}}} \diff \Phi_n \left[ \sum_{\ijt,\kt \in \{f\}}
  \sum_{f_j} \diff \Phi_1^{ij,k} D^S_{ij,k}
  (r_{\ijt,\kt} \changed{\{\tilde{a}\}}) \right],
\end{equation}
\changed{where $\{a\}, \{\tilde{a}\}$ are particle configurations, and $r_{\ijt,\kt}$ is 
the momentum mapping, eq.\ \eqref{eq:intro:PSmappingInverse}.}
In a parton shower, our goal is to generate the one-particle 
phase space $\diff \Phi_1^{ij,k}$. We parametrise it in three 
variables $t$, $z$ and the azimuthal angle $\phi$,
\begin{equation}
  \diff \Phi_1^{ij,k} = \frac{1}{16\pi^2} \diff t 
  \diff z \frac{\diff \phi}{2\pi} J(t,z,\phi),
\end{equation}
where $J(t,z,\phi)$ is the Jacobian. Now we also use 
factorisation of the soft and collinear limits of 
particle radiation, and factorise out the Born  
differential cross section $\diff \sigma_B$ from the 
radiative part, 
\begin{equation}
  \diff \sigma
  = \diff \sigma_B \sum_{\ijt,\kt \in \{f\}} \sum_{f_i}
  \frac{\mathcal{L}(r_{\ijt,\kt}(\{a\}))}{\mathcal{L}(\changed{\{\tilde{a}\}})}
  \,\frac{D^S_{ij,k}(r_{\ijt,\kt} (\{a\}))}
  {B (\changed{\{\tilde{a}\}})},
\end{equation}
where we have also included the particle luminosity factors
$\mathcal{L}$, since this is a differential cross section. 
\changed{The particle luminosity} is given by 
\begin{equation}
  \mathcal{L}(\{a\},\mu_F^2) = x_1 f_{f_1}^A(x_1,\mu_F^2) \,x_2 f_{f_2}^B(x_2,\mu_F^2),
\end{equation}
where $x_i$ is the momentum fraction of $f_i$ in $A$ or $B$, 
as appropriate. For final-state radiation, since the 
initial-state flavours and momenta do not change, the 
ratio of particle luminosities is 1.
This is the essence of our parton shower factorisation. 
We will give the explicit forms of the splitting functions 
$D^S_{ij,k}$ below. However, as mentioned above, applying 
this without modification would violate unitarity, neither 
resulting in the correct \ac{LO} cross section nor the 
correct \ac{NLO} one. 
Near the \ac{IR} limit, the logarithmic terms 
are dominant, and these exactly cancel between the virtual 
corrections and real-emission 
contributions due to the \ac{KLN} theorem.
We assume, in our \ac{LO}-accurate parton shower, that 
virtual corrections exactly cancel the finite (non-logarithmic) 
parts of the real corrections. This
leads to the addition of an unresolved-emission term 
\begin{equation} \label{eq:qedps:sudakov}
  \Omega^{\ijt,\kt}(t,t') = \exp{\left(-\frac{1}{16\pi^2}
  \sum_{f_i} \int_t^{t'} \diff t \int \diff z \int\frac{\diff \phi}{2\pi}\,
  \frac{1}{2}\, 
  \frac{\mathcal{L}(r_{\ijt,\kt}(\{a\}))}{\mathcal{L}(\changed{\{\tilde{a}\}})}
  \,\frac{D^S_{ij,k}(r_{\ijt,\kt} (\{a\}))}{B(\changed{\{\tilde{a}\}})}  
  \right)}.
\end{equation}
The product $\Omega(t,t')=\prod_{\{\ijt,\kt\}}\Omega^{\ijt,\kt}(t,t')$ 
gives the total no-emission probability of the parton shower,
which is clearly seen by extending the sum over emitter-spectator
pairs into the exponential. 
Since $\Omega(t,t')$ is a Poisson distribution in $\log{t}$, 
the probability of a 
branching at a scale $t$, starting from scale $t'$, is given by 
\begin{equation} \label{eq:qedps:probability}
  \mathcal{P}(t,t') = \frac{\diff \Omega(t,t')}{\diff \log{t}}.
\end{equation}
In a final-state parton shower, the no-emission probability 
$\Omega=\Delta$, the Sudakov form factor, which is defined to 
solve the \ac{DGLAP} equations. However, where initial-state 
particles are involved, $\Omega$ also contains a ratio of 
particle luminosities. A parton shower uses our knowledge of 
the form of $\mathcal{P}(t,t')$ using \ac{QCD} or \ac{QED} 
splitting functions to solve eq.\ \eqref{eq:qedps:probability}
for $t$, the splitting scale.

The forms of the emission and no-emission probabilities will 
inform our construction of the parton shower algorithm in 
section \ref{sec:qedps:veto} below. First, however, we 
will give specific expressions for the parton shower 
splitting functions.
\changed{The subtraction terms $D^S_{ij,k}$ have different forms 
depending on whether $\ijt$ and $\kt$ are in the initial or 
final state.}

\subsection{Splitting functions} \label{sec:qedps:SFs}

\paragraph*{Final-final} For a final-state splitter $\ijt$ and 
a final-state spectator $\kt$, 
\begin{equation} \label{eq:qedps:dipoleFF}
  D^S_{ij,k} = -\frac{1}{(p_i + p_j)^2-m_{\ijt}^2} 
  \,\mathbf{Q}_{\ijt\kt}^2 \,\langle \dots, \ijt, \dots, \kt, \dots 
  | \mathbf{V}_{ij,k} | \dots, \ijt, \dots, \kt, \dots \rangle,
  % \Theta (\alpha_{FF}-y_{ij,k})
\end{equation}
\changed{where $|\dots,\ijt,\dots,\kt,\dots\rangle$ is a vector in 
colour and helicity space, and thus the product indicates a 
squared matrix element, summed over final-state colours and spins.}

The charge correlator is defined as 
\cite{Yennie:1961ad,Dittmaier:1999mb,Dittmaier:2008md,Kallweit:2017khh,Schonherr:2017qcj}
\begin{equation} \label{eq:qedps:charge_correlator}
    \mathbf{Q}_{\ijt\kt}^2 = 
    \begin{cases}
        \frac{Q_\ijt \theta_\ijt Q_\kt \theta_\kt}{{Q_\ijt}^2}, & \text{for } \ijt \neq \gamma \\
        \kappa_{\ijt\kt}, & \text{for } \ijt = \gamma,
    \end{cases}
\end{equation}
where the $Q$ are the charges of the particles and $\theta = 1\,(-1)$ if the 
particle is in the final (initial) state.
Clearly $Q_{\kt} = Q_k$, since the spectator does not change 
flavour, and $\theta_{\kt} = \theta_k$. 
The $\kappa_{\ijt\kt}$ are parameters that can be chosen 
to implement a spectator-assigning scheme but are subject to the 
constraint 
\begin{equation}
    \sum_{\kt \neq \ijt} \kappa_{\ijt\kt} = -1 \quad \forall \ijt=\gamma,
\end{equation}
in order that the sum over all possible spectators adds up to the 
correct collinear limit. There are various schemes for assigning 
spectators to photon splittings, 
discussed further in chapter \ref{chapter:yfs} and in ref.
\cite{Schonherr:2018jva}.

The parton shower splitting functions are the spin-averaged 
forms of $\mathbf{V}_{ij,k}$, in 4 dimensions, divided by 
the Born contribution,
\begin{equation}
  S_{\ijt (\kt)\to i j (k)} = \frac{\langle s | \mathbf{V}_{ij,k} | s \rangle|_{\epsilon=0}}
  {B},
\end{equation}
in terms of the splitting variable $y_{ij,k}$ and light-cone momentum 
fraction $z_i$, which are defined as
\begin{equation} \label{eq:qedps:zy}
  y_{ij,k} = \frac{p_i p_j}{p_i p_j + p_i p_k + p_j p_k}
  \qquad\text{and}\qquad
  z_i = \frac{p_i p_k}{p_i p_k + p_j p_k}
  \;.
\end{equation}
We will usually suppress the splitting indices and write 
$y$ to mean $y_{ij,k}$ and $z$ instead of $z_i$.
From the definition of $z_i$ it is clear 
that $z_j = 1-z_i$, and so under $i\leftrightarrow j$, 
each splitting function is given by the exchange 
$z \leftrightarrow 1-z$.

Then the splitting functions are 
\begin{equation} \label{eq:qedps:FF:SFs}
  \begin{split}
    S_{f_\ijt(\kt)\to f_i\gamma_j(k)}
    \,=\;
    S_{\bar{f}_\ijt(\kt)\to \bar{f}_i\gamma_j(k)}
    \,=&\;
      8 \pi \,\alpha\,
      \left[\frac{2}{1-z+zy}-\frac{\tilde{v}_{\ijt,\kt}}{v_{ij,k}}\left(1+z+\frac{m_i^2}{p_ip_j}\right) \right] \\
    S_{\gamma_\ijt(\kt)\to f_i\bar{f}_j(k)}
    \,=&\;
      8\pi \, \alpha\,
      \left[
        1-2z(1-z)-z_+ z_- \vphantom{\frac{m_i^2}{p_ip_j}}
      \right].
  \end{split}
\end{equation}
The relative velocities $v,\tilde{v}$ are introduced to facilitate the 
analytic integration for the case of massive partons. $z_\pm$ 
are the phase space boundaries. Explicit expressions for these are 
given in the \Sherpa parton shower literature, e.g. ref. \cite{Schumann:2007mg}.
In chapter \ref{chapter:yfs} we use different solutions for 
the relative velocities and phase space boundaries, which are 
detailed there.

%%%%%%%%%%%%%%%%%%%%%%%%%%%%%%%%%

\paragraph*{Final-initial} For a 
final-state splitter $\ijt$ and a massless
initial-state spectator $\at$,\footnote{
The splitting functions for the case of a massive spectator are
given in sec. \ref{sec:yfs:fidip}.}
\begin{equation} \label{eq:qedps:dipoleFI}
  D^S_{aj,k} = -\frac{1}{(p_i + p_j)^2-m_{\ijt}^2} \,\frac{1}{y_{ij,a}}
  \,\mathbf{Q}_{\ijt\at}^2 \,\langle \dots, \ijt, \dots, \at, \dots 
  | \mathbf{V}_{ij,a} | \dots, \ijt, \dots, \at, \dots \rangle, 
\end{equation}
where $y_{ij,a}$ is the splitting variable defined below in eq.\ 
\eqref{eq:qedps:FIxz}.

The splitting variables are 
\begin{equation} \label{eq:qedps:FIxz}
  y_{ij,a} = 1 - \frac{p_i p_j - \frac{1}{2} \left( m_\ijt^2 -m_i^2 -m_j^2 \right)}
  {p_i p_a + p_j p_a} 
  \qquad\text{and}\qquad
  z_i = \frac{p_i p_a}{p_i p_a + p_j p_a}
  \;.
\end{equation}
The initial-state splitting variable is often called $x_{ij,a}$, 
but we will retain the label $y_{ij,a}=y$, to distinguish it from 
the momentum fraction $x$ of the beam which is carried by a 
given parton.
% but it is not to be confused with the momentum 
% fraction of the beam particle which is carried by a given parton.

Then, analogously to the final-final case,
\begin{equation} \label{eq:qedps:FI:SFs}
  \begin{split}
    S_{f_\ijt(\at)\to f_i\gamma_j(a)}
    \,=\;
    S_{\bar{f}_\ijt(\at)\to \bar{f}_i\gamma_j(a)}
    \,=&\;
      8 \pi \,\alpha\,
      \left[\frac{2}{2-z-y} -(1+z) - \frac{m_i^2}{p_i p_j}\right] \\
    S_{\gamma_\ijt(\at)\to f_i\bar{f}_j(a)}
    \,=&\;
      8\pi \, \alpha\,
      \left[
        1-2(z_+ + z)(z_- - z)
      \right].
  \end{split}
\end{equation}

%%%%%%%%%%%%%%%%%%%%%%%%%%%%%%
\paragraph*{Initial-final} For a massless 
initial-state splitter $\ajt$ and a final-state spectator $\kt$, 
\begin{equation} \label{eq:qedps:dipoleIF}
  D^S_{aj,k} = -\frac{1}{2 p_a p_j} \,\frac{1}{z_{aj,k}}\,
  \mathbf{Q}_{\ajt\kt}^2 \,\langle \dots, \ajt, \dots, \kt, \dots 
  | \mathbf{V}_{aj,k} | \dots, \ajt, \dots, \kt, \dots \rangle, 
\end{equation}
where the splitting variables are defined by 
\begin{equation} \label{eq:qedps:IFxz}
  z_{aj,k} = 1 - \frac{p_j p_k}{p_j p_a + p_k p_a}
  \qquad\text{and}\qquad
  u_i = \frac{p_i p_a}{p_i p_a + p_k p_a}
  \;.
\end{equation}

Then the splitting functions are given by 
\begin{equation} \label{eq:qedps:IF:SFs}
  \begin{split}
    S_{f_\ajt(\kt)\to f_a\gamma_j(k)}
    \,=\;
    S_{\bar{f}_\ajt(\kt)\to \bar{f}_a\gamma_j(k)}
    \,=&\;
      8 \pi \,\alpha\,
      \left[\frac{2}{2-u-z} -(1+u)\right] \\
    S_{\gamma_\ajt(\kt)\to f_a\bar{f}_j(k)}
    \,=&\;
      8\pi \, \alpha\,
      \left[
        1-2z(1-z)
      \right].
  \end{split}
\end{equation}

%%%%%%%%%%%%%%%%%%%%%%%%%%%%%%%%%

\paragraph*{Initial-initial} For a massless initial-state splitter 
$\ajt$ and a massless spectator $\bt$, 
\begin{equation} \label{eq:qedps:dipoleII}
  D^S_{aj,b} = -\frac{1}{2p_a p_j} \frac{1}{z_{aj,b}}
  \mathbf{Q}_{\ajt\bt}^2 \langle \dots, \ajt, \dots, \bt, \dots 
  | \mathbf{V}_{aj,b} | \dots, \ajt, \dots, \bt, \dots \rangle, 
  % \Theta (\alpha_{FI}-y_{ij,k})
\end{equation}
where the splitting variable $z_{aj,b}$ is now given by 
\begin{equation} \label{eq:qedps:IIx}
  z_{aj,b} = 1 - \frac{p_j p_a + p_j p_b}{p_a p_b}
  \;.
\end{equation}

Then, as before, the splitting functions are 
\begin{equation} \label{eq:qedps:II:SFs}
  \begin{split}
    S_{f_\ijt(\at)\to f_i\gamma_j(a)}
    \,=\;
    S_{\bar{f}_\ijt(\at)\to \bar{f}_i\gamma_j(a)}
    \,=&\;
      8 \pi \,\alpha\,
      \left[\frac{2}{2-z} -(1+z)\right] \\
    S_{\gamma_\ijt(\at)\to f_i\bar{f}_j(a)}
    \,=&\;
      8\pi \, \alpha\,
      \left[
        1-2z(1-z)
      \right].
  \end{split}
\end{equation}
Only one splitting variable is needed to specify the splitting 
functions in this case, but for the full kinematics a second 
variable $v_i = p_i p_a/p_a p_b$ must be defined \cite{Schumann:2007mg}. 

%%%%%%%%%%%%%%%%%%%%%%%%%%%%%%%
\newpage
These splitting functions, for the four dipole cases discussed,
reproduce both the soft eikonal limit 
and, when massless fermions are involved,
the collinear Altarelli-Parisi limit \cite{Catani:1996vz}.

%%%%%%%%%%%%%%%%%%%%%%%%%%%%%%

\subsection{The dipole picture and QED}

Most \ac{QCD} parton showers are built on the large-$N_C$ limit,
or the leading-colour approximation. This is justified by the 
fact that the next-to-leading term in the expansion in $N_C$ goes 
as $1/N_C^2$, which for $N_C=3$ is approximately a 10\% correction 
on what is already formally a 10\% effect, since $\alpha_s \approx 0.1$.
Many newer showers go beyond this limit and include 
subleading colour effects \cite{Platzer:2012np,Hamilton:2020rcu,
Forshaw:2021mtj,Scyboz:2022lvm}. However, in \ac{QED}, there 
is no leading-colour approximation since $N_C=1$. This reflects 
the fact that photon emission is properly described by coherent 
radiation from a whole charged multipole, rather than individual 
dipole terms. A parton shower can still be built using a dipole 
picture, however, but all charge dipoles must be included. This 
necessarily includes either same-charge dipoles or opposite-charge 
initial-final interference, which in both cases will give a 
negative contribution as seen from the charge correlator defined 
in eq.\ \eqref{eq:qedps:charge_correlator}.

% Maybe include something about this?
% Now we will show how subsequent emissions using the splitting 
% functions defined above,
% ordered in a suitable ordering parameter defined later, 
% resum the leading soft and collinear logarithms by exponentiating 
% them into a Sudakov factor. 

Now we will show how a `hit and miss' algorithm, ordered 
in a suitable ordering parameter defined later, reproduces 
the correct distribution of radiation according to the 
splitting functions defined above. Hence, by iterating 
these splittings, we resum the leading soft and collinear 
logarithms by exponentiating them into a Sudakov factor.
The problem of negative splitting 
functions will be solved by an analytic weighting method.

\subsection{The veto algorithm} \label{sec:qedps:veto}

The veto algorithm \cite{Bengtsson:1986et,Seymour:1994df} 
is a method for generating 
a parton splitting at a scale $t$ (the parton shower evolution variable) 
according to a 
probability distribution $f(t)$ 
when the integral of the distribution, $F(t)$ is unknown. 
To simplify the notation, the presence of only one splitting function 
is assumed here but the extension to multiple functions, including 
flavour changing splittings, is straightforward. This description is 
based on many excellent arguments in the literature, for example 
refs. \cite{Sjostrand:2006za,Platzer:2011dq,Lonnblad:2012hz}.

The differential probability for generating an emission at scale
$t$, starting from an upper scale $t'$, is given by eq.\ 
\eqref{eq:qedps:probability}, i.e.
\begin{equation} \label{eq:qedps:probabilityveto}
  \mathcal{P}(t,t') = f(t) \exp{\left(-\int_t^{t'} \diff \Tilde{t} \, f(\Tilde{t}) \right)},
\end{equation}
and the new scale is generated according to the distribution 
of $f(t)$:
\begin{equation}
  t = F^{-1}(F(t') + \log{r}),
\end{equation}
where $r$ is a random number between zero and one, $F(t)$ is the 
indefinite integral of $f(t)$, and $F^{-1}(x)$ is 
the inverse of $F(t)$. 

If $F(t)$ is unknown, an overestimate 
$g(t) \geq f(t)$ is defined, where the integral $G(t)$ of the 
overestimate is known. 
Then we generate the new scale instead using the 
integral of the overestimate $G(t)$, 
\begin{equation} \label{eq:qedps:generatet}
  t = G^{-1}(G(t') + \log{r}),
\end{equation}
and the new scale (and splitting) is accepted with probability 
$f(t)/g(t)$. Then the probability of a splitting being accepted, with 
$n$ intermediate rejections, is 
\begin{align*} \label{eq:qedps:veto2}
  \mathcal{P}_n(t,t') = \frac{f(t)}{g(t)} \, &g(t) \exp{\left( -\int_t^{t_1} \diff \Tilde{t} 
    \, g(\Tilde{t}) \right)} \\
  &\times \prod_{i=1}^n \left[ \int_{t_{i-1}}^{t_{n+1}} \diff t_i 
    \left( 1-\frac{f(t_i)}{g(t_i)} \right) g(t_i) \, \exp{\left( -\int_{t_i}^{t_{i+1}} 
    \diff \Tilde{t} \, g(\Tilde{t}) \right)} \right] , \numberthis
\end{align*}
where $t_0=t$ and $t_{n+1}=t'$. Note that $n=0$ is allowed here:
in that case $t_1 = t'$ and the second line of eq.\ \eqref{eq:qedps:veto2}
gives a factor of unity.

To reproduce eq.\ \eqref{eq:qedps:probabilityveto}, we use \changed{that}
\begin{equation}
  \mathcal{P}(t,t') = \sum_{n=0}^\infty \mathcal{P}_n(t,t'),
\end{equation}
i.e. we cannot distinguish the number of rejections which occur 
before a splitting. 
Note that the product in eq.\ \eqref{eq:qedps:veto2} can be simplified 
using properties of exponentials and integrals,
\begin{align*}
    \prod_{i=1}^n \exp{\left( - \int_{t_i}^{t_{i+1}} \diff \Tilde{t} \, g(\Tilde{t}) 
    \right)} &= \exp{\left( - \sum_{i=1}^n \int_{t_i}^{t_{i+1}} \diff \Tilde{t} \, 
    g(\Tilde{t}) \right)} \\
    &= \exp{\left(-\int_{t_1}^{t_{n+1}} \diff \Tilde{t} \, g(\Tilde{t}) \right) }. \numberthis
\end{align*}
When summing over all $n$, using symmetry under $t_i \leftrightarrow t_j$, 
we also recognise the infinite series 
\begin{align*} \label{eq:qedps:veto5}
  \sum_{n=0}^\infty \prod_{i=1}^n \int_{t_{i-1}}^{t_{n+1}} \diff t_i \left(1-\frac{f(t_i)}{g(t_i)} 
  \right) g(t_i) &= \sum_{n=0}^\infty \frac{1}{n!} \left[ \int_{t_0}^{t_{n+1}} \diff \Tilde{t} 
  \left(1-\frac{f(\Tilde{t})}{g(\Tilde{t})} \right) g(\Tilde{t}) \right]^n \\
  & = \exp{\left( \int_{t}^{t'} \diff \Tilde{t} \left(1-\frac{f(\Tilde{t})}
  {g(\Tilde{t})} \right) \right)}. \numberthis
\end{align*}
Therefore, eq.\ \eqref{eq:qedps:probabilityveto} follows immediately
upon summation of eq.\ \eqref{eq:qedps:veto2} over all numbers of 
rejections $n$.

In equation \eqref{eq:qedps:probabilityveto}, the term
\begin{equation}
    \Delta(t,t') = \exp{\left( - \int_t^{t'} \diff \Tilde{t} f(\Tilde{t}) \right)},
\end{equation}
which resums an infinite series, is defined as the Sudakov form factor and 
interpreted as the probability 
for no branching to occur between the two scales $t'$ and $t$. In the case 
of \iac{NLO}-matched parton shower, it will acquire additional meaning as 
the exponential of the \ac{IR} subtraction terms at \ac{NLO}.

When we apply a lower cutoff $t_c$ to the algorithm, we will encounter 
events which do not generate a splitting with $t>t_c$. In those cases, 
we no longer generate eq.\ \eqref{eq:qedps:probabilityveto}, but only the 
exponential term, as expected. These events are rare in a \ac{QCD} parton 
shower, but much more common in a \ac{QED} parton shower due to the 
smallness of $\alpha$.

We have shown that the veto algorithm correctly produces 
the probability distribution for one or zero emissions
to occur, after any number of rejected splittings. To 
extend the argument to a complete parton shower, therefore,
we must consider the fact that all evolution is strictly 
ordered in the variable $t$. Since $g(t)\geq f(t)$, and 
therefore the no-emission probability according to 
$g(t)$ is smaller than that from $f(t)$, once we have 
generated no emissions in a $t$ region, it is not 
necessary to revisit this region. In this way, nested 
integrals from subsequent emissions decouple, and the 
full parton shower evolution can be generated by 
repeated application of the veto algorithm in descending 
$t$.

\subsubsection*{The weighted veto algorithm} \label{sec:qedps:weightedveto}

When describing \ac{QED} radiation in a dipole picture, we must be 
mindful that unlike in \ac{QCD}, there is no leading-colour limit.
Thus, all dipoles must be included, some of which have negative 
splitting functions due to the product of splitter and spectator 
charges. This would destroy the probabilistic interpretation of the 
splitting functions. To restore this, we can introduce an additional 
overestimate, $h(t)$. We sample using the 
positive-definite $h(t)$, veto with probability $0\leq f(t)/g(t) \leq 1$, 
and then apply the additional weight $g(t)/h(t)$ analytically.

Using this method, the modified probability 
of a splitting being accepted after $n$ rejections is 
\begin{align*}
    \mathcal{P}_n(t,t') = \frac{f(t)}{g(t)} \, &h(t) \exp{\left( -\int_t^{t_1} 
      \diff \Tilde{t} \, h(\Tilde{t}) \right)} \\
    &\times \prod_{i=1}^n \left[ \int_{t_{i-1}}^{t_{n+1}} \diff t_i 
    \left( 1-\frac{f(t_i)}{g(t_i)} \right) h(t_i) \exp{\left( -\int_{t_i}^{t_{i+1}} 
    \diff \Tilde{t} \, h(\Tilde{t}) \right)} \right], \numberthis
\end{align*}
and the correct distribution $\mathcal{P}(t,t')$ is reproduced after 
applying the weight 
\begin{equation} \label{eq:qedps:weight}
    w_n(t,t_1,...,t_n) = \frac{g(t)}{h(t)} \prod_{i=1}^n \frac{g(t_i)}{h(t_i)} 
    \frac{h(t_i) - f(t_i)}{g(t_i) - f(t_i)},
\end{equation}
and summing over $n$ from 0 to infinity.
To see that this is the correct weight, we follow the same procedure as in 
equations \ref{eq:qedps:veto2}-\ref{eq:qedps:veto5} to find
that the weighted 
probability of acceptance after $n$ rejections becomes 
\begin{align*}
    w_n \mathcal{P}_n(t,t') = f(t) \prod_{i=1}^n &\left[ \int_{t_{i-1}}^{t_{n+1}}
    \diff t_i \left(1-\frac{f(t_i)}{g(t_i)} \right) h(t_i) \frac{g(t_i)}{h(t_i)} \, 
    \frac{h(t_i) - f(t_i)}{g(t_i) - f(t_i)} \right] \\
    &\times \exp{\left( -\int_{t_0}^{t_{n+1}} \diff \Tilde{t} \, h(\Tilde{t}) \right)}. \numberthis
\end{align*}
Simplifying and summing over all $n$, 
\begin{align*}
    w \mathcal{P}(t,t') &= \sum_{n=0}^\infty \left \{ f(t) \frac{1}{n!} 
    \left[ \int_{t_0}^{t_{n+1}} \diff \Tilde{t} (h(\Tilde{t})-f(\Tilde{t})) \right]^n 
    \exp{\left(-\int_{t_0}^{t_{n+1}} \diff \Tilde{t} \, h(\Tilde{t}) \right)} \right \} \numberthis\\
    &= f(t) \exp{\left( -\int_t^{t'} \diff \Tilde{t} \, h(\Tilde{t}) \right)} 
    \exp{\left( \int_t^{t'} \diff \Tilde{t} \, (h(\Tilde{t}) - f(\Tilde{t})) \right)} \\
    &= f(t) \exp{\left( -\int_t^{t'} \diff \Tilde{t} \, f(\Tilde{t}) \right)},
\end{align*}
which reproduces eq.\ \eqref{eq:qedps:probabilityveto}.
This works for all choices of $h(t)$,
\changed{and the primary reason for its use in this thesis is 
to correctly model photon emission from same-sign dipoles.
In this case, when there are negative \ac{QED} splitting 
functions, we let $h(t)=-g(t)$.
The weighted veto algorithm can also be used to probe 
statistically unlikely regions with better precision.}
For example, with a weight $h(t)=10 g(t)$, the emission 
of a photon from a quark could be considered roughly as often 
as the emission of a gluon. This is extremely useful in 
an interleaved shower when studying observables which depend on 
\ac{QED} radiation.

%%%%%%%%%%%%%%%%%%%%%%%%%%%%%

\subsection{The generating functional of the parton shower}

We will now put all the ingredients together and express 
the full parton shower evolution as a generating functional.
This will allow us to write down the effect of the parton 
shower on a general \ac{IR-safe} observable. We will also 
refer to this formalism when we discuss matching in chapter 
\ref{chapter:mcatnlo}.

The generating functional is recursively defined in terms 
of the splitting functions $D^{ij,k}$ as 
\begin{align*} \label{eq:qedps:generatingFunctional}
  \mathcal{F}_n(\Phi_n,O) =\, &
    \underbrace{\Delta(\mu_Q^2,t_c) \, O(\Phi_n)}_{\text{virtual + unresolved}} \\
    &+ \sum_{\{\ijt,\kt\}} \sum_f 
    \int \diff \Phi_1^{ij,k} \,\Theta(t_{ij,k}-t_c)\,
    \underbrace{S_{ij,k} \,\Delta(\mu_Q^2,t) \,D^{ij,k}}_{\text{resolved}} \\
    &\times \mathcal{F}_{n+1}(\Phi_{n+1},O), \numberthis
\end{align*}
where $S_{ij,k}$ is a ratio of symmetry factors, and 
$\mathcal{F}_{n+1}$ is the result of the parton shower 
acting on the new $(n+1)$-particle configuration.
Then the expectation value of an observable $O$ is 
given by 
\begin{equation}
  \langle O \rangle^\text{PS} = \sum_f \intdphi{n} 
  B \, \mathcal{F}_n(\Phi_n,O),
\end{equation}
where $B$ contains the appropriate parton luminosity 
factors.

%%%%%%%%%%%%%%%%%%%%%%%%%%%%%

\subsection{Kinematics}

To obtain the kinematics of the individual splittings, we 
invert the phase space factorisation which emerges from 
the subtraction scheme. For details, see ref. \cite{Schumann:2007mg}.
It is important to note that initial-state spectators cannot 
absorb any transverse recoil, so the whole final state undergoes a
Lorentz transformation. In this way, the initial-state radiation 
induces logarithmic corrections to the transverse masses of 
intermediate resonances.
This process has been improved with a new recoil scheme, described 
in ref. \cite{Hoeche:2009xc}.

%%%%%%%%%%%%%%%%%%%%%%%%%%%%%

\subsection{Details of the implementation}
\label{sec:qedps:methods:implementation}

There are a few remaining subtleties in implementing an interleaved 
QCD+QED parton shower, which are detailed below.

\paragraph*{Evolution variable.}
The choice of the evolution variable $t$ is an important factor in 
defining a parton shower algorithm. Here we follow the default choice made 
in the \ac{QCD} dipole shower in \Sherpa, namely that final-state
fermions emit 
vectors (gluons and photons) with the modified transverse momentum
\begin{equation} \label{eq:qedps:kTordering}
  t = \bar{k}_T^2 = (Q^2-m_i^2-m_j^2-m_k^2)\,y\,(1-z),
\end{equation}
and that photon splitting into fermions is described by the modified 
virtuality 
\begin{equation}
  t = \tilde{q}^2 = (Q^2-m_i^2-m_j^2-m_k^2)\,y.
\end{equation}
This ensures colour and charge coherence, and angular ordering where 
appropriate. More details are given in section \ref{sec:yfs:methods:photonsplit}.
Different choices of evolution variable are made for initial-state
evolution, which are described in the \Sherpa user manual.

In addition, the generation of evolution variables can be simplified.
Throughout the description of the veto algorithm, sec. \ref{sec:qedps:veto},
we treated $f(t)$ and $g(t)$ as (in principle) 
different functions of $t$, subject only to the constraint $g(t)\geq f(t)$.
However, in the case of parton shower splitting functions, the $t$ dependence 
is typically straightforward. As can be seen in equations 
\eqref{eq:qedps:dipoleFF}-\eqref{eq:qedps:dipoleII}, 
the $t$ dependence enters only through the virtuality, and hence all splitting 
functions $f(t) \propto 1/t$. Therefore $g(t)\propto 1/t$ and $F(t) \propto 
G(t) \propto \log{t}$. Using this fact, and defining an overestimate 
$G'$ of $S_{\ijt (\kt)\to i j k}$, eq.\ \eqref{eq:qedps:generatet} is usually 
implemented as 
\begin{equation} \label{eq:qedps:practicalGeneratet}
  t = t' \, r^{2\pi/G'}.
\end{equation}

\paragraph*{Infrared cutoff.}
For \ac{QCD} evolution, the quarks and gluons are not 
physical degrees of freedom at scales below a few GeV due to 
confinement. For this reason, the \ac{IR} cutoff of a \ac{QCD} parton 
shower is usually around 1-2 GeV. However, photon emission has 
no such intrinsic cutoff, and photon splittings into electrons 
are kinematically allowed to occur while the photon has virtuality 
greater than twice the electron mass. This means that the evolution 
variable can in principle take values of the order of the 
electron mass. In \ac{QED} 
there is no process like hadronisation to take care of the 
effects of softer radiation, so to properly model \ac{QED} 
radiation we must allow the parton shower to evolve to lower 
scales. Here we choose a cutoff $t_c = 10^{-6} \,\GeV^2$ for 
final-state evolution of leptons and photons. For initial-state 
evolution of leptons, the \ac{IR} cutoff is coupled with the 
upper cutoff on the \ac{PDF} or structure function momentum 
fraction, so the discussion is postponed to sec. \ref{sec:qedps:ISmethods}.

\paragraph*{Efficiency.}
In the case where many charged particles are present in the 
shower evolution, for example as a result of gluon splittings 
into quark-antiquark pairs, considering all dipoles as possible 
photon emitters prohibitively impacts efficiency, due to 
the abundance of negative weighted events. However, since 
these $q\bar{q}$ pairs are usually close in phase space due to 
the collinear enhancement in their production, soft 
photon emissions will not resolve their individual charges if 
emitted outside the cone spanned by the dipole. Since the dominant 
contribution to the radiation pattern is from soft and collinear 
photons, the effect of all dipoles except that spanned by the 
$q\bar{q}$ pair in question is to pairwise approximately cancel. 
This applies to all cases where there are many final-state charge 
dipoles, and the obvious efficiency improvement is to identify 
the dominant dipole in each case. The solution we propose is 
to identify as spectator only the \ac{OSSF} particle which 
results in the smallest dipole invariant mass\footnote{
This scheme is similar to the one implemented in \Pythia, where 
new $f\bar{f}$ pairs are tracked manually.}. This scheme  
must be used with consideration of the process in question. 
In particular, whenever initial-final dipoles are important, 
this scheme will fail. It will also fail for any process 
involving decay products of $W$ bosons, since these are not 
\ac{OSSF} pairs. However, this scheme performs well in most 
non-resonant situations, and in addition, where there are 
neutral resonances this scheme will preserve their virtuality 
in the shower. 
We have verified that in all processes presented in section 
\ref{sec:qedps:FSresults}, the transverse momentum distributions
and the resonance lineshapes 
are statistically in agreement between the unaltered shower and 
the efficiency-improved shower. The relevant settings to effect 
these options are detailed in appendix \ref{app:settings}.

Finally, we will discuss another important factor in any 
\ac{EW} calculation, which has not yet been systematically addressed 
for a \ac{QED} shower.

%%%%%%%%%%%%%%%%%%%%%%%%%%%%%%%%%%%%%%%%%%%%%%%%%%%%%%%%%%%%%%%%%%

\section{$\alpha$ and electroweak input schemes}
\label{sec:qedps:runningAlpha}

Before presenting results from the \ac{QED} parton shower, we 
briefly discuss \ac{EW} input schemes and the treatment of 
$\alpha$. For the evaluation of \acp{ME} in hard 
processes, it is most appropriate to use a value of $\alpha$
which resums higher-order corrections.
Using $\alpha(0)$ for processes at scales of the \ac{EW} 
gauge bosons or higher leads to logarithms of the type 
$\log(m_f^2/m_{Z/W}^2)$, which are large for light flavours $f$.
To absorb these photonic vacuum polarisation corrections
to the coupling, 
a commonly used choice is the $\alpha(m_Z)$ scheme.
Alternatively, the $G_\mu$ scheme absorbs higher-order 
corrections to the renormalisation of the weak mixing angle.
In the $G_\mu$ scheme, $\alpha$ is a derived parameter and 
does not run, so it will be denoted $\alpha_{G_\mu}$, without 
any scale dependence.
However, since the difference between $\alpha$ 
in different schemes is formally subleading, we have the 
freedom to define a different scheme for radiative corrections
\cite{Denner:2019vbn}. 
In the case of \ac{NLO} calculations, of course, this must 
be consistent in all parts of the \ac{NLO} calculation to 
ensure cancellation of singularities. For the case of the 
shower, which is unitary, we are free to use a running 
$\alpha$ in the shower, analogously to QCD. 

For photon emissions, since most will become 
long-distance photons (i.e. they will not split again), their \ac{QED}
coupling should be evaluated in the Thomson limit, so we choose 
$\alpha(0)$. This is because the photonic wavefunction 
renormalisation already exactly cancels the light-fermion 
logarithms in the renormalisation of the coupling constant.
For photon splittings into fermions, 
we choose $\alpha(t)$, where $t$ is the splitting variable,
here the modified virtuality.

To implement these choices, and to allow for freedom in 
choosing the most appropriate \ac{EW} input scheme in many 
different cases, some changes to the computation 
of the running $\alpha$ in \Sherpa were necessary.

The scale dependence of $\alpha$ is given, in terms 
of the Thomson limit, by
\begin{equation} \label{eq:qedps:runningalpha:from0}
  \alpha(Q^2) = \frac{\alpha(0)}{1-\frac{\alpha(0)}{3\pi}
  \left[\Pi_\text{lep}(Q^2)+\Pi_\text{top}(Q^2)\right]-\Pi_\text{had}(Q^2)},
\end{equation}
where each $\Pi_f(Q^2)$ term is the vacuum polarisation 
induced by species $f$.
$\Pi_\text{had}$ is fitted from data \changed{and implemented 
using a series of thresholds. Meanwhile, the analytically known }
$\Pi_\text{lep}$ and $\Pi_\text{top}$ have the form 
\begin{equation} \label{eq:qedps:runningalpha:pi}
  \Pi_f(Q^2) = \frac{1}{3}-\left(1+2\frac{m_f^2}{Q^2}\right)
  \left(\sqrt{1-4\,\frac{m_f^2}{Q^2}} \,\log\left(
    \frac{1-\sqrt{1-4m_f^2/Q^2}}{1+\sqrt{1-4m_f^2/Q^2}}
    \right)\right).
\end{equation}
However, when one defines $\alpha$ at a scale $\mu > 0$, 
the running must be computed from this 
defined value and scale. To achieve this, we first compute 
$\alpha(0)$ in the chosen scheme by rearranging eq.\ 
\eqref{eq:qedps:runningalpha:from0}, 
\begin{equation} \label{eq:qedps:runningalpha:to0}
  \alpha(0) = \frac{\alpha(\mu^2)(1-\Pi_\text{had}(\mu^2))}
  {1+\frac{\alpha(\mu^2)}{3\pi} \left[\Pi_\text{lep}(\mu^2)+\Pi_\text{top}(\mu^2)\right]},
\end{equation}
then use this value of $\alpha(0)$ to compute all necessary
values of $\alpha(Q^2)$.
Note that while it appears that the hadronic vacuum polarisation 
is needed as input here, in fact the dependence on $\Pi_\text{had}$
disappears when the $\alpha(\mu^2)$ input scheme is used for the 
calculation of a hard process at the scale $\mu$.

\begin{figure}
  \centering
  \includegraphics[width=0.7\textwidth]{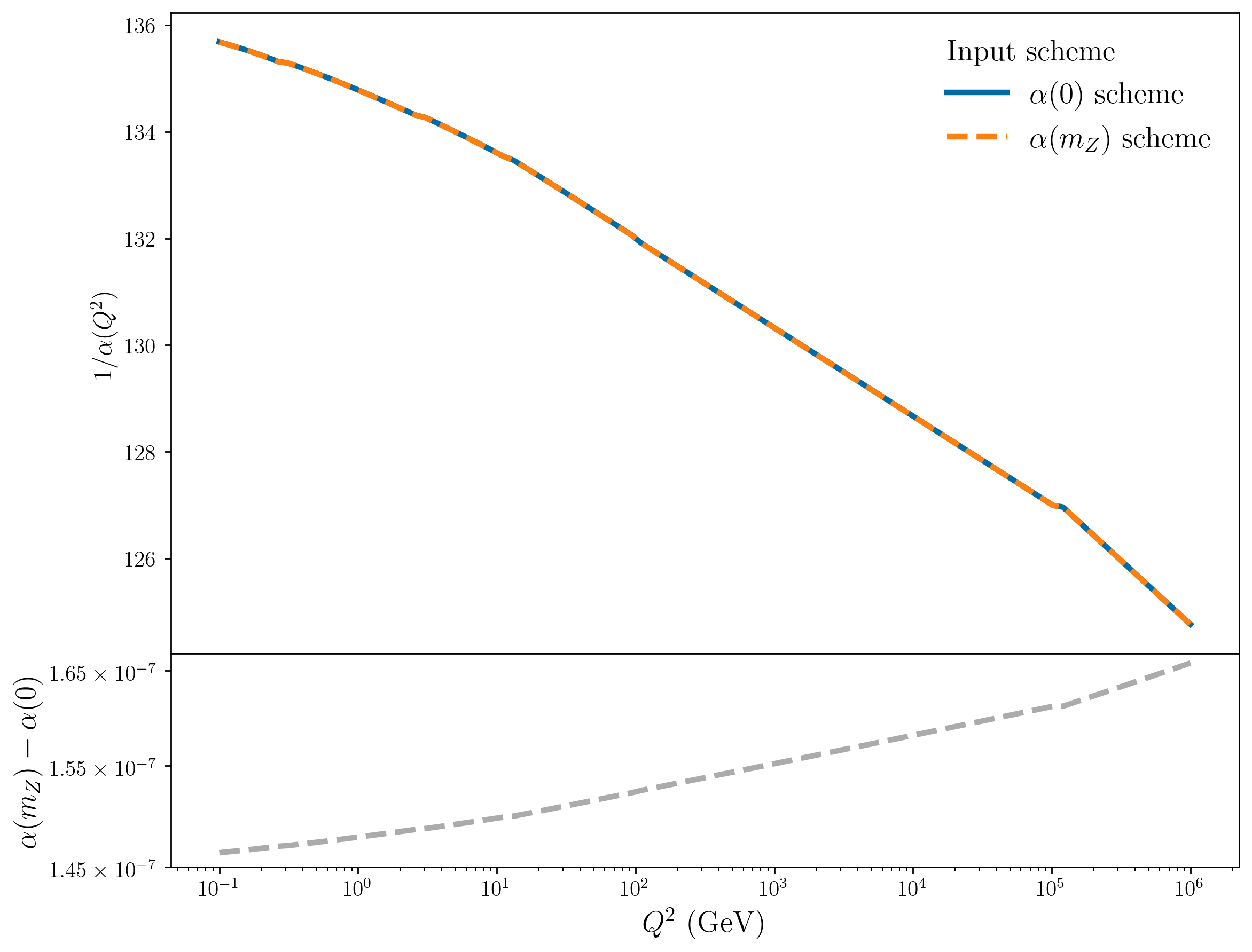}
  \caption{The running of $\alpha$ with energy,
  displayed as $1/\alpha$, in the $\alpha(0)$ scheme and 
  in the $\alpha(m_Z)$ scheme.
  \label{fig:qedps:runningAlpha}
  }
\end{figure}

Figure \ref{fig:qedps:runningAlpha} shows the scheme 
dependence of the running $\alpha$. From the residual plot, 
it can be seen that taking the measured 
value at the Thomson limit and running to $Q^2=m_Z^2$ using
eq.\ \eqref{eq:qedps:runningalpha:from0} induces a small 
correction to the value of $\alpha$ as measured at the 
$Z$ boson mass and evolved from there. However, the schemes 
are consistent to a high degree, showing that this method 
of modelling the running (in particular the hadronic 
vacuum polarisation) is valid.

For all the \ac{LO} hard process calculations discussed in 
this thesis, we use the $G_\mu$ input scheme which takes 
$\{G_F, m_W, m_Z, m_H\}$ and the decay widths as inputs. In 
this scheme, $\alpha$ is a derived parameter and does not 
run. However, as discussed above, we use a running
$\alpha$, defined in the $\alpha(0)$ scheme, for the 
radiative corrections. Wherever an external photon is 
present, it is considered to couple with $\alpha(0)$.

\subsubsection{Note: processes with external Born photons}

The discussion above is predicated on the Born-level process 
under consideration having no external photons. 
\changed{For a pure \ac{EW} process at order $n$ with $\ell$ external photons, 
the correct power counting would 
therefore be} $\alpha(0)^\ell 
\alpha(m_Z^2)^{n-\ell}$, or similarly $\alpha(0)^\ell 
\alpha_{G_\mu}^{n-\ell}$ if the $G_\mu$ scheme is preferred.
\changed{The factor $\alpha^{n-\ell}$ should be replaced by the 
appropriate power counting for the process under consideration.}

%%%%%%%%%%%%%%%%%%%%%%%%%%%%%%%%%%%%%%%%%%%%%%%%%%%%%%%%%%%%%%%
\section{Final-state results} \label{sec:qedps:FSresults}

In this section, we will present results produced using 
the \ac{QED} parton shower.
More work is needed to adapt the shower
for the case of an electron-positron collider due to 
the electron structure function (see sec. \ref{sec:qedps:ISmethods}),
so here we focus on showing its applicability to charged lepton 
final states. 
We first use a test process $\nu_\mu \bar{\nu}_\mu \to e^+ e^-$
at centre-of-mass energies of 91.2 GeV and 500 GeV, which are 
proposed energies of future 
$e^+ e^-$ colliders, where leptonic final states will be of 
great importance for precision measurements \cite{Freitas:2019bre}. 
By colliding muon neutrinos in this case study, we fully 
isolate the final state to validate the method. 
Then, in section \ref{sec:qedps:results:ggHleptons}
we will study the leptonic decays of a Higgs boson. We will 
again isolate the \ac{QED} final state by considering the Higgs 
to be produced through gluon fusion.

Throughout this section we will compare the \ac{QED} 
parton shower with the \ac{YFS} soft-photon resummation.
To produce the following results, 
\Sherpa's \Photons module was used \cite{Schonherr:2008av}. 
This produces exclusive photons in the soft approximation 
using eikonal factors, which can be corrected either 
using collinear splitting functions, or
exact higher-order soft-subtracted \acp{ME} can 
be used, if available. 
\changed{While the \ac{YFS} framework can incorporate higher-order 
corrections to any order, and \ac{NNLO} \ac{QED} corrections have 
been implemented in \Sherpa \cite{Krauss:2018djz}, here we use 
the publicly available \ac{NLO} \ac{EW} corrections in \Photons}
for the resonance decays presented here.
Note that in all cases the total cross section is not changed 
from the \ac{LO} cross section, so the results are comparable
with the \ac{LO} unitary parton shower without needing to account 
for differences in the total cross section.
In addition, the extension of the \ac{YFS} algorithm to 
charged particle pair production is included \cite{Flower:2022iew}
(see chapter \ref{chapter:yfs} for a detailed description of 
the \ac{YFS} formalism and the extension).
All charged particles are considered massive in the \ac{YFS} framework, 
while in the parton shower, all charged leptons and the $b$ and $t$ 
quarks are massive, but other quarks are treated massless.
For all results, \Amegic was used for the tree-level 
\ac{ME} generation \cite{Krauss:2001iv}.

\subsection{Case study: $\nu_\mu \bar{\nu}_\mu \to e^+ e^-$}
\label{sec:qedps:results:neutrinos}

We will first 
look at a process $\nu_\mu \bar{\nu}_\mu \to e^+ e^-$. 
In this section we will present results from this process 
on the $Z$ pole (with a centre-of-mass energy of 91.2 GeV)
and at higher energy, $\sqrt{s} = 500$ GeV.

To analyse our \ac{QED} parton shower in a way that is comparable 
to standard \ac{QCD} parton showers, we will use jet observables.
We define our \ac{QED} jets as objects produced using a $k_T$
jet algorithm (sometimes referred to as the Durham jet algorithm)
\cite{Catani:1993hr,Ellis:1993tq}. The distance parameter is given by 
\begin{equation}
  d_{ij} = \min(p_{Ti}^2,p_{Tj}^2) \, \frac{\Delta R_{ij}^2}{R^2},
\end{equation}
where $\Delta R_{ij}$ is the distance in pseudorapidity and azimuthal 
angle, $\Delta R_{ij}^2 = \Delta \eta_{ij}^2 + \Delta \phi_{ij}^2$.
We use a radius parameter $R=1$.
We choose to include electrons, muons and photons as input to 
the algorithm. This means we will miss contributions from 
photon splittings into $q\bar{q}$, pairs of light hadrons, 
or $\tau^+ \tau^-$,
but these are very rare (see chapter \ref{chapter:yfs}). 
Using a jet algorithm to 
define a final state means that we are 
inclusive with respect to flavour, analogously to \ac{QCD} 
parton showers. A $k_T$ algorithm was chosen since it effectively 
reverses the parton shower evolution, which is approximately 
ordered in $k_T$ (eq.\ \eqref{eq:qedps:kTordering}). 
The differential jet rate $d_{n,n+1}$ allows us to 
see at what scale the $(n+1)^\text{th}$ jet is formed, i.e. 
at what scale the emission happened. We will also study some 
observables, namely the photon multiplicity with different 
energy cuts, and the electron transverse momentum, which do 
not depend on this jet algorithm but similarly characterise 
our parton shower.

\begin{figure}
  \centering
  \includegraphics[width=0.45\textwidth]{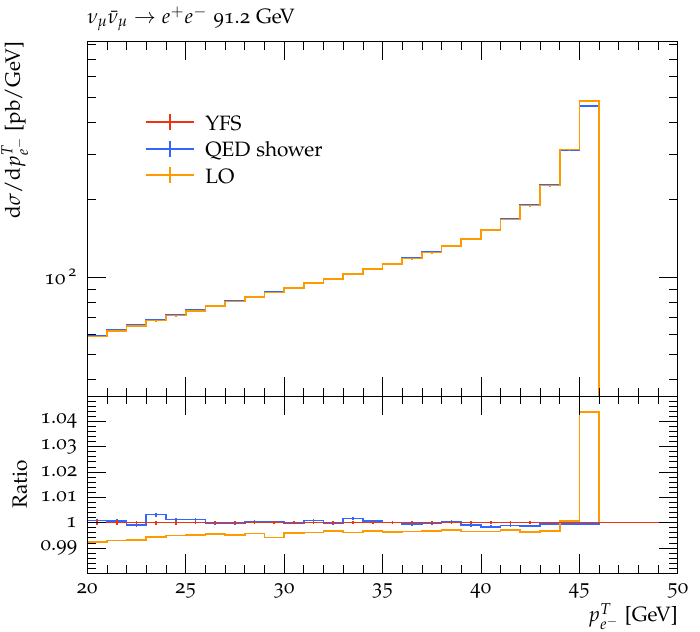}
  \includegraphics[width=0.45\textwidth]{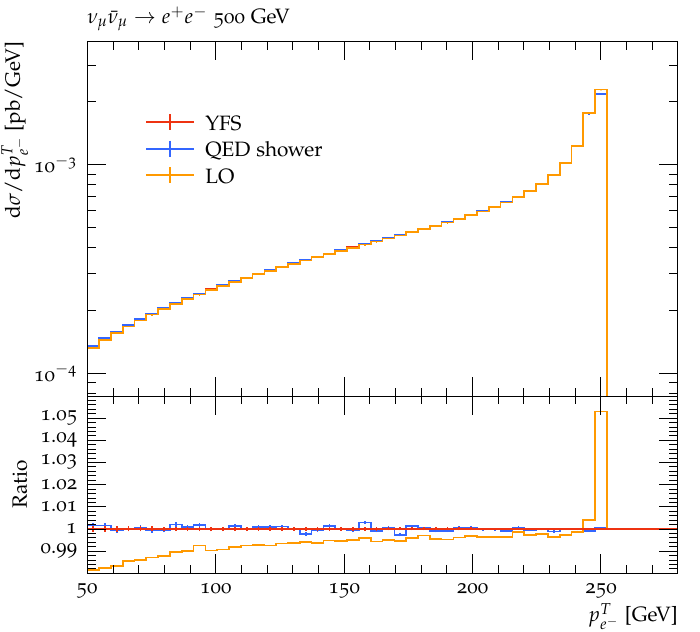}
  \caption[The electron transverse 
  momentum in $\nu_\mu \bar{\nu}_\mu \to e^+ e^-$, 
  comparing the YFS soft-photon resummation 
  with the QED shower prediction.]{ 
  The electron transverse 
  momentum in $\nu_\mu \bar{\nu}_\mu \to e^+ e^-$, 
  comparing the YFS soft-photon resummation with 
  the QED shower prediction and the 
  fixed-order LO distribution, for two 
  different collider energies. 
  \textbf{Left:} $\sqrt{s} = $ 91.2 GeV. 
  \textbf{Right:} $\sqrt{s} = $ 500 GeV. 
  The ratio plot shows the ratio to the YFS prediction.
  \label{fig:qedps:electronkT}
  }
\end{figure}

Figure \ref{fig:qedps:electronkT} shows the cross section 
differential in the transverse momentum of the hardest electron
for the process $\nu_\mu \bar{\nu}_\mu \to e^+ e^-$, comparing 
the prediction from different methods of modelling \ac{QED} 
radiation. In addition, we show the total size of the 
\ac{QED} corrections by comparing to the fixed-order 
\ac{LO} prediction for this observable.
The left plot shows a centre-of-mass energy of 
91.2 GeV, where it is clear that the \ac{YFS} prediction and 
the shower agree to high precision. On the right, for $\sqrt{s}=500$
GeV, the shower and \ac{YFS} agree perfectly. 
For both collider 
energies, the radiative corrections are approximately 
5\% in the last bin and up to 1\% throughout the 
distribution.

\begin{figure}
  \centering
  \includegraphics[width=0.45\textwidth]{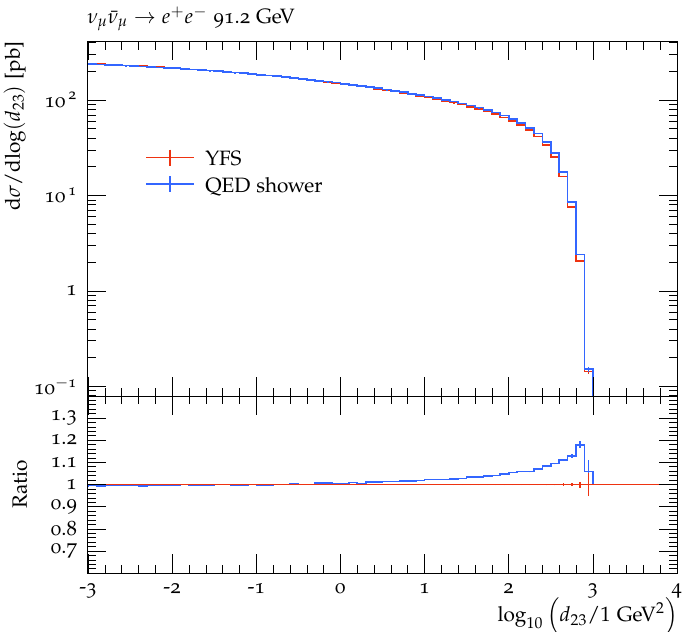}
  \includegraphics[width=0.45\textwidth]{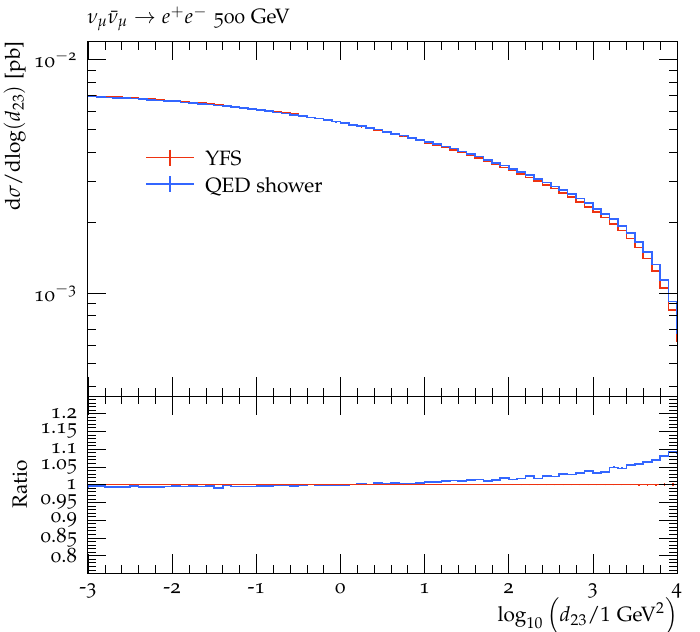}
  \caption[The 2-3 Durham jet rate in $\nu_\mu \bar{\nu}_\mu \to e^+ e^-$,
  comparing the YFS soft-photon resummation 
  with the QED shower prediction.]{
  The 2-3 Durham jet rate in $\nu_\mu \bar{\nu}_\mu \to e^+ e^-$,
  comparing the YFS method with the LO QED shower 
  prediction for 
  two different collider energies.
  \textbf{Left:} $\sqrt{s} = $ 91.2 GeV. 
  \textbf{Right:} $\sqrt{s} = $ 500 GeV.
  The ratio plot shows the ratio to the YFS prediction.
  \label{fig:qedps:jetrate}
  }
\end{figure}

\begin{figure}
  \centering
  \includegraphics[width=0.45\textwidth]{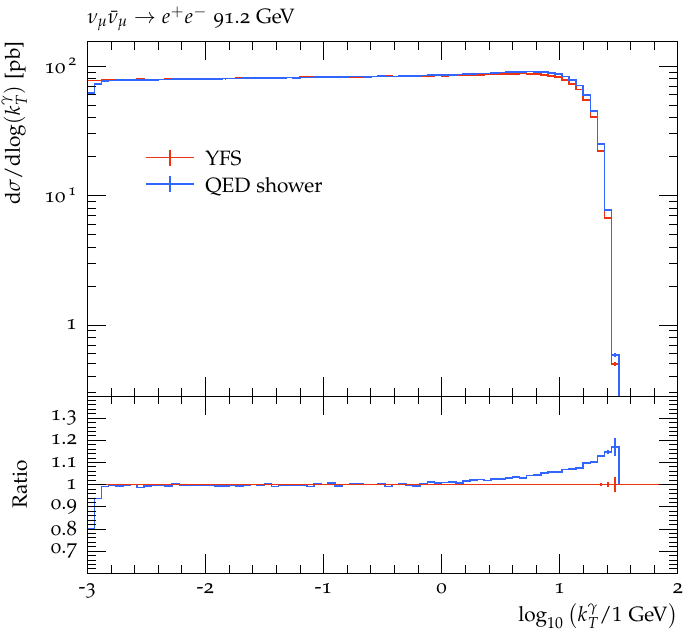}
  \includegraphics[width=0.45\textwidth]{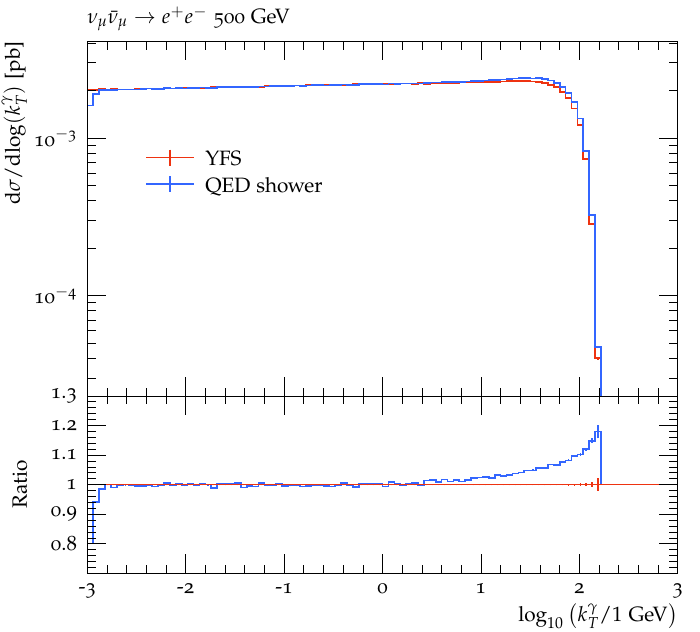}
  \caption[The third jet transverse momentum in $\nu_\mu \bar{\nu}_\mu \to e^+ e^-$,
  comparing the YFS soft-photon resummation 
  with the QED shower prediction.]{
  The third jet transverse momentum in $\nu_\mu \bar{\nu}_\mu \to e^+ e^-$,
  comparing the YFS method with the LO QED shower 
  prediction for 
  two different collider energies.
  \textbf{Left:} $\sqrt{s} = $ 91.2 GeV. 
  \textbf{Right:} $\sqrt{s} = $ 500 GeV.
  The ratio plot shows the ratio to the YFS prediction.
  \label{fig:qedps:thirdkT}
  }
\end{figure}

On the other hand, fig.\ \ref{fig:qedps:jetrate} shows 
$\diff \sigma/\diff \log(d_{23})$, the differential cross section in 
the 2-3 Durham jet rate, as described above. This characterises 
the hardest emission from the final state, which in the shower 
picture is the first emission. Compared to the \ac{YFS} 
prediction, for both collider energies,
the parton shower overestimates hard emissions.
This is due to resumming the collinear logarithms, without 
including the interference effects at higher orders, which are 
negative. Fig.\ \ref{fig:qedps:thirdkT} 
similarly shows the third jet transverse momentum, where the 
jets are ordered in transverse momentum. The shape differences 
in this observable are very similar to the $d_{23}$ plot, and
are also up to 20\% in size for the shower compared to the 
\ac{YFS}.

\begin{figure}
  \centering
  \includegraphics[width=0.45\textwidth]{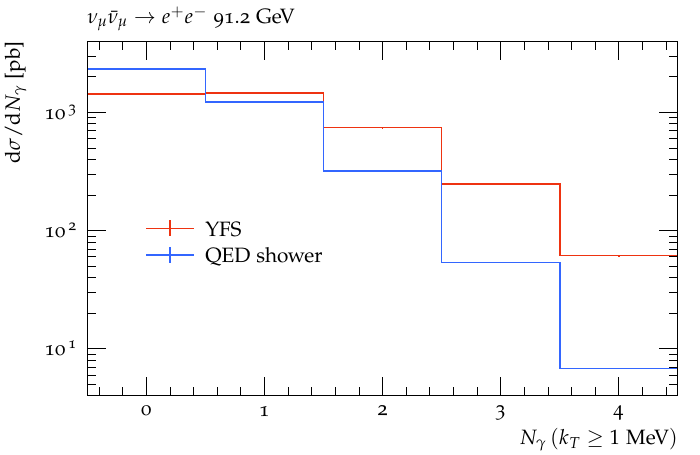}
  \includegraphics[width=0.45\textwidth]{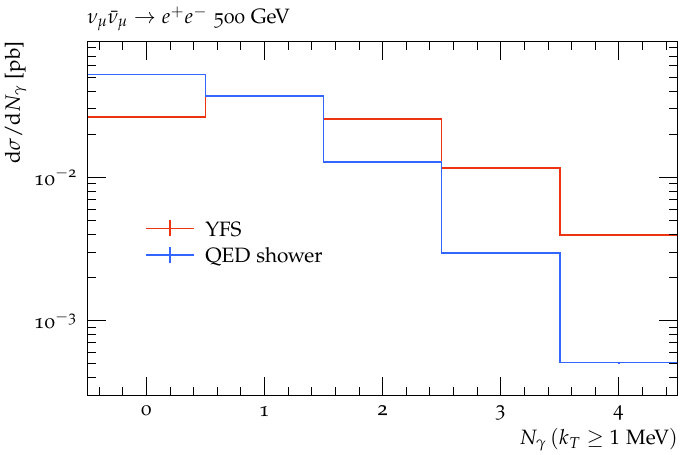}
  \caption[The multiplicity of photons 
  with $k_T \geq 1$ \MeV\, in $\nu_\mu \bar{\nu}_\mu \to e^+ e^-$,
  comparing the YFS soft-photon resummation 
  with the QED shower prediction.]
  {The multiplicity of photons with $k_T \geq 1$ \MeV\,
  in $\nu_\mu \bar{\nu}_\mu \to e^+ e^-$,
  comparing the YFS method with the LO QED shower 
  prediction for 
  two different collider energies.
  \textbf{Left:} $\sqrt{s} = $ 91.2 GeV. 
  \textbf{Right:} $\sqrt{s} = $ 500 GeV.
  \label{fig:qedps:multiplicity}
  }
\end{figure}

\begin{figure}
  \centering
  \includegraphics[width=0.45\textwidth]{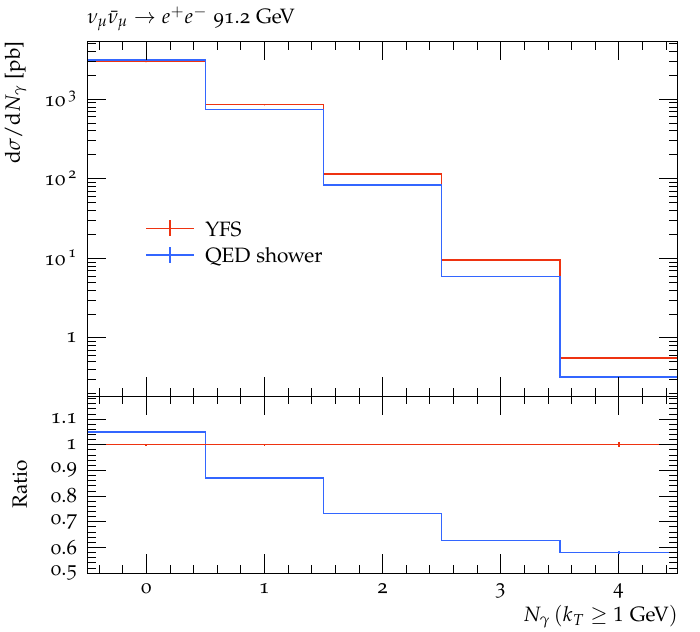}
  \includegraphics[width=0.45\textwidth]{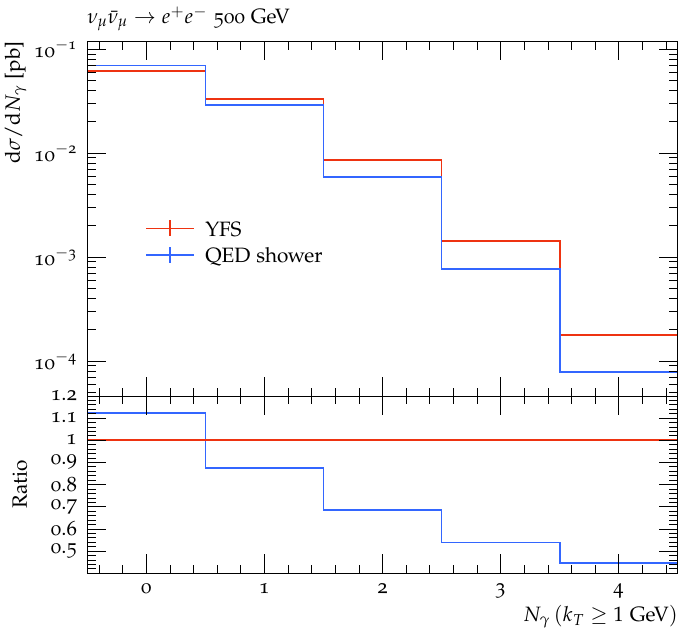}
  \caption[The multiplicity of photons 
  with $k_T \geq 1$ GeV in $\nu_\mu \bar{\nu}_\mu \to e^+ e^-$,
  comparing the YFS soft-photon resummation 
  with the QED shower prediction.]
  {The multiplicity of photons with $k_T \geq 1$ GeV
  in $\nu_\mu \bar{\nu}_\mu \to e^+ e^-$,
  comparing the YFS method with the LO QED shower 
  prediction for 
  two different collider energies.
  \textbf{Left:} $\sqrt{s} = $ 91.2 GeV. 
  \textbf{Right:} $\sqrt{s} = $ 500 GeV.
  The ratio plot shows the ratio to the YFS prediction.
  \label{fig:qedps:multiplicity1GeV}
  }
\end{figure}

Next, we study the multiplicity of photons produced by each 
method. We consider two different transverse momentum cuts, 
1 \MeV\, and 1 GeV. The former is shown in figure \ref{fig:qedps:multiplicity},
which clearly shows that the \ac{YFS} method produces far more 
photons (with $k_T \geq 1$ \MeV) than the shower does. This is a result 
of the resummation of all soft logarithms, where the number of 
photons increases as the energy decreases. While the shower splitting 
kernels do capture the leading soft divergences, the resummation is 
primarily of collinear logarithms and the veto algorithm, with 
its imposition of ordered emissions, does not produce as many 
soft photons as the unordered \ac{YFS} approach. However, 
fig.\ \ref{fig:qedps:multiplicity1GeV} shows the photon multiplicity 
with a higher cutoff of $k_T \geq 1$ GeV. This considers only the 
semi-soft and hard photons which have a considerable impact on 
recoil of leptons and other inclusive observables. We can see that 
the shower prediction is much closer to the \ac{YFS} prediction 
in this case, although large differences still emerge. The fact 
that both methods produce vastly different numbers of photons
(especially soft photons) but agree for a large part of kinematic 
observables validates the \ac{QED} parton shower as a description 
of higher-order \ac{QED} radiation.

\subsection{Leptonic Higgs decay}
\label{sec:qedps:results:ggHleptons}

In this section, we will study the processes $gg \to H \to \mu^+ \mu^-$
and $gg \to H \to e^+ e^- \mu^+ \mu^-$ in the \ac{HEFT}. 
The \ac{HEFT} is an effective field theory in which we obtain 
a direct effective coupling of gluons to the Higgs, 
\begin{equation}
    \mathcal{L} = \mathcal{L}_\text{SM} + g_\text{HEFT}\, G^{\mu\nu}_a G_{\mu\nu}^a H + \dots,
\end{equation}
by integrating out the top quark 
in the \ac{SM} loop-induced production of a Higgs via gluon fusion. 
In the \ac{SM}, other quarks also contribute to the loop, but since 
the top Yukawa coupling is much greater than the other quark 
Yukawa couplings due to its large mass,
only this contribution must be considered in the \ac{HEFT}.

We study gluon-induced Higgs production at the \ac{LHC},
where the colliding protons have a centre-of-mass energy 
of 13 \TeV. We use the \ac{PDF} 
set {\sc Pdf4Lhc21} from the \LHAPDF library \cite{Buckley:2014ana}. 
\Sherpa's default parton shower, {\sc Csshower}, was used for the 
initial-state \ac{QCD} shower \cite{Schumann:2007mg}. Beam remnants, 
hadronisation, and multiple interactions were not modelled.

\subsubsection{$gg \to H \to \mu^+ \mu^-$}

The \ac{LO} cross section for $gg \to \mu^+ \mu^-$ at the 13 \TeV\, \ac{LHC} in 
the \ac{HEFT} is $0.0028662(1)$ pb. 
In this section we present differential 
cross sections in various kinematical observables for bare muons 
and dressed muons. The muons are dressed with photons in a cone of 
radius $\Delta R=0.1$, where $\Delta R = \sqrt{\Delta \eta^2 + \Delta \phi^2}$.
The primary muons, whether bare or dressed, are subject to cuts 
on transverse momentum, $p_\mu^T > 10\,\GeV$, and rapidity, 
$\abs{y_\mu} < 2.5$.
We compare the interleaved QCD+QED shower presented in this chapter 
with the scenario where no \ac{QED} radiation is included (QCD shower 
only) in addition to the \ac{YFS} soft-photon resummation 
supplemented with exact \ac{NLO} corrections.

\begin{figure}
  \centering
  \includegraphics[width=0.45\textwidth]{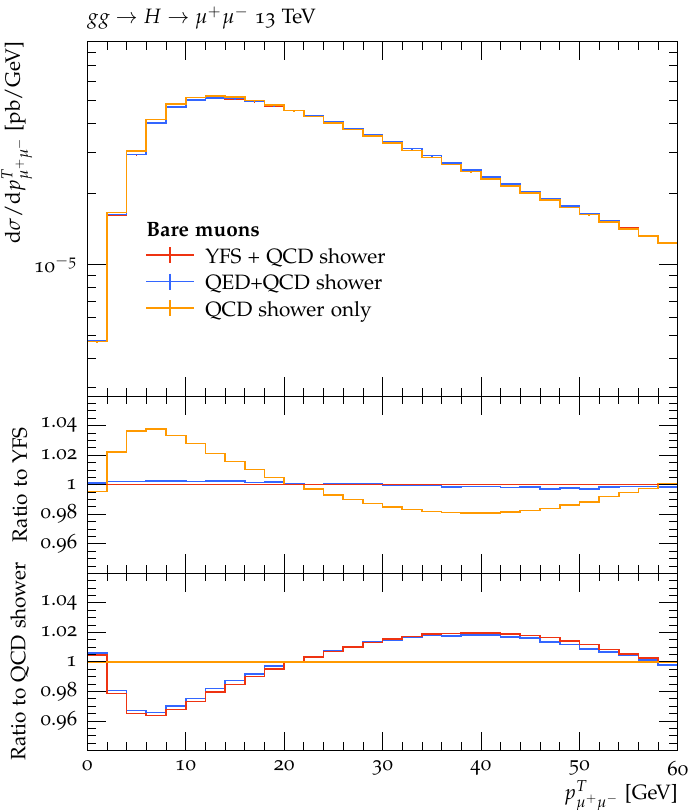}
  \includegraphics[width=0.45\textwidth]{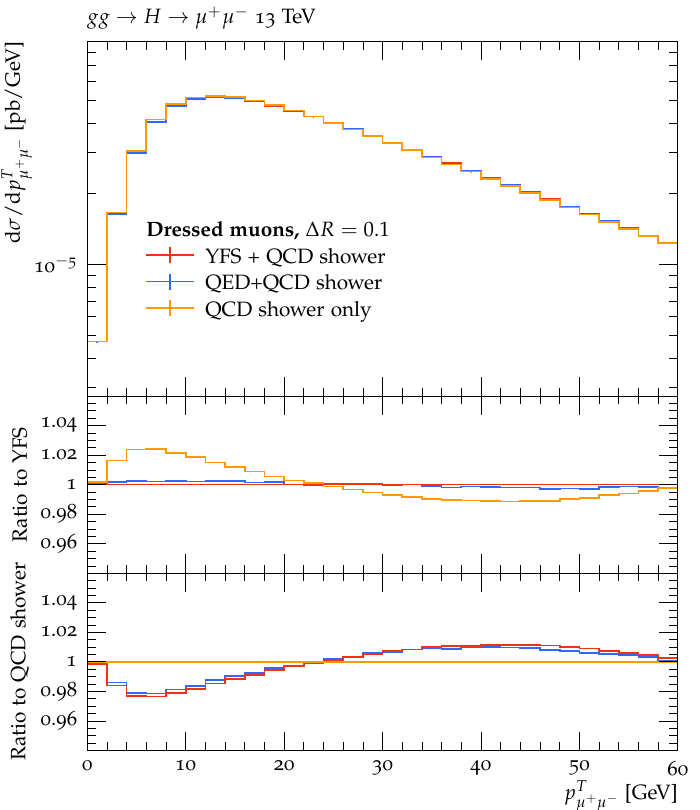}
  \caption[The dimuon transverse momentum 
  in $gg\to \mu^+\mu^-$, comparing the 
  YFS prediction with 
  the interleaved QCD+QED shower. The pure 
  QCD shower prediction is also included for comparison.]
  {The dimuon transverse momentum 
  in the $gg\to \mu^+\mu^-$ process, comparing the 
  YFS prediction with 
  the interleaved QCD+QED shower. The pure 
  QCD shower prediction is also included for comparison. 
  \textbf{Left:}
  bare muons, \textbf{Right:} photon-dressed muons with 
  a cone size $\Delta R = 0.1$.
  \label{fig:qedps:ggmumu:dilepton_pt}
  }
\end{figure}

First, figure \ref{fig:qedps:ggmumu:dilepton_pt} shows the dimuon 
transverse momentum distribution. The shape of the distribution 
is dominated by \ac{QCD} initial-state radiation, which induces
transverse recoil on the whole final state. However, there 
are small effects from \ac{QED} radiation recoil, as can be seen 
in the lower ratio plot. As can be seen from the upper ratio plot,
in both bare muons and dressed 
muons, the interleaved shower agrees very well with the 
\ac{YFS} prediction.

\begin{figure}
  \centering
  \includegraphics[width=0.45\textwidth]{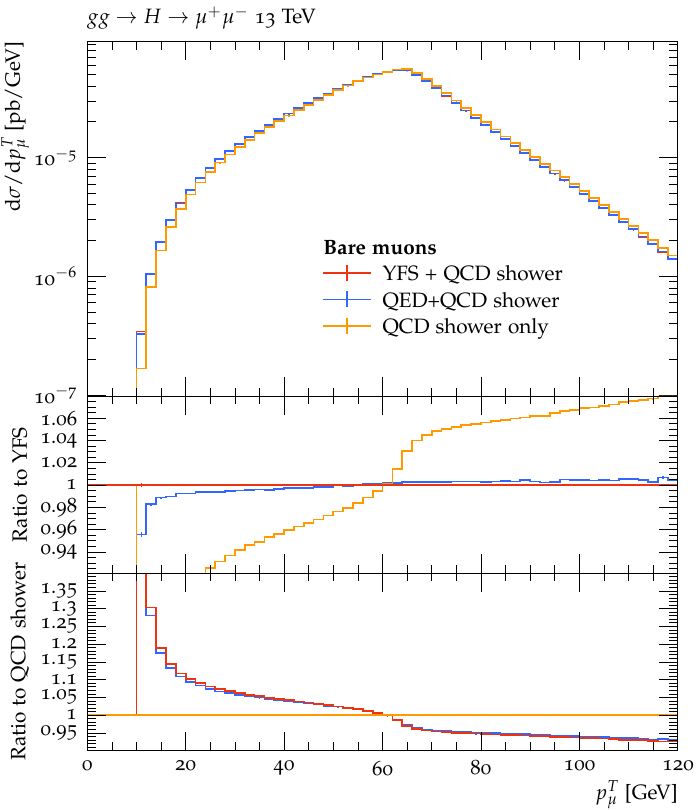}
  \includegraphics[width=0.45\textwidth]{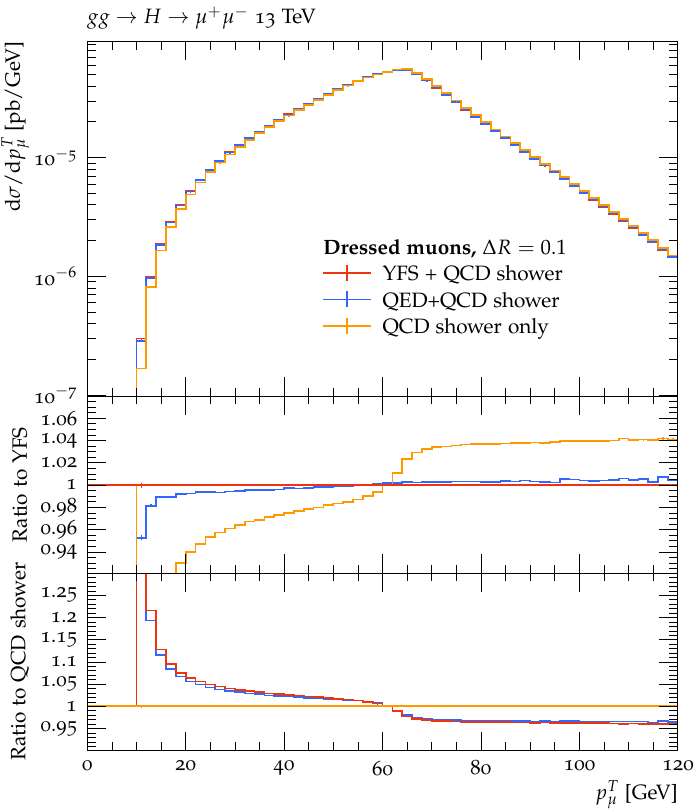}
  \caption[The hardest muon transverse momentum 
  in $gg\to \mu^+\mu^-$, comparing the 
  YFS prediction with
  the interleaved QCD+QED shower and the pure 
  QCD shower prediction.]{The hardest muon transverse momentum 
  in the $gg\to \mu^+\mu^-$ process, comparing the 
  YFS prediction with
  the interleaved QCD+QED shower and the pure 
  QCD shower prediction. We show ratios to 
  YFS (upper ratio plot) and to the QCD shower only (lower
  ratio plot).
  \textbf{Left:}
  bare muons, \textbf{Right:} photon-dressed muons with 
  a cone size $\Delta R = 0.1$.
  \label{fig:qedps:ggmumu:one_lepton_pt}
  }
\end{figure}

The distribution of the muon transverse momentum, shown in fig.\ 
\ref{fig:qedps:ggmumu:one_lepton_pt}, is given already at 
\ac{LO} by the kinematics of the Higgs decay. Even without 
any transverse initial-state radiation, there is a peak at 
$p_\mu^T=m_H/2$. The distribution including only initial-state 
\ac{QCD} radiation is shown in the plot. The addition of \ac{QED}
final-state radiation results in a softer hardest muon, as can 
be seen from the lower ratio plot. Comparing the two \ac{QED} 
radiation methods in the upper ratio plots, we see that the full 
\ac{YFS} including exact \ac{NLO} corrections produces the 
largest recoil corrections, though the soft logarithms account
for the bulk of these. The interleaved QED+QCD shower is in good 
agreement with \ac{YFS} for both bare and dressed muons, 
with better than 1\% precision.

\begin{figure}
  \centering
  \includegraphics[width=0.45\textwidth]{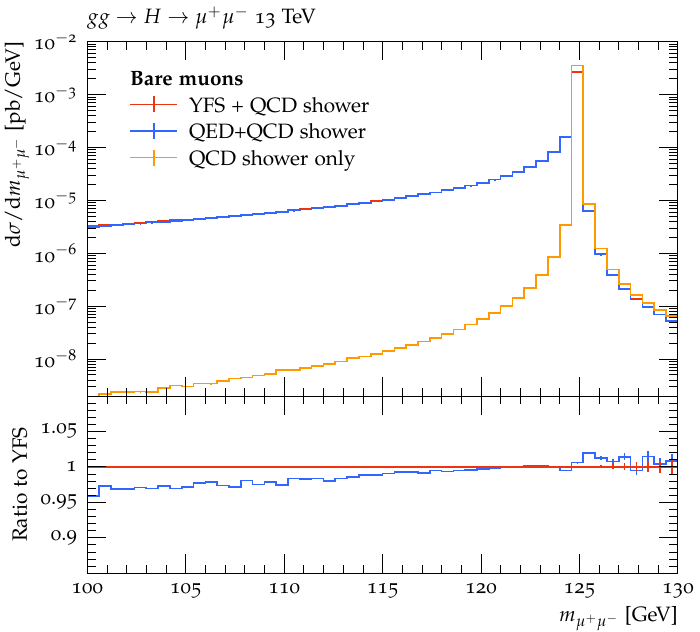}
  \includegraphics[width=0.45\textwidth]{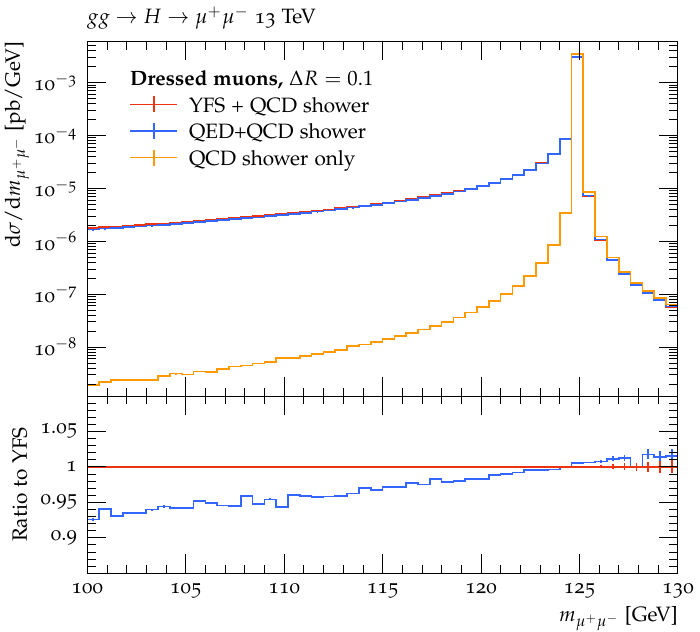}
  \caption[The dimuon invariant mass
  in $gg\to \mu^+\mu^-$, comparing the 
  full YFS prediction with 
  the interleaved QCD+QED shower. The pure 
  QCD shower prediction is also included for comparison.]{
  The dimuon invariant mass
  in the $gg\to \mu^+\mu^-$ process, comparing the 
  full YFS prediction with 
  the interleaved QCD+QED shower. The pure 
  QCD shower prediction is also included for comparison. 
  \textbf{Left:}
  bare muons, \textbf{Right:} photon-dressed muons with 
  a cone size $\Delta R = 0.1$.
  \label{fig:qedps:ggmumu:invariant_mass}
  }
\end{figure}

Finally, we study the Higgs lineshape in figure \ref{fig:qedps:ggmumu:invariant_mass}.
This plot shows the cross section differential in the dimuon 
invariant mass, for bare muons (left) and cone-dressed muons (right)
with a cone size of $\Delta R = 0.1$. We can see that the \ac{QCD}
shower alone produces very few events below the resonant peak, since 
if the final state is produced on the resonance, it cannot radiate 
further. The ratio plot shows the interleaved shower compared to the 
\ac{YFS} prediction. 
For bare muons, we see that the shower is in very good agreement 
with \ac{YFS} in the vicinity of the resonance. Below the resonance, 
the shower underestimates the cross section with respect to \ac{YFS}, 
since it does not contain \ac{NLO}-accurate hard photon 
emissions. This disagreement is accentuated for the case of 
dressed muons, since the kinematical impact of collinear radiation 
is decreased when muons are cone-dressed and hard wide-angle 
radiation contributes more strongly to the lineshape. 

\subsubsection{$gg \to H \to \mu^+ \mu^- e^+ e^-$}

The \ac{LO} cross section for $gg \to \mu^+ \mu^- e^+ e^-$ 
at the 13 \TeV\, \ac{LHC} in 
the \ac{HEFT} is $0.00084(1)$ pb. 
Due to the complex final state, 
simulations for this process are significantly more expensive than
the processes we have studied so far. However, we have found that 
the interleaved shower is faster or the same speed as the \ac{QCD}
shower plus \ac{YFS} simulations, when the one-spectator
efficiency option 
described in section \ref{sec:qedps:methods:implementation} and 
in the appendix is enabled. We have validated that even for this 
more complex final state, the one-spectator option reproduces the
results of the full dipole shower but with much higher statistical 
precision.
In this section, we will present distributions in the four-lepton 
invariant mass $m_{4\ell}$, the hardest $Z$ boson transverse 
momentum $p^T_Z = \max ((p_{e^+}+p_{e^-})^T,(p_{\mu^+}+p_{\mu^-})^T)$,
and the hardest lepton transverse momentum $p_\ell^T$.
The primary electrons and muons are subject to cuts 
on transverse momentum, $p_\ell^T > 10 \,\GeV$, and rapidity, 
$\abs{y_\ell} < 2.5$.
We compare the interleaved QCD+QED shower presented in this chapter 
with the \ac{YFS} soft-photon resummation including \changed{collinearly-approximated} 
\ac{ME} corrections.

\begin{figure}
  \centering
  \includegraphics[width=0.45\textwidth]{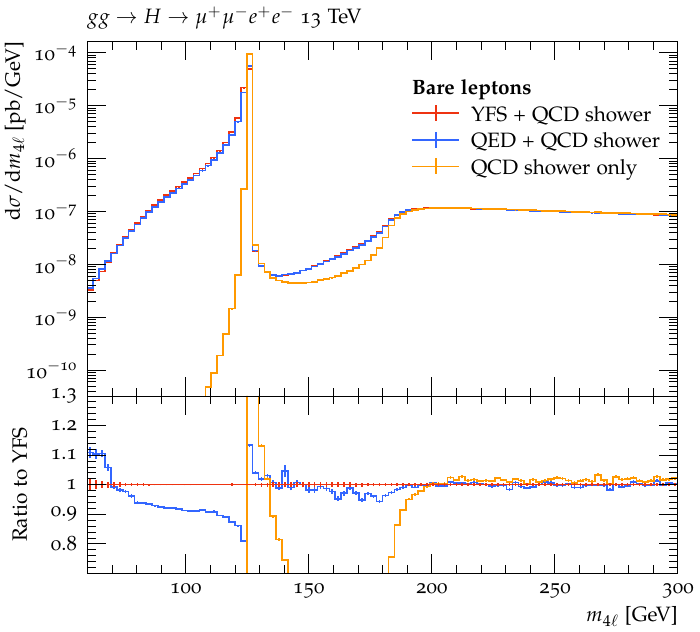}
  \includegraphics[width=0.45\textwidth]{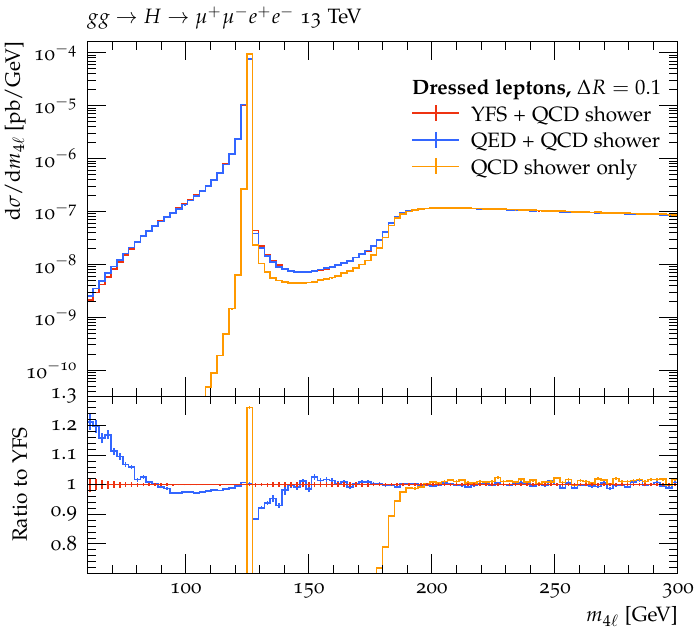}
  \caption{The four-lepton bare invariant mass
  in $gg\to \mu^+\mu^- e^+ e^-$, comparing the
  the QCD shower and YFS prediction 
  with the prediction from an interleaved QCD+QED shower.
  The pure QCD shower prediction is also included for comparison. 
  \textbf{Left:} bare leptons, \textbf{Right:} photon-dressed leptons 
  with a cone size $\Delta R = 0.1$.
  \label{fig:qedps:gg4l:invariant_mass}
  }
\end{figure}

Figure \ref{fig:qedps:gg4l:invariant_mass} shows the four-lepton 
invariant mass differential cross section. The left plot shows 
the observable for bare leptons, while the right plot shows the 
invariant mass of four primary photon-dressed leptons.
The main features of 
the lineshape are the Higgs resonance at 125 GeV and the 
two on-shell $Z$ threshold at 180 GeV. Above the threshold, the 
interleaved shower is in perfect agreement with the \ac{YFS} 
approach combined with the \ac{QCD} shower. Between the Higgs resonance 
and the di-$Z$ threshold, the interleaved shower produces fewer 
events for the bare-lepton observable. This is due to the shower producing 
fewer soft and collinear photon emissions than \ac{YFS}, as can 
be seen in the dressed lepton plot where this feature disappears.
Of more note are the differences in the two methods at the Higgs 
mass and just below it. 
The \ac{YFS} approach correctly resums all the soft 
logarithms and hence produces more events just below the Higgs 
resonance, where the leptons have only lost energy through relatively 
soft radiation. This effect is clearly mitigated in the dressed 
lepton plot, where recoil from photon emissions is minimised.
The \ac{QED} shower, by contrast, leaves more bare-lepton 
events exactly on the Higgs pole.
We can clearly see that both \ac{QED} radiation methods produce 
a very different distribution from that predicted by the \ac{QCD}
shower alone, which is clearly lacking modelling of final-state 
radiation.

\begin{figure}
  \centering
  \includegraphics[width=0.45\textwidth]{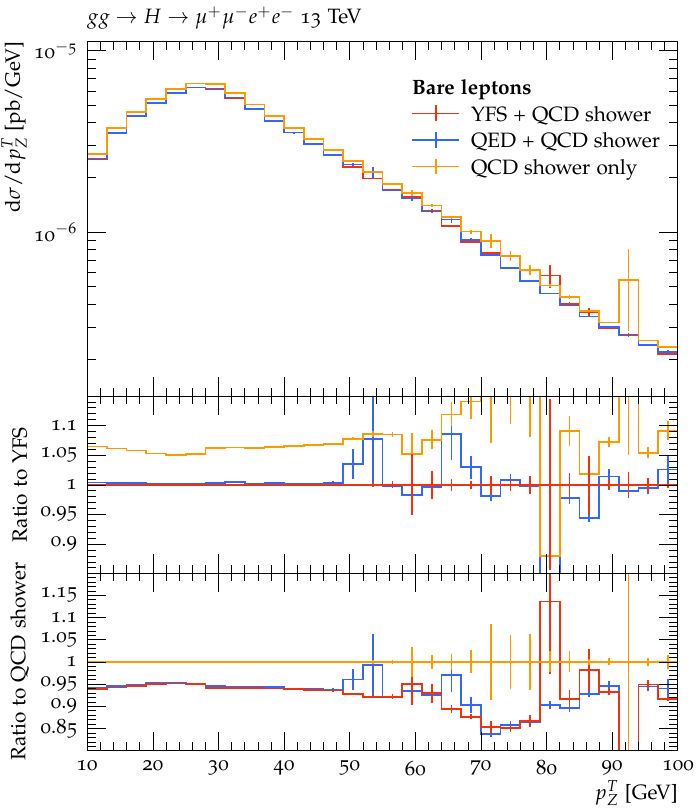}
  \includegraphics[width=0.45\textwidth]{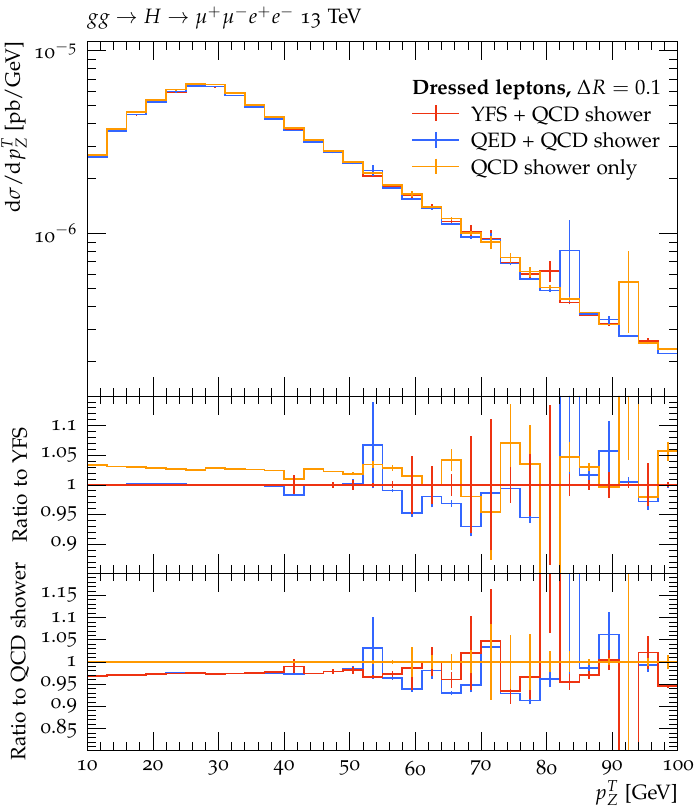}
  \caption[The hardest $Z$ boson transverse 
  momentum
  in $gg\to \mu^+\mu^- e^+ e^-$, comparing
  the QCD shower and YFS prediction 
  with the prediction from an interleaved QCD+QED shower.
  The pure QCD shower prediction is also included for comparison.]
  {The hardest $Z$ boson transverse momentum
  in $gg\to \mu^+\mu^- e^+ e^-$, comparing
  the QCD shower and YFS prediction 
  with the prediction from an interleaved QCD+QED shower.
  The pure QCD shower prediction is also included for comparison. 
  The upper ratio plot shows the ratio to the YFS + QCD shower 
  prediction, while the lower shows the ratio with respect to 
  the QCD shower alone.
  \textbf{Left:} bare leptons, \textbf{Right:} photon-dressed leptons 
  with a cone size $\Delta R = 0.1$.
  \label{fig:qedps:gg4l:ZpT}
  }
\end{figure}

Next, we study the Higgs decay to $Z$ bosons by reconstructing the 
momentum of each $Z$ and plot the hardest $Z$ transverse momentum 
in fig.\ \ref{fig:qedps:gg4l:ZpT}, using the flavour identification 
of bare leptons (left) or dressed leptons (right).
Here, we expect better agreement between the \ac{QED} radiation 
methods, since \ac{QED} radiation is already only a 3-6\% correction 
to the distribution predicted by the \ac{LO} \ac{QCD} shower
(as seen from the lower ratio plots).
Indeed, we see perfect agreement between the \ac{YFS} approach and the 
\ac{QED} shower approach for both bare and dressed leptons, 
when each is combined with the initial-state 
\ac{QCD} shower. Unlike the $H \to \mu^+\mu^-$ case 
in fig.\ \ref{fig:qedps:ggmumu:one_lepton_pt}, there is no peak at 
$m_H/2$ because the decay preferentially proceeds with one $Z$ 
on-shell (since both cannot be on-shell). Note that since the 
$p_T$ cut is applied to the leptons, the reconstructed $Z$ bosons 
can have zero transverse momentum.

\begin{figure}
  \centering
  \includegraphics[width=0.45\textwidth]{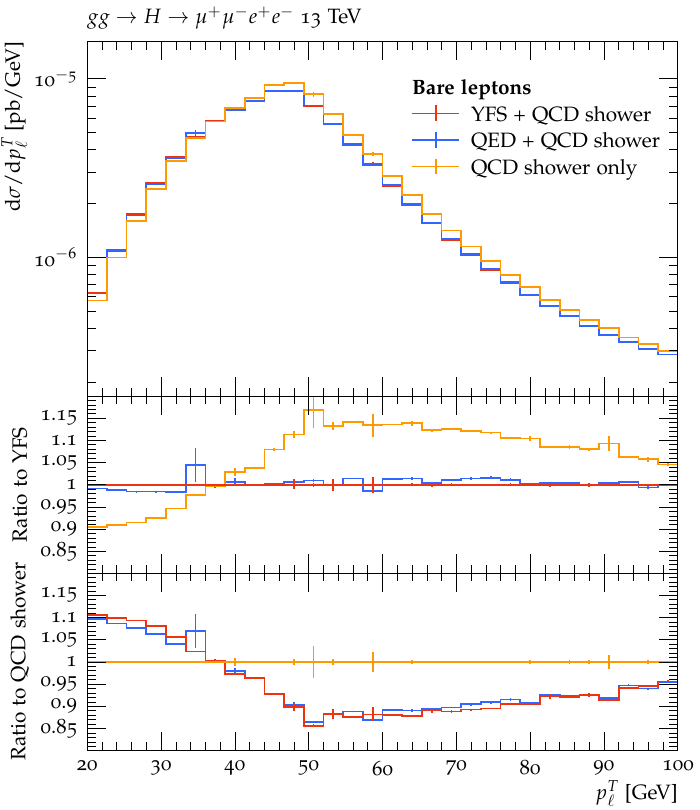}
  \includegraphics[width=0.45\textwidth]{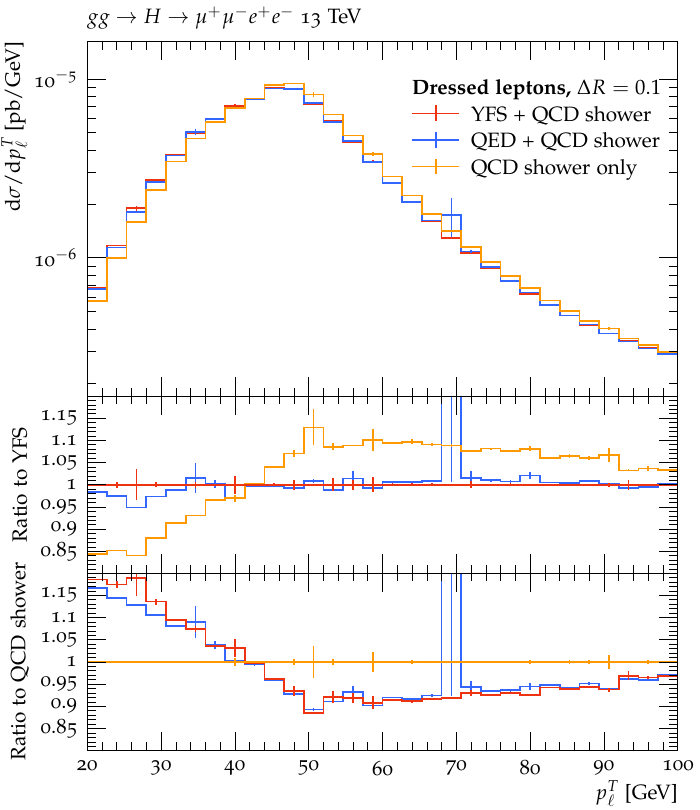}
  \caption[The hardest lepton transverse 
  momentum 
  in $gg\to \mu^+\mu^- e^+ e^-$, comparing
  the QCD shower and YFS prediction 
  with the prediction from an interleaved QCD+QED shower.
  The pure QCD shower prediction is also included for comparison.]
  {The hardest lepton transverse momentum 
  in $gg\to \mu^+\mu^- e^+ e^-$, comparing
  the QCD shower and YFS prediction 
  with the prediction from an interleaved QCD+QED shower.
  The pure QCD shower prediction is also included for comparison.
  The upper ratio plot shows the ratio to the YFS + QCD shower 
  prediction, while the lower shows the ratio with respect to 
  the QCD shower alone. 
  \textbf{Left:} bare leptons, \textbf{Right:} photon-dressed leptons 
  with a cone size $\Delta R = 0.1$.
  \label{fig:qedps:gg4l:leptonpT}
  }
\end{figure}

In figure \ref{fig:qedps:gg4l:leptonpT} we show the 
cross section differential in the hardest lepton transverse 
momentum. The left plot shows the hardest bare lepton $p_T$, 
while the right plot shows the hardest dressed lepton $p_T$.
The spectrum has a peak at $m_Z/2$ due to the preference 
for on-shell $Z$ production in the Higgs decay. There is very 
good agreement between the \ac{YFS} prediction and the interleaved 
shower. Both methods 
predict a different shape distribution than the \ac{LO} \ac{QCD}
shower, with differential cross section differences up to 15\%
for both bare and dressed leptons.
Note that the \ac{YFS} prediction here includes exact \ac{NLO} 
\ac{QED} corrections to the radiation pattern, and so produces more 
hard photonic radiation than the shower. This results in the 
interleaved shower predicting a slightly higher hardest lepton $p_T$,
which is visible when photonic radiation is not recombined.

\begin{figure}
  \centering
  \includegraphics[width=0.45\textwidth]{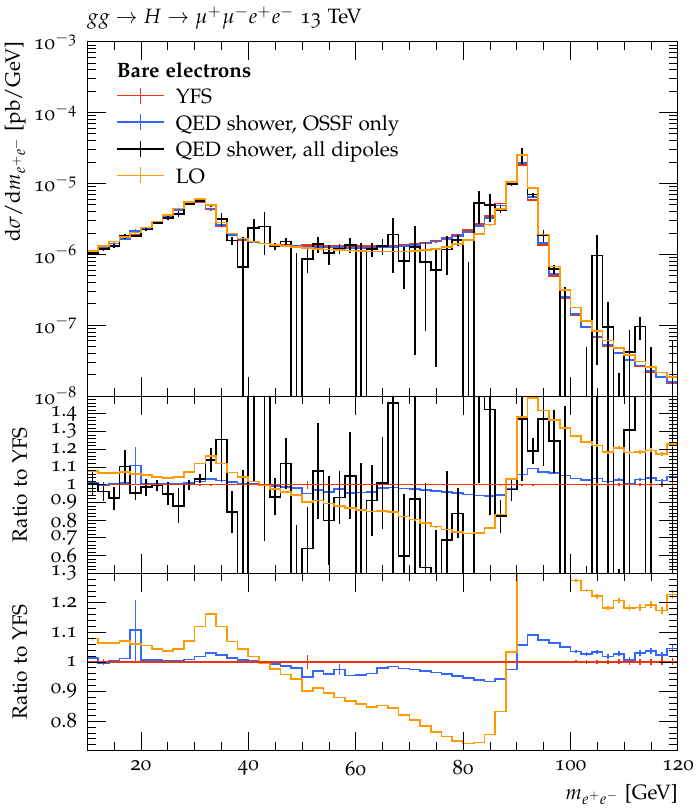}
  \includegraphics[width=0.45\textwidth]{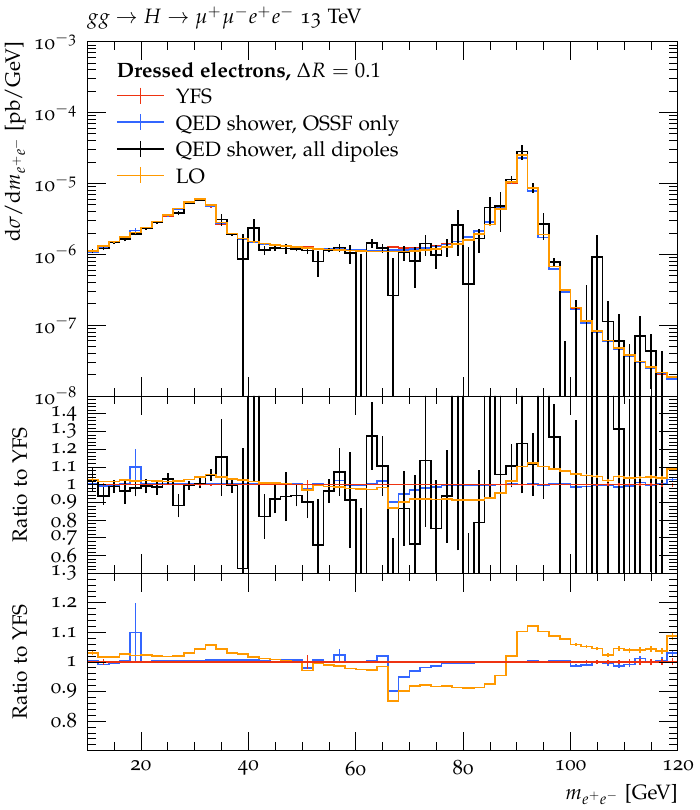}
  \hfill 
  \includegraphics[width=0.45\textwidth]{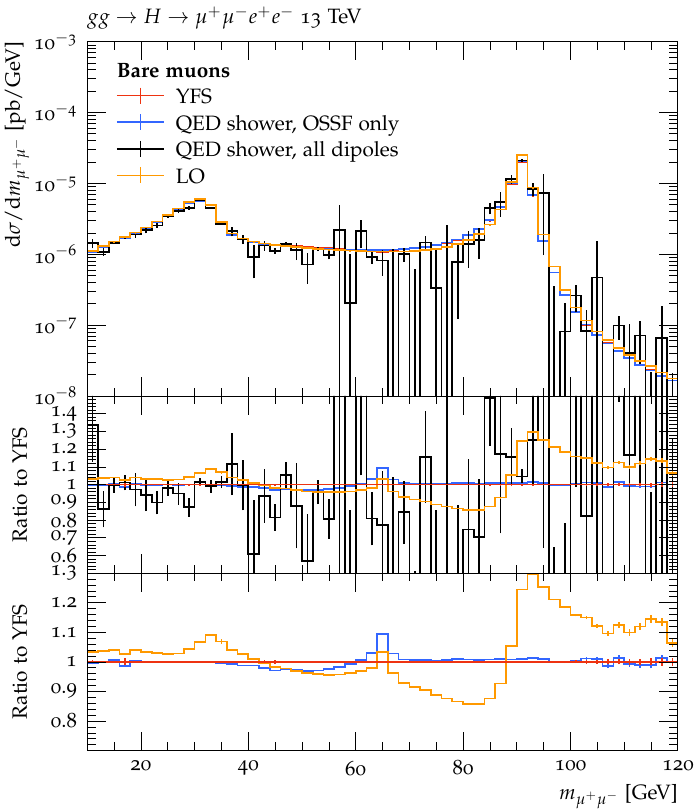}
  \includegraphics[width=0.45\textwidth]{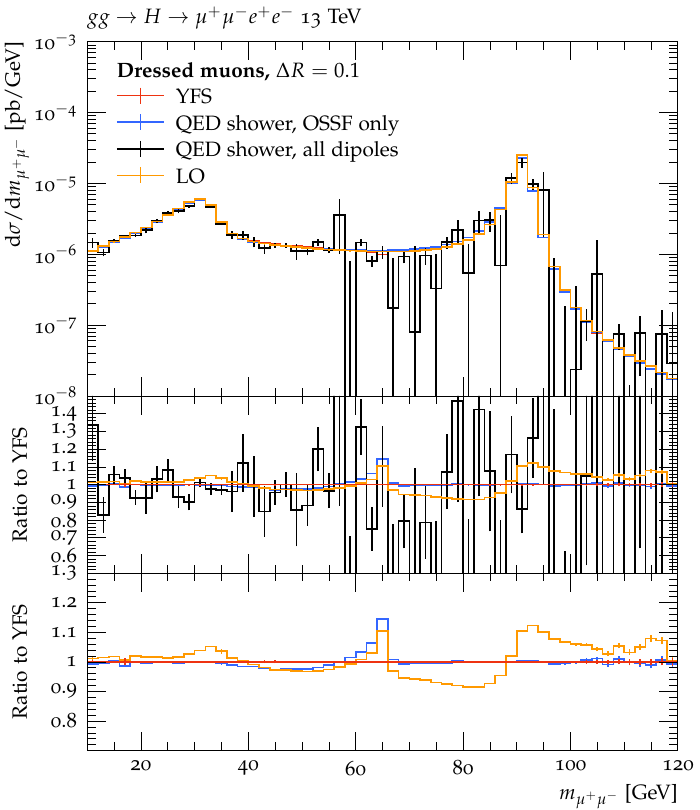}
  \caption[The dilepton invariant mass
  in $gg\to \mu^+\mu^-e^+e^-$, comparing the 
  QED shower either with OSSF dipoles only, or with all dipoles.
  The YFS resummation and LO prediction are also shown for 
  comparison.]{
  The dilepton invariant mass
  in the $gg\to \mu^+\mu^-e^+e^-$ process, comparing the 
  QED shower either with OSSF dipoles only, or with all dipoles.
  The YFS resummation and LO prediction are also shown for 
  comparison. 
  Both ratio plots are with respect to the YFS resummation, 
  but the all-dipoles case is not shown in the lower ratio plot.
  \textbf{Top:} electrons, \textbf{Bottom:} muons;
  \textbf{Left:}
  bare leptons, \textbf{Right:} photon-dressed leptons with 
  a cone size $\Delta R = 0.1$.
  \label{fig:qedps:gg4l:mee}
  }
\end{figure}

Figure \ref{fig:qedps:gg4l:mee} shows the impact of the 
efficiency improvement 
described in section \ref{sec:qedps:methods:implementation}.
In this approximation, rather than including all possible 
charged dipoles for photon emission, we identify \ac{OSSF}
pairs and consider only the dipoles spanned by these pairs to 
radiate photons. In addition to improving efficiency, this 
is a form of resonance-awareness which can help to avoid 
unphysical exchanges of momentum between decaying resonances.
Resonance-identification methods have been shown to reduce 
the tendency of real-radiation modelling methods to distort 
lineshapes \cite{Kallweit:2017khh,Carli:2010cg,Hoeche:2014lxa}.
In these plots we show the invariant mass of the $Z$ boson, 
reconstructed from particular flavours, for various methods in
$gg \to H \to \mu^+ \mu^- e^+ e^-$. Only \ac{QED} radiation is 
included, since the initial-state \ac{QCD} radiation affects 
all methods in the same way. The top-left and 
top-right plots show the $e^+ e^-$ invariant mass for bare 
and dressed electrons respectively, while the bottom two 
plots show the $\mu^+ \mu^-$ invariant mass. The \ac{LO}
prediction is shown for comparison, which is the case where no 
radiation effects are included.
We see that the all-dipoles case, as 
compared to the \ac{OSSF} dipoles only case, performs far 
worse in terms of statistics. This plot shows the result 
of simulating $10^9$ weighted events for the \ac{YFS}, 
\ac{OSSF}-only and \ac{LO} predictions, and $10^{10}$ 
weighted events for the all-dipoles shower. Despite this,
we can clearly see the impact of 
large numbers of negative weights. The statistical precision 
is not sufficient to verify whether the unconstrained 
momentum transfer distorts the $Z$ invariant mass spectrum, 
however, we can clearly see from these plots that the \ac{OSSF}
approximation is a very good one.

The lower ratio plot of each plot in fig.\ 
\ref{fig:qedps:gg4l:mee} shows the impact of lepton definitions 
on the invariant mass distribution. We see that the muon 
invariant mass is much less impacted by \ac{QED} radiation
on the bare level, due to its higher mass/charge ratio.
The differences between \ac{YFS} and the shower are also 
greater for the bare electrons than for the bare muons, 
which is due to differences 
in how the \ac{IR} cutoff is implemented in each case.
It is clear that dressing the leptons with photons 
eliminates most of the differences between the \ac{YFS} and 
\ac{QED} shower methods, since only the kinematic impact 
of the hardest photons (or semi-soft wide-angle photons) remains.
We will study the interplay of \ac{QED} radiation and lepton 
definitions further in chapter \ref{chapter:yfs}.

In this section we have presented results from the final-state \ac{QED} 
parton shower. An initial-state \ac{QED} shower is not usually 
phenomenologically relevant at the \ac{LHC}, although as 
described in section \ref{sec:qedps:construction}, all initial-initial 
and initial-final dipoles are implemented in the shower 
presented here. However, initial-state \ac{QED} radiation 
is of the utmost importance in lepton-lepton colliders.
Before concluding this chapter, we outline important 
considerations when implementing the initial-state \ac{QED} parton 
shower for electron-positron colliders.

%%%%%%%%%%%%%%%%%%%%%%%%%%%%%%%%%%%%%%%%%%%%%%%%%%%%%%%%%%%%%%%%%%

\section{The initial-state QED parton shower for $e^+ e^-$ colliders}
\label{sec:qedps:ISmethods}

For electron-positron colliders, it is essential to model 
initial-state radiation both for luminosity measurements and for 
processes of interest. For example, at low-energy $e^+ e^-$ 
colliders, the measured rate of $e^+ e^-$ to hadrons is used to 
extract the hadronic vacuum polarisation via the optical theorem
\cite{Aoyama:2020ynm}.
Understanding these processes requires precise modelling of the 
\ac{QED} radiation produced from the initial and final states.
At future high-energy $e^+ e^-$ colliders such as the FCC-ee, 
the precision will be high enough such that a simple structure 
function approach is 
unlikely to be sufficient to describe the makeup of the initial 
states. In both these cases, the \ac{YFS} approach can be used
\cite{Krauss:2022ajk}. 
However, we stand to benefit significantly from the resummation of 
collinear logarithms, which are formally of the same size as soft 
logarithms. We have seen in this chapter that they can be very 
important in practice, too, especially when the centre-of-mass energy 
is far from a resonance (e.g. some low-energy colliders and 
any future collider above the $Z$ mass). \ac{NLO} corrections, and 
possibly even higher orders, are also necessary to reach the precision
of future experiments. There have been many tools developed for 
\ac{NLO} parton shower matching for \ac{QCD} initial states, from which 
we can take inspiration. These ideas will be discussed further in 
chapter \ref{chapter:mcatnlo}, but first we discuss the necessary 
prerequisite of a parton shower for electron-positron initial 
states.

\subsection{The electron structure function} 
\label{sec:qedps:electronSF}

In \ac{QCD}, to obtain the \acp{PDF} for hadrons, they
must be fitted to data and then evolved numerically using the 
\ac{DGLAP} equations. In \ac{QED}, obtaining a
structure function for the electron is more straightforward.
The \ac{DGLAP} equations for \ac{QED} can be solved exactly to 
\ac{LL} accuracy, using the initial condition 
\begin{equation}
  f_e(x,0) = \delta(1-x),
\end{equation}
for the electron or positron (in pure \ac{QED} they are identical
to all orders; in the \ac{EW} theory they are identical at \ac{LO}).
With the inclusion of approximate higher-order corrections and 
a modification of the collinear exponentials to take into account 
soft radiation, this results in the \ac{LL} structure function
\begin{equation} \label{eq:qedps:electronSF}
  f_e(x,Q^2) = \beta \,\frac{\exp{\left(-\gamma_E \beta + 
  \frac{3}{4} \beta_S \right)}}{\Gamma(1+\beta)} (1-x)^{\beta-1} 
  + \beta_H \sum_{n=0}^\infty \beta^n_H \mathcal{H}_n(x),
\end{equation}
where $\beta = \frac{\alpha}{\pi} \left(\log{\left(\frac{Q^2}{m_e^2}\right)} - 1\right)$.
This form of $\beta$ comes from integration over all phase space 
available for the emission of soft photons in the \ac{YFS} framework,
thus allowing the resummation of soft radiation effects in the 
collinearly-derived structure function.
The soft photon residue $\beta_S$ can 
be set either to $\beta$, or to $\eta = \frac{\alpha}{\pi} 
\log{\left(\frac{Q^2}{m_e^2}\right)}$. The $\beta_H$ logarithms 
which come with the hard coefficients $\mathcal{H}_n$ have the 
same freedom: one can choose $\beta_H=\beta_S$, or define a 
mixed scheme in which $\beta_S=\beta,\, \beta_H=\eta$
\cite{Montagna:1994qu,Krauss:2022ajk}. 

Note that beyond \ac{LO}+\ac{LL} accuracy, the initial 
condition is a mixed electron-photon state. The \ac{NLO} \acp{PDF} 
have been recently derived in the one-flavour \ac{QED} case
\cite{Frixione:2012wtz,Bertone:2019hks,Frixione:2019lga}. 
These are expected to be implemented in a future version of 
\Madgraph \cite{Alwall:2014hca}. When these \acp{PDF} are extended to the case of 
multiple fermions, and can be implemented in an event generator 
in a similar way to hadron \acp{PDF}, this will allow for fully 
\ac{NLO}-accurate simulations of events at $e^+ e^-$ colliders.
With the present \ac{LO}+\ac{LL} structure function, any
calculation is formally limited by the structure function accuracy.
This will be an active area of future research.

The \ac{LL} structure function, eq.\ \eqref{eq:qedps:electronSF},
has an integrable singularity at $x=1$, since \changed{$0<\beta<1$}. 
This means that provided the 
observable of interest does not diverge as $x \to 1$, the result 
after integration over $x$ will be finite, despite the singularity.
However, \ac{MC} integration and event generation rely on sampling 
from $f_e(x)$,
and any numerical algorithm must have an upper limit $x_{\mathrm{max}}$.
For convenience, in the following, we introduce a small positive number 
$\epsilon$, defined as $x_{\mathrm{max}}=1-\epsilon$.
However, no matter how small $\epsilon$ is taken, significant contributions 
to the total cross section are missed, due to the singularity in
$f_e(x)$. 

This problem is traditionally solved by using the analytic knowledge 
of the singularity structure to rescale the structure function 
near the numerical upper limit, to ensure the correct total 
cross section is obtained. We define a second small number $\delta$,
such that $\delta > \epsilon$. Then, assuming that the partonic 
cross section is roughly flat in $x$ for $x>1-\delta$, we can write 
\begin{equation}
  \int_{1-\delta}^{1-\epsilon} \diff x \,\lambda(\epsilon,\delta) \, f_e(x,Q^2)
  = \int_{1-\delta}^1 \diff x \,f_e(x,Q^2),
\end{equation}
introducing the scale factor $\lambda(\epsilon,\delta)$. Using the 
asymptotic form of eq.\ \eqref{eq:qedps:electronSF} and solving 
for $\lambda$, we obtain 
\begin{equation}
  \lambda(\epsilon,\delta) = \frac{(\delta/\epsilon)^\beta}
  {(\delta/\epsilon)^\beta - 1},
\end{equation}
where $\beta$ is evaluated at $Q^2 = s$. Then the structure function 
$W^\lambda_e(x,Q^2)$ used in \ac{MC} integration is 
\begin{equation} \label{eq:qedps:weightedSF}
  W_e^\lambda(x,Q^2) =
  \begin{cases}
      f_e(x,Q^2) & 0 \leq x \leq 1-\delta \\
      \lambda(\epsilon,\delta) f_e(x,Q^2) & 1-\delta < x < 1-\epsilon \\
      0 & \text{else.}
  \end{cases}
\end{equation}

\begin{figure}
  \centering
  \includegraphics[width=0.7\textwidth]{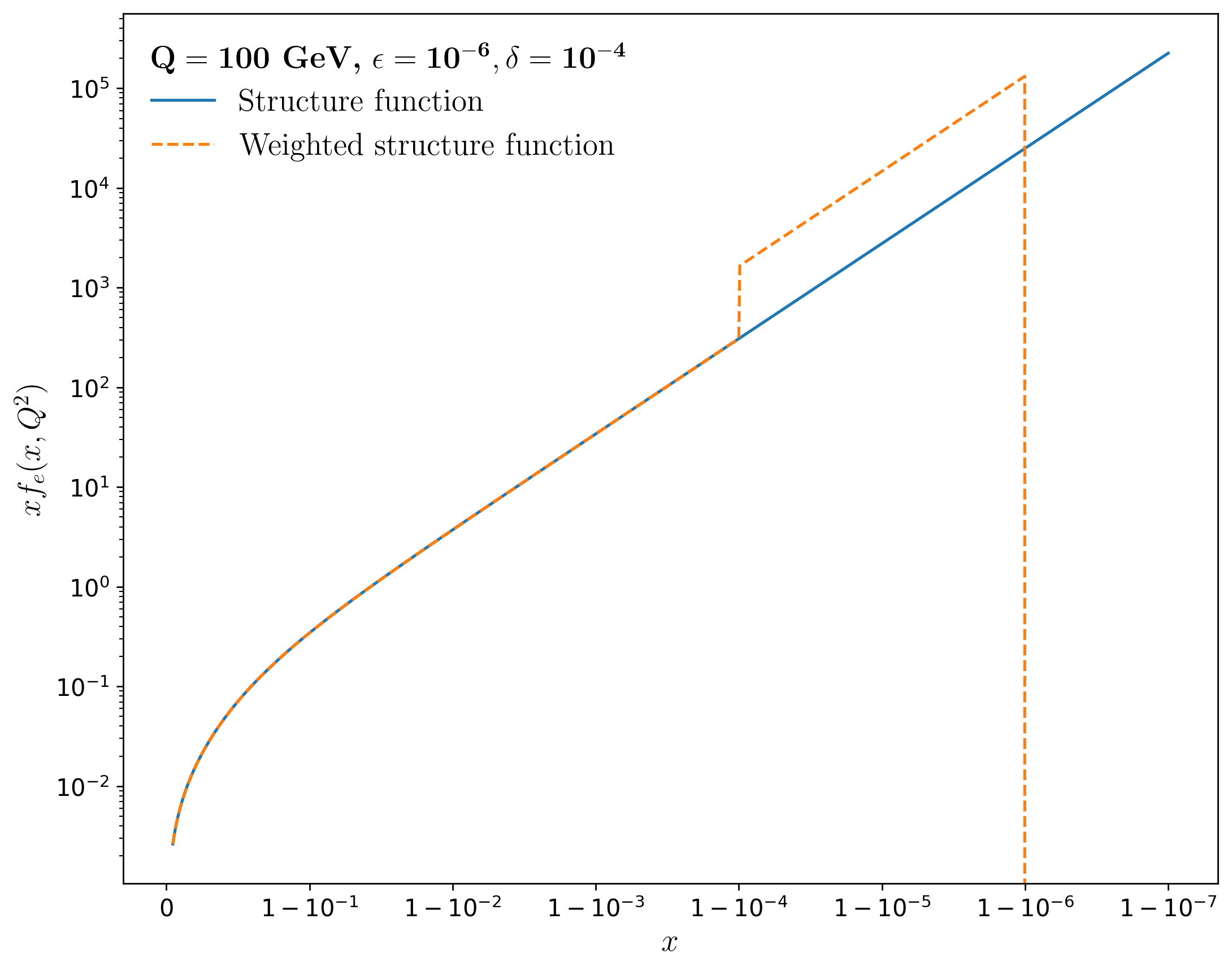}
  \caption
  {The LL electron structure function
  at a scale 100 GeV, compared to the weighted structure 
  function usually used for MC integration and event 
  generation.
  \label{fig:qedps:electronSF}
  }
\end{figure}

Figure \ref{fig:qedps:electronSF} shows the logarithmic 
behaviour of the structure function as $x \to 1$, at 
a scale $Q=100 \,\GeV$.
The weighted structure function,
eq.\ \eqref{eq:qedps:weightedSF}, is also shown, for 
the choices $\epsilon=10^{-6},\delta=10^{-4}$. We can 
see that the weighted structure function is discontinuous 
at the point $x = 1-\delta$, and therefore is no longer a valid 
solution of the \ac{DGLAP} equations. If the weight 
$\lambda(\epsilon,\delta)$ is also applied in the parton 
shower, this will lead to an overestimate of emissions 
into the region $1-\delta < x < 1-\epsilon$, which is 
unphysical.

We can solve some of these problems by instead considering 
the unweighted structure function with a cutoff $\epsilon$,
and defining a component proportional to $\delta(1-x)$ 
which contains the missing cross section contributions. 
This is defined as 
\begin{equation} \label{eq:qedps:deltaSF}
  W_e^\delta(x,Q^2) =
  \begin{cases}
      f_e(x,Q^2) & 0 \leq x \leq 1-\epsilon \\
      0 & 1-\epsilon < x < 1 \\
      C \delta(1-x) & x=1.
  \end{cases}
\end{equation}
Solving for $C$ in the same way as above, again 
approximating $Q^2=s$,
we obtain $C = \epsilon^{\beta/2}$.

This method, rather than smearing the missing cross section 
contributions across the high-$x$ region, adds it into the 
$x=1$ region. This has the advantage that a whole kinematical 
region is not affected by the overweighting. It is also more 
physically meaningful when considering photon emissions, since 
we define all emissions within the $1-\epsilon<x<1$ region to 
be unresolvable, both in terms of detectability and in their 
kinematical effects. Therefore, they are combined with the 
$x=1$ no-emission piece. This implementation is less convenient, 
however, since the phase space has four distinct regions in 
$e^+ e^-$ collisions: 
\begin{itemize}
  \item $x_1<1-\epsilon,\,x_2<1-\epsilon$
  \item $x_1<1-\epsilon,\,x_2=1$
  \item $x_1=1,\,x_2<1-\epsilon$
  \item $x_1=1,\,x_2=1$.
\end{itemize}
Therefore, the event generation must be completed in 
four phases. In \Sherpa, this process has not been automated.

Both methods for dealing with the integrable divergence 
in the electron structure function work well for \ac{LO}
computations. However, neither solve some of the problems 
which arise from a unitary parton shower acting on the 
electron-positron initial state.

\subsection{Implementation of a QED initial-state shower}

There are a few difficulties with implementing the \ac{QED} 
parton shower, as described in section \ref{sec:qedps:construction},
for the case of an $e^+ e^-$ collider.

One challenge lies in embedding the structure function into the 
veto algorithm (\see sec. \ref{sec:qedps:veto}). 
The parton shower splitting function 
includes the ratio of particle luminosities 
$\mathcal{L}(r_{\ijt,\kt}(\{a\}))/\mathcal{L}(\changed{\{\tilde{a}\}})$. In the 
initial-state splitting $e^{\pm} \to e^{\pm} \gamma$, both the 
numerator and the denominator of this fraction contain the \ac{LL}
electron structure function:
\begin{equation} \label{eq:qedps:luminosityRatio}
  \frac{\mathcal{L}(r_{\tilde{aj},\tilde{b}}(\{a\}))}{\mathcal{L}(\changed{\{\tilde{a}\}})}
  = \frac{\frac{x}{z_{aj,b}}f_e\left(\frac{x}{z_{aj,b}},Q^2\right)}
  {xf_e(x,Q^2)},
\end{equation}
where $z_{aj,b}$ is the splitting variable defined in eq.\ 
\eqref{eq:qedps:IIx}.
We can see from this equation that the splitting function 
diverges for $z_{aj,b} \to x$. In addition, we can see that as $x \to 1$, 
the emission phase space becomes restricted to very soft emissions, i.e. 
$z_{aj,b}$ must be very close to 1, for the argument of the structure 
function to be less than unity. 

The fact that this ratio contains an additional divergence is 
not the only difficulty. The veto algorithm requires an 
overestimate $g(t)$ of the splitting function, which is needed 
before any splitting kinematics are computed. This means 
we need an expression which overestimates eq.\ \eqref{eq:qedps:luminosityRatio}
which is independent of $z_{aj,b}$. The solution to this appears
straightforward: applying the same $x_\mathrm{max}$ logic as before
leads to an $x$-dependent overestimate given by
\begin{equation} \label{eq:qedps:overestimate}
  g(t) \supset \frac{x_\mathrm{max} f_e(x_\mathrm{max},Q^2)}
  {xf_e(x,Q^2)}.
\end{equation}
We would like to include the effect of radiation which, in a 
backwards-evolution picture, takes the initial electron/positron 
line into the $x_\mathrm{max} < x < 1$ region. This radiation is 
not restricted to be soft, so may be very relevant 
to a range of observables. We can achieve this 
to a first approximation by allowing the generation of $z_{aj,b}>x/x_\mathrm{max}$.
However, if the current $x$ is above $x_\mathrm{max}$, further 
emissions are forbidden. These further emissions must be soft, since 
$z_{aj,b}$ is restricted to be very close to unity, so are unlikely 
to be relevant phenomenologically.
This choice allows the dominant effects 
of emission into the kinematically favoured $x \to 1$ region to 
be included. Eq.\ \eqref{eq:qedps:overestimate}
is not always strictly an overestimate in this case, but if 
this is problematic in practice, it is always possible to 
reweight it.

Correctly overestimating the splitting function introduces 
another difficulty, however.
The veto algorithm requires the new scale $t$ to be calculated
according to the distribution of $g(t)$,
as in eq.\ \eqref{eq:qedps:generatet} and eq.\ \eqref{eq:qedps:practicalGeneratet}.
For any sensible (small) value of $\epsilon$, this calculation is 
likely to occasionally produce a new scale which is, in floating-point 
arithmetic, indistinguishable from the current scale. When this happens, 
the algorithm will stall. 
By imposing a limit on the size of the overestimate 
used to generate $t$, one can produce radiation using the shower. 
As expected, however, a dependence on the form of the overestimate 
and on $\epsilon$ remains, so the results cannot be trusted.

To avoid the divergence as $z_{aj,b} \to x$, we can flatten the 
function which is sampled using standard \ac{MC} techniques
\cite{Sjostrand:2006za}. 
By sampling logarithmically in $x-z_{aj,b}$, we flatten the first-order 
$1/(x-z_{aj,b})$ pole.

To deal with large overestimates in the generation of new scales, 
which will be necessary even
with the above modification since $f_e(x_\mathrm{max},Q^2)$ is 
very large numerically,
we can use the method of analytic weights (\see section 
\ref{sec:qedps:weightedveto}). However, this method has not been used 
before for large overweights and so should be stringently tested.
It will also have a negative impact on \ac{MC} statistics, as 
with any method which introduces large weight fluctuations between 
events.

The results of these new developments in modelling the $e^+ e^-$
initial state radiation will be detailed in a 
future publication, and made available in a future release of 
\Sherpa.

%%%%%%%%%%%%%%%%%%%%%%%%%%%%%%%%%%%%%%%%%%%%%%%%%%%%%%%%%%%%%%%%%%
\section{Conclusions}
\label{sec:qedps:conclusion}

In this chapter we have presented novel results from a
QCD+QED final-state parton shower. After outlining the 
construction of a parton shower and detailing QED-specific 
considerations, we validated our 
implementation using a test process 
$\nu_\mu \bar{\nu}_\mu \to e^+ e^-$ at two different 
energies: just above the $s$-channel resonance and far 
above it. We saw that the parton shower produces a radiation 
pattern which is in good agreement with that produced 
by the \ac{YFS} soft-photon resummation, when the latter is 
supplemented with exact \ac{NLO} corrections.

We also presented results showing the applicability of 
our method to phenomenological studies. We studied the 
production of a Higgs boson via gluon fusion and its 
decay to two or four leptons. We found that for $H \to \mu^+ \mu^-$,
the \ac{QED} shower predictions agreed with the \ac{YFS}
predictions to within 1\% for transverse observables and 
within 5\% for the Higgs invariant mass. Similarly, for 
the four-lepton final state, the transverse momentum 
distributions of the hardest lepton and hardest $Z$ were 
in very good agreement. The differences in the Higgs invariant 
mass were larger for this decay, but localised, as expected, 
just below the resonant $H$ pole. The interleaved shower 
presented in this chapter thus produces useful predictions, for 
a range of processes, which are consistent with other methods for 
calculating \ac{QED} radiative corrections.
We also validated the use of a primary dipole identification 
method to preserve resonance lineshapes and improve efficiency.

Finally, we discussed the differences which arise when 
considering an electron-positron collider. We presented the 
situations in which the current algorithm is insufficient, 
and outlined methods to solve these difficulties. This will 
be an important topic of future research, becoming ever more relevant as 
future collider plans are confirmed.

In the next chapter, we extend our interleaved QCD+QED shower 
by matching it to \ac{NLO}. We outline the proof of the \MCatNLO
method for \ac{QED} and for QCD+QED, and discuss details of the 
implementation. We will show that matching to \ac{NLO} improves 
the accuracy of our shower perturbatively while retaining the 
resummation accuracy of the parton shower which we have 
demonstrated in this chapter.

The interleaved dipole shower was implemented 
in the event generator \Sherpa and will be incorporated 
in a future release in the \Sherpa 3 series. All analyses 
and plots were made using \Rivet \cite{Buckley:2010ar,Bierlich:2019rhm}.

%% file: text/mcatnlo.tex
\chapter{QED MC@NLO}
\label{chapter:mcatnlo}

\section{Introduction to NLO matching} \label{mcatnlo:sec:intro}

As was described in section \ref{sec:intro:mceg:NLO}, there are 
various difficulties in computing \ac{NLO} observables numerically, 
but these have been solved by using phase-space slicing or 
subtraction schemes. In this chapter we will review the solution 
to the additional problem of incorporating these \ac{NLO} calculations 
into an event generator, in particular, of combining \iac{NLO}
calculation with a parton shower.

Around the turn of the 21st century, \ac{LO} \ac{QCD} plus parton shower 
predictions were insufficient to describe the multi-jet final states 
being produced at colliders. As a result, \ac{LO} parton shower 
merging was developed \cite{Catani:2001cc,Lonnblad:2001iq}, which 
allows one to describe each additional jet with its own full matrix 
element correction to the parton shower approximation. However, for 
many processes including Higgs production via gluon fusion, large 
$K$-factors were observed in comparisons to data, indicating that 
\ac{NLO} virtual corrections are non-negligible in many cases. This 
led to the development of matching methods. 

The idea of \ac{NLO} parton shower matching is to produce a prediction 
for an observable $\langle O \rangle$ which contains the parton 
shower factor (containing emission probabilities and Sudakov form 
factors), but which reproduces the correct \ac{NLO} value for the 
observable, i.e.
\begin{equation}
    \langle O \rangle^\text{Matched} = \langle O \rangle^\text{NLO}
    + \mathcal{O}(\alpha^{m+2}),
\end{equation}
for an observable which is $\mathcal{O}(\alpha^m)$ at \ac{LO}.

As was explained in detail in chapter \ref{chapter:qedps}, a parton 
shower resums the leading logarithms associated with the emission of 
soft or collinear radiation, whilst assuming perfect cancellation 
of both the \ac{IR} singularities and the finite parts of the 
\ac{NLO} contribution to an observable. Contrary to analytic resummation, 
a parton shower produces exclusive final states containing a 
number of additional particles. The radiation pattern is, however, 
formally only correct in the soft and collinear limits, although the 
choice of finite parts of the splitting functions can improve the 
description away from these limits. \Iac{NLO} matched
parton shower, on the other hand, describes the first emission 
radiation pattern at \ac{LO} everywhere in phase space. This 
is interpolated with the parton shower resummed prediction in the 
soft and collinear regions. Importantly, in contrast to LO merging 
predictions, the total cross section is the correct \ac{NLO} one and 
the \ac{NLO} $K$-factor is present in observable predictions. 
\changed{The $K$-factor is defined by $\sigma^\mathrm{LO}(1+K)=\sigma^\mathrm{NLO}$,
sometimes called a global $K$-factor, and usually quoted as a percentage.} 
In \ac{MC} event generators, \iac{NLO} matched parton shower allows us to 
carry out a full simulation of events at this accuracy, including 
hadronisation, hadron decays, multiple interactions and underlying 
event.

The \MCatNLO \cite{Frixione:2002ik} and \Powheg \cite{Nason:2004rx} 
methods were developed to solve the problem of double-counting 
the first emission, while exponentiating the appropriate parton 
shower factors. Both methods utilise so-called additive matching, 
but some newer matching methods use multiplicative matching, 
for example {\sc KrkNlo} \cite{Jadach:2015mza}, or 
a combination of both types. In this chapter, we will focus 
exclusively on the \MCatNLO method, since it 
only exponentiates the \ac{IR}-singular parts of the real 
correction. This is an advantage over \Powheg, which sometimes 
suffers from artificial local
$K$-factors caused by exponentiating large hard real corrections. 
However, \MCatNLO has the disadvantage that negative 
event weights occur when the parton shower overestimates 
the \ac{NLO} real-emission pattern. This most often occurs when 
the shower is leading-colour and is matched to full-colour \ac{NLO}
\cite{Danziger:2021xvr}. However, 
we will use weighted events to produce distributions shown 
here; in addition, we do not need to pass events through a detector simulation, 
and more events can be generated to overcome the statistical 
inefficiencies caused by negative weights. For a detailed review of the 
strengths and drawbacks of \Powheg and \MCatNLO, see 
ref. \cite{Hoeche:2011fd}.

In the following section, we will outline the \MCatNLO method and 
show that the predictions it produces have the desired properties.
We will then summarise some subtleties in the implementation of 
this method, in particular with regards to \ac{QED}. We present test results 
from a hypothetical neutrino collider in sec. \ref{sec:mcatnlo:results:neutrinos}.
In section \ref{sec:mcatnlo:results:ggHleptons} we present results 
for a QCD+QED \MCatNLO for Higgs production via gluon fusion, where 
the \ac{EW} corrections are effected to the Higgs decay to 
leptons. Finally, we conclude in sec. \ref{sec:mcatnlo:conclusions}.

\section{Methods}

\subsection{The \MCatNLO method in detail}

In this section we will show how, by applying a parton 
shower to a subset of \iac{NLO} calculation, we can generate 
a formula for the observable which reproduces its \ac{NLO} value.
This calculation follows the original \MCatNLO paper, ref. \cite{Frixione:2002ik}.
We will begin by outlining the process symbolically, suppressing 
flavour sums, splitting indices, and phase space mappings. The latter 
are given in equations \eqref{eq:intro:PSmapping} and \eqref{eq:intro:PSmappingInverse}.

The expectation value of \iac{IR-safe} observable at \ac{NLO} 
can be written, with subtraction terms $D^S$, as 
\begin{align*} \label{eq:mcatnlo:Onlo}
    \Oangle{NLO} = &\intdphi{n} \left[B+\tilde{V}+I^S\right] O(\Phi_n) \\
    &+ \intdphi{n+1} \left[ R \, O(\Phi_{n+1})-D^S \,O(\Phi_n)\right]. \numberthis
\end{align*}
\changed{Before we can extend this to \ac{NLO} plus parton shower 
(NLOPS) accuracy, it will be helpful to reformulate eq.\ \eqref{eq:mcatnlo:Onlo}}
by adding and subtracting an additional set of subtraction 
terms $D^A$, as 
\begin{align*} \label{eq:mcatnlo:expNLO}
    \Oangle{NLO} = \intdphi{n} \bar{B} \,O(\Phi_n) 
    + \intdphi{n+1} \left[ R \, O(\Phi_{n+1})- D^A\,O(\Phi_n)\right], \numberthis
\end{align*}
where \changed{we define a $\bar{B}$ term containing the Born and subtracted 
virtual terms, in addition to the finite combination of both subtraction 
terms $D^A-D^S$, i.e.}
\begin{equation} \label{eq:mcatnlo:bbar}
    \bar{B} = B+\tilde{V}+I^S + \intdphi{1} \left[D^A-D^S\right].
\end{equation}
As was argued in ref. \cite{Frixione:2002ik}, generating the 
$R$ and $D^A$ terms in eq.\ \eqref{eq:mcatnlo:expNLO} independently 
results in double counting, so we combine them into a 
subtracted real \ac{ME} squared, $R-D^A$, and 
separate out the observable dependence:
\begin{align*} \label{eq:mcatnlo:expNLOocorr}
    \Oangle{NLO} = &\intdphi{n} \bar{B} \,O(\Phi_n) 
    + \intdphi{n+1} \left[ R - D^A\right] O(\Phi_{n+1}) \\
    &+ \intdphi{n+1} D^A \left[O(\Phi_{n+1}) - O(\Phi_n)\right]. \numberthis
\end{align*}
We will see that the new subtraction terms $D^A$ must capture 
the soft and collinear parts of 
the real emission, which should be exponentiated by the parton
shower. 

The parton shower factor $\mathcal{F}_n$ acting on an 
$n$-particle state contribution to an observable $O$
can be written recursively as (eq.\ \eqref{eq:qedps:generatingFunctional})
\begin{equation}
    \mathcal{F}_n(\Phi_n,O) = \Delta_n(\mu_Q^2,t_c) \, O(\Phi_n)
    + \intdphi{1} \Delta_n(\mu_Q^2,t) \,\mathcal{K}_n \,\mathcal{F}_{n+1}(\Phi_{n+1},O),
\end{equation}
where $\mathcal{K}_n$ are the splitting kernels and 
\begin{equation} \label{eq:mcatnlo:sudakov}
    \Delta_n(\mu_Q^2,t') = \exp{\left(-\int_{t'}^{\mu_Q^2} \diff \Phi_1
    \mathcal{K}_n \right)}
\end{equation}
is the Sudakov form factor. 

As an aside,
note that throughout this 
section the phase space element $\diff \Phi$ is written 
symbolically as a single integral. 
Practically, we write 
the single-particle phase space element $\diff \Phi_1$ 
in terms of a quantity $t$ with units of energy squared, 
a dimensionless variable $z$ and the azimuthal angle $\phi$.
As in chapter \ref{chapter:qedps}, the phase space element 
is given by 
\begin{equation}
    \diff \Phi_1 = \frac{1}{16\pi^2} \diff t 
    \diff z \frac{\diff \phi}{2\pi} J(t,z,\phi),
\end{equation}
where $J(t,z,\phi)$ is the Jacobian. The variable 
$t$ was referred to as the parton shower evolution variable
in chapter \ref{chapter:qedps}.
The limits on the integral in the Sudakov form factor, eq.\ 
\eqref{eq:mcatnlo:sudakov}, refer to the integral over $t$
for the given emission.
$t_c$ is the \ac{IR} cutoff
of the parton shower, and $\mu_Q^2$ is the shower starting 
scale, sometimes called the resummation scale.

We would like to examine the effect of the parton shower 
on the \ac{NLO} observable. We will do this by expanding 
the formulae in $\alpha$, defining our observable $O$ to 
be $\mathcal{O}(\alpha^m)$ at \ac{LO}.
First, we expand the Sudakov factor
in $\alpha$,
\begin{equation}
    \Delta_n(\mu_Q^2,t) = 1 - \intdphi{1}
    \mathcal{K}_n + \mathcal{O}(\alpha^2),
\end{equation}
and therefore the parton shower factor expands as 
\begin{align*}
    \mathcal{F}_n(\Phi_n,O) = \; &O(\Phi_n) - \intdphi{1} 
    \mathcal{K}_n \, O(\Phi_n) \\
    &+ \intdphi{1} \left(1-\intdphi{1}
    \mathcal{K}_n \right) \, \mathcal{K}_n \left[ O(\Phi_{n+1})
    -\intdphi{1} \mathcal{K}_{n+1} \, O(\Phi_{n+1}) + \dots \right] \\
    = \; &O(\Phi_n) - \intdphi{1}\mathcal{K}_n \,O(\Phi_n)
    + \intdphi{1}\mathcal{K}_n \,O(\Phi_{n+1}) + \mathcal{O}(\alpha^2). \numberthis
\end{align*}

To relate this to the \ac{NLO} observable defined above, we let 
the splitting kernels be the newly-defined subtraction 
terms, $\mathcal{K}_n = D^A/B$.
Then the parton shower is applied to 
eq.\ \eqref{eq:mcatnlo:expNLOocorr},
\begin{equation} \label{eq:mcatnlo:mcatnloF}
    \Oangle{NLOPS} = \intdphi{n} \bar{B} \, \mathcal{F}_n(\Phi_n,O)
    + \intdphi{n+1} \left[ R-D^A \right] \mathcal{F}_{n+1}(\Phi_n,O). 
\end{equation}
Recalling that 
$\bar{B} = B + \mathcal{O}(\alpha^{m+1})$ and expanding 
in $\alpha$, we obtain 
\begin{align*} \label{eq:mcatnlo:nlopsOalpha}
    \Oangle{NLOPS} = &\intdphi{n} \bar{B} \, O(\Phi_n) + \intdphi{n+1} 
    D^A \left[ O(\Phi_{n+1})-O(\Phi_n) \right] \\ 
    & + \intdphi{n+1} \left[R-D^A \right] O(\Phi_{n+1}) + 
    \mathcal{O}(\alpha^{m+2}), \numberthis
\end{align*}
i.e. the parton shower generates the third term of 
eq.\ \eqref{eq:mcatnlo:expNLOocorr}. Eq.\ \eqref{eq:mcatnlo:nlopsOalpha} 
is therefore accurate to \ac{NLO}.

To obtain a clearer physical picture, we can expand 
the parton shower factor in eq.\ \eqref{eq:mcatnlo:mcatnloF}
in terms of its Sudakov factor $\bar{\Delta}^A$,
\begin{align*} \label{eq:mcatnlo:nlops}
    \Oangle{NLOPS} = &\intdphi{n} \bar{B} \, \left[ \bar{\Delta}^A 
    \, O(\Phi_n) + \intdphi{1} \bar{\Delta}^A \, \frac{D^A}{B} \,
    O(\Phi_{n+1}) \right] \\ 
    &+ \intdphi{n+1} \left[ R-D^A \right] O(\Phi_{n+1}). \numberthis
\end{align*}
This is the main idea of the \MCatNLO method. However, there 
are many subtleties, and to see these it will be important 
to include the splitting sums and indices which we have 
neglected so far. The Sudakov factor $\bar{\Delta}^A$,
unlike the standard parton shower Sudakov factor, must be 
integrated over all phase space. It is given by 
\begin{equation}
    \bar{\Delta}^A(t') = \prod_{\ijt,\kt} \bar{\Delta}^A_{\ijt,\kt}(t'),
\end{equation}
where
\begin{equation}
    \bar{\Delta}^A_{\ijt,\kt}(t') = \exp{\left(-\sum_f \int \diff \Phi_1^{ij,k}
    \,\Theta(t_{ij,k}-t')\, S_{ij,k} \,\frac{D^A_{ij,k}}{B}\right)}.
\end{equation}
$S_{ij,k}$ is a ratio of symmetry factors, explained 
in ref. \cite{Hoche:2010pf}, and we have also included the sum 
over flavours. This encodes whether \ac{QCD} and/or \ac{QED} 
subtraction terms are being considered. 
Eq.\ \eqref{eq:mcatnlo:nlops} 
is, more precisely, 
\begin{align*} \label{eq:mcatnlo:mcatnlo}
    \Oangle{NLOPS} = &\sum_f \intdphi{n} \bar{B}\, \left[
    \bar{\Delta}^A(t_c) 
    \, O(\Phi_n) \right.\\
    &\left.+ \sum_{\ijt,\kt}\sum_f \int \diff \Phi_1^{ij,k}
    \,\Theta(t_{ij,k}-t_c) \,S_{ij,k} \,\bar{\Delta}^A_{\ijt,\kt}(t) \, 
    \frac{D^A_{ij,k}}{B} \,
    O(\Phi_{n+1}) \right] \\ 
    &+ \intdphi{n+1} \left[ R- \sum_{ij,k} D^A \right] O(\Phi_{n+1}). \numberthis
\end{align*}
The $\Theta$-functions ensure the uniqueness of the first 
emission and define a starting scale for the 
parton shower. 

The \MCatNLO method as implemented within \Sherpa,
referred to henceforth as the \SMCatNLO method, makes the 
choice $D^A_{ij,k}=D^S_{ij,k} \,\Theta(\mu_Q^2-t_{ij,k})$. 
Modified versions of the \ac{CS}
subtraction terms are used both for the \ac{NLO} subtraction 
and the parton shower evolution. While in \ac{QED} the 
weighted veto algorithm (see sec. \ref{sec:qedps:weightedveto}) is 
used in the ordinary shower to deal with negative charge 
correlators, in the \SMCatNLO method it is needed even in 
\ac{QCD} to deal with the exponentiation of subtraction terms 
which are negative due to subleading colour or spin effects.
However, beyond the first emission, the usual spin-averaged 
and leading-colour shower can be used to generate further 
emissions. The \SMCatNLO method has the advantage that to 
generate the unresolved emission term proportional to 
$\bar{B}$, the parton shower kernels do not need to be 
integrated over the one-particle phase space. Integrating 
the remaining $D^S$ is then easy since the $I^S$ are known
analytically.

In the next section we will briefly consider the implementation 
of the \SMCatNLO method. We will introduce terminology and 
discuss the modifications needed for \iac{QED} matching, and for 
mixed matching to \ac{NLO} QCD+QED.

\subsection{Implementation} \label{sec:mcatnlo:implementation}

In an event generator, to generate events according to the 
\MCatNLO equation \eqref{eq:mcatnlo:mcatnlo}, we use 
eq.\ \eqref{eq:mcatnlo:mcatnloF}, shown in the previous 
section to be \ac{NLO} accurate. A seed event is generated 
either using $\bar{B}$ (a so-called soft event or $\mathbb{S}$-event)
or $R-D^A$ (a hard or $\mathbb{H}$-event). $\mathbb{S}$-events are 
passed to a 
one-step parton shower where the splitting kernels are 
either exactly equal to $D^A/B$, or are reweighted to this value
using the weighted veto algorithm. An emission may or may not 
occur, generating the first and second terms of eq.\ \eqref{eq:mcatnlo:mcatnlo}
respectively. The event is then passed to the standard parton 
shower to generate further emissions, starting from the scale 
of any emission that occurred. If an $\mathbb{H}$-event is selected, 
it already has real-emission kinematics, and is thus passed 
directly to the standard shower. This generates the 
third term of eq.\ \eqref{eq:mcatnlo:mcatnlo}.
For further details of the implementation of the \MCatNLO method 
in \Sherpa, see refs. \cite{Hoeche:2011fd,Hoeche:2009xc}.
However, in this section we will discuss some of the 
implementation details which are relevant for \ac{QED}.

\paragraph*{Scale choice.} The choice of scales in 
\ac{QED} \MCatNLO is relevant despite the slow running of 
$\alpha$ compared to that of $\alpha_s$. 
As has been argued \cite{Hoeche:2011fd,Nagy:2003tz}, 
a lack of phase-space restrictions on the exponentiated
($D^A$) part of the real emission can lead to 
logarithmic contributions to \MCatNLO of $\alpha \log^2{(q^2/s)}$,
where $q^2$ is the virtuality of the emitter particle, 
instead of the correct parton shower contribution of 
$\alpha \log^2{(q^2/\mu_F^2)}$. This implies that where 
$\mu_F^2 \ll s$, a careful choice of scale is needed. 
The scale chosen must be soft- and collinear-safe, so 
the invariant mass of bare charged particles cannot be used
to define the scale.

\paragraph*{Electron PDF.} The electron structure 
function and the full electron \ac{PDF} both provide 
an obstacle to initial-state showers and \MCatNLO, 
due to their divergent behaviour. Some techniques to 
control this behaviour have been described in section \ref{sec:qedps:electronSF}.
These involve using a cutoff to produce an 
overestimate which can be used for shower evolution,
and generating shower variables logarithmically to 
avoid the divergence.
The under-production of hard collinear radiation when 
a cutoff is used can cause additional problems in 
\MCatNLO compared to \iac{LO} parton shower, 
however. When the subtracted real correction is sizeable 
and negative, there is no room for an underestimate 
of the first shower emission, since the difference can 
become negative for observables which are \ac{LO}
in the first emission. Work is currently underway to minimise 
this issue.

\paragraph*{Running $\alpha$ and EW input scheme.} 
As described in sec. \ref{sec:qedps:runningAlpha}, 
external photons 
(whether produced in the hard interaction or by a \ac{QED} 
parton shower) should couple with $\alpha(0)$. This 
is completely consistent with our matching procedure, since 
differences between definitions of $\alpha$ enter beyond 
\ac{NLO}.
Here we make the choice to use a running $\alpha$ for the 
$\mathcal{O}(\alpha^{m+1})$ parts of the \ac{NLO} calculation 
and for further shower emissions, while using the $G_\mu$ 
scheme for the $\mathcal{O}(\alpha^m)$ parts.
Specifically, however, we use the coupling $\alpha(0)$ for 
photon emissions, since the majority of emitted photons 
will become long-distance objects, and hence should have 
the appropriate long-distance coupling. 
Correcting the \ac{NLO} calculation thus 
involves rescaling the virtual-Born interference term, 
the $D^A-D^S$ term, and the subtracted real correction 
with a factor $\alpha(0)/\alpha_\mu \approx 0.96$.

\section{Results}

In this section, we will present results produced using 
the \ac{QED} \MCatNLO method for charged lepton final 
states. More work is needed to adapt the \MCatNLO method 
for the case of an electron-positron collider due to 
the electron structure function, so here we will focus 
on validating the 
method for matching a final-state shower. To this end, 
we first use a test process $\nu_\mu \bar{\nu}_\mu \to e^+ e^-$
at 91.2 GeV and 500 GeV, then in section \ref{sec:mcatnlo:results:ggHleptons}
we will study the leptonic decays of a Higgs boson.

Throughout this section, we will compare the \ac{QED} 
\MCatNLO method with the \ac{LO} \ac{QED} parton 
shower and with the \ac{YFS} soft-photon resummation.
To produce the following results, 
\Sherpa's \Photons module was used \cite{Schonherr:2008av}. 
Exclusive $N$-photon 
final states are produced in the soft approximation 
using eikonal factors, and can be corrected either 
using collinear splitting functions or
exact higher-order soft-subtracted \acp{ME},
if the latter are specifically implemented. 
\changed{While the \ac{YFS} framework can incorporate higher-order 
corrections to any order, and \ac{NNLO} \ac{QED} corrections have 
been implemented in \Sherpa \cite{Krauss:2018djz}, here we use 
the publicly available \ac{NLO} \ac{EW} corrections in \Photons}
for the resonance decays presented here.
Note that in all cases the total cross section \changed{in the \ac{YFS} 
prediction} is not changed, 
and so the \ac{NLO} $K$-factor is not present.
Also, the extension of the \ac{YFS} algorithm to 
charged particle pair production is included \cite{Flower:2022iew}
(see chapter \ref{chapter:yfs} for a detailed description of 
the \ac{YFS} formalism and the extension).
It is also important to note that in the \ac{YFS} framework, 
all fermion masses must be taken into account to regulate 
the collinear divergences. The parton shower results presented 
here were also produced with finite lepton masses. The \MCatNLO
method can in principle also take into account all fermion masses. 
However, to aid in efficiency of calculating the virtual 
contributions, we have here set all fermion masses 
to zero in the \ac{NLO} parts of the calculation. The resummation 
includes the lepton masses.
For all results, \Amegic was used for the Born
\ac{ME} generation \cite{Krauss:2001iv} and \Comix was used 
for the real-emission \acp{ME} \cite{Gleisberg:2008fv}. 
The one-loop \acp{ME} for \MCatNLO were provided 
by \OpenLoops \cite{Buccioni:2019sur,Denner:2016kdg,Ossola:2007ax,vanHameren:2010cp}.
We use the $G_\mu$ 
\ac{EW} input scheme for the Born \ac{ME} calculation, and the 
running $\alpha$ scheme described in sec. \ref{sec:qedps:methods:implementation}
for the shower and the \ac{YFS} resummation. The \MCatNLO 
\ac{EW} scheme is described in sec. \ref{sec:mcatnlo:implementation}.

\subsection{Case study: $\nu_\mu \bar{\nu}_\mu \to e^+ e^-$}
\label{sec:mcatnlo:results:neutrinos}

As in chapter \ref{chapter:qedps}, we first isolate the 
\ac{QED} \MCatNLO from other parts of event generation by 
looking at a process $\nu_\mu \bar{\nu}_\mu \to e^+ e^-$. 
In this section we will present results from this process 
on the $Z$ pole (with a centre-of-mass energy of 91.2 GeV)
in addition to at higher energy, $\sqrt{s} = 500$ GeV.
\changed{In the latter case, where the 
invariant mass of the electron-positron pair (before 
any radiation) is far from the $Z$ mass, the on-shell $Z$
decay \ac{ME} is not usually appropriate. Hence, the \Photons 
module used to produce the \ac{YFS} prediction usually employs 
a collinearly-approximated \ac{NLO} correction. This would 
reduce its formal accuracy to the same as the parton shower.
However, it was observed that the collinearly-approximated 
correction to this process suffers from an overestimate of hard 
radiation, and does not reproduce the correct real-emission 
cross section to the desired accuracy. Therefore,
we use the $Z$ decay 
\ac{ME} to describe the process at 500 GeV.
This is justified here, since in this process the initial 
state completely factorises, there is no photon exchange 
diagram, and there are no three-vector-boson vertices.}

The majority of the cross section for this process resides at 
$\sqrt{s}=91.2$ GeV. The \ac{LO} 
cross section is  $\sigma_\text{LO} = 3950.30(2)$ pb and the 
\ac{NLO} total cross section is 
$\sigma_\text{NLO} = $ 3985.1(3) pb
resulting in a positive $K$-factor of 0.9\%.
At a centre-of-mass energy 
of 500 GeV, far above the $Z$ pole, the Born cross section
is much smaller, $\sigma_\text{LO} = 0.10524(1)$ pb, 
and there is a negative \ac{NLO} $K$-factor of 7.6\%
since $\sigma_\text{NLO} = $ 0.972197(1) pb.

\begin{figure}
    \centering
    \includegraphics[width=0.45\textwidth]{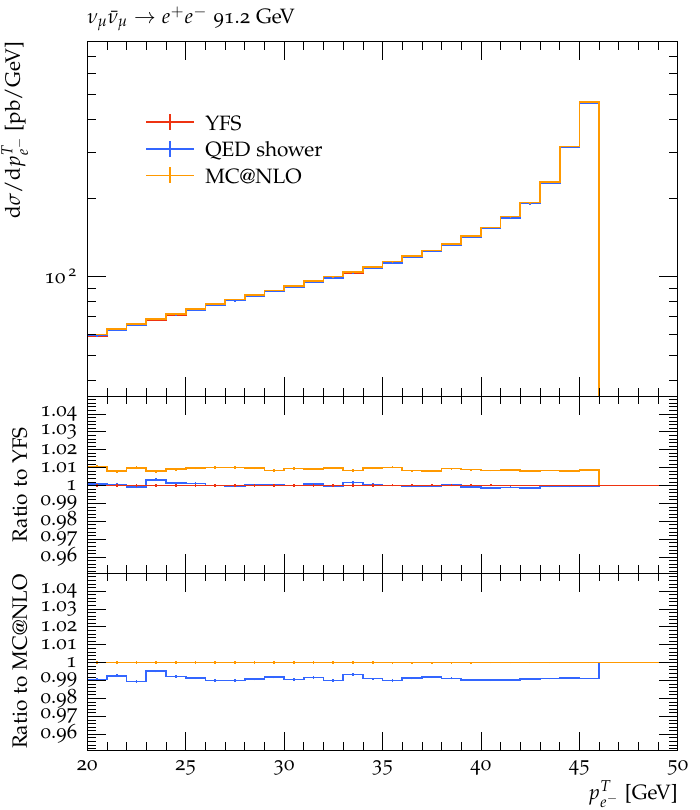}
    \includegraphics[width=0.45\textwidth]{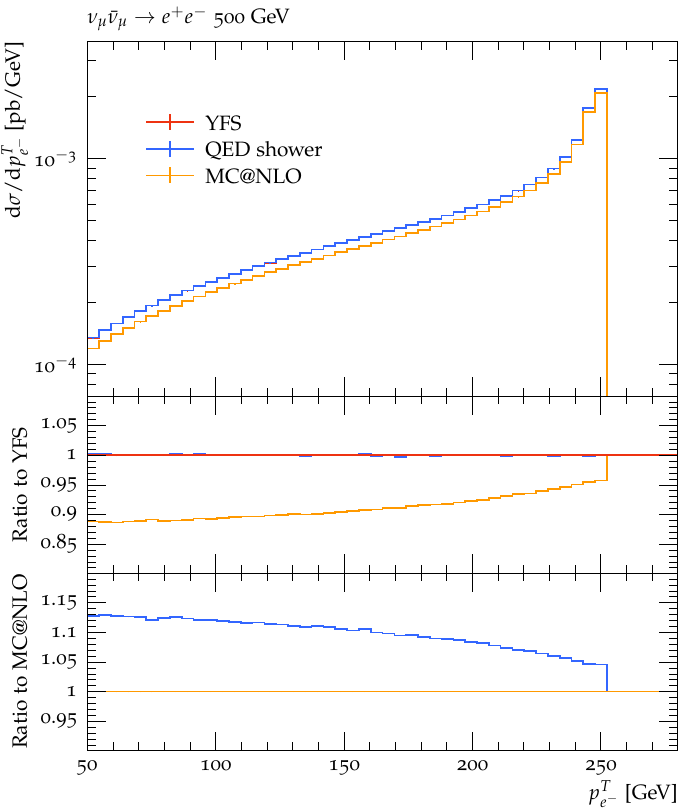}
    \caption[The electron transverse 
    momentum in $\nu_\mu \bar{\nu}_\mu \to e^+ e^-$,
    comparing the YFS soft-photon resummation with the LO 
    QED shower prediction and the QED \MCatNLO prediction.]{ 
    The electron transverse 
    momentum in $\nu_\mu \bar{\nu}_\mu \to e^+ e^-$, 
    comparing the YFS soft-photon resummation with the LO
    QED shower prediction and the QED \MCatNLO prediction for two 
    different collider energies. 
    \textbf{Left:} $\sqrt{s} = $ 91.2 GeV. 
    \textbf{Right:} $\sqrt{s} = $ 500 GeV. 
    The upper ratio plot shows the ratio to the YFS prediction 
    while the lower shows the ratio of the LO QED shower to the 
    QED \MCatNLO prediction.
    \label{fig:mcatnlo:electronkT}
    }
\end{figure}

First, fig.\ \ref{fig:mcatnlo:electronkT} shows a comparison 
of our \MCatNLO implementation with the \ac{QED} parton shower 
and with the \ac{YFS}
soft-photon resummation for the cross section differential in 
the electron transverse momentum. The left plot shows the 
electron $k_T$ for a collider energy of $\sqrt{s}=$ 91.2 GeV,
while the right plot is for $\sqrt{s}=$ 500 GeV.
We see that the only difference between the methods for 
the process on the $Z$ mass is the 
\ac{NLO} $K$-factor describing the difference in the 
total cross section. For 500 GeV, on the other hand, there 
is a shape difference between the methods, with the 
\MCatNLO producing comparatively more events with higher 
electron transverse momentum. Since this observable does 
not depend on real radiation except through small recoil 
effects, the \ac{YFS} predictions only have \ac{LO}
accuracy here. In the \ac{NLO} calculation, resonant 
structures are also present, for example due to resonant 
di-$Z$ production in a box diagram, but these are very small 
and not visible here.

\begin{figure}
    \centering
    \includegraphics[width=0.45\textwidth]{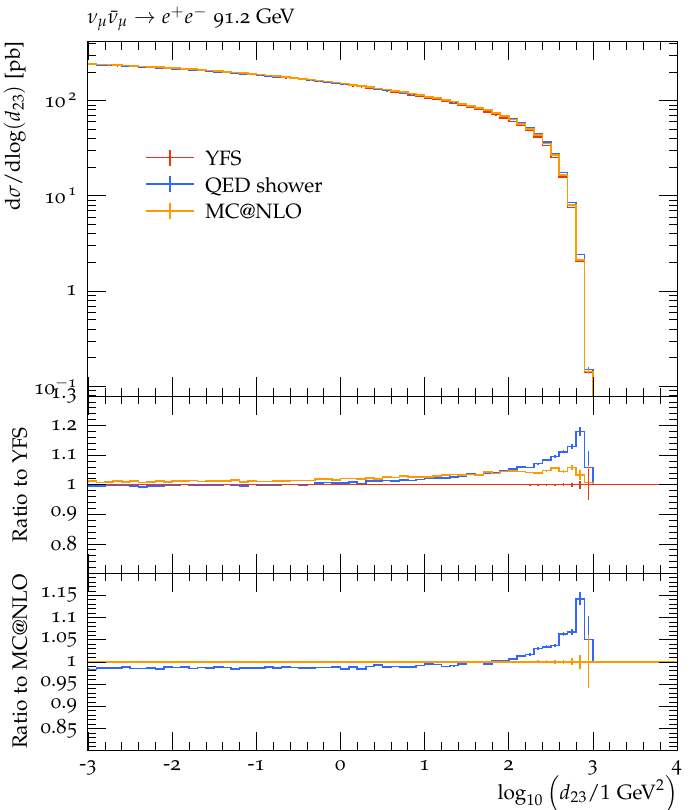}
    \includegraphics[width=0.45\textwidth]{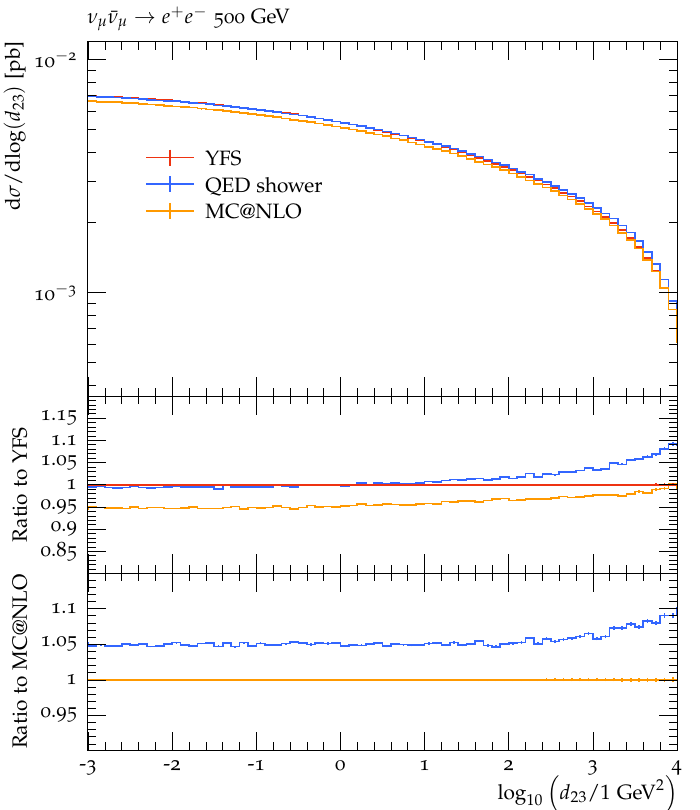}
    \caption[The 2-3 Durham jet rate in $\nu_\mu \bar{\nu}_\mu \to e^+ e^-$, 
    comparing the YFS method with the LO QED shower 
    prediction and the QED \MCatNLO prediction.]{
    The 2-3 Durham jet rate in $\nu_\mu \bar{\nu}_\mu \to e^+ e^-$, 
    comparing the YFS method with the LO QED shower 
    prediction and the QED \MCatNLO prediction for 
    two different collider energies.
    \textbf{Left:} $\sqrt{s} = $ 91.2 GeV. 
    \textbf{Right:} $\sqrt{s} = $ 500 GeV.
    The upper ratio plot shows the ratio to the YFS prediction 
    while the lower shows the ratio of the LO QED shower to the 
    QED \MCatNLO prediction.
    \label{fig:mcatnlo:jetrate}
    }
\end{figure}

Fig.\ \ref{fig:mcatnlo:jetrate} shows an observable 
which is \ac{LO} in the first emission, so here 
we expect to see significant shape differences between the four 
predictions. In this plot we show the cross section 
differential in the Durham $2\to 3$ jet rate, produced
by clustering \ac{QED} particles with a $k_T$ jet 
algorithm (for more details see sec. 
\ref{sec:qedps:results:neutrinos}).
We expect that the \MCatNLO result should 
give the full real-emission cross section for hard emissions,
and should tend to the shower-approximated result (multiplied 
by a $K$-factor) for soft or collinear emissions.
Comparing the \MCatNLO prediction 
with the \ac{YFS} prediction (upper ratio plot), 
there is agreement to within a few percent at all scales,
indicating that both predictions are \ac{NLO} accurate. However, 
the \MCatNLO approach produces slightly more hard radiation
than the \ac{YFS} method at both collider energies.
Looking at the lower ratio plot, we can see that the 
\MCatNLO corrects the overproduction of hard radiation 
that the \ac{LO} shower suffers from.

\begin{figure}
    \centering
    \includegraphics[width=0.45\textwidth]{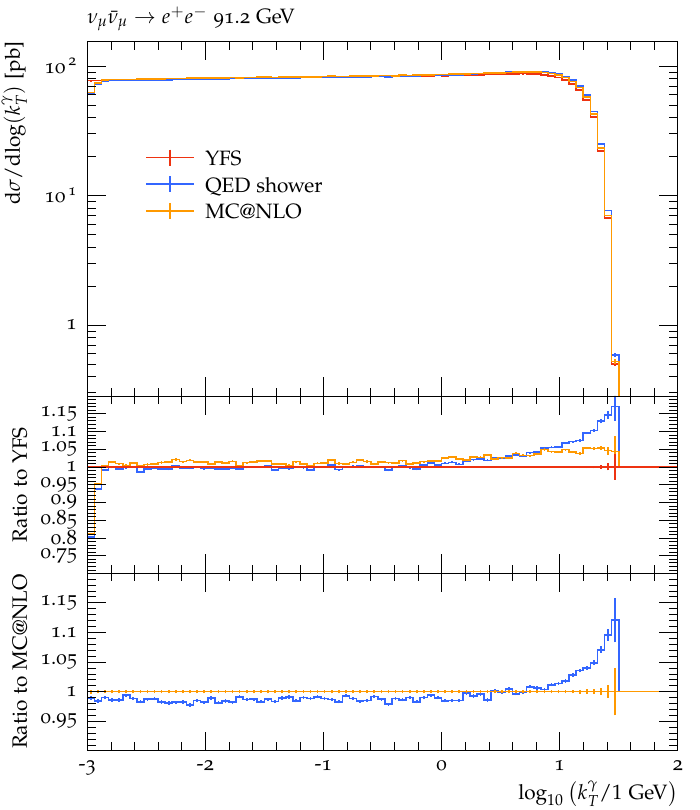}
    \includegraphics[width=0.45\textwidth]{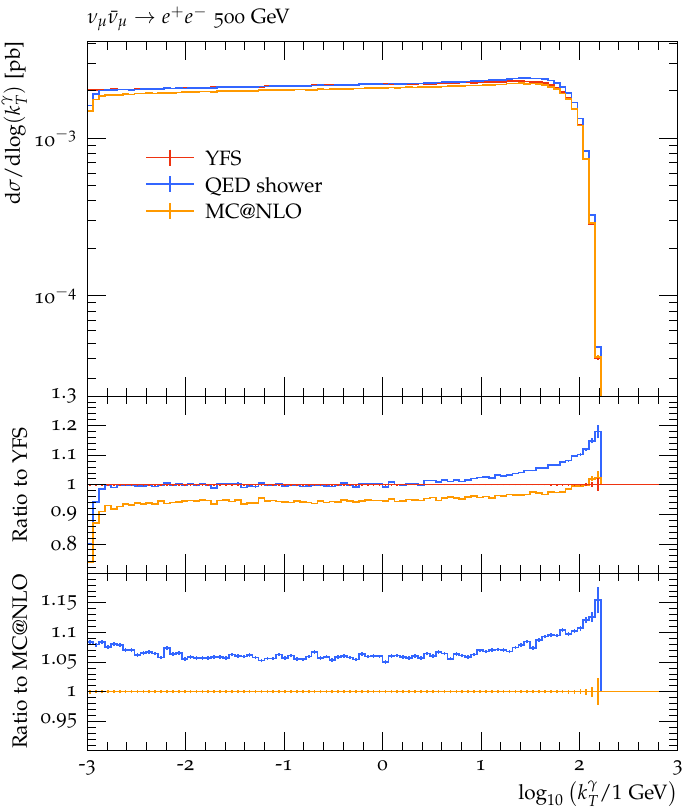}
    \caption[The third jet transverse momentum in $\nu_\mu \bar{\nu}_\mu \to e^+ e^-$, 
    comparing the YFS method with the LO QED shower 
    prediction and the QED \MCatNLO prediction.]{
    The third jet transverse momentum in $\nu_\mu \bar{\nu}_\mu \to e^+ e^-$, 
    comparing the YFS method with the LO QED shower 
    prediction and the QED \MCatNLO prediction for 
    two different collider energies.
    \textbf{Left:} $\sqrt{s} = $ 91.2 GeV. 
    \textbf{Right:} $\sqrt{s} = $ 500 GeV.
    The upper ratio plot shows the ratio to the YFS prediction 
    while the lower shows the ratio of the LO QED shower to the 
    QED \MCatNLO prediction.
    \label{fig:mcatnlo:thirdkT}
    }
\end{figure}

Similarly, fig.\ \ref{fig:mcatnlo:thirdkT} shows the 
third jet transverse momentum, produced using the same 
Durham jet algorithm. This observable has a strong 
probability of corresponding to the hardest photon transverse
momentum, so 
we denote it $k_T^\gamma$. Again, for both centre-of-mass energies,
both \ac{NLO}-accurate predictions agree up to a $K$-factor.
The shape differences caused by only including a 
collinear resummation, in the form of the \ac{QED} shower,
are corrected by the \MCatNLO method. It is clear that 
the \MCatNLO method 
reproduces the correct shape across a wide range of photon 
transverse momenta. The advantage of the \MCatNLO method over the 
matrix-element-corrected \ac{YFS} resummation is that 
off-the-shelf \ac{NLO} calculations can be used.

\begin{figure}
    \centering
    \includegraphics[width=0.45\textwidth]{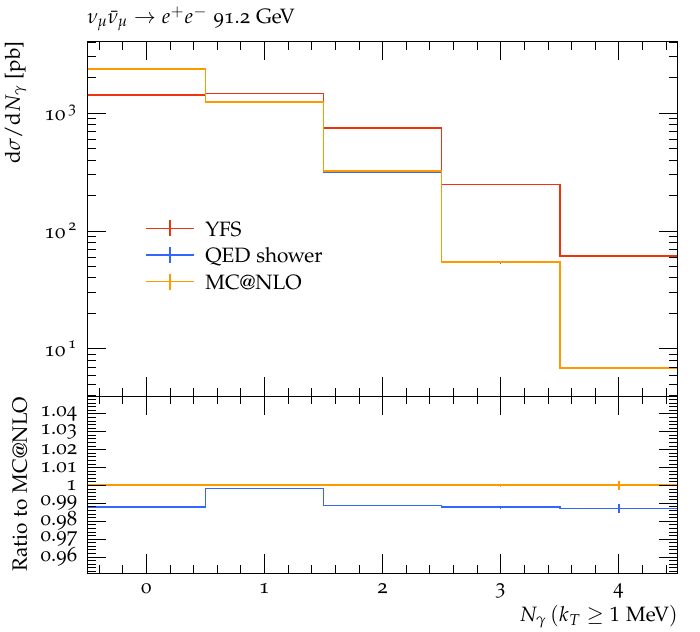}
    \includegraphics[width=0.45\textwidth]{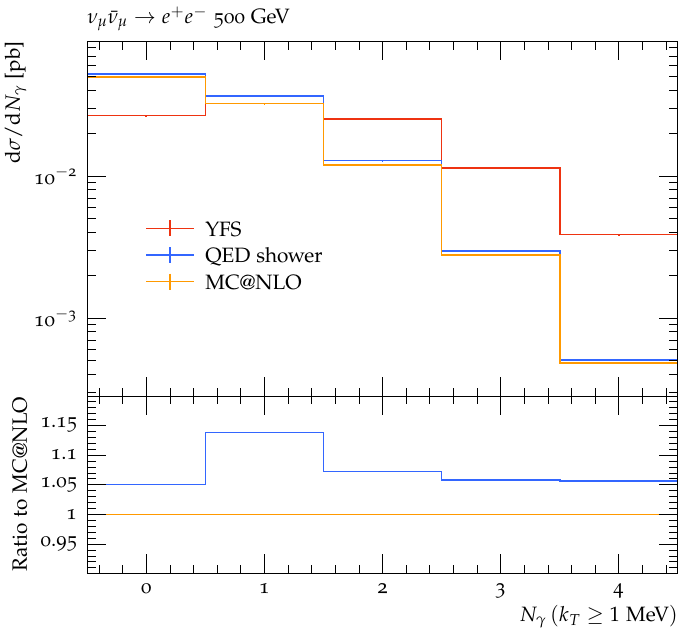}
    \caption[The multiplicity of photons 
    with $k_T \geq 1$ \MeV\, in $\nu_\mu \bar{\nu}_\mu \to e^+ e^-$, 
    comparing the YFS method with the LO QED shower 
    prediction and the QED \MCatNLO prediction.]
    {The multiplicity of photons with $k_T \geq 1$ \MeV\,
    in $\nu_\mu \bar{\nu}_\mu \to e^+ e^-$, 
    comparing the YFS method with the LO QED shower 
    prediction and the QED \MCatNLO prediction for 
    two different collider energies.
    \textbf{Left:} $\sqrt{s} = $ 91.2 GeV. 
    \textbf{Right:} $\sqrt{s} = $ 500 GeV.
    The ratio plot shows the ratio of the LO QED shower to the 
    QED \MCatNLO prediction.
    \label{fig:mcatnlo:multiplicity}
    }
\end{figure}

Fig.\ \ref{fig:mcatnlo:multiplicity} shows the number of photons 
with $k_T\geq1$ \MeV\, produced by each method. 
The \ac{YFS} resummation predicts a higher average number of 
photons per event, with the modal number of photons being 1, 
compared to the collinear resummation methods which do not 
produce a large number of soft photons, and retain a modal 
number of photons of 0. This difference is characteristic 
of the resummation method employed in each case. The 
\ac{YFS} approach primarily resums soft logarithms, and 
collinear ones are only included through higher-order 
corrections, meaning that many soft photons are produced. 
On the other hand, the parton shower is derived from collinear 
resummation, 
and only includes the eikonal factors through modifications
to the splitting functions.

The fact that the two classes 
of resummation produce vastly different numbers of photons 
per event, but agree up to a few percent for physical inclusive 
distributions, validates the methods as viable alternative 
descriptions of the same physical process.

\begin{figure}
    \centering
    \includegraphics[width=0.45\textwidth]{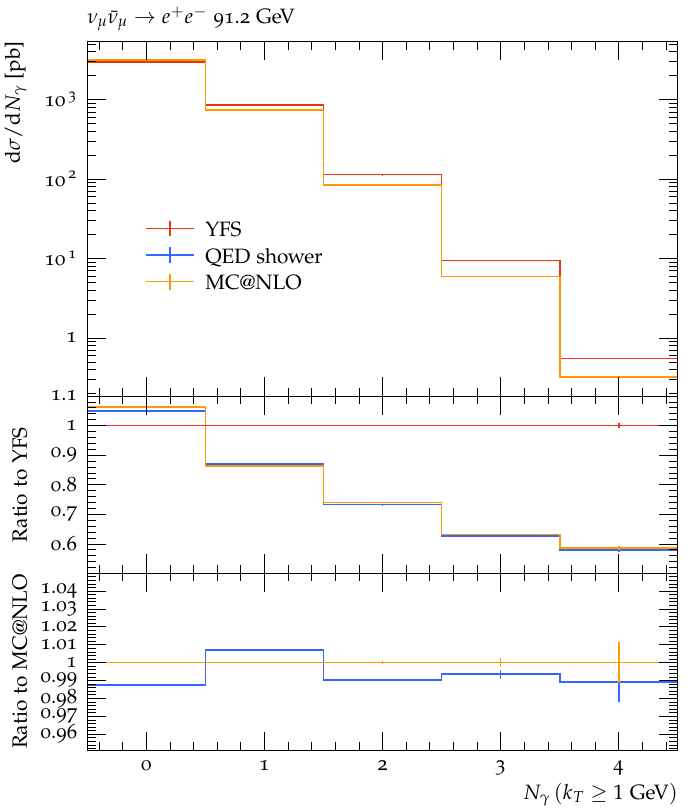}
    \includegraphics[width=0.45\textwidth]{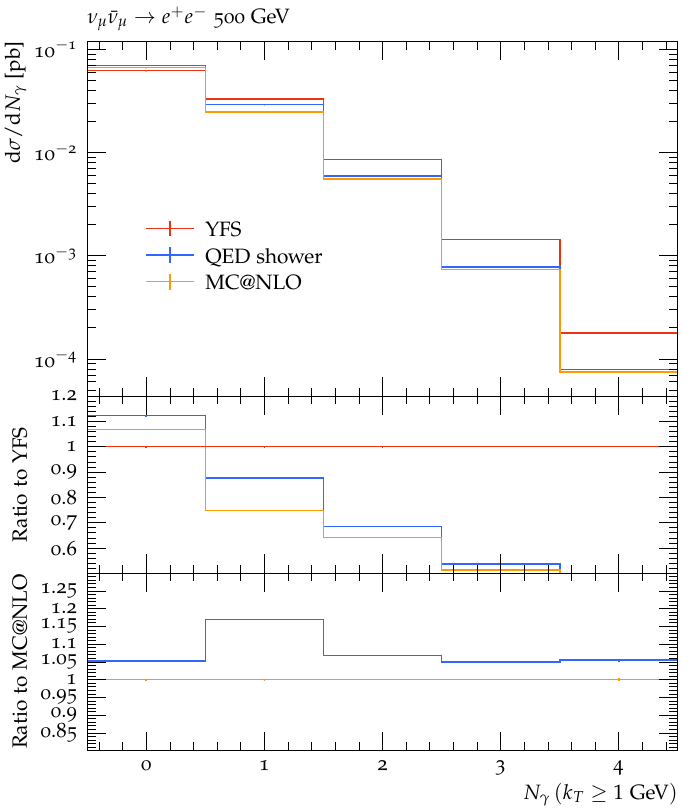}
    \caption[The multiplicity of photons 
    with $k_T \geq 1$ GeV in $\nu_\mu \bar{\nu}_\mu \to e^+ e^-$, 
    comparing the YFS method with the LO QED shower 
    prediction and the QED \MCatNLO prediction.]
    {The multiplicity of photons with $k_T \geq 1$ GeV
    in $\nu_\mu \bar{\nu}_\mu \to e^+ e^-$, 
    comparing the YFS method with the LO QED shower 
    prediction and the QED \MCatNLO prediction for 
    two different collider energies.
    \textbf{Left:} $\sqrt{s} = $ 91.2 GeV. 
    \textbf{Right:} $\sqrt{s} = $ 500 GeV.
    The upper ratio plot shows the ratio to the YFS prediction 
    while the lower shows the ratio of the LO QED shower to the 
    QED \MCatNLO prediction.
    \label{fig:mcatnlo:multiplicity1GeV}
    }
\end{figure}

To verify this statement, we show the number of photons 
produced by each method, but with a higher $k_T$ cut of 
1 GeV, in fig.\ \ref{fig:mcatnlo:multiplicity1GeV}. It is 
immediately clear that the two classes of resummation methods 
produce similar numbers of hard photons. These will provide the 
bulk of the recoil effects, in addition to being more detectable
experimentally. Differences in the zero-photon and one-photon 
bins still exist up to 15\% at the $Z$ mass and 25\% at 
500 GeV, but as we have seen from the $d_{23}$ and third 
jet $k_T$ plots above, the overall differences in the radiation 
pattern are much smaller than this. \changed{In particular, in 
the lower ratio plot comparing the shower and \MCatNLO methods, 
we see that with this harder photon energy cut the \MCatNLO method 
produces significantly more one-photon events than the shower. 
The 2-, 3- and 4-photon bins are unchanged compared to fig.\ 
\ref{fig:mcatnlo:multiplicity}, as expected.}

%%%%%%%%%%%%%%%%%%%%%%%%%%%%%%%%%%%%%%%%%%%%%%%%%%%%%%%%%%%%%%
\newpage
\subsection{Leptonic Higgs decay}
\label{sec:mcatnlo:results:ggHleptons}

\changed{One of the primary use cases of \ac{NLO} \ac{EW} matching 
for the \ac{LHC} is in Higgs decays to leptons, and their irreducible 
backgrounds in vector boson pair production. Recently 
the \ac{QCD}+\ac{QED} \ac{NLO} matched corrections were presented 
for $pp \to VV'$ processes where the vector bosons decay 
leptonically \cite{Chiesa:2020ttl}. However, the method 
presented is difficult to extend to Higgs decay processes due 
to \Powheg's exponentiation of large $K$-factors. Here we present 
the \ac{QCD}+\ac{QED} \MCatNLO method, which has no such issue.}

\begin{figure}
    \centering
    % Born
    \begin{subfigure}[b]{0.25\textwidth}
    \includegraphics[width=0.9\columnwidth]{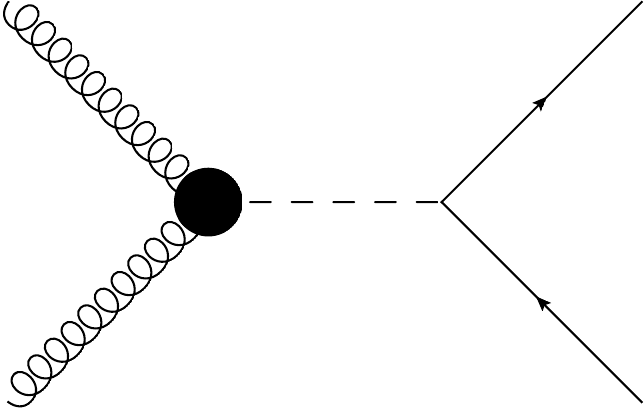}
    \caption{$\mathcal{O}(\alpha_s^2,\alpha^2)$ \label{subfig:a}}
    \end{subfigure}
    % Real QCD
    \begin{subfigure}[b]{0.25\textwidth}
    \includegraphics[width=0.9\columnwidth]{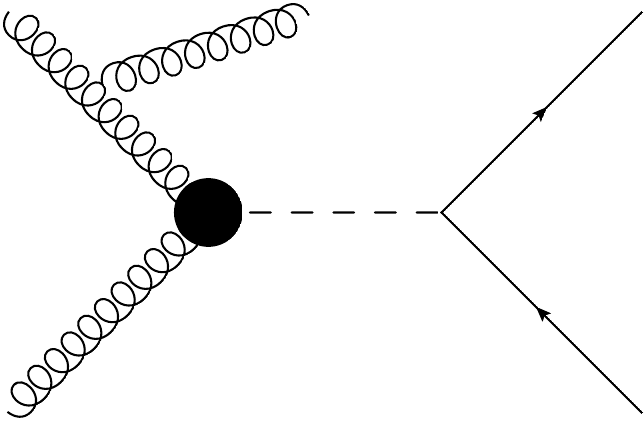}
    \caption{$\mathcal{O}(\alpha_s^3,\alpha^2)$ \label{subfig:b}}
    \end{subfigure}
    % Real QED
    \begin{subfigure}[b]{0.25\textwidth}
    \includegraphics[width=0.9\columnwidth]{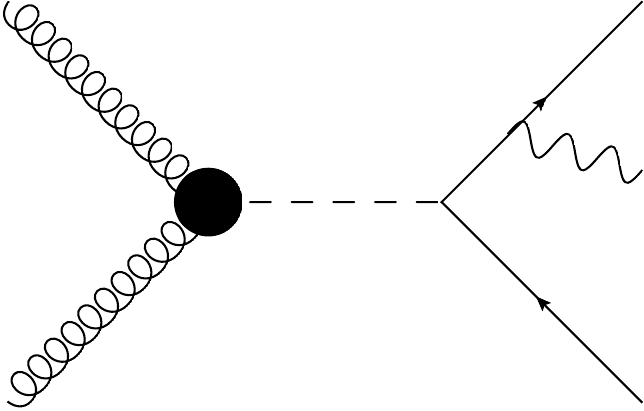}
    \caption{$\mathcal{O}(\alpha_s^2,\alpha^3)$ \label{subfig:c}}
    \end{subfigure}
    \hfill
    % Virtual QCD
    \begin{subfigure}[b]{0.45\textwidth}
    \includegraphics[width=0.9\columnwidth]{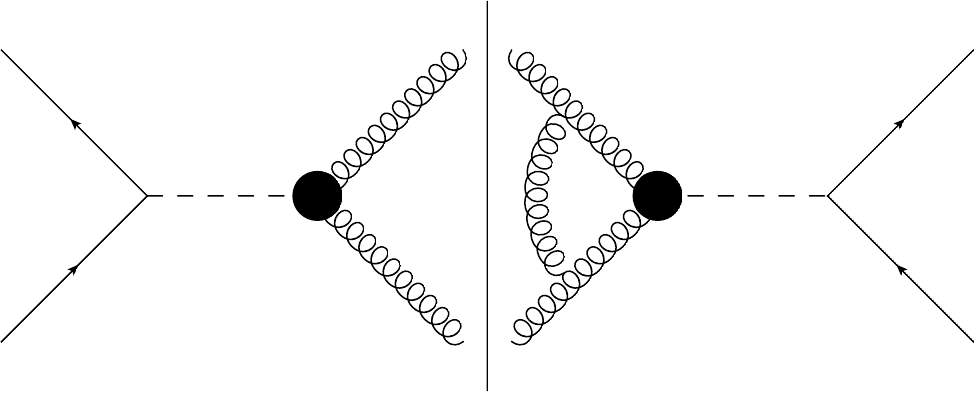}
    \caption{$\mathcal{O}(\alpha_s^3,\alpha^2)$ \label{subfig:d}}
    \end{subfigure}
    % Virtual QED    
    \begin{subfigure}[b]{0.45\textwidth}
    \includegraphics[width=0.9\columnwidth]{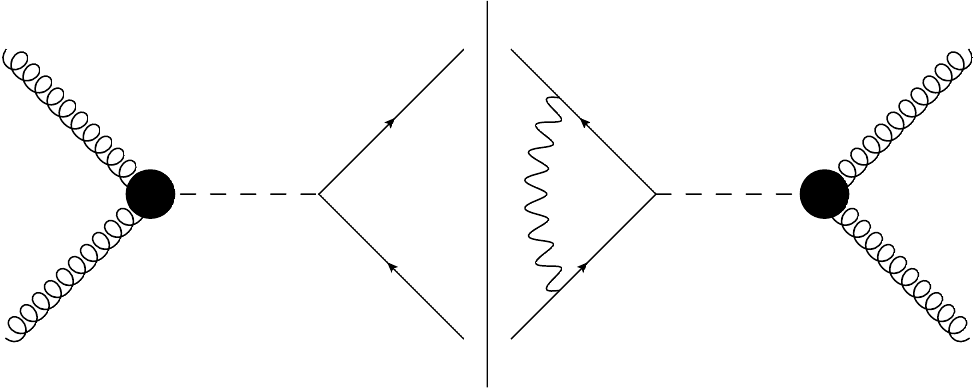}
    \caption{$\mathcal{O}(\alpha_s^2,\alpha^3)$ \label{subfig:e}}
    \end{subfigure}

    \caption[A representative subset of the diagrams contributing 
    at each order
    of the NLO
    QCD+QED calculation of the process $gg\to\mu^+\mu^-$.]{
    A representative subset of the diagrams contributing 
    at each order (labelled) 
    of the NLO 
    QCD+QED calculation of the process $gg\to\mu^+\mu^-$.
    Diagrams \ref{subfig:a}, \ref{subfig:b} and \ref{subfig:c}
    are individually squared, while the virtual correction diagrams 
    \ref{subfig:d} and \ref{subfig:e} are interfered with the 
    Born process as shown.
    \label{fig:mcatnlo:ggmumuFeynman}
    }
\end{figure}

\begin{figure}
    \centering
    % Born
    \begin{subfigure}[b]{0.25\textwidth}
    \includegraphics[width=0.9\columnwidth]{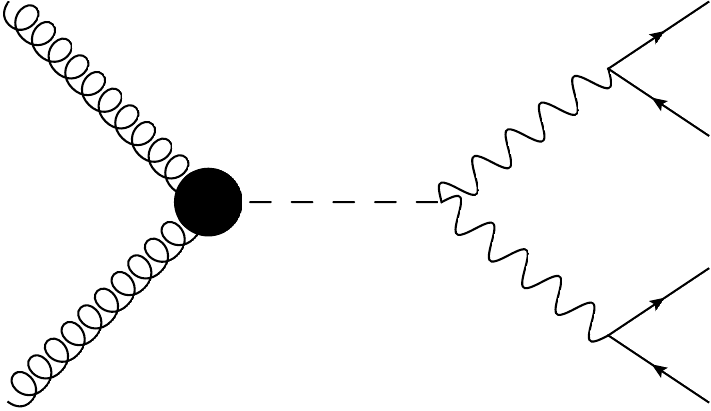}
    \caption{$\mathcal{O}(\alpha_s^2,\alpha^4)$ \label{subfig:gg4l:a}}
    \end{subfigure}
    % Real QCD
    \begin{subfigure}[b]{0.25\textwidth}
    \includegraphics[width=0.9\columnwidth]{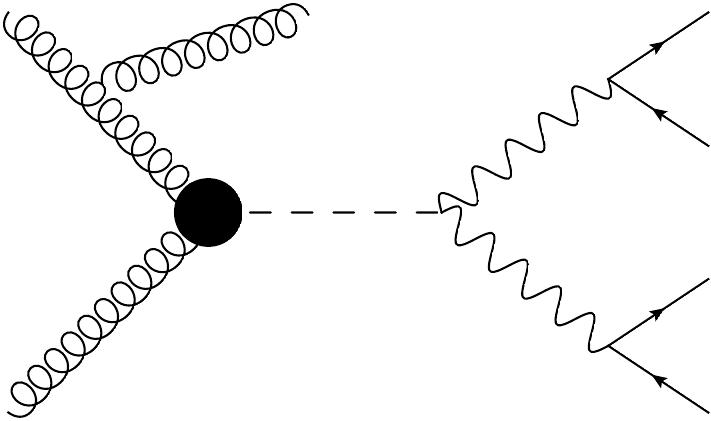}
    \caption{$\mathcal{O}(\alpha_s^3,\alpha^4)$ \label{subfig:gg4l:b}}
    \end{subfigure}
    % Real QED
    \begin{subfigure}[b]{0.25\textwidth}
    \includegraphics[width=0.9\columnwidth]{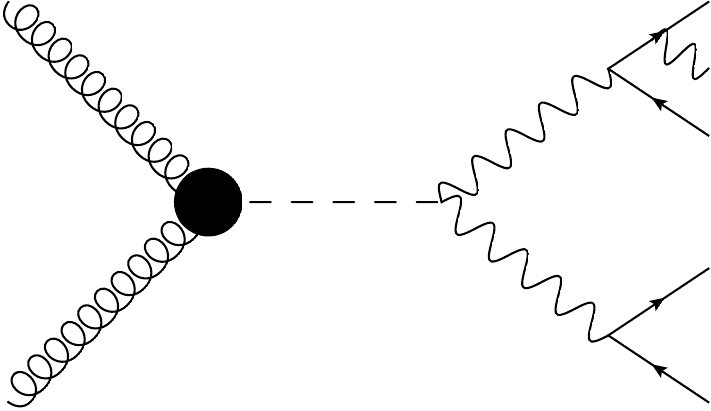}
    \caption{$\mathcal{O}(\alpha_s^2,\alpha^5)$ \label{subfig:gg4l:c}}
    \end{subfigure}
    \hfill
    % Virtual QCD
    \begin{subfigure}[b]{0.45\textwidth}
    \includegraphics[width=0.9\columnwidth]{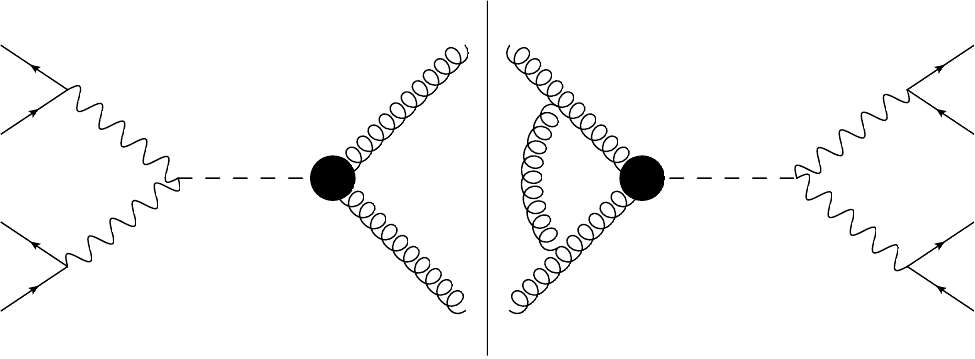}
    \caption{$\mathcal{O}(\alpha_s^3,\alpha^4)$ \label{subfig:gg4l:d}}
    \end{subfigure}
    % Virtual QED    
    \begin{subfigure}[b]{0.45\textwidth}
    \includegraphics[width=0.9\columnwidth]{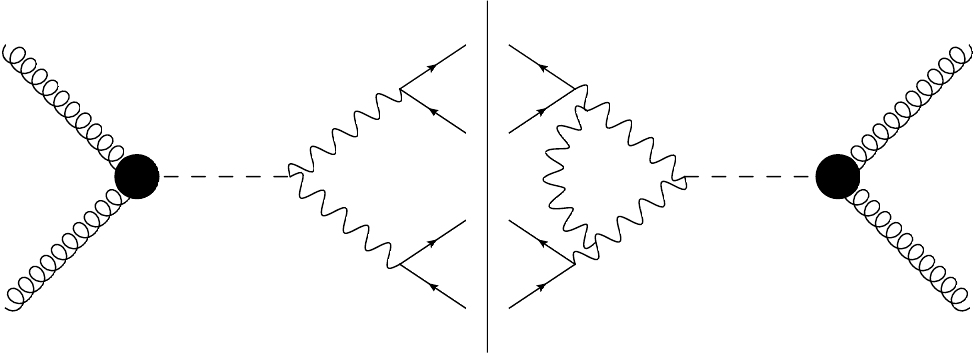}
    \caption{$\mathcal{O}(\alpha_s^2,\alpha^5)$ \label{subfig:gg4l:e}}
    \end{subfigure}

    \caption[A representative subset of the diagrams contributing
    at each order 
    of the NLO
    QCD+QED calculation of the process $gg\to\mu^+\mu^- e^+ e^-$.]{
    A representative subset of the diagrams contributing
    at each order (labelled) 
    of the NLO
    QCD+QED calculation of the process $gg\to\mu^+\mu^- e^+ e^-$.
    Diagrams \ref{subfig:gg4l:a}, \ref{subfig:gg4l:b} and \ref{subfig:gg4l:c}
    are individually squared, while the virtual correction diagrams 
    \ref{subfig:gg4l:d} and \ref{subfig:gg4l:e} are interfered with the 
    Born process as shown. While diagram \ref{subfig:gg4l:e}
    emphasises that pure weak corrections are included 
    at this order, the dominant virtual EW contribution 
    will be from photonic vertex corrections to the $Z$ decay.
    \label{fig:mcatnlo:gg4lFeynman}
    }
\end{figure}

In this section, we will study the processes $gg \to H \to \mu^+ \mu^-$
and $gg \to H \to e^+ e^- \mu^+ \mu^-$ in the \ac{HEFT}. 
We will present a general method for automating the matching 
of an interleaved QCD+QED parton shower with the 
the \ac{NLO} \ac{EW} corrections to the Higgs decay 
and the \ac{NLO} \ac{QCD} corrections to the production process. 
We refer to these \ac{NLO} corrections 
as NLO QCD+QED in the following. \changed{The method is 
process-independent, but to aid in the explanation we will 
refer to these Higgs decay processes throughout.}
For clarity, a subsection of the diagrams 
included in the computation of $gg \to H \to \mu^+ \mu^-$
are shown in fig.\ \ref{fig:mcatnlo:ggmumuFeynman}. Note 
that we do not include the mixed contribution, which enters at
$\mathcal{O}(\alpha_s^3,\alpha^3)$.
Similarly, for the four-lepton decay a selection of 
diagrams are shown in fig.\ \ref{fig:mcatnlo:gg4lFeynman}.
While diagram \ref{subfig:gg4l:e}
emphasises that pure weak corrections are included 
at this order, the dominant virtual \ac{EW} contribution 
will be from photonic vertex corrections to the $Z$ decay.

The \ac{HEFT} was introduced in section \ref{sec:qedps:results:ggHleptons},
but we will briefly recall it here since we now work at \ac{NLO}
QCD+QED.
In the \ac{HEFT}, we obtain 
a direct effective coupling of gluons to the Higgs, 
\begin{equation}
    \mathcal{L} = \mathcal{L}_\text{SM} + g_\text{HEFT}\, G^{\mu\nu}_a G_{\mu\nu}^a H + \dots,
\end{equation}
by integrating out the top quark 
in the \ac{SM} loop-induced production of a Higgs via gluon fusion. 
Of course, other quarks can run in the loop, but since 
the top Yukawa coupling $y_t$ is much greater than the other quark 
Yukawa couplings due to its large mass,
other contributions can safely be ignored.
Using this formalism, we neglect 
\ac{EW} corrections to the top quark loop, which would be 
dominated by \ac{EW} Sudakov virtual high-energy logarithms
(these have been implemented in various automated tools, 
\see refs. \cite{Bothmann:2020sxm,Pagani:2021vyk,Pagani:2023wgc,Lindert:2023fcu}).
In the diagrams in figs. \ref{fig:mcatnlo:ggmumuFeynman} and 
\ref{fig:mcatnlo:gg4lFeynman}, the $ggH$ vertex is shown 
as a black blob, since it is not \iac{SM} vertex. In the \ac{SM},
this effective coupling comes with a factor $g_s^2 y_t$.
The numerical 
value of the $ggH$ coupling used here 
is $g_\text{HEFT} \approx 5 \times 10^{-5}$.

Further to the discussion on 
the interleaved QCD+QED parton shower in section \ref{sec:qedps:FSresults},
we will briefly outline the \MCatNLO procedure before presenting 
results.

The \MCatNLO algorithm extends straightforwardly to match QCD+QED
\ac{NLO} in these processes. Separating \ac{QCD} and \ac{QED} contributions, eq.\ 
\eqref{eq:mcatnlo:mcatnlo} becomes 
\begin{align*} \label{eq:mcatnlo:QCDQED}
    \Oangle{NLOPS} = &\sum_f \intdphi{n} \Biggl[ \bar{B} \,
    \bar{\Delta}^A(t_c) 
    \, O(\Phi_n) \Biggr.\\
    &\Biggl.+ \sum_{\ijt,\kt}\sum_f \int \diff \Phi_1^{ij,k}
    \,\Theta(t_{ij,k}-t_c) \,S_{ij,k} \,\bar{\Delta}^A(t_{ij,k}) \, 
    \frac{D^A_{ij,k}}{B} \,
    O(\Phi_{n+1}) \Biggr] \\ 
    &+ \intdphi{n+1} \left[ R_\text{QCD} + R_\text{QED}
    - \sum_{ij,k} \left(D^A_\text{QCD} + D^A_\text{QED}\right) \right] 
    O(\Phi_{n+1}). \numberthis
\end{align*}
The $R_\text{QCD}$ and $R_\text{QED}$ terms are given by the matrix 
element squared of the diagrams in fig.\ \ref{subfig:b} and \ref{subfig:c},
respectively, and the equivalent diagrams where the gluon or 
photon is emitted from the other external leg,
for the $gg \to \mu^+ \mu^-$ process. For 
$gg \to \mu^+ \mu^- e^+e^-$, the diagrams \ref{subfig:gg4l:b}
and \ref{subfig:gg4l:c} contribute.
The $\bar{B}$ function is defined as 
\begin{align*}
    \bar{B} = \,&B+\tilde{V}_\text{QCD} + V_\text{EW}+I^S_\text{QCD} 
    +I^S_\text{QED} \\
    &+ \intdphi{1} \left[D^A_\text{QCD}-D^S_\text{QCD}\right]
    + \intdphi{1} \left[D^A_\text{QED}-D^S_\text{QED}\right]. \numberthis
\end{align*}
$\tilde{V}_\text{QCD}$ contains the collinear mass factorisation 
terms from the initial state, in addition to the interference term 
shown in fig.\ \ref{subfig:d} or \ref{subfig:gg4l:d}. 
$V_\text{EW}$ contains the interference 
term, \changed{e.g.} in fig.\ \ref{subfig:e} or \ref{subfig:gg4l:e}. 
The shower kernel $D^A_{ij,k}/B$ and the Sudakov factor 
$\bar{\Delta}^A$ are generated in the usual way with an interleaved 
shower, and therefore contain both $D^A_\text{QCD}$ and $D^A_\text{QED}$.

When an $\mathbb{H}$-event is selected, it is selected to be 
either \ac{QCD} or \ac{QED} according to the relative size of $R_\text{QCD}$ and 
$R_\text{QED}$. If, as is the case here, the \ac{QED} real emissions are 
of interest, the \ac{ME} can be overweighted by a factor 
$\alpha_s/\alpha$ so that \ac{QED} real-emission events are selected as 
often as \ac{QCD} real-emission events. A corrective weight is then 
applied as an analytic event weight. In a similar way, for 
$\mathbb{S}$-events
an analytic overweighting can be applied to the \ac{QED} splitting functions.
The weights in this case must also be applied to the Sudakov factors 
when no emission occurs, as described in the weighted veto algorithm 
(section \ref{sec:qedps:weightedveto}).

In this thesis, we are unable to present the full 
\ac{NLO} QCD+QED corrections as described, due to 
current structural limitations of the process handling
in \Sherpa. Care must be taken when defining 
the different parts of a mixed \ac{NLO} calculation 
to account for all the possible divergences which 
arise, especially for processes with four or more 
quarks at Born level. However, these results and 
considerations will be presented in a 
future publication.
Here, to demonstrate the validity 
of the method, we present the pure \ac{EW} 
\ac{NLO}-matched parton shower, having already demonstrated
the success of the interleaved QCD+QED evolution in 
chapter \ref{chapter:qedps}. We focus on distributions 
in invariant mass of final-state particles, since these 
are insensitive to \ac{QCD} radiation, and hence are 
\ac{NLO} accurate up to a $K$-factor.

We consider gluon-induced Higgs production at the \ac{LHC},
where the colliding protons have a centre-of-mass energy 
of 13 \TeV. We use the \ac{PDF} 
set {\sc Pdf4Lhc21} from the \LHAPDF library \cite{Buckley:2014ana}. 
Beam remnants, hadronisation, and multiple interactions were not modelled.

\subsubsection{$gg \to H \to \mu^+ \mu^-$}

The \ac{LO} cross section for $gg \to \mu^+ \mu^-$ at the 
13 \TeV\, \ac{LHC} in 
the \ac{HEFT} is $\sigma_\text{LO} = 0.0028662(1)$ pb. 
The \ac{NLO} \ac{EW} cross 
section is $\sigma_\text{NLO} = 0.0028747(4)$ pb, which results in a 
positive \ac{EW} $K$-factor of 0.3\%.
In this section we present the differential 
cross section in the muon invariant mass 
for bare muons and dressed muons. 
The muons are dressed with photons in a cone of 
radius $\Delta R=0.1$, where $\Delta R = \sqrt{\Delta \eta^2 + \Delta \phi^2}$.
The primary muons, whether bare or dressed, are subject to cuts 
on transverse momentum, $p_\mu^T > 10\,\GeV$, and rapidity, 
$\abs{y_\mu} < 2.5$.
We compare the \ac{QED} \MCatNLO
with the \ac{YFS} soft-photon 
resummation supplemented with exact \ac{NLO} corrections,
in addition to the \ac{QED} shower presented in the 
previous chapter.

\begin{figure}
    \centering
    \includegraphics[width=0.45\textwidth]{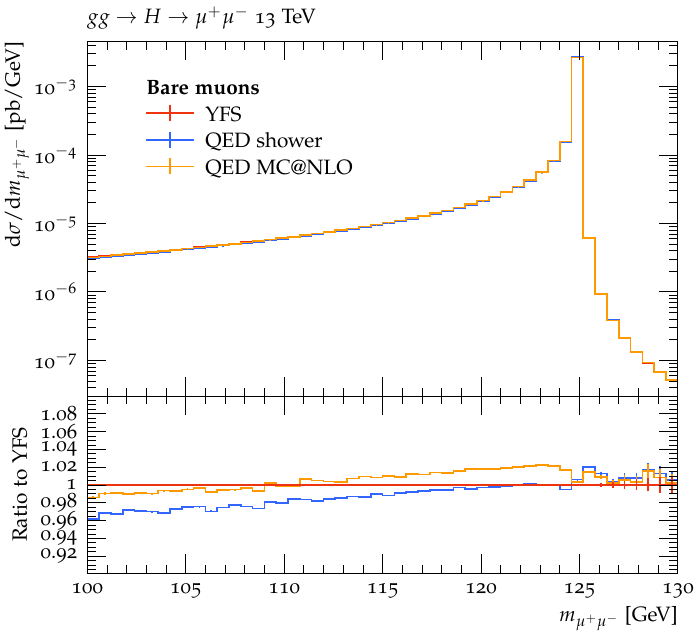}
    \includegraphics[width=0.45\textwidth]{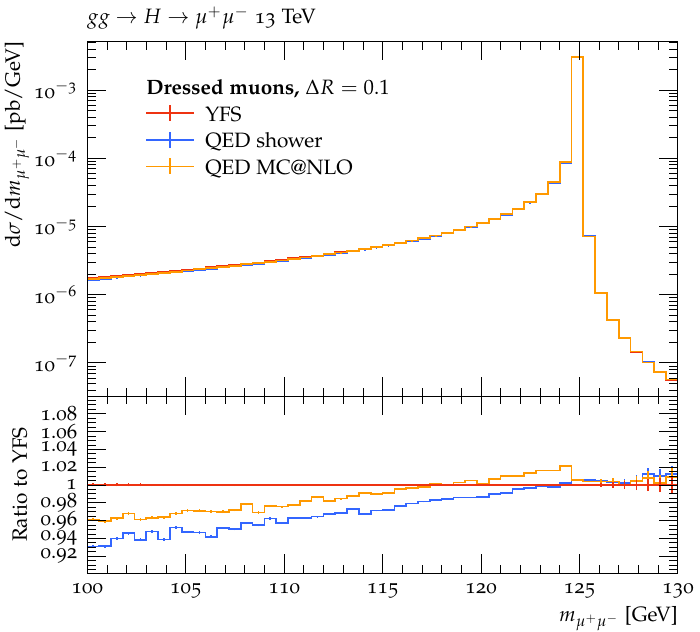}
    \caption[The dimuon invariant mass
    in $gg\to \mu^+\mu^-$, comparing the 
    QED \MCatNLO with the YFS prediction and the QED shower.]{
    The dimuon invariant mass
    in the $gg\to \mu^+\mu^-$ process, comparing the 
    QED \MCatNLO with the YFS prediction and the QED shower. 
    \textbf{Left:}
    bare muons, \textbf{Right:} photon-dressed muons with 
    a cone size $\Delta R = 0.1$.
    \label{fig:mcatnlo:ggmumu:invariant_mass}
    }
\end{figure}

Figure \ref{fig:mcatnlo:ggmumu:invariant_mass} shows the 
dimuon invariant mass in the $gg\to \mu^+ \mu^-$ process. 
No \ac{QCD} corrections to the initial state are effected, 
apart from the gluon \ac{PDF}. We compare our implementation 
of the \ac{QED} \MCatNLO with the \ac{QED} shower described in 
the previous chapter, and with the \ac{YFS} resummation. 
We can see that the \ac{YFS} and \MCatNLO approaches agree
to within 2\% for this observable when bare muons are used to 
reconstruct the Higgs, and to within 4\% when dressed muons are 
used. This shows very good 
agreement, since this invariant mass distribution is approximately 
a $\delta$-function at \ac{LO}.
The \MCatNLO prediction clearly has the same shape as the 
parton shower prediction, but integrates to the total 
\ac{NLO} cross section. Just below the Higgs mass, 
the \MCatNLO prediction has an 
enhancement of 2\% above the \ac{YFS} prediction for both 
bare and dressed muons, though the differential cross section 
in the immediate vicinity of the resonance is in excellent 
agreement when dressed muons are used.

\subsubsection{$gg \to H \to \mu^+ \mu^- e^+ e^-$}

The \ac{LO} cross section for the rare process $gg \to \mu^+ \mu^- e^+ e^-$ 
at the 13 \TeV\, \ac{LHC} in the \ac{HEFT} is 
$\sigma_\text{LO} = 0.00085(2)$ pb. 
The \ac{NLO} \ac{EW} cross section is $\sigma_\text{NLO} = 0.0097(6)$ 
pb, resulting in a positive $K$-factor of 13\%. To provide some 
context, the \ac{QCD} $K$-factor of around a factor of 2 
was one of the primary motivations for the development of 
matching methods. Here, we see that the \ac{EW} corrections 
are also enhanced, and must be included at the level 
of precision of modern experiments.

In this section, we will present distributions in the four-lepton 
invariant mass $m_{4\ell}$, as well as invariant masses of 
$Z$ bosons reconstructed via the primary electrons and muons. 
As before, the primary electrons and muons are subject to cuts 
on transverse momentum, $p_\ell^T > 10 \,\GeV$, and rapidity, 
$\abs{y_\ell} < 2.5$.
We compare the \ac{QED} \MCatNLO
with the \ac{YFS} soft-photon 
resummation supplemented with \changed{collinearly-approximated}
\ac{NLO} corrections,
in addition to the \ac{QED} shower presented in the 
previous chapter.

\begin{figure}
    \centering
    \includegraphics[width=0.45\textwidth]{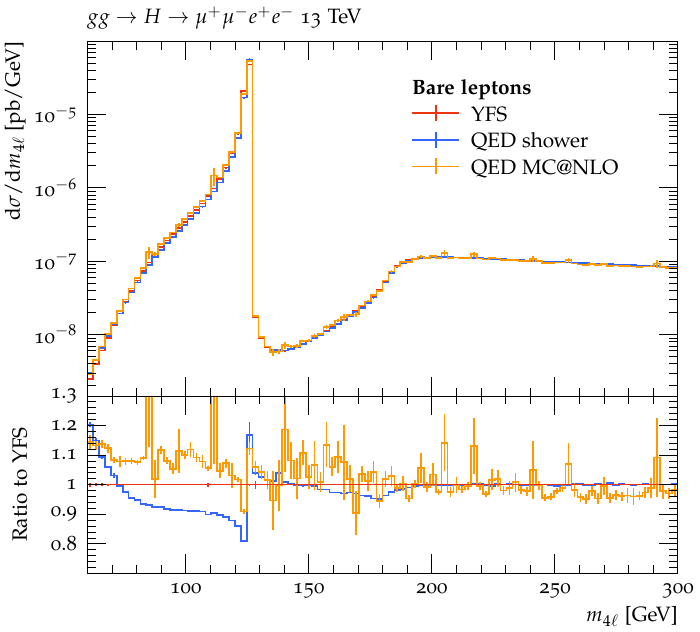}
    \includegraphics[width=0.45\textwidth]{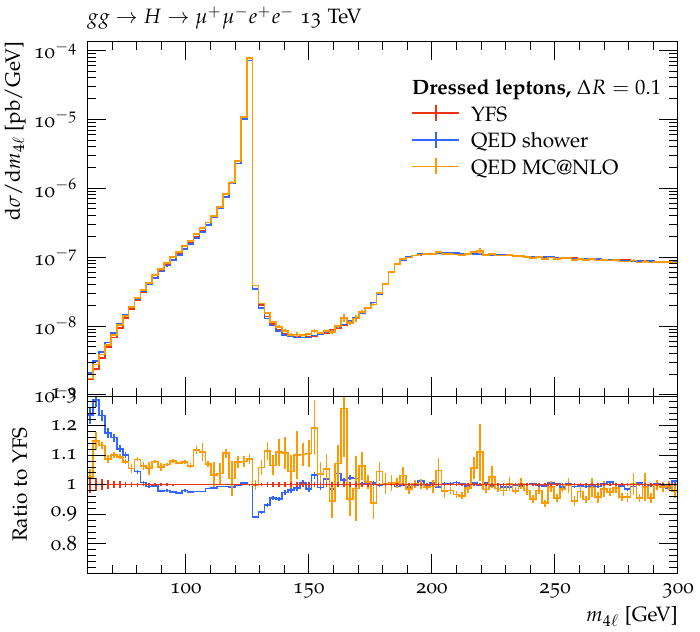}
    \caption{The four-lepton invariant mass
    in $gg\to \mu^+\mu^- e^+ e^-$, comparing the
    QED \MCatNLO with the YFS prediction and the QED shower.
    \textbf{Left:} bare leptons, \textbf{Right:} photon-dressed leptons 
    with a cone size $\Delta R = 0.1$.
    \label{fig:mcatnlo:gg4l:invariant_mass}
    }
\end{figure}

Figure \ref{fig:mcatnlo:gg4l:invariant_mass} shows the four-lepton 
invariant mass differential cross section. The main features of 
the lineshape are the Higgs resonance at 125 GeV and the 
two on-shell $Z$ threshold at 180 GeV. The left plot shows 
the predictions for the invariant mass distribution reconstructed 
from bare leptons, while dressed leptons are used in the right 
plot. We use a cone dressing with a size $\Delta R = 0.1$ to 
recombine photons into primary leptons.
Above the di-$Z$ threshold,
all three predictions are in excellent agreement. Between the 
resonance and the threshold, for bare leptons,
the parton shower approach underestimates 
the cross section, while the \ac{YFS} and the \MCatNLO approaches 
agree up to statistical uncertainty. This feature is not 
present in the dressed lepton distribution. 
At the Higgs resonance, the 
\MCatNLO shares the parton shower feature of roughly 15\% more 
events on the resonance compared to just below it, 
compared to the \ac{YFS} prediction. On the dressed level,
the \ac{YFS} and shower agree perfectly at the resonance, and 
the \MCatNLO prediction has the same shape, but is modulated 
by the \ac{NLO} cross section contribution in this region.
At lower invariant masses, however, the \MCatNLO calculation 
corrects the underproduction of events which the shower predicts,
which is necessary for a good description of this process at 
the bare level.
Overall, the \MCatNLO and \ac{YFS} predictions agree within a 
few percent.
However, we can see in both plots that the 
\MCatNLO suffers from worse statistics, since all dipoles are 
considered in the \MCatNLO $\mathbb{S}$-events, and the 
same-sign dipoles result in negative weights. This is a 
motivation to include the \ac{OSSF} dipole identification efficiency 
improvement, described in sec. \ref{sec:qedps:methods:implementation},
in the \MCatNLO method. This would not impact the \ac{NLO}
accuracy of the simulation, and has been shown in sec. \ref{sec:qedps:results:ggHleptons}
not to impact the parton shower logarithmic accuracy.

\begin{figure}
    \centering
    \includegraphics[width=0.45\textwidth]{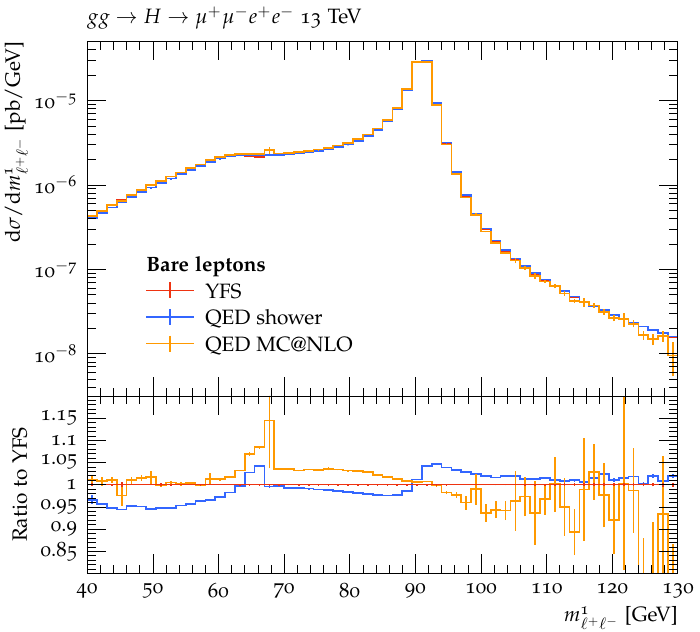}
    \includegraphics[width=0.45\textwidth]{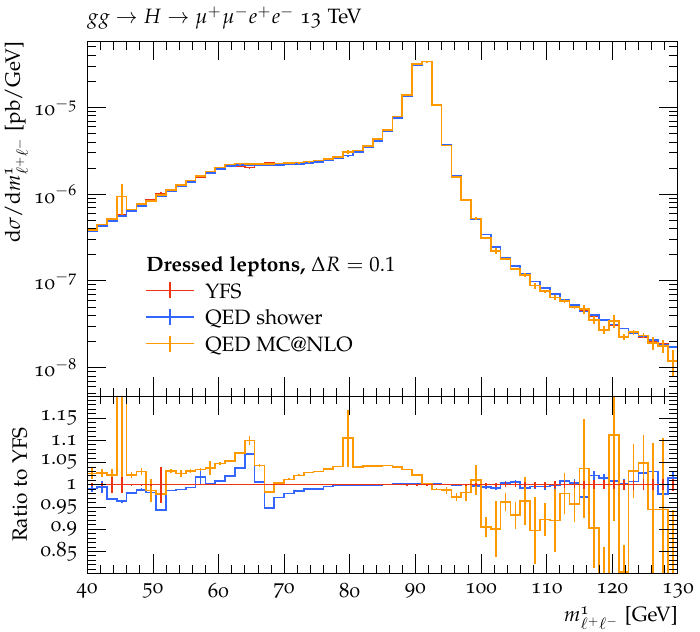}
    \caption{The hardest $Z$ boson invariant mass
    in $gg\to \mu^+\mu^- e^+ e^-$, comparing
    QED \MCatNLO with the YFS prediction and the QED shower.
    \textbf{Left:} bare leptons, \textbf{Right:} photon-dressed leptons 
    with a cone size $\Delta R = 0.1$.
    \label{fig:mcatnlo:gg4l:Z1minv}
    }
\end{figure}

\begin{figure}
    \centering
    \includegraphics[width=0.45\textwidth]{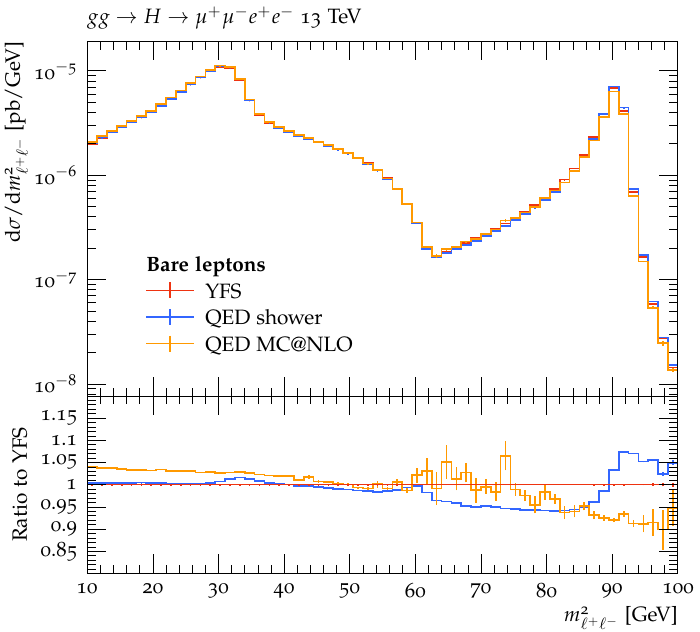}
    \includegraphics[width=0.45\textwidth]{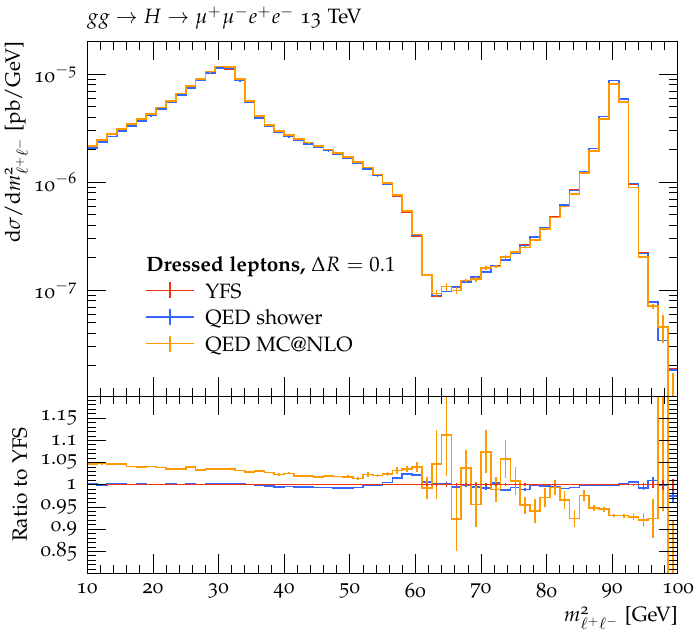}
    \caption{The second $Z$ boson invariant mass
    in $gg\to \mu^+\mu^- e^+ e^-$, comparing
    QED \MCatNLO with the YFS prediction and the QED shower.
    \textbf{Left:} bare leptons, \textbf{Right:} photon-dressed leptons 
    with a cone size $\Delta R = 0.1$.
    \label{fig:mcatnlo:gg4l:Z2minv}
    }
\end{figure}

We study two more invariant mass distributions, in figs. 
\ref{fig:mcatnlo:gg4l:Z1minv} and \ref{fig:mcatnlo:gg4l:Z2minv}.
$Z$ bosons were reconstructed from the primary $e^+ e^-$ and 
$\mu^+ \mu^-$ pair, identified by their energy. The left 
plots in both figures show the $Z$ invariant mass reconstructed
from bare leptons, while the right plots show the same 
observable using photon-dressed leptons with a cone size 
$\Delta R = 0.1$. 

Fig.\ \ref{fig:mcatnlo:gg4l:Z1minv} shows the 
larger of the two $Z$ invariant masses, comparing the 
\ac{QED} \MCatNLO prediction with the shower and the \ac{YFS}
predictions. We see that the shape is well-reproduced in all 
cases, with the two main features being a small local maximum 
at $m_H/2$, and a large peak at $m_Z$. Compared to the \ac{YFS}
prediction, the \MCatNLO predicts an excess of events on the $m_H/2$ 
peak compared to just below it, amounting to differences 
of up to 10\%. The \MCatNLO prediction improves on the 
behaviour of the shower near the $Z$ resonance for bare 
leptons, but for dressed leptons the main advantage of the 
\MCatNLO method is the correct \ac{NLO} cross section, mostly 
concentrated in the region 60-90 GeV.

On the other hand, fig.\ \ref{fig:mcatnlo:gg4l:Z2minv} shows 
the smaller of the $Z$ invariant masses. We see this time 
that the peak at $m_H/2$ is non-existent, though the $Z$ 
resonance is still clearly visible, but that there is a 
peak at around 30 GeV. This is produced by an approximately
on-shell Higgs decaying to an on-shell $Z$, leaving the other 
with the remaining 35 GeV invariant mass. We do not 
see a difference in the position of this peak for bare 
and dressed leptons, suggesting that any \ac{QED} radiation 
effects are small in this region. In both plots, we see that 
compared to the 
\ac{YFS} distribution, the \MCatNLO prediction is skewed 
towards smaller $Z$ invariant masses. This suggests that the 
leptons have lost more energy to photon radiation and secondary 
charged particles.

Overall, for the invariant mass distributions which are 
characterised by \ac{QED} radiation in this process, the 
\MCatNLO method produces distributions consistent with the \ac{YFS}
prediction at a level of a few percent, and the correct 
\ac{NLO} cross section.

\section{Conclusions} \label{sec:mcatnlo:conclusions}

In this chapter we have presented novel results from a
QCD+QED \ac{NLO} matched parton shower. After arguing that 
the \MCatNLO method is \ac{NLO}-accurate and that it extends to 
\ac{QED} and mixed \ac{NLO} corrections, we validated our 
implementation of the method using a test process 
$\nu_\mu \bar{\nu}_\mu \to e^+ e^-$ at two different 
energies: just above the $s$-channel resonance, and far 
above it. We saw that the \MCatNLO method performs well 
in both cases, producing results highly consistent with 
the widely used \ac{YFS} soft-photon resummation in the 
former case. In the latter case, collinear logarithms 
play a larger role in the differential cross section 
determination, since the electron mass is comparatively 
smaller. The \ac{YFS} approach only includes these 
logarithms at fixed order, while the \MCatNLO includes 
the exact \ac{NLO} result and the resummation of all orders 
of leading collinear logarithms. For this reason, we 
expect the \MCatNLO method to provide a more accurate 
description at this energy. 

We also demonstrated the applicability of 
our method to phenomenological studies. We studied the 
production of a Higgs boson via gluon fusion and its 
decay to two or four leptons. By matching the \ac{NLO} 
\ac{QCD} corrections to the production process to a 
\ac{QCD} parton shower, and the \ac{NLO} \ac{EW} corrections 
to the decay to a final-state \ac{QED} parton shower, 
we produced \iac{NLO} QCD+QED matched parton shower. 
As a first step, before the necessary process handling 
is available, we presented results from the \ac{NLO}
\ac{EW} matched parton shower. We studied the Higgs 
lineshape and found large \ac{QED} radiative corrections, 
as expected. These agreed well with the \ac{YFS} 
predictions (with \ac{ME} corrections). In addition, 
we studied the lineshapes of the intermediate $Z$ bosons.
We again found large \ac{QED} corrections which were 
well-modelled by the matched shower. We found that 
an advantage of the \MCatNLO approach is the freedom to 
use massless leptons in the \ac{ME} calculation, while 
keeping the mass dependence in the resummed parts of the 
calculation. In this way, collinear logarithms are resummed 
without relying on small lepton masses to regularise them. 
This is an advantage since \ac{NLO} calculations using massless 
fermions are more widely available than those with full 
mass dependence. We also studied 
the impact of lepton dressing on various invariant mass 
observables. We showed that dressed lepton observables have a 
reduced dependence on the radiative correction methods 
used, and hence a smaller theoretical uncertainty, than 
bare leptons.

Overall, our results show that in the precision era 
of particle physics, higher-order \ac{EW} and \ac{QED} 
corrections need to be included, and their accuracy 
quantified through the comparison of independent methods.
Our implementation of the \MCatNLO method allows a new 
class of methods to be explored, including future 
extensions of the method to electron-positron colliders.
This would allow for another cross-check of predictions 
for the $e^+ e^- \to$ hadrons cross section, which is a 
vital input to the calculation of $(g-2)_\mu$, through 
the determination of the hadronic vacuum polarisation. 

The interleaved dipole shower and \MCatNLO were implemented 
in the event generator \Sherpa, and will be incorporated 
in a future release in the \Sherpa 3 series. One-loop 
matrix elements were provided by \OpenLoops 
\cite{Buccioni:2019sur,Denner:2016kdg,Ossola:2007ax,vanHameren:2010cp}.
All analyses 
and plots were made using \Rivet \cite{Buckley:2010ar,Bierlich:2019rhm}.

%% file: text/yfs.tex
\chapter{Photon splitting corrections to YFS}
\label{chapter:yfs}

\section{Introduction}
\label{sec:yfs:intro}

Precision measurements of the \ac{SM} continue to stress-test
our understanding of particle physics at an unprecedented level.
In particular, charged and neutral Drell-Yan production at hadron colliders
like the \ac{LHC} are used as standard candles due to their large cross
sections and exceedingly small experimental uncertainties, often
below the percent level.
However, these electroweak precision observables have also been 
brought to the forefront of searches for new physics, in the form of 
measured deviations from the \ac{SM} prediction. 
For example, the recent extraction of the $W$ boson mass, performed by the CDF 
experiment on legacy \Tevatron data \cite{CDF:2022hxs}, is in
apparent tension with the world average \cite{ParticleDataGroup:2022pth}
and previous hadron and lepton collider measurements \cite{LHCb:2021bjt,
  ATLAS:2017rzl,CDF:2013dpa,CDF:2013bqv,D0:2012kms,CDF:2012gpf,D0:2009yxq,
  CDF:2007mxw,CDF:2007cmy,CDF:2003tdi,CDF:2000gwd,D0:2002fhu,D0:1999rsi,
  D0:1999ivw,DELPHI:2008avl,ALEPH:2006cdc,L3:2005fft,OPAL:2005rdt},
as well as measurements of other fundamental \ac{EW} parameters in $Z$ production
\cite{ATLAS:2016rnf,ATLAS:2012ewf,CMS:2015cyj,LHCb:2022tbc,CDF:2011ksg,
  CDF:2013uau,ALEPH:2005ab,CDF:2005qwt,ATLAS:2012au,CMS:2011kaj}.
Measurements such as this motivate 
precise theoretical input with uncertainties in the permille
range or lower.
At this level of required accuracy,  
higher-order \ac{QCD} and \ac{EW} corrections in vector-boson production 
must be supplemented with additional sources of theoretical precision.
In addition to a consideration of the structure functions 
that describe 
the make-up of the incident particles, a detailed description of the
vector boson's decay is paramount.
Special emphasis must lie on the precise phase space distribution and flavour
composition of the accompanying radiation, in order to be able to precisely
model the detector response.
With this chapter we contribute to the effort to determine the size and
uncertainty of higher-order \ac{QED} corrections in the description of the
decay of massive vector bosons.

Higher-order corrections to Drell-Yan processes are known to
\ac{NLO} in the complete \ac{EW} \ac{SM}
\cite{Wackeroth:1996hz,Baur:1997wa,Baur:1998kt,Baur:2001ze,Baur:2004ig,
  Andonov:2004hi,Dittmaier:2001ay,Dittmaier:2009cr,Li:2012wna,Alioli:2016fum}.
The recent advances at \ac{NNLO} \ac{QCD}-\ac{EW} mixed calculations \cite{Bonciani:2021zzf,
  Buonocore:2021rxx,Armadillo:2022bgm,Delto:2019ewv,
  Buccioni:2020cfi,Bonciani:2020tvf,Bonciani:2021iis,Behring:2020cqi,
  Dittmaier:2014qza,Dittmaier:2015rxo},
though an impressive achievement in their own right, have not
increased the perturbative accuracy of the description of \ac{EW} or
\ac{QED} radiative corrections themselves.
Alternatively, universal \ac{QED} corrections can be resummed to all orders
either in traditional \ac{QED} parton showers 
by means of the \ac{DGLAP} equation \cite{Seymour:1991xa} 
(see chapter \ref{chapter:qedps}),
or through the soft-photon resummation devised by Yennie, Frautschi,
and Suura \cite{Yennie:1961ad}. These resummations can of course 
be matched to the fixed-order calculations mentioned above (see 
chapter \ref{chapter:mcatnlo}). 
As was described in chapter \ref{chapter:qedps},
\iac{QED} parton shower is available in all major general-purpose
Monte-Carlo event generators:
\Herwig \cite{Bellm:2019zci,Bellm:2015jjp},
\Pythia \cite{Bierlich:2022pfr,Sjostrand:2014zea}, and
\Sherpa \cite{Hoeche:2009xc,Gleisberg:2008ta,Bothmann:2019yzt}, 
while the \ac{YFS} approach is implemented in \Herwig \cite{Hamilton:2006xz}
and \Sherpa \cite{Schonherr:2008av,Krauss:2018djz} for particle decays.
The implementation in \Sherpa has recently been extended to also
resum initial-state soft-photon radiation in $e^+e^-$ collisions
\cite{Krauss:2022ajk}. In addition, dedicated \ac{MC} programs such as 
\Photos \cite{Barberio:1990ms, Barberio:1993qi,Golonka:2005pn,
Davidson:2010ew} are widely used to add \ac{QED} final-state radiation to 
any process.

To reach the necessary precision to make full use of the existing and
future experimental datasets, the \ac{QED} effects impacting the leptonic
final state of the Drell-Yan process have to be understood in detail.
These effects are driven by soft and collinear photon radiation.
They can be resummed to all orders, and be further improved order by
order in perturbation theory.
Such calculations, matching to at least \ac{NLO} \ac{EW} corrections and sometimes
even including \ac{NNLO} \ac{QED} ones, have been implemented using \ac{QED} parton showers
in \Horace \cite{CarloniCalame:2001ny,CarloniCalame:2003ux,
  CarloniCalame:2005vc,CarloniCalame:2006zq,CarloniCalame:2007cd,
  CarloniCalame:2016ouw}
and \Powheg \cite{Bernaciak:2012hj,Barze:2012tt,Muck:2016pko,Barze:2013fru},
using the structure function approach in \Rady
\cite{Dittmaier:2001ay,Dittmaier:2009cr}, and through a \ac{YFS}-type
soft-photon resummation in \Winhac/\Zinhac \cite{Placzek:2003zg}, \Herwig
\cite{Hamilton:2006xz} and \Sherpa \cite{Schonherr:2008av,Krauss:2018djz}.
In addition, the \Photos Monte-Carlo provides an
algorithm based on both soft-photon resummation and \ac{ME} corrections.
Dedicated comparisons between \Sherpa's \ac{YFS}-type resummation and
\Photos \cite{Gutschow:2020cug}, between \Horace and \Photos
\cite{Kotwal:2015bfa}, as well as \Horace and \Winhac
\cite{CarloniCalame:2004fza} have yielded very good agreement.

A key element in the description of final state radiative corrections,
however, has only been sporadically and not very systematically addressed:
the possible splitting of the radiated bremsstrahlungs photons
into secondary charged-particle pairs.
These corrections only enter at a relative $\order(\alpha^2)$ in Drell-Yan
processes, but the production of light flavours may
be enhanced logarithmically and thus gain relevance.
In addition, and in contrast to \ac{QCD}, photons and light charged flavours
like electrons, muons, or pions, are experimentally distinguishable --
such conversions alter the visible make-up of the final state and are thus
of importance at the envisaged theoretical precision.
It is also important to consider here the usual experimental and 
phenomenological practice of dressing charged leptons with photon radiation. 
While definitions of \ac{QCD} jets have been constantly refined, there has 
been little discussion of dressed lepton algorithms since the adoption 
of cone-dressing strategies where all photons within a certain radius 
of the lepton are absorbed. Considering higher-order corrections in the 
form of photons splitting into charged particles has the potential to 
spoil the physically meaningful definition of a lepton dressed with 
photons. The treatment of charged leptons in the presence of secondary 
charged flavours must therefore be handled with care. 
Thus, while a first implementation of pair-production
corrections exists in \Photos \cite{Arbuzov:2012dx,Antropov:2017bed}, 
it only covers photon splittings into electrons and
muons, and their theoretical and phenomenological impact has not been
rigorously appraised.
In this chapter, we address this issue by introducing 
a rigorous independent
framework to calculate these corrections and study the resulting theoretical
and phenomenological implications.

This chapter proceeds as follows: We begin by providing a brief summary
of the \ac{YFS} soft-photon resummation as implemented in \Sherpa before
providing a comprehensive description of the photon splitting implementation,
including a detailed examination of their interplay and the splitting
properties in sec.\ \ref{sec:yfs:methods}.
We then present a detailed discussion of possible extensions of the
standard lepton dressing algorithm to cope with the presence of
secondary pairs of (light) charged particles, and 
quantify their effect on $Z\to e^+e^-$ decays in sec.\ \ref{sec:yfs:dress}.
Finally, we offer some concluding remarks in sec.\ \ref{sec:yfs:conclusions}.

%%%%%%%%%%%%%%%%%%%%%%%%%%%%%%%%%%%%%%%%%%%%%%%%%%%%%%%%%%%%%%%%%%%%%%%%%%%%%%%%%%%%%%%%%%%%%%%%%%%%%%

\section{Soft-photon resummation and photon splittings}
\label{sec:yfs:methods}

Incorporating photon-splitting processes alongside photon
emissions are straightforwardly implemented when both are described
in a common parton shower framework, as was discussed in 
chapter \ref{chapter:qedps}.
In this chapter, however, we base our
implementation on the existing and superior description of
photon emission corrections in the \ac{YFS} framework of 
the \Photons module in \Sherpa \cite{Schonherr:2008av},
including its inherent coherent-radiation formulation and
existing \ac{NNLO} \ac{QED} and \ac{NLO} \ac{EW} corrections \cite{Krauss:2018djz}.
In this section we thus start by providing a brief summary of the
\ac{YFS} soft-photon resummation and its
implementation in the \Sherpa event generator.
The remainder of this section discusses the construction
of the photon splitting algorithm in detail
before examining its properties.

\subsection{The YFS soft-photon resummation}
\label{sec:yfs:methods:yfs}

The work of Yennie, Frautschi and Suura \cite{Yennie:1961ad} 
describes the \ac{IR} singularities of \ac{QED} to all orders.
To achieve this, the \ac{YFS} approach considers all charged 
particles of the theory 
to be massive, and as a consequence only singularities associated with 
soft-photon emission are present.
In particular, all photon splittings are finite and thus do not
partake in the analysis of the \ac{IR} singular structure.
Using that knowledge, the \ac{YFS} algorithm reorders the perturbative
expansion of a scattering or decay \ac{ME}. 
This is achieved by separating the \ac{IR} divergences from the finite 
remainders to all orders. The IR divergent terms form a series which can 
be exponentiated, amounting to a resummation of soft-photon logarithms
in the enhanced real and virtual regions, leaving a perturbative expansion
in \ac{IR}-finite, hard photons (both real emissions and virtual exchanges).

In the implementation of the \ac{YFS} resummation in \Sherpa for
particle decays \cite{Schonherr:2008av}, the all-orders
soft-photon resummed differential decay rate is written as
\begin{equation}\label{eq:methods:yfsmaster}
  \begin{split}
    \done\Gamma^\text{YFS}
    =\;&
      \done\Gamma_0\cdot e^{\alpha Y(\omega_\text{cut})}\cdot
      \sum\limits_{n_\gamma=0}^\infty\frac{1}{n_\gamma!}
      \left[
        \prod\limits_{i=1}^{n_\gamma}\done \Phi_{k_i}\cdot\alpha\,
        \tilde{S}(k_i)\,\Theta(k_i^0-\omega_\text{cut})
        \cdot\mathcal{C}
      \right]\,,
  \end{split}
\end{equation}
wherein $\done\Gamma_0$ is the \ac{LO} differential decay rate and
the \ac{YFS} form factor $Y(\omega_\text{cut})$ contains the soft-photon
logarithms.
The decay rate is then summed over all possible additional photon
emissions with an energy larger than $\omega_\text{cut}$ 
with respect to the \ac{LO} decay.
Each emission is described through its eikonal $\tilde{S}$ and
corrected for hard emission effects up to a given order through
the correction factor $\mathcal{C}$.\footnote{
  The hard (real and virtual) photon-emission corrections $\mathcal{C}$
  are available up to \ac{NLO} \ac{EW} for leptonic $W$ decays and up to
  \ac{NNLO} \ac{QED} + \ac{NLO} \ac{EW} for leptonic $Z$ decays \cite{Krauss:2018djz}.
}

Unlike a conventional parton shower, where the resummation
is reliant on the factorisation of subsequent emissions when ordered in an
evolution variable, \ac{YFS} photons are unordered.
They are also emitted coherently from the charged multipole
through the radiator function $\tilde{S}$ and are thus not inherently
associated with a specific emitter particle.
Consequently, when the produced final state is to be further treated by
a dedicated photon-splitting parton shower, the existing configuration
must be interpreted in the parton shower's evolution and splitting
language before any further splittings take place.
Of course, care has to be taken so as to not compromise its 
\ac{LL} soft-photon resummation.
The effects of the algorithm we will describe in the following 
are completely beyond the scope of the \ac{YFS} formulation
without any potential overlap. Hence, this requirement amounts to 
ensuring the kinematic
recoil induced by a splitting photon on the primary charged particle ensemble
(and possibly other existing photons) vanishes in the limit that the
energy of the splitting photon vanishes.
While this is trivially true as all charged particles are treated
as massive, the recoil assignment performed in this study and described 
in section \ref{sec:yfs:methods:photonsplit} introduces corrections to the momenta of the primary 
charged particle ensemble which scale non-logarithmically with the photon 
energy and hence do not contribute to the \ac{LL} resummation.

\subsection{Photon splittings}
\label{sec:yfs:methods:photonsplit}

In this section we introduce the parton shower algorithm
which computes the photon splitting probabilities and kinematics,
while the principal user input commands to steer its behaviour
are described in appendix \ref{app:settings}.
We will use the usual notation associated with a \ac{CS}
dipole shower.
% A parton shower generates
% radiation off existing particles using the splitting functions, which are derived
% from the \ac{ME} in the \ac{IR} limit.
Since the \ac{YFS} algorithm requires massive charged particles,
it is necessary to include all masses in this shower for consistency.
There are therefore no \ac{IR} singularities associated with
our photon splittings, since the collinear pole is regulated
by the fermion and scalar masses.
However, the aim is still to capture the correct behaviour in
the quasi-collinear limit, accounting for the logarithmic
enhancement for collinear splitting into light flavours.
Throughout this section we will focus on configurations where
all relevant particles are in the final state of the
decay process, \ie decays of neutral resonances.
In section \ref{sec:yfs:fidip}, we discuss the 
modifications needed to describe the decay of charged resonances.

The key part of the parton shower algorithm is, as usual, the
veto algorithm (\see sec. \ref{sec:qedps:veto}).
This allows us to avoid the
problem of analytically integrating the splitting functions,
which are detailed below, by using an
overestimate to evaluate the cumulative emission probability,
and then vetoing emissions
with a probability which corrects for the overestimate.
The evolution begins at some starting scale $\tstart$ which is the 
highest possible scale for a splitting to take place;
we postpone its exact definition to the end of this section.
All splitting functions compete: a splitting scale is calculated
for each possible combination of splitter, splitting
products, and spectator.
Whichever splitting results in the highest splitting 
scale is selected.
If the splitting is accepted (not vetoed), a new particle is
created and the flavours and kinematics of 
existing particles are updated.
The whole process is repeated, starting from the selected splitting scale, 
and iterated until some \ac{IR} cutoff $t_c$ is reached. 
This cutoff is needed to regulate the divergence
of the splitting functions in the general case
where these appear.
For a \ac{QCD} shower, a physical choice for the cutoff is the
hadronisation scale, which is of $\order(1\,\text{GeV})$,
well above $\Lambda_\text{QCD}$ where \ac{QCD} dynamics
turn non-perturbative.
For a \ac{QED} shower, however, the splittings
which do not involve quarks can evolve to arbitrarily low scales.
In the algorithm presented here,
which contains only splitting functions of photon emissions off
charged scalars and fermions as well as of photons splitting into massive
fermions or pseudo-scalar hadrons,
the cutoff is dictated by the mass of the lightest
fermion, i.e. $t_c = 4m_e^2$ or lower.

As stated earlier, in the case of a photon splitting to a fermion or 
scalar particle-antiparticle pair, there is no soft divergence.
The collinear divergence present for massless splitting products
is converted into a logarithmic collinear
enhancement when masses are included;
hence, lighter particles will have a larger contribution
to photon splitting corrections.
Here we include all possible splittings up to a mass cutoff of
$2 m_i \lesssim 1 \,\GeV$ in addition to $\tau$ pair production which,
while rare, contributes
to some observables through the decays to lighter leptons or hadrons.
Since most splittings occur
near or below the hadronisation scale, we consider hadrons, not quarks,
to be the relevant \ac{QCD} degrees
of freedom.
Using this mass cutoff, the hadrons which can be produced are the charged pions
and kaons. They are pseudo-scalars, and their interaction with photons
is modeled using point-like scalar \ac{QED}, neglecting any substructure
effects.
We use the scalar splitting functions of ref. \cite{Catani:2002hc}.
Depending on the experimental environment, the kaons and $\tau$ leptons
might decay before hitting any detector.
This can be handled within the usual (hadron) decay treatment available
within the \Sherpa framework \cite{Bothmann:2019yzt,Hoche:2014kca}.

\paragraph*{Splitting functions and spectator assignment.}
In the usual parton shower notation, we use the following
dipole splitting functions \cite{Catani:2002hc,Schumann:2007mg,Schonherr:2017qcj}
\begin{equation} \label{eq:methods:SFs}
  \begin{split}
    S_{s_\ijt(\kt)\to s_i\gamma_j(k)}
    \,=&\;
    S_{\bar{s}_\ijt(\kt)\to \bar{s}_i\gamma_j(k)}
    \,=\;
      -\,\mathbf{Q}^2_{\ijt\kt}\;\alpha\,
      \left[\frac{2}{1-z+zy}-\frac{\tilde{v}_{\ijt,\kt}}{v_{ij,k}}\left(2+\frac{m_i^2}{p_ip_j}\right) \right] \\
    S_{f_\ijt(\kt)\to f_i\gamma_j(k)}
    \,=&\;
    S_{\bar{f}_\ijt(\kt)\to \bar{f}_i\gamma_j(k)}
    \,=\;
      -\,\mathbf{Q}^2_{\ijt\kt}\;\alpha\,
      \left[\frac{2}{1-z+zy}-\frac{\tilde{v}_{\ijt,\kt}}{v_{ij,k}}\left(1+z+\frac{m_i^2}{p_ip_j}\right) \right] \\
    S_{\gamma_\ijt(\kt)\to s_i\bar{s}_j(k)}
    \,=&\;
    S_{\gamma_\ijt(\kt)\to f_i\bar{f}_j(k)}
    \,=\;
      -\,\mathbf{Q}^2_{\ijt\kt}\; \alpha\,
      \left[
        1-2z(1-z)-z_+ z_- \vphantom{\frac{m_i^2}{p_ip_j}}
      \right],
  \end{split}
\end{equation}
for splittings involving the scalars $s$, fermions $f$, their antiparticles
$\bar{s}$ and $\bar{f}$, and a photon $\gamma$,
in terms of the splitting variable $y$ and light-cone momentum fraction $z$.
These are defined as
\begin{equation} \label{eq:methods:zy}
  y = \frac{p_i p_j}{p_i p_j + p_i p_k + p_j p_k}
  \qquad\text{and}\qquad
  z = \frac{p_i p_k}{p_i p_k + p_j p_k}
  \;.
\end{equation}
Further, $m_i$ is the mass of the splitting product $i$, and
$z_-$ and $z_+$ are the phase space boundaries
\begin{equation} \label{eq:methods:zlimits}
  z_\pm = \frac{2 \mu_i^2 + (1-\mu_i^2-\mu_j^2-\mu_k^2)\,y}{2(\mu_i^2+\mu_j^2 + (1-\mu_i^2-\mu_j^2-\mu_k^2)\,y)} (1 \pm v_{ij,i} \, v_{ij,k})\;,
\end{equation}
where the dimensionless rescaled masses $\mu_i^2=m_i^2/Q^2$ are introduced for convenience, and
$Q^2=(p_i+p_j+p_k)^2=(p_\ijt+p_\kt)^2$ is the invariant mass of the dipole.
The relative velocities $\tilde{v}_{\ijt,\kt}$, $v_{ij,k}$, and $v_{ij,i}$
are given by
\begin{equation} \label{eq:methods:relv}
  \begin{split}
    \tilde{v}_{\ijt,\kt} \,=&\; \frac{\sqrt{\lambda(1,\mu_{\ijt}^2,\mu_\kt^2)}}{1-\mu_{\ijt}^2-\mu_\kt^2}\;, \\
    v_{ij,i} \,=&\; \frac{\sqrt{(1-\mu_i^2-\mu_j^2-\mu_k^2)^2 \,y^2 -4\mu_i^2\mu_j^2}}{(1-\mu_i^2-\mu_j^2-\mu_k^2)\,y+2\mu_i^2}\;, \\
    v_{ij,k} \,=&\; \frac{\sqrt{(2\mu_k^2+(1-\mu_i^2-\mu_j^2-\mu_k^2)(1-y))^2-4\mu_k^2}}{(1-\mu_i^2-\mu_j^2-\mu_k^2)(1-y)}\;.
  \end{split}
\end{equation}
Finally, the charge correlator $\mathbf{Q}^2_{\ijt\kt}$ is defined in 
eq.\ \eqref{eq:qedps:charge_correlator}, repeated here for convenience:
\begin{equation} \label{eq:methods:chargecorrelator}
  \mathbf{Q}^2_{\ijt\kt} = \begin{cases}
     \;\;\frac{Q_{\ijt} Q_{\kt} \theta_{\ijt} \theta_{\kt}} {Q^2_{\ijt}} & \ijt \neq \gamma \\
     \;\;\kappa_{\ijt\kt} & \ijt=\gamma
  \end{cases}
  \qquad\text{with}\qquad
  \sum\limits_{\kt\neq\ijt}\kappa_{\ijt\kt}=-1\quad\forall\,\ijt=\gamma\;,
\end{equation}
where the $Q_{\ijt}$ and $Q_{\kt}$ are the charges of the splitter and
spectator respectively and their $\theta_{\ijt/\kt}$ are 1 ($-1$) if
they are in the final (initial) state.
The $\kappa_{\gamma\kt}$ must ensure that the splitting
functions are appropriately normalised such that the correct \ac{IR}
limit is found, but are otherwise unconstrained.
Here we choose
\begin{equation}
  \kappa_{\gamma\kt} = -\frac{1}{\mathcal{N}_\mathrm{specs}},
\end{equation}
where $\mathcal{N}_\mathrm{specs}$ is the chosen number of possible
spectators, \ie we choose to weigh all selected spectators $\kt$ equally.
The photon splittings themselves are free of soft divergences, hence the
spectator is only needed for momentum conservation and, in
principle, any other particle of the process may assume this role.
In the present context, we consider all primary charged decay products
as possible spectators of photon splittings as our default, but the choice
to consider only the splitting photon's originator particle (as
reconstructed, described below) has also been implemented, \see
appendix \ref{app:settings}.
While the other present \ac{YFS} photons and other neutral decay products
as well as the decaying particle itself are all valid spectators, the two
choices described above both guarantee that enough energy is available 
to allow photon splitting into heavier flavours to occur.
Limiting the number of spectators also helps to reduce the computational
complexity.
For further photon radiation off the products of a photon splitting,
the spectator assignment, and therefore the recoil, should be kept
local in that system.

\paragraph*{Evolution variable.} \label{par:methods:evovariable}
For the evolution variable $t$ used in the parton 
shower, the requirement of \ac{LL} accuracy 
means that any choice which preserves $\mathrm{d}t/t$ is formally  
correct in the \ac{IR} limit.
We consider two variants,
modified virtuality $\qbar^2$ and
transverse momentum $\kTsq$. We will usually refer to the 
modified virtuality simply as the virtuality, since as 
we see from the definition in eq.\ \eqref{eq:methods:startingscalevirt}
below, for photon emitters the definitions coincide.
% To be precise, with the help of the dipole invariant mass
% $Q^2$ introduced earlier, the virtuality is defined as
The virtuality is defined in terms of the dipole invariant mass $Q^2$
and the masses of the emitter $m_\ijt$, splitting products $m_{i/j}$,
and spectator $m_k$. The modified virtuality is given by
\begin{equation}\label{eq:methods:startingscalevirt}
  \qbar^2 = (Q^2-m_i^2-m_j^2-m_k^2)\,y + m_i^2 + m_j^2 - m_{\ijt}^2\;.
\end{equation}
Specifically, for the two relevant cases this translates to
\begin{equation}\label{eq:methods:startingscalevirt1}
  \qbar_{\mathrm{f}\to\mathrm{f}\gamma}^2
  = (Q^2-m_\mathrm{f}^2-m_k^2)\,y
\end{equation}
for photon emissions, where the flavour $\mathrm{f}$
can either be a scalar $s$, fermion $f$, or their
antiparticles $\bar{s}$ and $\bar{f}$, and
\begin{equation}\label{eq:methods:startingscalevirt2}
  \qbar_{\gamma\to\mathrm{f}\bar{\mathrm{f}}}^2
  = (Q^2-2m_\mathrm{f}^2-m_k^2)\,y + 2m_\mathrm{f}^2
\end{equation}
for photon splittings.
We see that, as stated earlier, photon emissions are possible at arbitrarily
low evolution scales while photon splittings can only occur
if the virtuality exceeds the pair-production threshold.
Likewise, the transverse momentum is given by 
\begin{equation}\label{eq:methods:startingscaleKt}
  \kTsq = (Q^2-m_i^2-m_j^2-m_k^2)\,y\,z(1-z) - m_i^2\,(1-z)^2 - m_j^2\,z^2\;.
\end{equation}
This is the momentum generated in the splitting which is transverse to 
the plane defined by the emitter-spectator dipole.
Again, for photon emissions this translates to
\begin{equation}\label{eq:methods:startingscaleKt1}
  \kTsq_{\,\mathrm{f}\to\mathrm{f}\gamma}
  = (Q^2 - m_\mathrm{f}^2-m_k^2)\,y\,z(1-z) - m_\mathrm{f}^2\,(1-z)^2
\end{equation}
and to
\begin{equation}\label{eq:methods:startingscaleKt2}
  \kTsq_{\,\gamma\to\mathrm{f}\bar{\mathrm{f}}}
  = (Q^2 - 2m_\mathrm{f}^2-m_k^2)\,y\,z(1-z) - m_\mathrm{f}^2\,(z^2+(1-z)^2)\;
\end{equation}
for photon splittings.
As discussed, photon emissions are possible down to $\kTsq=0$,
but in this case the photon-splitting threshold also lies at $\kTsq=0$.
As a result, the chosen \ac{IR} cutoff $t_c$ will induce a minimal
$\kT$, and thus a minimal opening angle, produced in 
the pair-creation process.
The pair's virtuality $\qbar^2$ automatically introduces a pair-production 
threshold, making it a viable candidate for the ordering variable for 
photon emissions. In addition, the \ac{BLM} argument \cite{Brodsky:1982gc}
states that since the only ultraviolet divergences in this process come from 
vacuum polarisation insertions in the photon propagator, the correct 
choice of scale to minimise higher-order corrections is the photon 
virtuality.

% Since this is a reconstructed splitting,
% we have the final momenta rather than the initial momenta and so must
% add the photon's momentum to that of the splitter-spectator pair
% under consideration. For the purposes of reconstructing the history
% and determining the starting scale, each photon is treated as though
% it is the last to be emitted (all other \ac{YFS} photons have already
% removed energy from the dipole). Care has been taken to define the
% virtuality such that $t=0$ for an on-shell particle (which occurs when $y=0$ and no splitting
% can take place).

It can be seen that the relation 
\begin{equation}\label{eq:methods:psdifferential}
  \frac{\mathrm{d}t}{t} = \frac{\mathrm{d}\kTsq}{\kTsq} = \frac{\mathrm{d}\bar{q}^2}{\bar{q}^2}
\end{equation}
holds, therefore both transverse momentum and virtuality are possible choices
of evolution variable and no Jacobian is needed to translate between them.

Using these definitions, there are three well-motivated choices for
the global evolution variable.
\begin{enumerate}
 \item
    As in most \ac{QCD} parton showers, $t=\kTsq$ is a viable choice.
    In analogy to \ac{QCD},
    ordering photon emissions by transverse momentum
    results in the inclusion of charge coherence effects \cite{Amati:1980ch},
    but there is no particular motivation to use $t=\kTsq$ as the evolution
    variable for photon splittings into charged-particle pairs.
  \item
    Choosing $t=\qbar^2$ is an equally valid option.
    Due to the $s$-channel nature of photon splittings, the photon virtuality
    is expected to be a good ordering variable here, as outlined in 
    the \ac{BLM} approach to determining scales. In addition,
    the gluon virtuality is
    commonly used in \ac{QCD} parton showers to describe the $g\to q \bar{q}$ 
    splitting, as described in the \Pythia manual \cite{Bierlich:2022pfr}.
    But since it does not implement angular ordering natively, it is not
    expected to yield the best description of soft-photon emissions.
  \item
    Following eq.\ \eqref{eq:methods:psdifferential},
    we are free to interpret the evolution variable differently in
    different splitting processes as long as $\done t/t$ is invariant.
    As our default, we thus choose to interpret the evolution
    variable $t$ as $\kTsq$ in photon emissions and as $\qbar^2$
    in photon splittings. We will refer to this the ``mixed scheme'' in
    later sections.
\end{enumerate}
All three choices are implemented, \see appendix \ref{app:settings},
and some of their respective consequences will be explored in
sec.\ \ref{sec:yfs:methods:diagnostics}.

\paragraph*{Generation of splitting variables.}
While the evolution variable $t$ is generated as usual in the veto
algorithm, the light-cone momentum fraction $z$ has to be generated within
its allowed range $[z_-,z_+]$.
The integration limits $z_\pm$ are defined in 
eq.\ \eqref{eq:methods:zlimits}, but in order
to generate a Sudakov factor we work with the evolution variable $t$.
We generate a trial emission using the integral of the
overestimate of the splitting function, for which the
$z$ limits are necessary, but
we do not yet know the value of the kinematic variable
$y$ (eq.\ \eqref{eq:methods:zy}).
Using a change of variables to replace 
$y$ with the evolution parameter yields usable $z$ ranges at this stage. 
Hence in the $\kT$ ordered scheme, the $z$ limits are \cite{Schumann:2007mg}
\begin{equation}
  z_{\pm,\kT} = \text{min/max}{\left[\frac{1}{2} \left( 1 \pm \sqrt{1-\frac{4t_c}{Q^2}} \right),\,z_\pm\right]}.
\end{equation}
Note that the $z_\pm$ are not yet known, but can be overestimated by 0 and 1,
respectively.
The above expression thus gives an overestimate of the true phase space
available.
The number of splittings rejected as a result is very small, however,
and does not have a large impact on the efficiency of the algorithm.

On the other hand, in the virtuality ordered scheme $\qbar^2$ has no $z$ dependence.
This means that $y$ can be determined independently of the
light-cone momentum fraction $z$ as well, $y(t,z)=y(t)$, by
solving eq.\ \eqref{eq:methods:startingscalevirt} for $y$.
This implies that the $z$ limits can be written as 
\begin{align*}
  z_{\pm,\qbar}
  = \text{min/max}
    &\Biggl[
      \frac{q^2+m_i^2-m_j^2}{2q^2} \Biggr. \\
      &\times \Biggl. \left(
        1
        \pm \sqrt{1-\frac{4m_i^2 q^2}{\left(q^2+m_i^2-m_j^2\right)^2}} \,
            \sqrt{1+\frac{4m_k^2q^2}{\left(Q^2-q^2-m_k^2\vphantom{m_j^2}\right)^2}}
      \right),\,z_\pm\Biggr] \numberthis
\end{align*}
where $q^2=\qbar^2+m_\ijt^2$.
Again, however, it is only a small price in efficiency to use larger and simpler limits at 
the trial emission stage. In the results that follow, $z_{-,\bar{q}^2}=0$ and 
$z_{+,\bar{q}^2}=1$ have been used. 

\paragraph*{Kinematics.}
% With the above definitions of the evolution and splitting variable,
% $t$ and $z$, or dipole variables $y$ and $z$, respectively, 
With the above definitions of the dipole variables $y$ and $z$ or, 
alternatively, with the evolution variable $t$ and the splitting variable $z$,
and the
uniformly distributed azimuthal splitting angle $\phi$, we can now
build the kinematics of the splitting products $i$ and $j$ and the
spectator $k$ after the emission process.
The new momenta are given by an inversion of the momentum maps
of ref. \cite{Catani:2002hc}, and in their construction we follow
ref. \cite{Schumann:2007mg}.
In particular, for the final-final dipoles discussed here,
they are given in sec.\ 3.1.1 eq.\ (49)--(58) of ref. \cite{Schumann:2007mg}.
Note that this redistribution of momenta is \ac{IR} safe and
does not spoil the \ac{LL} accuracy of the \ac{YFS} resummation, since
it introduces non-logarithmic corrections only.

\paragraph*{Starting conditions.}
Having defined the evolution and splitting variables as well as
the splitting functions and kinematic mappings above, we now
need to specify the initial conditions to fully define the algorithm.
As the photon emissions are already generated by the \ac{YFS} soft-photon
resummation, the existing distribution has to be reinterpreted as if
it was generated by our shower algorithm.
Then, the missing photon-splitting corrections can be embedded into
the existing calculation.
By not allowing further photon-radiation off the primary
charged-particle ensemble, double counting is avoided.

To determine the scale at which each existing photon has been
produced, we calculate the emission probabilities according to
the splitting functions $S_{f_\ijt(\kt)\to f_i\gamma_j(k)}$ and
$S_{s_\ijt(\kt)\to s_i\gamma_j(k)}$, respectively, for every
existing soft-photon $\gamma_j$.
Therein, every primary charged particle (all existing charged particles
of the process at the this stage) can act as possible emitter $\ijt$
or spectator $\kt$.
One of those possible splitting functions is then selected either
according to its probability $S_{\ijt\kt\to ijk}/\sum_{\ijt\kt} S$
(default), or by selecting the one with the largest splitting
probability (\see appendix \ref{app:settings}).
Its reconstructed evolution variable $t$ is then set as the starting
scale $t_{\text{start},j}$ of the further evolution for photon $\gamma_j$.
The above parton shower algorithm is then started at the largest
of all photons' starting scale, $\tstart=\max[t_{\text{start},j}]$,
but each individual photon's evolution is only active for
$t\leq t_{j,\text{start}}$.

%%%%%%%%%%%%%%%%%%%%%%%%%%%%%%%%%%%%%%%%%%%%%%%%%%%%%%%%%%%%%%%%%%%%%%%%%%%%%

\subsection{Properties of the photon-splitting algorithm}
\label{sec:yfs:methods:diagnostics}

Having the algorithm to calculate photon splitting probabilities
at hand, we can now examine its properties and assess the
consequences of the specific algorithmic choices discussed above.
To be precise, we use the example of an on-shell $Z$ boson
decaying to an $e^+e^-$ pair (maximising the number of radiated
photons).
Hence, as we are not in a collider environment, we use
a spherical coordinate system to measure relative radial
distances $\Delta\Theta$ in the following.

We begin by presenting a detailed look into the conditions under
which the photons generated through the \ac{YFS} soft-photon resummation
split.
As discussed, in a first step, the existing distribution of photons
and primary emitters must be clustered in order to assign individual
starting scales to the evolution of each photon.
This assignment is of course dependent on the choice of evolution
variable for photon emissions off charged particles as well as the
choice of spectator scheme.

\begin{figure}
  \centering
  \includegraphics[width=0.47\textwidth]{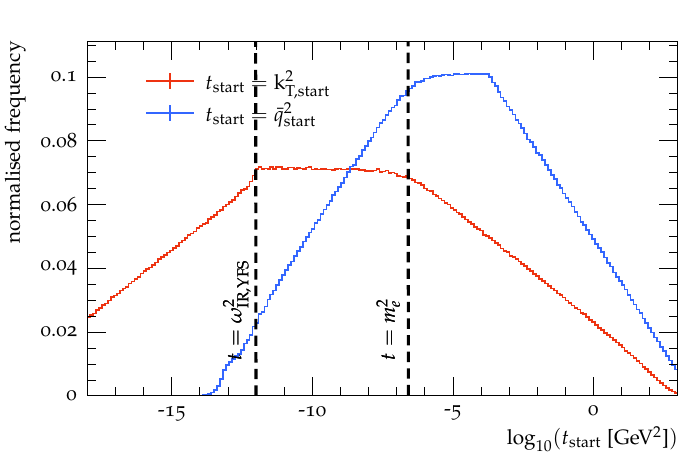}
  \hfill
  \includegraphics[width=0.47\textwidth]{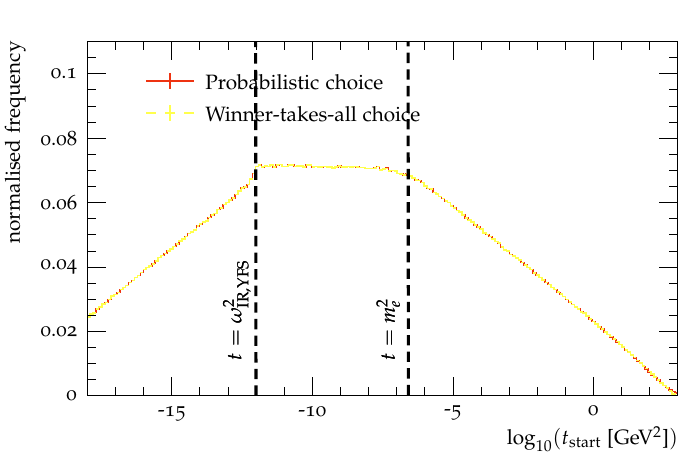}
  \caption[A comparison of the algorithmic choices made 
    in reconstructing the initial $e^\pm\to e^\pm\gamma$ splitting
    generated by \Photons.
    \textbf{Left:} A comparison of the frequency of the reconstructed starting scales \tstart
    using two choices for the evolution variable $t$, $\kTsq$ or $\qbar^2$,
    in the reconstructed initial $e^\pm\to e^\pm\gamma$ splitting.
    \textbf{Right:} A comparison of the frequency of the reconstructed starting scales
    $t_\text{start}=\mathrm{k}_{\mathrm{T},\text{start}}^2$
    using either a probabilistic determination of the emitter lepton
    or a winner-takes-all
    in the reconstructed initial $e^\pm\to e^\pm\gamma$ splitting.]{
      A comparison of the algorithmic choices made 
    in reconstructing the initial $e^\pm\to e^\pm\gamma$ splitting
    generated by \Photons.
    \textbf{Left:} A comparison of the frequency of the reconstructed starting scales \tstart
    using two choices for the evolution variable $t$, $\kTsq$ or $\qbar^2$,
    in the reconstructed initial $e^\pm\to e^\pm\gamma$ splitting.
    \textbf{Right:} A comparison of the frequency of the reconstructed starting scales
    $t_\text{start}=\mathrm{k}_{\mathrm{T},\text{start}}^2$
    using either a probabilistic determination of the emitter lepton
    or a winner-takes-all
    in the reconstructed initial $e^\pm\to e^\pm\gamma$ splitting.
    The threshold for photons splitting into charged particle pairs is
    $t>4m_e^2$, and $\omega_{\text{IR},\text{YFS}}^2$ is the IR
    cutoff of the YFS-style algorithm which generates the photons.
    \label{fig:yfs:results:startingscale}
  }
\end{figure}

Fig.\ \ref{fig:yfs:results:startingscale} shows the distribution
of starting scales when the photon emissions are reconstructed with the inverse
emission kernels.
In the left plot, the ordering variable is interpreted as
either a relative transverse momentum, $t=\kTsq$ (red) or a
virtuality, $t=\qbar^2$ (blue).
In the transverse momentum ordering scheme we observe an approximately
logarithmic rise in the abundance of starting scales, starting at
the kinematic limit of $\kTsq\simeq\tfrac{1}{4}\,m_Z^2$.
This reflects the photon spectrum produced by the soft-photon resummation.
This logarithmic rise levels out at $\kTsq\approx m_e^2$, formed by
the reconstructed $\kTsq$ of ultra-soft photons of the event.
This plateau ends at the soft-photon cutoff $\omega^2_\text{IR,YFS}$
 used in the soft-photon resummation.
In contrast, in the virtuality ordering
scheme we see that the majority of events have starting scales above $10^{-6}$ GeV.
In both cases the characteristic scale at $t=m_e^2$ is induced by both the splitter and the spectator masses of the primary decay.
The effect of the \ac{IR} cutoff is
straightforward in the $t=\kTsq$ case, as shown by the labelled black dashed
line. In the $t=\qbar^2$ case the cutoff does not dictate the turning point
of the frequency plot, but, indirectly, the point at which the frequency falls
to zero. Due to normalisation this has the effect of increasing the frequency
above the electron mass. This appears as a flattening off of the plot just above
the electron mass squared, before the frequency falls towards zero at $m_Z^2$
independent of the IR cutoff.
We note that in the mixed ordering scheme, which we choose as our default
ordering variable scheme, $\tstart = \mathrm{k}_{\mathrm{T},\text{start}}^2$.

On its right-hand side, fig.\ \ref{fig:yfs:results:startingscale} shows
the distribution of starting
scales $\tstart=\kTsq$ when the reconstructed emission of a \ac{YFS} photon from
one of the final-state charged particles is chosen either probabilistically
according to the relative sizes of the splitting functions (red, our default) or by simply
choosing the more likely emitter (yellow, dashed). There is no significant difference
between the two schemes. A very small difference occurs at the high $\tstart$ end.
In the winner-takes-all
scheme, large starting scales are less likely because the emitter with
the largest splitting function is always chosen; the chosen emitter is
the particle which results
in a smaller starting scale, due to the soft divergence and collinear enhancement
of the splitting functions.

\begin{figure}
  \centering
  \includegraphics[width=0.45\textwidth]{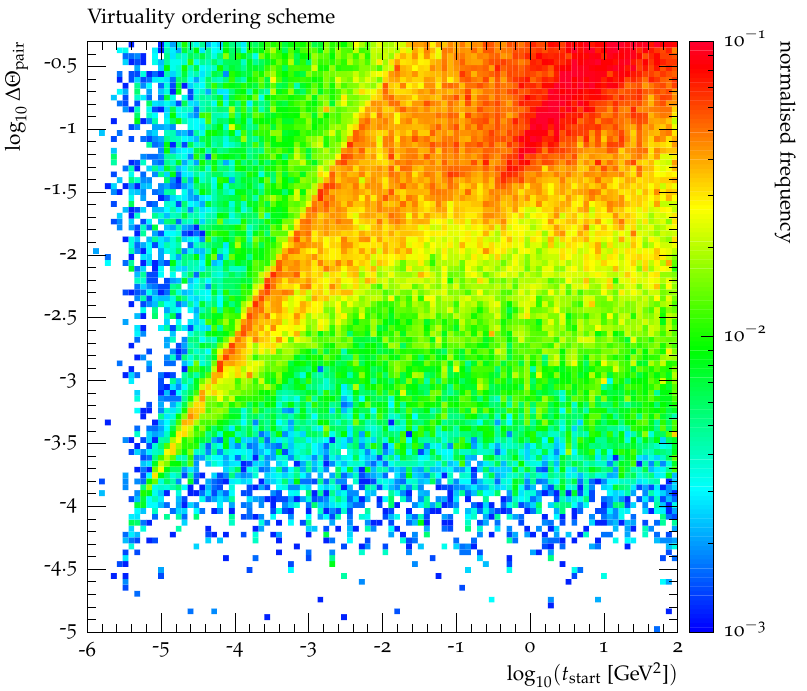}
  \includegraphics[width=0.45\textwidth]{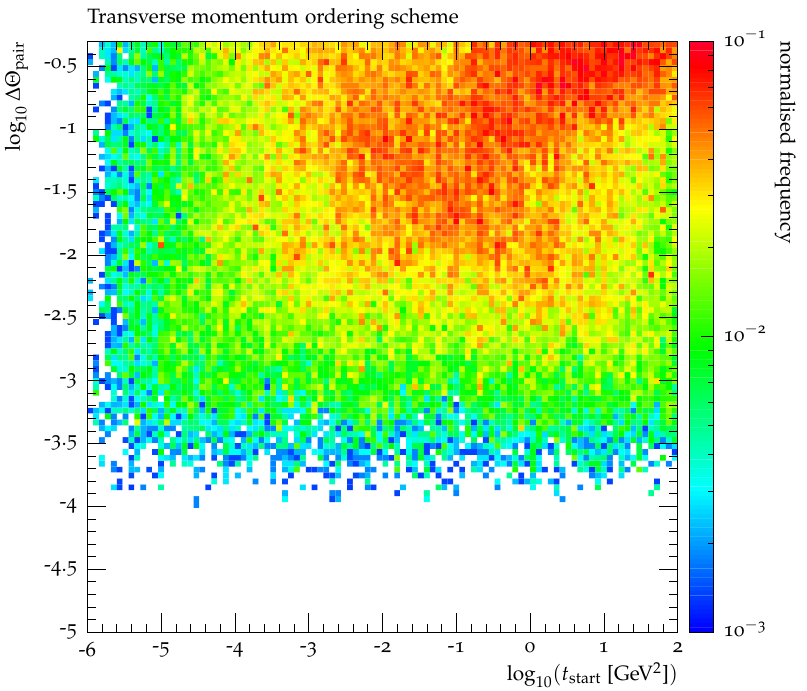}
  \hfill
  \includegraphics[width=0.45\textwidth]{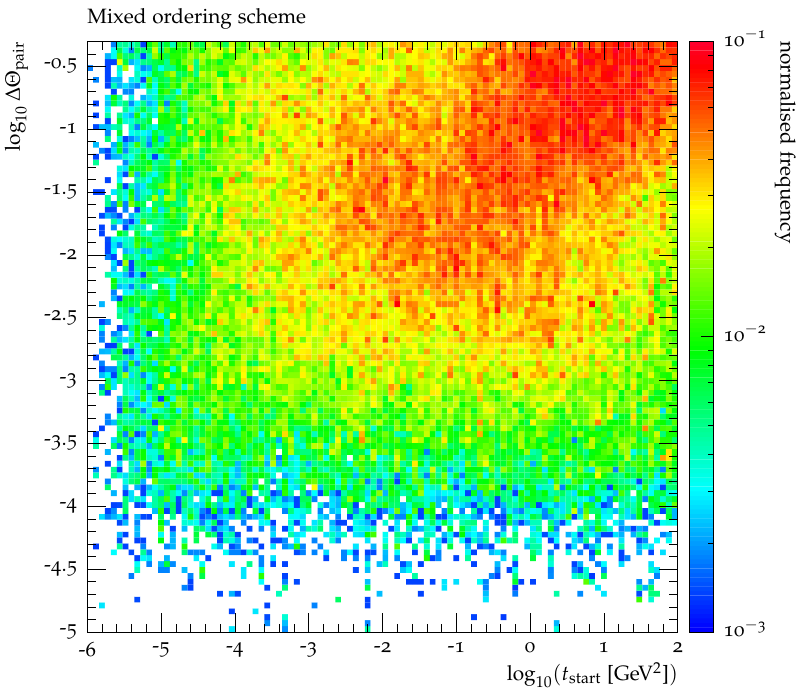}
  \caption[The interdependence of the starting scale $\tstart$ of a photon
  and the angular separation between the particles produced
  in its splitting.]{
    The interdependence of the starting scale $\tstart$ of a photon
    and the angular separation between the particles produced
    in its splitting, $\Delta \Theta_\mathrm{pair}$, in the
    in the $\qbar$-ordered scheme (left), the $\kT$-ordered
    scheme (centre), and the mixed ordering scheme (right).
    \label{fig:yfs:results:tdR}
  }
\end{figure}

Having established the starting conditions of the photons' evolution
we can now examine their splitting process into pairs of charged
particles, and the interplay of the choice of interpretation of the
evolution variable in either splitting process.
Therefore, fig.\ \ref{fig:yfs:results:tdR} depicts the correlation of the
starting scale $t_{\text{start},j}$ of a photon and the collinearity,
or opening angle $\Delta\Theta_\text{pair}$, of 
its splitting products (mainly $e^+e^-$ pairs),
for all three different choices of interpretation of the evolution
variable $t$: $\qbar^2$, $\kTsq$, or mixed.
In the virtuality-ordered scheme where $t=\qbar^2$,
photons can only split if $t$ exceeds the pair-creation threshold of
the lightest charged species, $t\geq 4\,m_e^2\approx 10^{-6}\,\text{GeV}^2$.
Furthermore, there is also a strong correlation between
the starting scale of the evolution and the eventual splitting angle,
which is mainly a consequence of the identification of the starting
scale $\tstart$ for each photon.
As already anticipated in section \ref{sec:yfs:methods:photonsplit} above,
the evolution scale in the $\kT$-ordered case is only constrained to be
above the \ac{IR} cutoff $t_c$, which is also chosen to be
$t_c=4\,m_e^2$ here.
This constrains the opening angle of the pair of splitting 
products to be $\Delta\Theta_\text{pair} \gtrsim 10^{-4}$.
The mixed scheme, in interpreting $t=\kTsq$ in reconstructing the photon
emission to define $\tstart$ and $t=\qbar^2$ in photon splittings,
combines aspects of these two schemes, producing a smooth
distribution in the whole $(\tstart,\Delta \Theta_\text{pair})$ space independent
of $t_c$, as long as $t_c\leq 4\,m_e^2$.
The opening angle becomes relevant when studying the recombination
properties of the splitting product into a dressed primary charged
particle, \see sec.\ \ref{sec:yfs:results:dress}.

\begin{figure}
  \centering
  \includegraphics[width=0.45\textwidth]{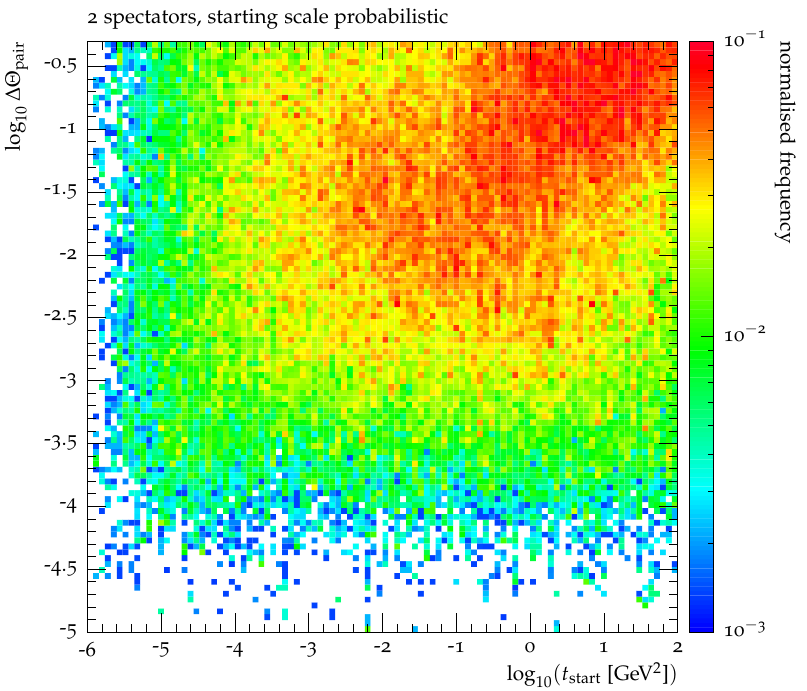}
  \hspace*{0.05\textwidth}
  \includegraphics[width=0.45\textwidth]{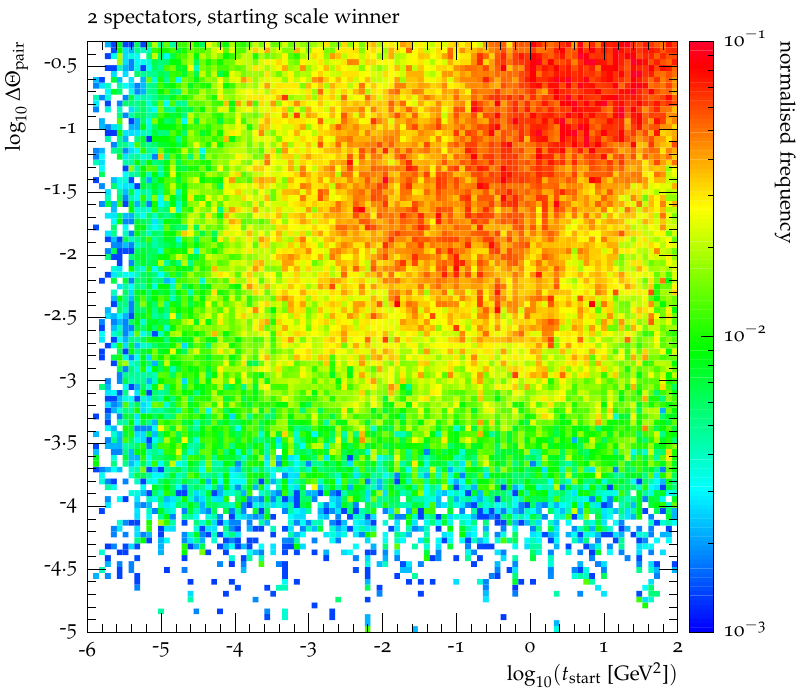}
  \\[2mm]
  \includegraphics[width=0.45\textwidth]{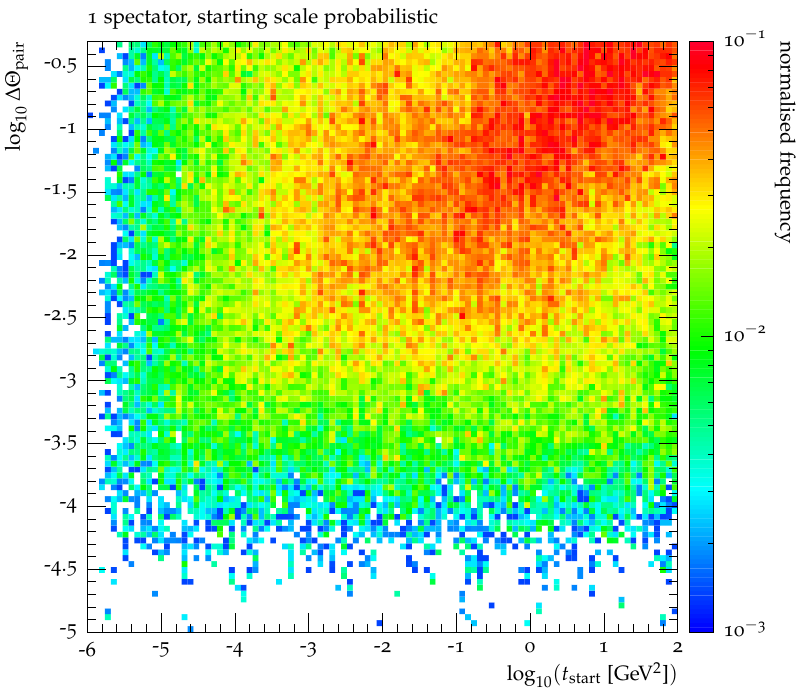}
  \hspace*{0.05\textwidth}
  \includegraphics[width=0.45\textwidth]{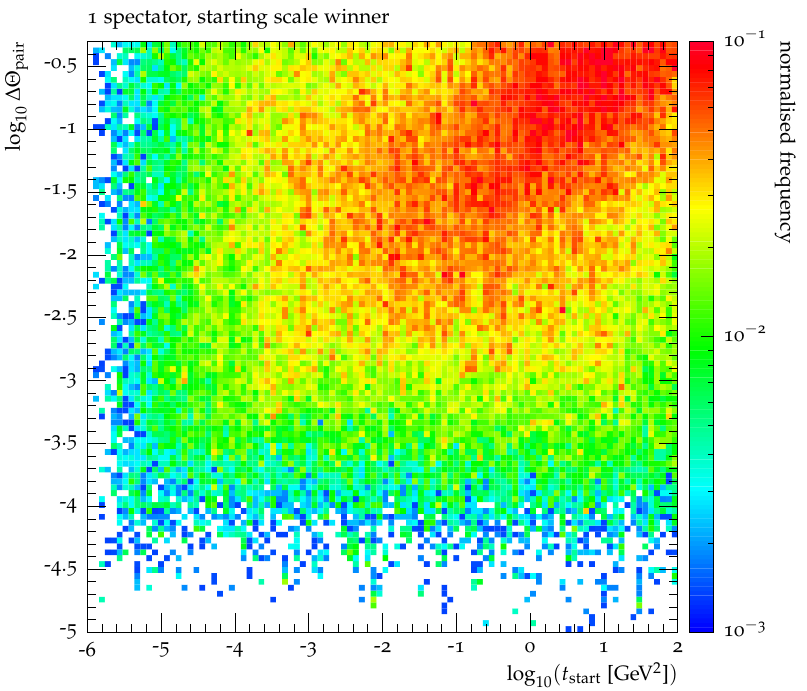}
  \caption[The interdependence of the starting scale $\tstart$ of a photon
  and the angular separation between the particles produced
  in its splitting, $\Delta \Theta_\mathrm{pair}$, in the mixed ordering
  scheme with different choices of kinematic spectators of the
  photon splitting.]{
    The interdependence of the starting scale $\tstart$ of a photon
    and the angular separation between the particles produced
    in its splitting, $\Delta \Theta_\mathrm{pair}$, in the mixed ordering
    scheme with different choices of kinematic spectators of the
    photon splitting: both charged primary leptons (top row) or
    only the primary lepton the splitting photon was reconstructed to
    have been emitted from (bottom row). We also show the dependence 
    on the way in which the
    starting scale of the evolution is chosen: probabilistically
    (left column) or by always choosing the winning dipole (right column).
  }
  \label{fig:yfs:results:systematicstdR}
\end{figure}

To further investigate the effects of our results on specific
algorithmic choices, fig.\ \ref{fig:yfs:results:systematicstdR} focuses on 
the same observable familiar from the previous figure:
the interdependence of the starting scale
$\tstart$ and the opening angle $\Delta \Theta_\text{pair}$.
As in fig.\ \ref{fig:yfs:results:startingscale}, we see here that 
the effect of a winner-takes-all choice of starting scale as opposed 
to our default probabilistic starting scale definition is not 
significant. The winner-takes-all choice results in the distribution 
of starting scales being skewed to slightly smaller values, as discussed above,
which correlates loosely with a more collinear splitting. 
This results in a slight extension of the high-frequency (red) 
region of the plot towards the small-$\tstart$ small-angle corner 
in the lower two plots of fig.\ \ref{fig:yfs:results:systematicstdR}
compared to the upper two plots. 
We also show the spectator
scheme dependence in the photon splitting, i.e. whether we allow both
primary leptons to be spectators, or only the lepton that the photon
was reconstructed to have been emitted from.
Since in photon splittings the spectator's only role is to absorb
recoil to guarantee momentum conservation, it is physically well
motivated for the splitting photon's progenitor to be the sole particle
to absorb its gained virtuality necessary for the splitting
process.
Note that this choice does not affect the value of $\tstart$, only the energy
available in the splitting, which affects the overall splitting probability
and the allowed opening angle of the splitting products. Fig.\ 
\ref{fig:yfs:results:systematicstdR} shows that this choice has negligible 
effect on the distribution of splitting events in the
$(\tstart,\Delta\Theta_\text{pair})$ plane.

\begin{figure}
  \centering
  \includegraphics[width=0.65\textwidth]{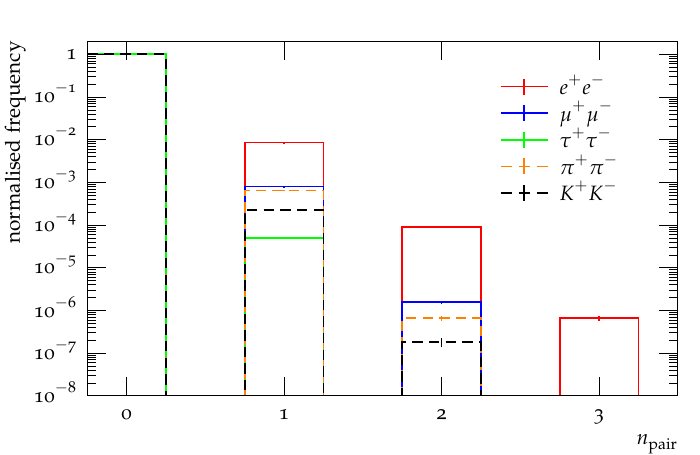}
  \caption{
    The relative abundance of secondary pairs of each species
    of charged particle produced in photon splittings in
    the mixed ordering scheme.
    \label{fig:yfs:results:number}
  }
\end{figure}

To conclude this section, fig.\ \ref{fig:yfs:results:number} shows the
relative frequency of photons splitting into different species of
charged lepton and hadron.
As the driving factor is the produced particle species' mass,
electron-positron pairs are most commonly produced,
around an order of magnitude more commonly the products of
photon splittings than muons or charged pions.
The probability of producing a second pair of a given species
roughly follows the na\"ive expectation of being the square
of the probability of producing one pair.
In a more detailed consideration one finds a factor of
$\alpha^2\log(m_Z/E_\gamma)\,\log(\tstart/m^2)$ associated
with each secondary pair production.
Therein, $E_\gamma$ is the energy of the bremsstrahlungs photon
that subsequently splits into the pair of particles of mass $m$,
and $\tstart$ is its reconstructed starting scale.
Hence, we observe a single-logarithmic suppression
of heavier flavours, modulo possible minor differences in the splitting
function itself.
This is well-reproduced by our algorithm.
In fact, in the current example, the drop in frequency of
producing an additional pair of
particles of the same flavour is between 2.5 and 4.5 orders of magnitude.

%%%%%%%%%%%%%%%%%%%%%%%%%%%%%%%%%%%%%%%%%%%%%%%%%%%%%%%%%%%%%%%%%%%%%%%%%%%%%%%%%%%%%%%%%%%%%%%%%

\section{Lepton dressing beyond photons}
\label{sec:yfs:dress}

In this section we analyse the final states produced by our algorithm,
and in particular the consequences of further
resolving the photons produced by the standard soft-photon resummation
into charged-particle pairs.
We will continue to use the decay of an on-shell $Z$ boson
into an $e^+e^-$ pair as a testbed for our algorithm.
We will analyse the corrections induced by photon splittings 
on a number of physical properties that are related to the charged
particle content of the radiation cloud surrounding the primary
decay products.

We continue to use
a spherical coordinate system to measure relative radial
distances $\Delta\Theta$.
Further, please note that for this study we turn off kaon and $\tau$ decays
for the greatest accuracy in identifying primary final-state particles.
By default, however, kaon and $\tau$ lepton decays would be handled
as normal in \Sherpa \cite{Bothmann:2019yzt}, including various
state-of-the-art parametrisations of all known decay channels and
including their own respective \ac{QED} corrections.

\subsection{Dressing strategies in the presence of photon splittings}
\label{sec:yfs:results:dress}

Lepton dressing is commonly used to define \ac{IR-safe} observables
through recombining a primary bare lepton with its surrounding
radiation cloud, in analogy with the jet clustering of \ac{QCD}.
While lepton dressing is essential when massless leptons are used
in a calculation due to the presence of collinear singularities,
the inclusion of a lepton mass renders both dressed and bare lepton
definitions physical.
Nonetheless, bare leptons suffer from large corrections that are
logarithmic in the lepton's mass, making them particularly
relevant for electrons.
Hence, a dressed lepton definition is also advantageous in calculations
with massive, but light, leptons.

In practice there are two common methods for lepton dressing,
analogous to jet definitions in \ac{QCD}: cone dressing and sequential
recombination dressing.
While a sequential recombination algorithm typically uses either
the anti-$k_t$ or Cambridge-Aachen algorithm \cite{Cacciari:2008gp},
the cone-based dressing uses the bare lepton to define the cone axis and,
at variance with historical \ac{QCD} cone algorithms, keeps the cone
itself stable throughout the recombination procedure, rendering it
collinear safe with respect to photon emissions.
In either case, unlike in \ac{QCD}, 
the algorithm is not completely blind to particle flavour
since (at least) the primary bare lepton is used as the dressing initiator
and defines the flavour of the resulting dressed lepton.
As long as only photon radiation is considered as a higher-order
correction to lepton production, which is the current standard in both
\ac{YFS} based soft-photon resummations \cite{Hamilton:2006xz,Schonherr:2008av}
and \Photos \cite{Barberio:1993qi,Davidson:2010ew}, both algorithms work
very straightforwardly by subsequently combining the primary lepton
with the surrounding photon cloud using the respective distance
measure.

When photon splittings are included in the \ac{QED} corrections to lepton
production as well, the radiation cloud surrounding the primary
lepton becomes flavour-diverse.
Considering the underlying physical process, these photon-splitting
corrections are simply resolving the structure of the photons
constituting the above photon cloud.
While these corrections are \ac{IR}-finite when all lepton masses
are considered, large logarithmic effects can be expected in particular
when branching into the lightest species, electrons, occurs. Further,
the splitting into electrons is the most probable branching for a photon 
emitter.
Thus, while continuing to dress the primary leptons with photons only is
\ac{IR-safe}, it is natural to demand that the resulting dressed lepton
definition does not strongly depend on whether or not we include further 
photon splittings.
We will thus investigate the following choices for the flavour set
$\fdress$ which is used to dress the primary lepton:
\begin{tabular}{p{0.25\textwidth} p{0.7\textwidth}}
  $\boldsymbol{\{\gamma\}}$
  & We continue to use only photons to dress the primary charged lepton. \\
\end{tabular}
\begin{tabular}{p{0.25\textwidth} p{0.7\textwidth}}
  $\boldsymbol{\{\gamma,e\}}$
  & In addition to the mandatory dressing with the surrounding photons,
    we also include the lightest charged particle, the electron, in the
    dressing procedure of the primary lepton.
    This is not only motivated by the fact that splittings into
    $e^+e^-$ pairs give the largest corrections, but also that experimentally
    both electrons and photons are measured similarly in the calorimeter.
    Of course, the presence of a magnetic field between the interaction
    and the calorimeter does in principle decorrelate the direction of
    their respective momentum vectors. \\
\end{tabular}
\begin{tabular}{p{0.25\textwidth} p{0.7\textwidth}}
  $\boldsymbol{\{\gamma,e,\pi,K\}}$
  & We also include the lightest hadronic splitting products in the
    dressed lepton definition.
    Such a definition is a compromise between theoretical inclusivity
    and experimental feasability. \\
\end{tabular}
\begin{tabular}{p{0.25\textwidth} p{0.7\textwidth}}
  $\boldsymbol{\{\gamma,e,\pi,K,\mu,\tau\}}$ 
    & We include all species produced in our photon-splitting implementation
      in order to be completely inclusive.
      It has to be noted though, that in realistic 
      experimental environments
      muons are well distinguishable even at low muon 
      energies, and $\tau$
      leptons of course decay further before detection rendering their
      inclusion in any realistic dressing algorithm highly non-trivial. \\
\end{tabular}

\begin{figure}[b!]
  \centering
  \includegraphics[width=0.47\textwidth]{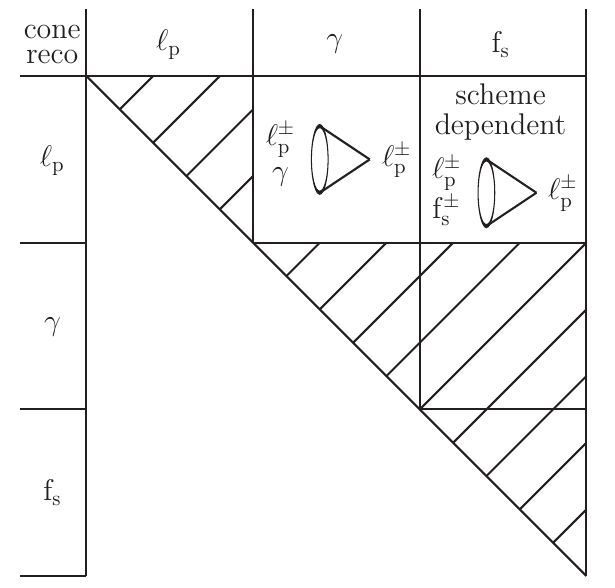}
  \hfill
  \includegraphics[width=0.47\textwidth]{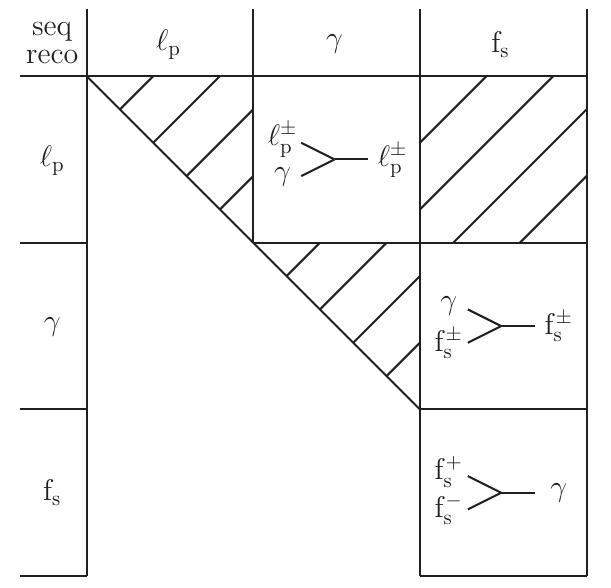}
  \caption[Recombination matrices of lepton dressing strategies beyond photon radiation.]{
    Recombination matrices of lepton dressing strategies beyond photon radiation.
    While $\gamma$ labels the photon, $\ellp$ and $\ells$ denote
    the primary leptons and secondary flavours, respectively.
    \label{fig:yfs:results:flavourmatrix}
  }
\end{figure}

A schematic of how both the cone and sequential recombination
dressing algorithms in the presence of photon splittings proceed
is given in fig.\ \ref{fig:yfs:results:flavourmatrix}.
In the case of cone dressing, the primary leptons
should be identified and dressed with all \ac{QED} radiation
that surrounds them, including other leptons and hadrons.
In particular, the flavour of the dressed lepton does not change
even if flavours other than a photon are included in it as it
is determined entirely by the primary lepton.
Thus, in consequence, the cone-dressed lepton may have a net
charge that is different from that of its assigned flavour when
not all photon-splitting products are recombined into the same
dressed lepton.
We will use this algorithm for the remainder of this study.

Nonetheless, a diagram of a flavour recombination matrix for sequential
recombination dressing is shown on the right-hand side of
fig.\ \ref{fig:yfs:results:flavourmatrix}.
Here it is possible to recombine a secondary (and hence soft/collinear)
lepton-antilepton pair into a photon, while allowing for even
softer or more collinear surrounding photons to be combined with
these charged leptons first.
On the level of primary leptons, then, they are only dressed
with photons, either from the final state or from previous
secondary-lepton clusterings.
This has the obvious advantage that the charge and flavour
of the primary lepton matches that of the dressed lepton.
However, it is schematically more intricate and does not
always yield circular dressed leptons, which are favoured
experimentally. The investigation of sequential recombination 
dressing is left to a future study.

\begin{figure}
  \centering
  \includegraphics[width=0.47\textwidth]{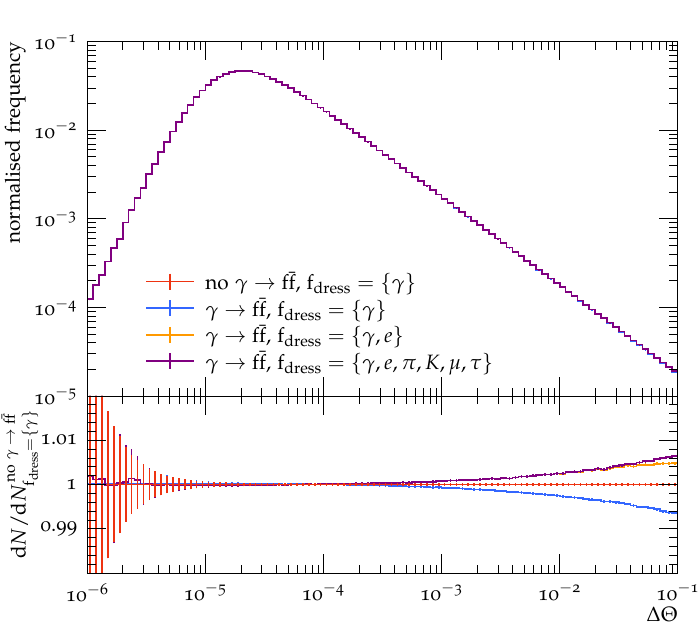}
  \hfill
  \includegraphics[width=0.47\textwidth]{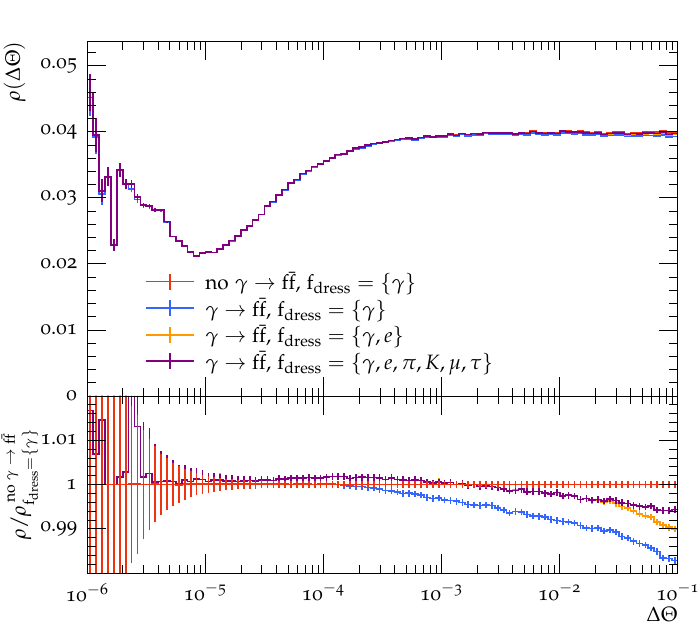}
  \caption[\textbf{Left:} The differential distribution of the dressed lepton
  constituents, including radiated photons with
  $E_\gamma>0.1\,\text{MeV}$, in dependence on the angular
  distance $\Delta\Theta$ from the primary lepton.
  \textbf{Right:} The energy density $\rho$ within the dressed lepton
  as a function of the angular distance $\Delta \Theta$ from the primary
  lepton.]{
    \textbf{Left:} The differential distribution of the dressed lepton
    constituents, including radiated photons having
    $E_\gamma>0.1\,\text{MeV}$, dependent on the angular
    distance $\Delta\Theta$ from the primary lepton.
    \textbf{Right:} The energy density $\rho$ within the dressed lepton
    as a function of the angular distance $\Delta \Theta$ from the primary
    lepton.
    Shown are the predictions without accounting for photon splittings (red),
    compared to the predictions allowing photons to split: dressed
    with photons only (blue), photons and electrons (orange) or all particles (violet).
    \label{fig:yfs:results:dtheta_rho}
  }
\end{figure}

The first observables we examine offer closer looks into the
substructure of the cone-dressed leptons produced by different
dressing strategies.
We introduce the legend notation:
either photon splittings to charged flavours $\mathrm{f}$ are
present ($\gamma \to \ffbar$) or they are
not (no $\gamma \to \ffbar$);
the dressing algorithm is specified by the set of particle
flavours \fdress\ which are included in the dressing.
% In fig. \ref{fig:yfs:results:dtheta_rho}, the plots extend to the cone size of
% $\dRdress = 0.1$ used in the dressing.

The left-hand side plot of fig.\ \ref{fig:yfs:results:dtheta_rho}
displays the angular distance $\Delta\Theta$ of the cone-dressed
lepton constituent from the primary lepton.
To ensure \ac{IR} safety, only photons with $E_\gamma>0.1\,\text{MeV}$
are included. A cutoff just below the electron mass has been selected 
to ensure that all electrons are included in the analysis.
Besides observing the primary lepton's dead cone for
$\Delta\Theta\lesssim 2\times 10^{-5}$,
we find that for $\Delta\Theta\lesssim 10^{-4}$ the constituent
multiplicity when including photon splittings, irrespective of the
dressing scheme used, coincides with the multiplicity when omitting
such splitting.
This corroborates our earlier expectation that collinear photons
largely lack the necessary virtuality to split into a charged-particle
pair.
At larger angles, where the required virtuality can be more easily
gained, a photon's probability to split increases.
In consequence, when including photon-splitting effects in the calculation,
but not accounting for the splitting products in the dressing, a drop
in multiplicity can be observed.
Including electrons as well as photons in the dressing reincorporates
most splitting products into the dressed lepton definition (\see fig.\
\ref{fig:yfs:results:number}). We find an increase above the reference 
of approximately the same number of constituents that are lost in the 
$\gamma\to \ffbar$, $\fdress=\{\gamma\}$ case. 
The completely flavour-inclusive dressing definition then shows
the same effects scaled to the production of heavier secondary species:
larger virtualities, and thus larger $\Delta\Theta$, are needed for 
a non-zero splitting probability, so fewer photons actually split into
these heavier flavours. This leads to a much smaller effect of these 
splittings, concentrated at the outside of the cone. We expect out-of-cone 
effects to be small, since the frequency spectrum falls steeply towards 
the edge of the cone. More generally, it appears that the splitting products 
are close to collinear with the progenitor photon, at least on average. 

On the other hand, the right-hand side plot of
fig.\ \ref{fig:yfs:results:dtheta_rho} shows the
distribution of energy within the dressed lepton as
a fraction of the energy of the entire dressed lepton.
Resolving photons into other species, \ie pairs of
charged particles, but continuing to dress the
primary lepton with photons only naturally decreases
the energy radial density of the dressed lepton.
The fact that this energy density loss is not constant
but rather increases with the radial distance to the
primary lepton is again a result of the increasing
possible off-shellness at larger $\Delta\Theta$, and
therefore the increased splitting probability.
Even when the photon splitting products are part of
the dressing procedure (either secondary electrons
only or the set $\{e,\pi,K,\mu,\tau\}$) the energy density $\rho$
falls below the reference at some distance from the
primary lepton, showing that a non-negligible number of more energetic 
splitting products end up outside the dressing cone radius.

\begin{figure}
  \centering
  \includegraphics[width=0.65\textwidth]{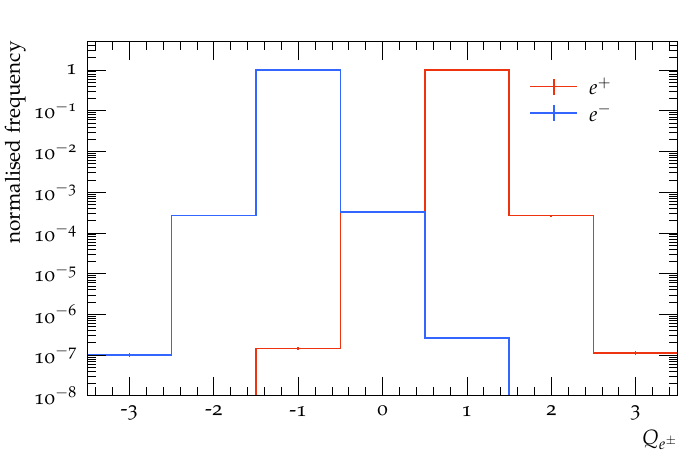}
  \caption{
    The total charge of the cone-dressed electron and positron with
    $\dRdress=0.1$ and including all secondary flavours,
    \ie $\fdress=\{\gamma,e,\pi,K,\mu,\tau\}$.
    \label{fig:yfs:results:charge}
  }
\end{figure}

As mentioned above, it is possible for the charge of the
dressed lepton to be different from the charge of its
primary constituent.
This is shown in fig.\ \ref{fig:yfs:results:charge} for the
case of cone dressing with $\dRdress = 0.1$.
Fewer than a thousandth of the dressed leptons are neutral
or doubly charged, while a fraction of $10^{-7}$ of them
are either triply charged or appear to be their own
antiparticle (a dressed electron having a charge of $+1$
or a dressed positron having a charge of $-1$).
Again, this is a consequence of only partially capturing
the photon splitting products.

In the next section we will look at the separate and combined effects of photon splittings 
and flavour-aware lepton dressing on physical observables in the decay of an on-shell 
$Z$ boson.

%%%%%%%%%%%%%%%%%%%%%%%%%%%%%%%%%%%%%%%%%%%%%%%%%%%%%%%%%%%%%%%%%%

\subsection{Case study: \texorpdfstring{$Z$}{Z} boson decay}
\label{sec:yfs:results:Zdecay}

Next, we look at the decay of an on-shell $Z$ boson into an
$e^+ e^-$ pair and investigate the impact of the photon splitting
corrections introduced in this chapter on physical observables.
To be precise, we present the effects of including $\gamma \to \ffbar$
splittings and the consequences of (not) using flavour-aware dressing algorithms
on the decay rate, differential with respect to the invariant mass
$m_{\ell\ell}$ of the primary electron-positron pair.

\begin{figure}
  \centering
  \includegraphics[width=0.65\textwidth]{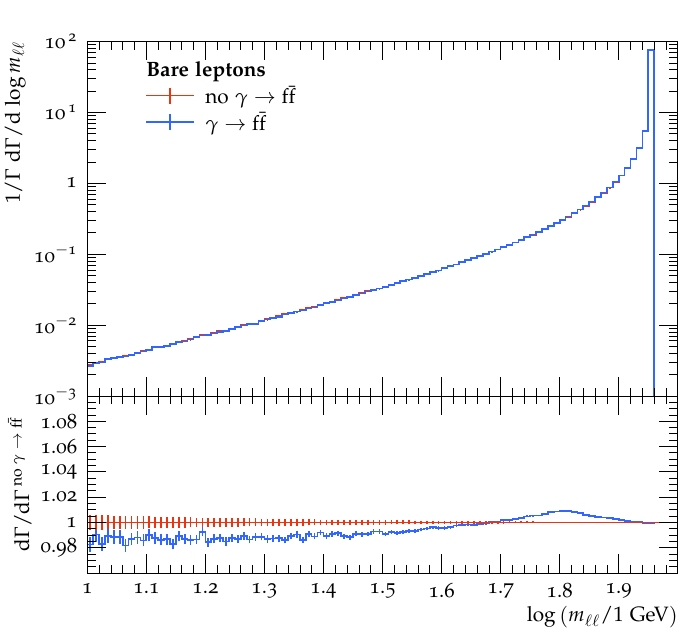}
  \caption[The bare dilepton invariant mass $m_{\ell\ell}$ in on-shell $Z$ decay
  as described by the
  YFS soft-photon resummation only or additionally resolving the
  photons further into pairs of charged particles, in the mixed
  ordering scheme.]
  {
    The bare dilepton invariant mass $m_{\ell\ell}$ in on-shell $Z$ decay
    as described by the
    YFS soft-photon resummation only (red) or additionally resolving the
    photons further into pairs of charged particles (blue), in the mixed
    ordering scheme.
    \label{fig:yfs:results:mllCompareBare}
  }
\end{figure}

We begin by examining the bare differential decay rate, \ie the invariant
mass of the primary lepton pair that is not dressed with the radiation
around it, in order to quantify the kinematic effect of photon splittings 
on the primary leptons themselves without confusing this effect with the 
intricacies of the dressing algorithm.
We note that bare leptons are theoretically well-defined as all lepton masses
are fully accounted for.
To this end, fig.\ \ref{fig:yfs:results:mllCompareBare} isolates the
effect of allowing \ac{YFS} photons to split by presenting the bare
invariant mass of the two most energetic leptons of opposite charge,
one electron and one positron.
In the overwhelming majority of cases these are expected to be the
primary electron-positron pair generated in the on-shell $Z$ decay.
The largest deviation from the pure \ac{YFS} prediction without photon splittings,
which is taken as the reference, is about $1\%$ in the region of
most interest.
It occurs just below the $Z$ mass, at about $60-70\,\GeV$.
It is driven by extracting additional momentum from the primary
leptons to accommodate the necessary virtuality for
photon splittings to occur.
Although barely visible, this is fueled by a minute reduction
of the much larger differential decay rate closer to the
nominal $Z$ mass itself.
Although of less interest due to the smaller absolute decay rate,
the opposite effect is seen at very small invariant masses, below
$50\,\GeV$. In this regime, through the same mechanism, the decay rate is
diminished by about $1-2\%$ as the slope of the distribution is
shallower but the momentum extraction is similar in magnitude to that 
at larger invariant masses. 

Finally, a change of the precise definition of the ordering
variable, both for the reconstructed starting scale of the evolution
and the splitting scale of the eventual photon splitting, generally
increases the size of the corrections for this observable.
While using $t=\kTsq$ for all splittings only increases the observed
corrections slightly, due to the increased photon splitting probability
as $\kT<\qbar$ throughout, using $t=\qbar^2$ almost doubles the size
of the corrections as now the starting scales of each photon's
evolution reconstruct to much larger values, \see
fig.\ \ref{fig:yfs:results:startingscale}.
This is a consequence of the different properties of these 
ordering variables as discussed in section \ref{sec:yfs:methods:photonsplit},
although \textit{a priori} all choices have the same formal accuracy.

\begin{figure}
  \centering
  \includegraphics[width=0.47\textwidth]{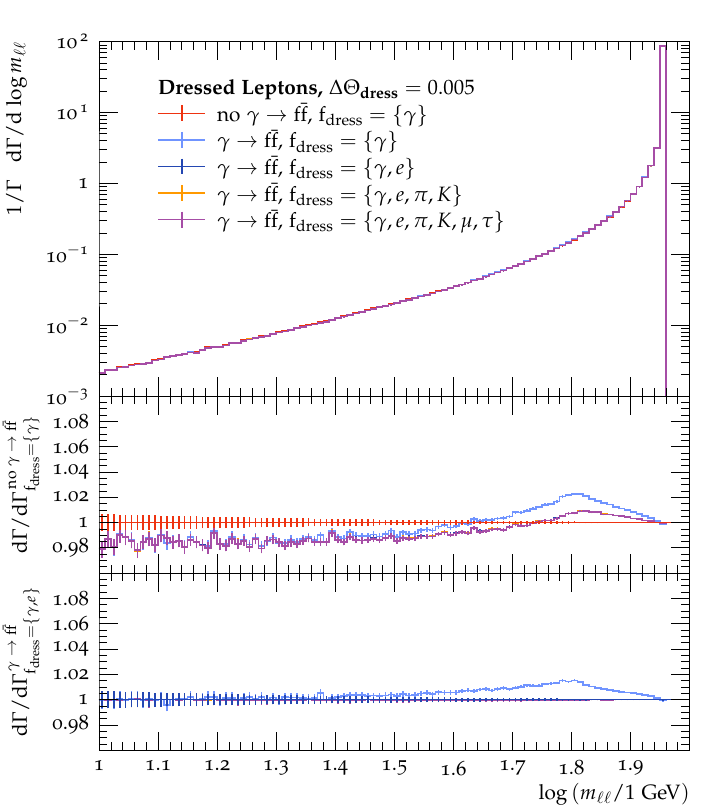}
  \hfill
  \includegraphics[width=0.47\textwidth]{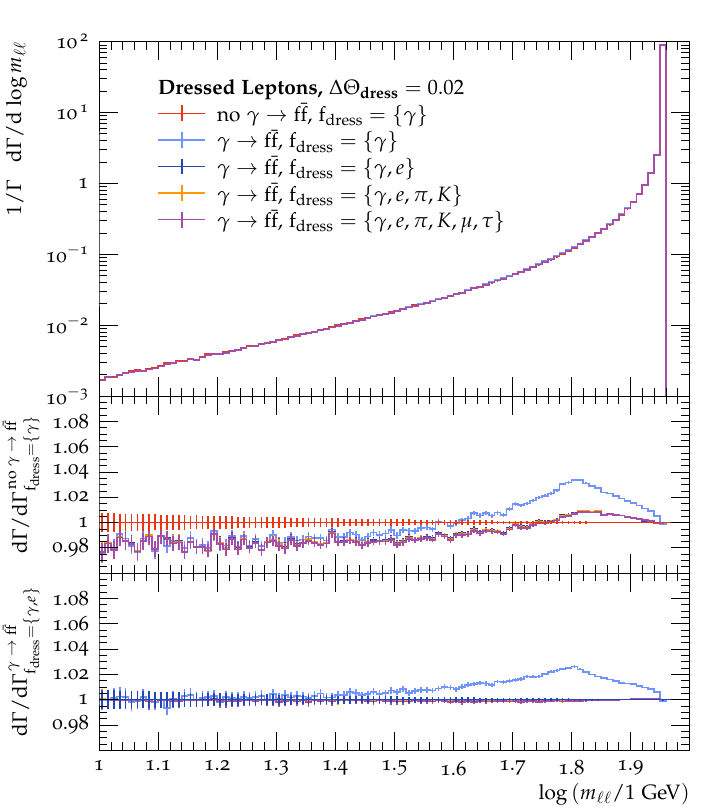}

  \includegraphics[width=0.47\textwidth]{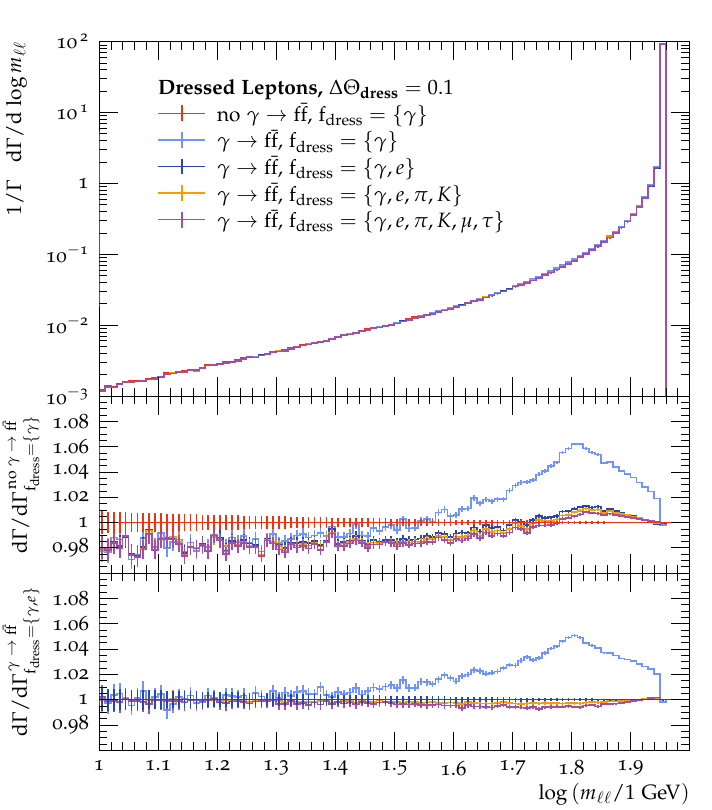}
  \hfill
  \includegraphics[width=0.47\textwidth]{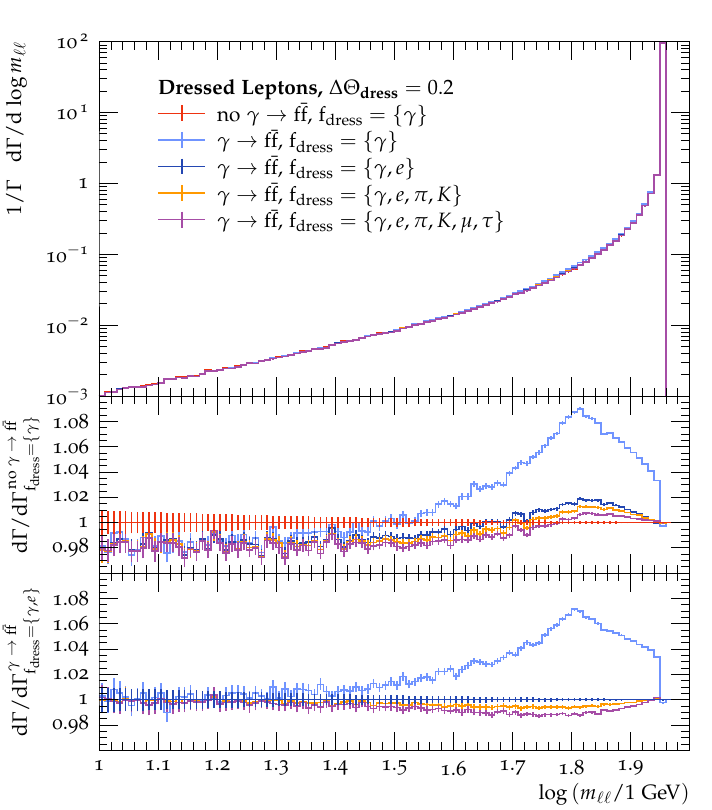}
  \caption[The dressed dilepton invariant mass $m_{\ell\ell}$ in on-shell $Z$ decay
  as described by the
  YFS soft-photon resummation only or additionally resolving the
  photons further into pairs of charged particles for four different
  dressing cone sizes.]{
    The dressed dilepton invariant mass $m_{\ell\ell}$ in on-shell $Z$ decay 
    as described by the
    YFS soft-photon resummation only (red) or additionally resolving the
    photons further into pairs of charged particles for four different
    dressing cone sizes, $\dRdress=0.005$ (top left), $0.02$ (top right),
    0.1 (bottom left), and 0.2 (bottom right), in the mixed ordering scheme.
    We differentiate various different dressing strategies, recombining
    photons only (light blue), photons and electrons (dark blue), photons,
    electrons and charged hadrons (orange), and all charged particles
    (violet) within the dressing cone with the primary charged lepton.
    Two ratios are presented, either with reference to the soft-photon resummation
    without photon splittings (upper), or the soft-photon resummation
    including photon splittings and dressing the primary leptons with
    photons as well as secondary electrons (lower).
    \label{fig:yfs:results:mllCompare3}
  }
\end{figure}

Having assessed the basic kinematic effects on the bare primary
leptons, we now turn to dressed leptons.
We will investigate the impact the different dressing strategies
discussed in sec.\ \ref{sec:yfs:results:dress} have, once the radiation cloud around
the primary leptons is not comprised of only photons but is resolved
further into various different flavours of secondary charged particles.
To this end, fig.\ \ref{fig:yfs:results:mllCompare3} contrasts the pure \ac{YFS}
soft-photon resummation without further photon splittings with a range of
dressing strategies when photon splittings are included.
Four different cone sizes are considered, from a minimum $\dRdress = 0.005$ to
a maximum $\dRdress = 0.2$.
The upper ratio illustrates the deviation of each prediction
from the pure \ac{YFS} case, due to both the presence of photon splittings
and the details of the dressing algorithm.
The lower ratio isolates the effect of the dressing strategy
by showing the deviation with respect
to the photon-only-dressed events. In particular, this shows 
which secondary flavours are recombined with the primary
lepton into the dressed lepton.
We observe that when the photon radiation off the primary electrons
is further resolved into charged-particle pairs but the primary
electrons are still only dressed with only the photons of their
surrounding radiation cloud, large effects are manifest.
They range from slightly over 2\% for $\dRdress = 0.005$ to 6\%
for the commonly used $\dRdress = 0.1$, and up to 9\% for the more inclusive
cone radius of $\dRdress = 0.2$.
This difference originates in the fact that as long as only photons
are included in the dressing, every photon lost by resolving it into
a charged-particle pair cannot be recombined into the dressed lepton,
which then ends up with less energy simply because higher-order
corrections have been included.
The observation that our algorithm reconstructs higher starting
scales for hard wide-angle photons
than either soft or collinear ones, and thus these are 
more likely to possess the necessary virtuality to split into
charged-particle pairs, explains the dressing-cone-size dependence.
However, when more inclusive dressing algorithms are considered,
the effect of photon splittings on the differential decay rate is
reduced, as is the $\dRdress$ dependence.
As photons predominantly resolve into $e^+e^-$ pairs, their inclusion
in the dressed lepton definition already captures the bulk of the
effect, in particular at smaller dressing cone radii.
Along the lines of the above argument, photons need to be sufficiently separated
from the primary lepton in order to gain enough
virtuality to split into the heavier particle species.
Thus, the inclusion of further secondary flavours in the dressing
algorithm only plays a role at larger dressing cones, with effects
ranging from 1\% at $\dRdress=0.1$ to 2\% at $\dRdress=0.2$.
The effect of changing the ordering scheme for the photon splitting 
algorithm on fig.\ \ref{fig:yfs:results:mllCompare3} is very similar to the effect
on fig.\ \ref{fig:yfs:results:mllCompareBare}. Again, using the transverse momentum
or virtuality ordered schemes increases the size of 
the corrections induced by photon splittings in a very similar way as before.
It is still the case that reincorporating splitting products in the dressing
recovers the bare-lepton level deviation from the pure \ac{YFS} prediction.
As above, it needs to be noted that such a change in the ordering
variable results in a suboptimal description of the physical process,
and is thus not recommended to be used as an estimator of the intrinsic
uncertainty.

\begin{figure}
  \centering
  \includegraphics[width=0.47\textwidth]{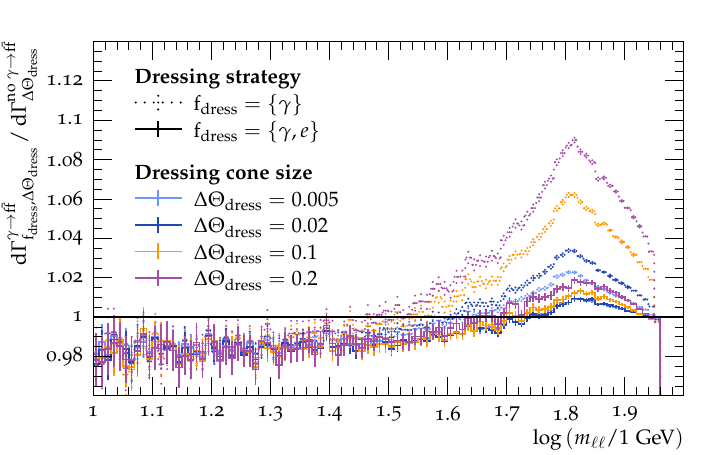}
  \hfill
  \includegraphics[width=0.47\textwidth]{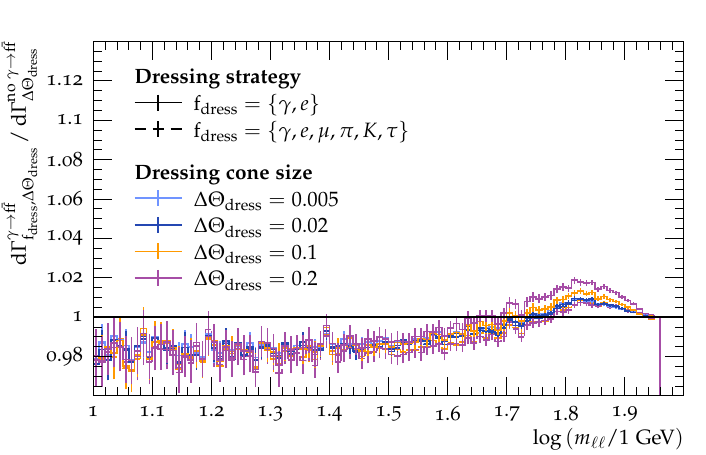}
  \caption[The cone size dependence of different dressing
  strategies in on-shell $Z$ decay. The differential decay rate
  $\mathrm{d}\Gamma^{\gamma \to \ffbar}_{\fdress,\dRdress}/
  \mathrm{d}\log{m_{\ell \ell}}$
  has been divided by the corresponding
  $\mathrm{d}\Gamma^{\mathrm{no}\,\gamma \to \ffbar}_{\dRdress}/
  \mathrm{d}\log{m_{\ell \ell}}$,
  in dependence of both the flavour set $\fdress$ included
  in the dressing and the dressing cone of size $\dRdress$.]{
    The cone size dependence of different dressing
    strategies in on-shell $Z$ decay. The differential decay rate
    $\mathrm{d}\Gamma^{\gamma \to \ffbar}_{\fdress,\dRdress}/
    \mathrm{d}\log{m_{\ell \ell}}$
    has been divided by the corresponding
    $\mathrm{d}\Gamma^{\mathrm{no}\,\gamma \to \ffbar}_{\dRdress}/
    \mathrm{d}\log{m_{\ell \ell}}$,
    in dependence of both the flavour set $\fdress$ included
    in the dressing and the dressing cone of size $\dRdress$.
    The \textbf{left} plot shows the difference case where only
    photons are used in the dressing (dotted) and using both
    photons and secondary electrons (solid),
    whereas the \textbf{right} plot shows the difference between
    a dressing strategy using only photons and electrons (solid)
    and all secondary flavours (dashed).
    \label{fig:yfs:results:mllDivided}
  }
\end{figure}

In fig.\ \ref{fig:yfs:results:mllDivided} we show more clearly the recovery 
of the pure soft-photon prediction using the two most relevant 
charged-particle-inclusive dressing strategies.
The figure shows the ratio of the differential
cross section including photon splittings to that without photon splittings 
for different dressing choices.
We find that
including charged particles in the cone dressing limits the effect of
photons splitting corrections to the 1\% level, irrespective of cone size.
Including electrons in the dressing similarly limits the corrections 
to 2\% even for the largest cone sizes considered here.

\subsection{Results: Drell-Yan production at hadron colliders}
\label{sec:yfs:results:DY}

In this section we present results for the phenomenologically 
relevant case of Drell-Yan at proton-proton colliders. We focus 
on the process $pp\to e^+ e^-$ at a centre-of-mass energy of 
13 TeV. We use \Sherpa's \Ahadic for hadronisation and the \ac{PDF} 
set {\sc Pdf4Lhc21} from the \LHAPDF library \cite{Buckley:2014ana}. 
These results were produced with a pre-release 
version of \Sherpa 3.0. \Sherpa's default parton 
shower, {\sc Csshower}, was used for the initial-state \ac{QCD} shower
\cite{Schumann:2007mg}. \Comix was used for \ac{ME} generation 
\cite{Gleisberg:2008fv}.

As before, the effect of photon splittings is seen most clearly 
in the dilepton invariant mass for a range of dressing strategies. 
In the hadron collider environment, it is clear that dressing 
leptons with light hadrons is not appropriate due to the large 
number of hadrons produced by unrelated \ac{QCD} processes, such as 
initial-state radiation, multiple interactions and pile-up.
As a result, in this section 
we focus on the modified cone dressing where secondary electrons 
are included in addition to photons.

\begin{figure}
  \centering
  \includegraphics[width=0.65\textwidth]{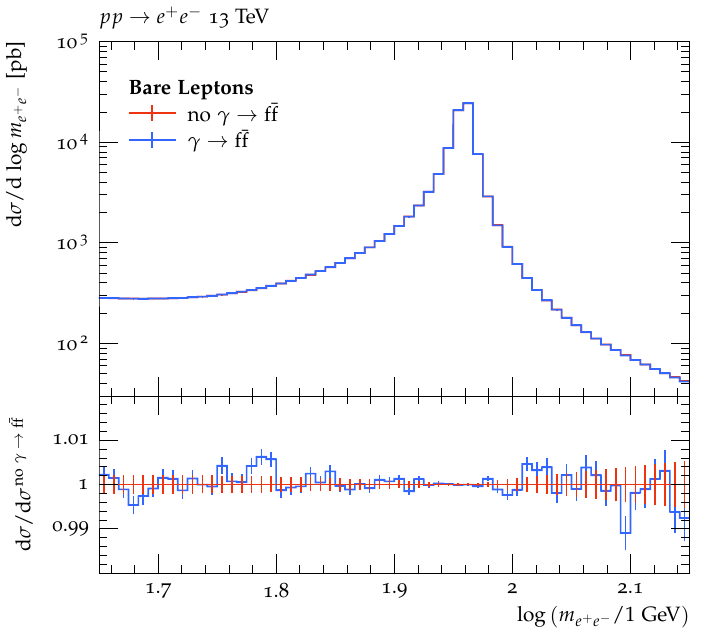}
  \caption{
    The bare dilepton invariant mass $m_{e^+ e^-}$ in Drell-Yan production
    as described by the
    YFS soft-photon resummation only (red) or additionally resolving the
    photons further into pairs of charged particles (blue), in the mixed
    ordering scheme.
    \label{fig:yfs:results:mllDYbare}
  }
\end{figure}

Fig.\ \ref{fig:yfs:results:mllDYbare} shows the dielectron invariant 
mass for bare electrons identified, as before, as the highest-energy
opposite-flavour pair. Whilst for the case of on-shell $Z$ decay the 
$m_{e^+e^-}$ spectrum was \iac{LO} observable in the \ac{QED}
radiative correction, in Drell-Yan production the spectrum is 
given at \ac{LO} by the $x$ distribution of the initial partons. 
The \ac{QED} radiative corrections are large, however: they are around 
50\%. Hence we expect the photon splitting corrections 
to this observable to be much smaller than in the previous case study,
but still non-negligible. 
This can be seen in the plot, where there is only a small region 
in which photon splitting corrections contribute a statistically significant 
correction. Below the $Z$ peak, at an invariant mass of around 
60 GeV, there is a positive correction of 0.5\% to the cross section. 
While statistically significant on an event-generation level, the 
correction to this observable 
is not significant physically since scale-variation uncertainties
in the spectrum are much larger than the \ac{MC} statistical errors 
shown here. However, as before, these corrections are dependent on 
the lepton definition used in the observable.

\begin{figure}
  \centering
  \includegraphics[width=0.65\textwidth]{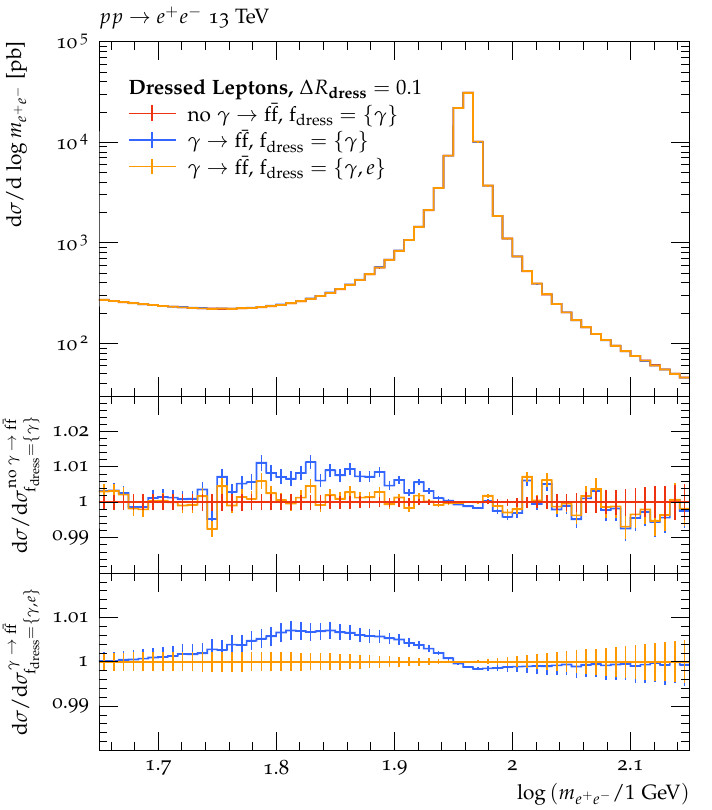}
  \caption[The dressed dilepton invariant mass $m_{e^+ e^-}$ in Drell-Yan production
  as described by the
  YFS soft-photon resummation only (red) or additionally resolving the
  photons further into pairs of charged particles for a dressing cone 
  size of $\Delta R_\mathrm{dress}=0.1$,
  in the mixed ordering scheme.]{
    The dressed dilepton invariant mass $m_{e^+ e^-}$ in Drell-Yan production
    as described by the
    YFS soft-photon resummation only (red) or additionally resolving the
    photons further into pairs of charged particles for a dressing cone 
    size of $\Delta R_\mathrm{dress}=0.1$
    in the mixed ordering scheme.
    We differentiate a dressing strategy recombining
    photons only (blue), and recombining photons and electrons (orange)
    within the dressing cone with the primary charged lepton.
    Two ratios are presented, either taking the soft-photon resummation
    without photon splittings (upper ratio), or the soft-photon resummation
    including photon splittings and dressing the primary leptons with
    photons as well as secondary electrons (lower ratio),
    as the reference.
    \label{fig:yfs:results:mllDYdressed}
  }
\end{figure}

To see this, we also investigate the effect of photon splitting corrections 
on dressed leptons. fig. \ref{fig:yfs:results:mllDYdressed} shows the 
dressed-electron invariant mass for a dressing cone size of 
$\Delta R_\mathrm{dress}=0.1$. We compare the usual photon-dressed 
electrons with those where secondary electrons have also been included 
in the dressing. The upper ratio plot shows the deviation with respect 
to the case where no photon splittings have been included. There is a 
clear 1\% correction in a large kinematical region below the $Z$ peak
introduced by including $\gamma\to\ffbar$. However, as in the previous 
section, we see that this deviation is almost eliminated when 
secondary electrons are recombined into primary electrons in dressing 
cones. The lower ratio plot shows more clearly the effect of changing 
the dressing algorithm. These results echo the findings from 
the $Z$ decay case study in a realistic collider environment.

In this section we have demonstrated the effectiveness of the 
proposed photon-splitting algorithm in modelling higher-order 
\ac{QED} radiation in $Z$ decays, including in neutral-current Drell-Yan 
processes. Before concluding this chapter, we briefly discuss 
the modifications to the algorithm in order to model corrections 
to $W$ decays.

%%%%%%%%%%%%%%%%%%%%%%%%%%%%%%%%%%%%%%%%%%%%%%%%%%%%%%%%%%%%%%%%%%%%%%%%%%%
\section{Charged resonances}
\label{sec:yfs:fidip}

In this section we give the definitions for final-initial dipoles needed
for the description of photon splittings in the \ac{QED} corrections
of charged particle decays, such as $W\to\ell\nu$.

The notation used to describe these dipoles is for the most part 
consistent with the final-final dipoles described above.
We consider a charged resonance $\at$ ($a$)
decaying to a charged particle $\ijt$ ($i$) and a recoiling system
$\{\tilde{n}\}$ ($\{n\}$):\, $\at\to\ijt\,\{\tilde{n}\}$ before and
$a\to i\,j\,\{n\}$ after the emission of a photon $j$, respectively.
Ordinarily the recoil from the splitting would be absorbed locally by
either the spectator $a$ when $i$ is the emitter, or vice versa.
Since $a$ is the decaying particle, however, we have chosen
to keep its momentum unchanged and redistribute the recoil effectively to the
particle(s) $\{n\}$.
This allows us to use a single momentum map for 
both situations and to
combine both emitter-spectator designations into a single dipole
splitting function.
This not only simplifies its description, but also removes problems
with the positivity of the partial-fractioned
\ac{CS} splitting functions in situations where the mass correction
is larger than the (quasi-)collinear emission term.
Hence, we follow the treatment in refs. \cite{Basso:2015gca,Dittmaier:1999mb} 
to construct the splitting functions and kinematic variables.

As described in sec.\ \ref{sec:yfs:methods}, the first step in the photon
splitting algorithm is to determine the starting scale
of each photon by reconstructing its emission history.
In principle, emission of a photon can occur from the decaying particle $\at$
or from its charged decay product $\ijt$. However, since the former splitting is 
suppressed by the decaying particle's mass, it is much more likely to act
as spectator.
Therefore, as discussed above, we employ a single splitting
function which contains the initial-state emission term in addition to 
the final-state emission.
As a consequence, in the soft limit the full eikonal is recovered
and the dipole radiates coherently,
but splitting from the initial-state particle is never kinematically 
considered when building the required single-emitter history in the
collinear interpretation of our parton shower.

After calculating the starting scale, the photons' evolution begins and 
photon splittings are considered. Since a photon is never considered to be 
emitted from the initial-state charged particle, and the decaying particle 
has a restricted phase space for absorbing recoil in any case, the spectator 
in all photon splittings is chosen to be the final-state particle $i$.
For this reason, we do not need the kinematic mappings for a final-initial dipole.
The splitting function and evolution variable definitions are detailed 
below.

\paragraph*{Splitting functions.}

The dipole invariant mass $Q^2$ is defined as 
\begin{equation}
  Q^2 = (\tilde{p}_{\ijt}-\tilde{p}_{\at})^2 = (p_i+p_j-p_a)^2
\end{equation}
for the case of a dipole with final-state emitter $i$ and initial-state 
spectator $a$.

For convenience, we define the quantity 
\begin{equation}
  \bar{Q}^2 = m_a^2 - m_i^2 -m_j^2 - Q^2.
\end{equation}

The kinematic variables $z$ and $y$ are defined differently from the final-final
case; they are given by 
\begin{equation}
  y = \frac{p_i p_j}{p_a p_i + p_a p_j - p_i p_j -2m_i^2 -2m_j^2}\,,
  \quad
  z = \frac{p_i p_a - p_i p_j -m_i^2}{p_i p_a + p_j p_a -2p_i p_j -m_i^2-m_j^2}\,.
\end{equation}

In terms of these variables the splitting functions are given by 
\begin{align*}
  S_{f_\ijt(\at)\to f_i\gamma_j(a)}
  \,=\;
  S_{\bar{f}_\ijt(\at)\to \bar{f}_i\gamma_j(a)}
  \,=\;
    -\,\mathbf{Q}^2_{\ijt\at}\;\alpha\,
    \left[ \frac{2}{1-z(1-y)}
    \left(1+\frac{2m_i^2}{\bar{Q}^2}\right)
    -(1+z) \right. & \\ 
    -\frac{m_i^2}{p_i p_j} 
    - \left.\frac{(p_i p_j)}{\bar{Q}^2} \,\frac{m_a^2}{\bar{Q}^2}\,
    \frac{4}{[1-z(1-y)]^2} \right]&, \\
  S_{s_\ijt(\at)\to s_i\gamma_j(a)}
  \,=\;
  S_{\bar{s}_\ijt(\at)\to \bar{s}_i\gamma_j(a)}
  \,=\;
    -\,\mathbf{Q}^2_{\ijt\at}\;\alpha\,
    \left[ \frac{2}{1-z(1-y)}
    \left(1+\frac{2m_i^2}{\bar{Q}^2}\right)
    -2-\frac{m_i^2}{p_i p_j} \right. & \\
    - \left.\frac{(p_i p_j)}{\bar{Q}^2} \,\frac{m_a^2}{\bar{Q}^2}\,
    \frac{4}{[1-z(1-y)]^2} \right].& \numberthis
\end{align*}
The additional factor $(1+2m_i^2/\bar{Q}^2)$ is needed to recover the soft 
eikonal limit by cancelling some of the mass dependence of the variables 
$z$ and $y$. Note that $m_j=0$ needs to be taken for the soft limit, so it is 
not present in this additional factor.

\paragraph*{Evolution variable.} As before, we consider two choices of 
evolution variable, virtuality and transverse momentum. The form of these 
variables in terms of the dipole invariant mass $Q^2$ and the masses of 
the particles in the process are very similar to those for final-final dipoles. 

The virtuality is given by 
\begin{equation}
  \qbar^2 = (m_a^2-m_i^2-m_j^2-Q^2)\,y+m_i^2+m_j^2-m_{\ijt}^2,
\end{equation}
while the transverse momentum can be written 
\begin{equation}
  \kTsq = (m_a^2-m_i^2-m_j^2-Q^2) \,y\, z(1-z) - m_i^2\,(1-z)^2 - m_j^2\,z^2.
\end{equation}
As before, the default scheme for the evolution variable is the mixed 
scheme, where the transverse momentum is computed as the starting scale 
for photon evolution but is interpreted as a virtuality thereafter. The pure 
transverse momentum and virtuality schemes are implemented as well.

Having discussed the details of our photon splitting algorithm 
for the case of $W$ decays, we conclude this chapter.

%%%%%%%%%%%%%%%%%%%%%%%%%%%%%%%%%%%%%%%%%%%%%%%%%%%%%%%%%%%%%%%%%%%%%%%%%%%%%%%%%%%%%%%%%%%%%%%%%%%%%

\section{Conclusions}
\label{sec:yfs:conclusions}

In this chapter we detailed an extension to the soft-photon resummation
in the Yennie-Frautschi-Suura framework, which incorporates higher \ac{QED}
corrections originating from photon splittings into charged-particle pairs.
These photon-splitting corrections, which resolve the substructure of 
the newly produced photons, are larger than suggested by 
their fixed-order accuracy. In particular, they can be 
logarithmically enhanced with the ratio of the lightest charged 
particle, the electron, to the possible virtuality of the 
splitting photon. 

Using the decay $Z\to e^+e^-$, we found that that the limit on
the virtuality of the photon bremsstrahlung off a primary lepton 
is strongly correlated with the angular distance to this primary lepton, and
thus also to the probability of that photon to split.
We also investigated the systematics of our photon-splitting algorithm 
and found that algorithmic choices do not have a large
impact on results. 
We found that the frequency of occurrence of different species 
agreed with theoretical expectations, showing a logarithmic dependence 
on the mass of the produced particles.

As a consequence of our extension, the \ac{MC} prediction for 
the cloud of \ac{QED} radiation 
surrounding the primary leptons of a hard decay contains an 
array of particle flavours, not solely photons.
Hence, the standard dressing algorithms
to define \ac{IR-safe} dressed leptons were found to develop
a strong sensitivity to further resolving the initial soft-photon
cloud, in particular for larger dressing-cone radii.
We therefore developed a novel set of flavour-aware strategies
for dressing charged leptons and investigated their respective
properties.
We found that including secondary electrons as a minimal addition
in the dressing procedure already substantially reduces this
dependence on photon resolution, while an inclusion of all possible 
secondary flavours minimises it.

Using the example of the $Z\to e^+e^-$ decay rate, we investigated
the dilepton invariant mass in detail.
We found corrections of around 1\% from photon splittings, 
with respect to
the previous standard of not further resolving the initial photon
radiation on the bare electrons.
In the more relevant case of leptons cone-dressed with photons only,
these could become much larger, namely up to 9\% for large dressing
cone radii of $\dRdress=0.1$ or $0.2$.
Introducing a flavour-aware dressing algorithm restored the
bare result to a large degree, however. The photon-splitting 
corrections were reduced to $1-2\%$, and their 
cone-size dependence was removed.

We then studied how the above effects
translate to the general off-shell production of a Drell-Yan
electron-positron pair at a hadron collider. We found very 
small corrections to bare leptons, as expected, while the 
corrections to photon-dressed leptons were up to 1\% for a 
cone size of $\Delta R_\mathrm{dress}=0.1$. Including 
secondary electrons in the dressing removes these corrections.
Unlike the academic case of on-shell $Z$ decay, 
it is not appropriate to include hadrons in lepton 
dressing in a hadron collider environment due to the abundance 
of their production from \ac{QCD} processes. These findings 
reinforce the need for consistent treatment of higher-order 
\ac{QED} corrections and lepton definitions in theory, 
event generators and experiment.

The photon splitting corrections were implemented in the
\Sherpa Monte-Carlo event generator and are incorporated in 
the \Sherpa 3.0.0 release.
All analyses and dressing strategies were implemented
using \Rivet's analysis tools \cite{Buckley:2010ar,Bierlich:2019rhm}.

%% file: text/conclusions.tex
\chapter{Conclusions}
\label{chapter:conclusions}

In modern particle phenomenology, the precise prediction of \ac{SM}
processes is of great importance, both for background subtraction in 
the search for new physics, and for correct fitting of non-perturbative 
models to data. Since the inception of \ac{QCD} 50 years ago, 
\ac{QCD} calculations 
have undergone a huge reduction in uncertainty, allowing for accurate 
predictions of exclusive final states and differential distributions. 
The \ac{EW} sector, despite being the source of many precision 
observables at hadron colliders, is often considered only to \ac{LO}, 
due to the small size of $\alpha$ compared to $\alpha_s$. 
In recent years, however, that has begun to change. In addition to 
the ever-increasing precision of experimental measurements at the 
\ac{LHC} and its upcoming high-luminosity upgrade, there are 
other factors to motivate higher precision in the \ac{EW} sector.
\changed{First, the extraction of the $W$ boson mass
from legacy \Tevatron data is in tension with previous measurements
and with measurements of other \ac{EW} parameters.
Second}, the anomalous magnetic moment of the muon has been 
measured to be in conflict with the current \ac{SM} prediction.
Finally, there are multiple proposals for future $e^+ e^-$ colliders, 
for which theoretical research and development must start 
imminently.

In this thesis, we presented three novel methods which increase the 
precision of \ac{QED} radiative corrections in an automated, 
process-independent way. 

We took inspiration from \ac{QCD} in chapters \ref{chapter:qedps}
and \ref{chapter:mcatnlo}, developing \iac{NLO}-matched parton shower.
This resums the leading logarithms associated both with the emission of 
a collinear or soft photon, and with the splitting of a photon into a collinear 
fermion pair. We demonstrated how the shower can be interleaved 
with a \ac{QCD} parton shower, respecting the hadronisation scale 
of \ac{QCD} while allowing pure \ac{QED} radiation to continue 
to lower scales. We validated the shower, and its matched \MCatNLO 
method, for the test case of a neutrino collider,
then presented results for the phenomenologically relevant case of 
Higgs production via gluon fusion and its subsequent leptonic decay. 
We showed that the shower and \MCatNLO are in excellent agreement 
with the \ac{YFS} approach. We showed that a dipole identification 
helps to improve efficiency by eliminating negative \ac{QED} shower 
weights in many processes.

We also discussed the implementation of a \ac{QED} parton shower 
(and hence an \MCatNLO implementation) for electron-positron colliders. 
Although this work is ongoing, we discussed properties of the 
\ac{LL} electron structure function, numerical techniques to tackle 
its integrable singularity, and other considerations when 
implementing a parton shower for $e^+ e^-$ initial states.

In chapter \ref{chapter:yfs}, we took a different approach, building 
on the hugely successful \ac{YFS} soft-photon resummation. We used the 
techniques developed in chapter \ref{chapter:qedps} to extend the 
formalism to resum the leading logarithms from charged particle 
pair production. We found effects of a few percent on  
invariant mass distributions of resonance decay products, 
corresponding to around a 1\% correction to the Drell-Yan 
process at the \ac{LHC}.
We studied the dependence of these 
observables on the flavour composition of the \ac{QED} radiation 
surrounding final-state leptons, and found situations in 
which the usual method of cone dressing with photons introduces 
unphysical effects. One example is the strong cone size dependence
of the higher-order corrections.
We suggested alternative lepton dressing strategies to mitigate 
these effects, namely cone dressing using secondary electrons 
and positrons in addition to photons, and flavour-aware 
recombination dressing.

We implemented all methods described in this thesis in public 
\ac{MC} code and analysis frameworks. The extension to the \ac{YFS}
formalism was released as part of \Sherpa 3.0, while the \ac{QED}
parton shower and \MCatNLO will be included in a future \Sherpa 3 
release. The lepton collider developments, once these are finalised 
and validated against predictions made using the \ac{YFS} approach
and other parton shower codes, 
will also be released as part of \Sherpa. All analyses used to produce 
plots in this thesis were implemented in \Rivet and can be obtained 
from the author on request, along with the \Sherpa runcards.
Public distribution of automated methods and new results is vital 
to the collaboration of the particle physics community and the 
continued progress in the development of our description of our 
universe.

%% file: text/appendix1.tex
\chapter{Usage and settings}
\label{app:settings}

In this appendix we list the available keywords and settings
which steer the calculation of the effects described in this 
thesis. \changed{All are correct at the time of publication, 
however, these are subject to change between versions 
of \Sherpa, so the 
reader is encouraged to consult the appropriate version of the manual.}

For the \ac{QED} parton shower, \changed{under the scoped setting 
\texttt{SHOWER},} the following settings can be changed:
\begin{description}
    \item[\texttt{EW\_MODE}] \
      This setting turns on splitting functions for the \ac{QED} parton shower.
      \begin{itemize}
        \item[\texttt{0}] no \ac{QED} splittings (default),
        \item[\texttt{1}] all \ac{QED} splittings,
        \item[\texttt{2}] photon emissions off fermions only.
      \end{itemize}
    \item[\texttt{WED\_FS\_PT2MIN}] \
      The infrared cutoff in $t$, the parton shower ordering variable, 
      for the \ac{QED} final-state evolution.
      Default is $10^{-8}\,\GeV^2$.
    \item[\texttt{QED\_IS\_PT2MIN}] \
      The infrared cutoff in $t$, the parton shower ordering variable, 
      for the \ac{QED} initial-state evolution.
      Default is $2.5\times 10^{-7}\,\GeV^2$.
    \item[\texttt{QED\_SPECTATOR\_SCHEME}] \
      This setting controls which charged particles may act as spectators 
      for the emission of a photon off a fermion.
      \begin{itemize}
        \item[\texttt{One}] only the nearest \ac{OSSF} particle considered (default),
        \item[\texttt{All}] all charged external legs in the event considered.
      \end{itemize}
      The nearest particle is selected by taking the minimum invariant mass.
      All charged particles are always considered as spectators for 
      photon splittings and are given equal weight, unlike in the 
      \ac{YFS} module described above. 
      For an $s$-channel process, this should generally be left at the 
      default value. For a $t$-channel process it is recommended that 
      the \texttt{All} setting is used.
    \item[\texttt{QED\_ALLOW\_FI}] \
      This setting controls whether to allow initial-final interference 
      in the \ac{QED} shower. Note that if switched on, 
      \texttt{QED\_SPECTATOR\_SCHEME} must be set to \texttt{All}.
      \begin{itemize}
        \item[\texttt{false}] no initial-final interference (default),
        \item[\texttt{true}] initial-final and final-initial dipoles included.
      \end{itemize}
    \item[\texttt{QED\_ALLOW\_II}] \
      This setting controls whether to allow initial-state particles to 
      participate in the \ac{QED} parton shower.
      \begin{itemize}
        \item[\texttt{false}] no initial-state \ac{QED} radiation,
        \item[\texttt{true}] initial-initial dipoles included (default).
      \end{itemize}
      This setting is included to facilitate comparisons with the 
      \ac{YFS} final-state \ac{QED} radiation, or for efficiency improvements
      for quark initial states.
\end{description}

To turn on the \ac{QED} \MCatNLO, the \texttt{NLO\_Order} parameter in the 
\texttt{Process} section of the runcard must be set appropriately. 
In addition, the subtraction must be set as follows:
\begin{description}
  \item[\texttt{NLO\_SUBTRACTION\_MODE}] \
    This setting controls which IR divergences are subtracted.
    \begin{itemize}
      \item[\texttt{QCD}] only QCD subtraction (default),
      \item[\texttt{QED}] only \ac{QED} subtraction,
      \item[\texttt{QCD+QED}] all SM IR divergences will be subtracted.
    \end{itemize}
\end{description}
The \MCatNLO is turned on in the usual way, using \texttt{NLO\_MODE: MC@NLO}
and the associated settings. All of the \ac{QED} shower and \MCatNLO 
settings will be available in a future release of \Sherpa 3, however, 
the exact settings and default values are subject to change. Any 
changes will be detailed in the relevant user manual.

For the modifications to the \ac{YFS} resummation, under the scoped setting 
\texttt{YFS}, the available settings are: 
\begin{description}
  \item[\texttt{PHOTON\_SPLITTER\_MODE}] \
    This setting governs which secondary flavours will be considered.
    \begin{itemize}
      \item[\texttt{0}] photons do not split,
      \item[\texttt{1}] photons split into electron-positron pairs,
      \item[\texttt{2}] muons,
      \item[\texttt{4}] tau leptons,
      \item[\texttt{8}] and/or light hadrons
                        (up to \texttt{YFS\_PHOTON\_SPLITTER\_MAX\_HADMASS}).
    \end{itemize}
    The settings are additive; the default is \texttt{15}.
  \item[\texttt{PHOTON\_SPLITTER\_MAX\_HADMASS}] \
    This setting sets the mass of the heaviest hadron which can be
    produced in photon splittings.
    Note that vector splitting functions are not currently implemented.
    Default is $0.5\,\GeV$.
  \item[\texttt{PHOTON\_SPLITTER\_ORDERING\_SCHEME}] \
    This setting defines the ordering scheme used.
    \begin{itemize}
      \item[\texttt{0}] transverse momentum ordering,
      \item[\texttt{1}] virtuality ordering,
      \item[\texttt{2}] mixed scheme (default).
    \end{itemize}
  \item[\texttt{PHOTON\_SPLITTER\_SPECTATOR\_SCHEME}] \
    This setting defines the allowed spectators for the photon splitting
    process.
    \begin{itemize}
      \item[\texttt{0}] all primary emitters may act as spectator (default),
      \item[\texttt{1}] only the photon's reconstructed emitter is eligible
                        as a spectator.
    \end{itemize}
  \item[\texttt{PHOTON\_SPLITTER\_STARTING\_SCALE\_SCHEME}] \
    This setting governs the determination of the starting scale.
    \begin{itemize}
      \item[\texttt{0}] starting scale is chosed probabilistically (default),
      \item[\texttt{1}] the starting scale is chosen using a
                        winner-takes-all strategy.
    \end{itemize}
\end{description}
These are available from the \Sherpa 3.0 release onwards.